\documentclass[journal,article,submit,pdftex,moreauthors]{Definitions/mdpi}

\usepackage{cancel}

\newcommand{\mi}{\mathrm{i}} 

\newcommand{\dfd}[3]{\hspace{-0.4em}\ensuremath{\frac{\mathrm{d}^{#1}#3}{(2\pi)^{#2}}}\,}
\newcommand{\dx}[1]{\hspace{-0.4em}\ensuremath{\mathrm{d}#1}\,}

\newcommand{\eqn}[1]{Eq.~(\ref{#1})}

\def\Kbar{\overline{K}}
\def\K0bar{\overline{K^0}}

\def\Kbar{\overline{K}}
\def\Dbar{\overline{D}}

\firstpage{1} 
\pubvolume{1}
\issuenum{1}
\articlenumber{0}
\pubyear{2024}
\copyrightyear{2024}
\datereceived{ } 
\daterevised{ } 
\dateaccepted{ } 
\datepublished{ } 
\hreflink{https://doi.org/} 

\usepackage{natbib}
\usepackage{graphicx}
\usepackage{dcolumn}
\usepackage{bm}
\usepackage{amsmath,amssymb}
\usepackage{float}
\usepackage{multirow}
\usepackage{slashed}
\usepackage{xcolor}
\usepackage{mathtools,braket}
\usepackage{physics}
\usepackage{multirow}
\usepackage[normalem]{ulem}
\usepackage{color}

\Title{Heavy-heavy and heavy-light mesons in cold nuclear matter}

\TitleCitation{Title}

\Author{J\'{e}sus J. Cobos-Martínez$^{1}$\orcidA{}, 
Guilherme N. Zeminiani $^{2}$\orcidB{}, 
and Kazuo Tsushima$^{* 2}$\orcidC{}}

\AuthorNames{Firstname Lastname, Firstname Lastname and Firstname Lastname}

\AuthorCitation{Lastname, F.; Lastname, F.; Lastname, F.}

\address{%
$^{1}$ \quad Departamento de Física, Universidad de Sonora, Boulevard Luis Encinas J. y Rosales, Colonia Centro, Hermosillo, Sonora 83000, México \\
$^{2}$ \quad Laboratório de Física Teórica e Computacional (LFTC),
Programa de P\'{o}sgradua\c{c}\~{a}o em Astrof\'{i}sica e F\'{i}sica Computacional,
Universidade Cidade de S\~ao Paulo (UNICID), 01506-000, S\~ao Paulo, SP, Brazil
}

\corres{Corresponding author: kazuo.tsushima@gmail.com, kazuo.tsushima@cruzeirodosul.edu.br}

\abstract{
We review the in-medium modifications of effective masses (Lorentz scalar potentials or
phenomenon of mass shift) of heavy-heavy and heavy-light mesons in symmetric nuclear matter
and their nuclear bound states.
We use a combined approach with the quark-meson coupling (QMC) model
and an effective Lagrangian.
As demonstrated by the cases of pionic and kaonic atoms,
studies of meson-nucleus bound state can provide us with important information
on chiral symmetry in dense nuclear medium.
In this review, we treat the mesons,
$K, K^*, D, D^*, B, B^*, \eta, \eta', \phi, \eta_c, J/\psi, \eta_b, \Upsilon$, and $B_c$,
where our emphasys is on the heavy mesons.
In addition, we also present some new results for the $B_c$-nucleus bound states.
}

\keyword{Hadrons; nuclear matter; strong interactions; chiral symmetry restoration in medium}
\PACS{}

\begin{document}
\nolinenumbers

\section{Introduction}

Quantum chromodynamics (QCD) is the theory of strong interactions at the fundamental level,
namely, at the level of quarks and gluons which compose the observed hadrons  
in the standard model (SM)~\cite{Accardi:2023chb,Brodsky:2015aia, Heinz:2015tua, Alkofer:2000wg,Brambilla:2014jmp,Bashir:2012fs,Cloet:2013jya}.
However, a quantitative understanding of the strong force and strongly interacting matter
from the underlying first principles of QCD is still limited, in particular when the hadrons
are under the circumstance of many nucleons such as emersed in nuclei and dense nuclear medium
The study of the interactions of heavy-heavy and heavy-light mesons with atomic nuclei
is an important tool for understanding the properties
of strongly interacting matter in vacuum and in extreme conditions of temperature
and density based on QCD~Refs.~\cite{Hosaka:2016ypm,Krein:2016fqh,Metag:2017yuh,Krein:2017usp,Hatsuda:1994pi,Leupold:2009kz,Hayano:2008vn,piAF:2022gvw}.
In this review we treat the zero temperature case.
The understanding of hadronic interactions with the nuclear
medium is imperative for studying the production of heavy mesons in high energy heavy
ion collisions~\cite{Schroedter:2000ek,Andronic:2015wma},
because the medium modifications of hadron properties
may have a significant impact on the experimental results.
Decay processes and decay rates involving mesons in a nuclear medium should also be modified.
For example, decays of the type $B_c \rightarrow D_s \ell^+ \ell^-$
involving a flavor-changing neutral-current process are
highly suppressed in the Standard Model (SM)~\cite{Li:2023mrj,Bird:2004ts,
Altmannshofer:2014rta,Buras:2014fpa,Descotes-Genon:2015uva,Dutta:2017xmj,CMS:2021hug},
thus being very important for investigating the physics beyond the SM.
Moreover, the suppression or enhancement in the production of mesons
such as $J/\psi$, $\Upsilon$, and $B_c$ makes them interesting probes of quark-gluon plasma
(QGP)~\cite{Akram:2013dhd,Lodhi:2007zz,Lodhi:2011zz,Wu:2023djn,CMS:2022sxl,
Harris:2023tti,Lin:2000ke,CMS:2018zza,ATLAS:2022exb,STAR:2022rpk}.

To calculate the in-medium (effective) masses of the mesons that contain light quarks,
such as $K$, $K^*$, $D$, $D^*$, $B$, $B^*$, $\eta$ and $\eta'$,
we use the quark-meson coupling (QMC) model~\cite{Guichon:1987jp}.
Because of the Okubo-Zweig-Iizuka (OZI) rule, the heavy quarkonium-nucleus interaction
via the exchange of mesons made of only light quarks is suppressed, so that
the quarkonium-nucleus interaction is primarily by a QCD van der Waals type
interactions~\cite{Brodsky:1997gh}.
For the mesons that do not contain light quarks, namely,
$\phi$, $\eta_c$, $J/\psi$, $\eta_b$, $\Upsilon$, and $B_c$,
we employ a combined approach, in which the mechanism for the meson
interact with the nuclear medium through the excitation of the
intermediate-state mesons which do contain light quarks, in the self-energy.
Where the in-medium masses of the intermediate-state mesons
are calculated by the QMC model, and the meson
self-energies are estimated with effective Lagrangians.

Partial restoration of chiral symmetry, and chiral symmetry itself in nuclear medium,
are another interesting phenomena that can be studied in an empirical sense, because
the (effective) mass reduction of the medium-modified hadron may be associated with a
signature of partial restoration of chiral symmetry
~\cite{Krein:2010vp,Tsushima:2011kh,Ko:1992tp,Asakawa:1992ht}.
This negative mass shift can be regarded as an attractive Lorentz scalar potentials,
that, if sufficiently attractive, can bind mesons to atomic nuclei.
Deeply bound pionic atoms were first discussed in 1985~\cite{Friedman:1984yg},
and later observed in the $^{208}$Pb($d$, $^3$He) reaction~\cite{Yamazaki:1996ch}.
In addition, studies of kaonic atoms were
performed~\cite{Davies:1979aj,Lee:1994jj,Ito:1998yi,Curceanu:2019uph,Hirenzaki:2000da}.
The studies of the pionic and kaonic atoms can provide us with very important
information on chiral symmetry in dense nuclear medium.
Furthermore, other possible meson-nucleus bound states were
proposed~\cite{Hayano:1998sy,Tsushima:1998qw,Tsushima:1998ru}.
Charmonium-nucleus systems were proposed in
1989~\cite{Brodsky:1989jd}, followed by many predictions
~\cite{Krein:2010vp,Tsushima:2011kh,Hosaka:2016ypm,Metag:2017yuh,Krein:2017usp,
Lee:2000csl,Krein:2013rha,Klingl:1998sr,Hayashigaki:1998ey,Kumar:2010hs,Belyaev:2006vn,
Yokota:2013sfa,Peskin:1979va,Kharzeev:1995ij,Kaidalov:1992hd,Luke:1992tm,deTeramond:1997ny,
Brodsky:1997gh,Sibirtsev:2005ex,Voloshin:2007dx,TarrusCastella:2018php,Cobos-Martinez:2020ynh}.
Lattice QCD have also predicted such states~\cite{Yokokawa:2006td,Kawanai:2010ev,Skerbis:2018lew},
and also $\phi$-nucleon bound states~\cite{Chizzali:2022pjd}.
In the bottom sector of quarkonia, strong nuclear bound states with various
nuclei~\cite{Zeminiani:2020aho,Zeminiani:2021vaq,Zeminiani:2021xvw,Cobos-Martinez:2022fmt}
were predicted for $\Upsilon$ and $\eta_b$.
In this article, we review the downward shift of meson masses in nuclear matter, and
the meson-nucleus bound states focusing on the heavy-heavy and heavy-light mesons.
We also comment on some new results for the $B_c$-nucleus bound states.

This review is organized as follows. In Section~\ref{qmcmodel} we present the
details of quark-meson coupling (QMC) model needed to understand better most of our results.
In Section~\ref{qmcmodelresults} we present the results for the mass shift of mesons with
heavy-light quark content using the quark-meson coupling (QMC) model.
Since heavier flavor quarks $Q = s, c, b$ do not directly
interact directly with the mean fields in a  nuclear medium, to compute the effective masses
for the mesons with a (heavy quark)-(heavy antiquark) content, we use a combined
approach using both the QMC model with and effective Lagrangians.
We describe this in Section~\ref{combinedapproach}.
In Section~\ref{nuclearpotentials}, using the {calculated amounts of the downward shift of masses}
for the mesons considered in this work, we present our results for the meson-nucleus
potentials for various nuclei in a wide range of nuclear masses.
Using the meson-nucleus potentials obtained in the previous section,
in Section~\ref{bsenergies} we present our results for
meson-nucleus bound state energies and widths for some mesons)
by solving the Klein-Gordon equation.
Finally, in Section~\ref{conclusions} we present a summary and conclusions.

\section{\label{qmcmodel} The quark-meson coupling (QMC) model}

The quark-meson coupling (QMC) model, the standard version we use, is a quark-based model for
nuclear matter and nuclei that describes the internal structure of the nucleons using the
non overlapping MIT bag model, and the binding of nucleons (nuclear matter)
by the self-consistent couplings of
the confined light quarks $u$ and $d$ to the Lorentz scalar-$\sigma$, 
Lorentz-vector-isoscalar-$\omega$ and Lorentz-vector-isovector-$\rho$ meson mean
fields generated by the confined light quarks in the nucleons~\cite{Guichon:1987jp}. 

In a nuclear medium, the hadrons with light quarks are expected to be modified
their properties predominantly, such as evidenced by the European Muon Collaboration 
effect~\cite{EuropeanMuon:1983wih,Geesaman:1995yd} and the modifications of bound proton
electromagnetic form factors~\cite{Dieterich:2000mu,POLE}.
Thus, one can expect that the nuclear medium can
modify the internal structure of nucleons and hadrons, and can affect
the interaction with nucleons.
Thus, to study such effects due to the hadron internal structure based on the
quarks and gluons, can make the QMC model a useful phenomenological tool to describe
the change of the internal structure of hadrons in a nuclear medium.

The QMC model has been successfully applied for the studies of various properties
of infinite nuclear matter and finite 
(hyper)nuclei~\cite{Saito:2005rv,Guichon:2008zz,Shyam:2019laf,Guichon:2018uew,Krein:2017usp}.
Here we briefly present the necessary details to understand our
results better. For more detailed discussions and some successful features of the model,
see Refs.~\cite{Saito:2005rv,Guichon:2008zz,Guichon:2018uew,Krein:2017usp},
and references therein.

We consider nuclear matter (NM) in its rest frame, where all the scalar and
vector mean field potentials, which are responsible for the nuclear 
many-body interactions, are constants in Hartree approximation.
We assume $SU(2)$ symmetry for the quarks ($m_q = m_u = m_d$ and $q = u$ or $d$).
Note that, the heavier quarks $Q = s,\; c,\; b$ (we will denote simply heavy quarks as $Q$ hereafter including the $s$ quarks)
are not affected directly by the mean field potentials in the standard QMC model.
Thus, when dealing with mesons composed of valence
(heavy quark)-(heavy antiquark) pairs, we will have to proceed in a different manner
for hidden flavor heavier mesons with $Q = s,\; c,\; b$ and
two-heavy-flavored mesons such as $B_c$ and $B_s$ mesons.
In this study, we will treat the $B_c$ and $B_c^*$ mesons only; however, for the other
two-heavy-flavored mesons $B_s, B_s^*, D_s$ and $D_s^*$ mesons, 
see Ref.~\cite{Zeminiani:2023gqc}.

The Dirac equations for the quarks and antiquarks  
in nuclear matter (neglecting the Coulomb force), in a bag of a hadron,
$h$, ($q = u$ or $d$, and $Q = s,c$ or $b$),
are given by  ($x=(t,\textbf{r})$ and for $|\textbf{r}|\le$
bag radius)~\cite{Tsushima:1997df,Tsushima:1998ru,Sibirtsev:1999js,Sibirtsev:1999jr,Tsushima:2002cc},
\begin{eqnarray}
\left[ i \gamma \cdot \partial_x -
(m_q - V^q_\sigma)
\mp \gamma^0
\left( V^q_\omega +
\dfrac{1}{2} V^q_\rho
\right) \right] 
\left( \begin{array}{c} \psi_u(x)  \\
\psi_{\bar{u}}(x) \\ \end{array} \right) &=& 0,
\label{Diracu}\\
\left[ i \gamma \cdot \partial_x -
(m_q - V^q_\sigma)
\mp \gamma^0
\left( V^q_\omega -
\dfrac{1}{2} V^q_\rho
\right) \right]
\left( \begin{array}{c} \psi_d(x)  \\
\psi_{\bar{d}}(x) \\ \end{array} \right) &=& 0,
\label{Diracd}\\
\left[ i \gamma \cdot \partial_x - m_{Q} \right] \psi_{Q} (x) = 0, ~~~~~~~~  
\left[ i \gamma \cdot \partial_x - m_{Q} \right] \psi_{\overline{Q}} (x) &=& 0,  
\label{DiracQ}
\end{eqnarray}
where, the mean field potentials are defined by, 
$V^q_\sigma \equiv g^q_\sigma \sigma$, 
$V^q_\omega \equiv g^q_\omega \omega$, and
$V^q_\rho \equiv g^q_\rho b$,
with $g^q_\sigma$, $g^q_\omega$, and
$g^q_\rho$ being the corresponding quark-meson coupling constants. 
We assume SU(2) symmetry, $m_q = m_{\bar{q}} \equiv m_{u,\bar{u}}=m_{d,\bar{d}} \equiv
m_{q,\bar{q}}$.
The Lorentz-scalar ''effective quark masses'' 
are defined by, $m^*_q=m^*_{u,\bar{u}}=m^*_{d,\bar{d}}=m^*_{q,\bar{q}} \equiv
m_q-V^q_{\sigma}$, and thus $m_q^*$ is dominated by
$-V^q_{\sigma}$ as baryon density increases, and can be negative,
but one should not demand the positivity of usual particle mass, since this is
nothing but the reflection of the strong attractive scalar potential.
Note that, $m_Q=m_Q^*$, since the $\sigma$ field does not couple 
to the heavier (''heavy'') quarks $Q=s,c,b$ in the QMC model.
Furthermore, when we consider symmetric nuclear matter (SNM) with Hartre approximation, 
the $\rho$-meson mean field becomes zero, $V^q_{\rho}=0$, in Eqs.~(\ref{Diracu}) 
and~(\ref{Diracd}), and we can ignore hereafter.
However, when we consider the meson-nucleus bound states, the isospin dependent
$\rho$-meson mean field, as well as the Coulomb potential in nuclei
will be included if necessary. In this study, only the Coulomb potentials
for the $B_c$-nucleus bound states will enter.

The static solution for the ground state quarks (antiquarks) in asymmetric nuclear matter (ANM) 
with a flavor $f (=u,\,d,\,s,\,c,\,b)$ is written as $\psi_{f}\left(x\right) =
N_{f}e^{-i\epsilon_{f}t/R^{*}_{h}}\psi_{f}\left(\textbf{r}\right)$, with the
$N_f$ being the normalization factor, and $\psi_{f}\left(\textbf{r}\right)$
the corresponding spin and spatial part of the wave function.

The eigenenergies for the quarks and antiquarks in a hadron $h$,  
in units of the in-medium bag radius of hadron h, $1/R^{*}_{h}$, are given by
\begin{eqnarray}
&&\begin{pmatrix}
        \epsilon_{u}\\
        \epsilon_{\overline{u}}
       \end{pmatrix} = \Omega^{*}_{q} \pm R^{*}_{h} 
\left(V^{q}_{\omega} + \frac{1}{2}V^{q}_{\rho}\right),  \\ 
&&\begin{pmatrix}
        \epsilon_{d}\\
        \epsilon_{\overline{d}}
       \end{pmatrix} = \Omega^{*}_{q} \pm R^{*}_{h} 
\left(V^{q}_{\omega} - \frac{1}{2}V^{q}_{\rho}\right),\\
&&\epsilon_{Q} = \epsilon_{\overline{Q}} = \Omega_{Q}, \hspace{3mm} 
Q= s,\; c,\; b.
\end{eqnarray}
The in-medium mass and bag radius of hadron $h$ in the nuclear medium,
$m^{*}_h$ and $R^{*}_h$ respectively, are determined from
\begin{equation}
\label{eqn:mhnm}
m_h^{*}= \sum\limits_{j=q,\;\overline{q}\;,Q,\;\overline{Q}}
\dfrac{n_j\Omega_j^{*}-z_h}{R_h^{*}} + {4\over 3}\pi R_\eta^{* 3} B_p,
\hspace{8mm}
\left.\frac{d m_h^*}
{d R_h}\right|_{R_h = R_h^*} = 0,
\end{equation}
in particular, for the mesons $h=K,\;K^{*},\;D,\; D^{*},\; B,\; B^{*}$, while for 
the $\eta$ and $\eta'$ mesons, to take into account flavor mixing, these 
are given by
\begin{eqnarray}
\label{eqn:metanm}
m_\eta^*
&=& \frac{2 [a_{P}^2\Omega_q^* + b_{P}^2\Omega_s] - z_\eta}{R_\eta^*}
+ {4\over 3}\pi R_\eta^{* 3} B_p, \quad\quad
\left.\frac{d m_h^{*}}{d R_j}\right|_{R_h = R_h^{*}} = 0, \quad\quad (h =\eta,\eta'), \\
a_{P} &\equiv& \sqrt{1/3} \cos\theta_{P}
- \sqrt{2/3} \sin\theta_{P}, \quad\quad
b_{P} \equiv  \sqrt{2/3} \cos\theta_{P}
+ \sqrt{1/3} \sin\theta_{P} \label{mix} \\
& &\hspace{-12ex}({\rm for}\hspace{1ex} \eta',\hspace{1ex}
\eta \to \eta',\, {\rm and}\hspace{1ex}
a_P \leftrightarrow b_P), \nonumber
\end{eqnarray}
where $m_{q}^{*}= m_{q}-V_{\sigma}^{q}$, $m_{Q}^{*}= m_{Q}$ (as already mentioned),
$\Omega^{*}_{q} = \Omega^{*}_{\overline{q}} = \left[x^{2}_{q} + \left(R^{*}_{h} m^{*}_{q}\right)^{2}
\right]^{1/2}$,
$\Omega^{*}_{Q} = \Omega^{*}_{\overline{Q}} = \left[x^{2}_{Q} + \left(R^{*}_{h}m_{Q}\right)^{2}
\right]^{1/2}$,
with $x_{q,Q}$ being the lowest mode bag eigenfrequencies; 
$B_p$ is the bag constant; $n_{q,Q}$ ($n_{\overline{q},\overline{Q}}$) are the lowest mode valence
quark (antiquark) numbers for the quark flavors $q$ and $Q$  in the corresponding mesons;  and
$z_{h}$ parameterize the sum
of the center-of-mass and gluon fluctuation effects and are 
assumed to be independent of density~\cite{Guichon:1995ue}. 
The MIT big parameters $z_N$ ($z_h$) and $B_p$ are fixed by fitting the
nucleon (hadron) mass in free space.

We choose the values ($m_q, m_s, m_c, m_b = (5, 250, 1270, 4200)$ MeV for the current quark
masses, and $R_{N} = 0.8\; \text{fm}$ for the free space nucleon bag radius.
(See Ref.~\cite{Tsushima:2020gun} for other values used  ($m_q, m_s = (5, 93, 1270, 4180)$ MeV
result.)
The quark-meson coupling constants, $g^{q}_{\sigma}$, $g^{q}_{\omega}$ and
$g^{q}_{\rho}$, for the light quarks were determined by the fit
to the saturation energy (-15.7 MeV) at the saturation density  
($\rho_0 = 0.15$ fm$^{-3}$) of symmetric nuclear matter for $g^q_\sigma$ and 
$g^q_\omega$, and by the bulk symmetry energy (35 MeV) for
$g^q_\rho$~\cite{Guichon:1987jp,Saito:2005rv}.
The obtained values for the quark-meson coupling constants are 
($g_\sigma^q$, $g_\omega^q$, $g_\rho^q$)= (5.69, 2.72, 9.33).

Finally, for the mixing angle $\theta_P$ appearing in Eq.~(\ref{mix}), we use the value
$\theta_P=-11.3^{\circ}$, neglecting any possible mass dependence and imaginary
parts~\cite{Workman:2022ynf,Tsushima:2020gun}.
Furthermore, we also assume that the value of the mixing angle does not change in the nuclear
medium.

\section{\label{qmcmodelresults} Results with the QMC model}

\begin{figure}[htb!]
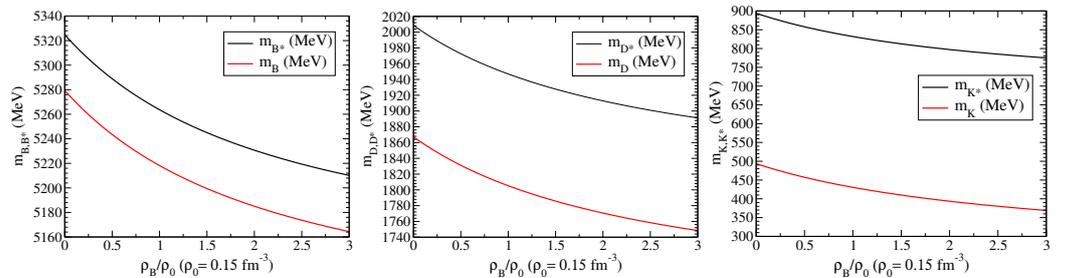
%
\centering
\includegraphics[width=4.5cm]{meson_BBs_mass.eps}
\includegraphics[width=4.5cm]{meson_DDs_mass.eps}
\includegraphics[width=4.5cm]{meson_KKs_mass.eps}
\caption{\label{fig:mh-nm} 
$B$ and $B^*$ (left panel), $D$ and $D^*$ (middle panel) and $K$ and $K^*$ (right panel) meson
Lorentz-scalar effective masses in symmetric nuclear matter versus baryon density
($\rho_B/\rho_0$), calculated with the QMC model.}%
\end{figure}

\begin{figure}[ht]
\centering
\scalebox{0.85}{
\begin{tabular}{cc}
 \includegraphics[scale=0.30]{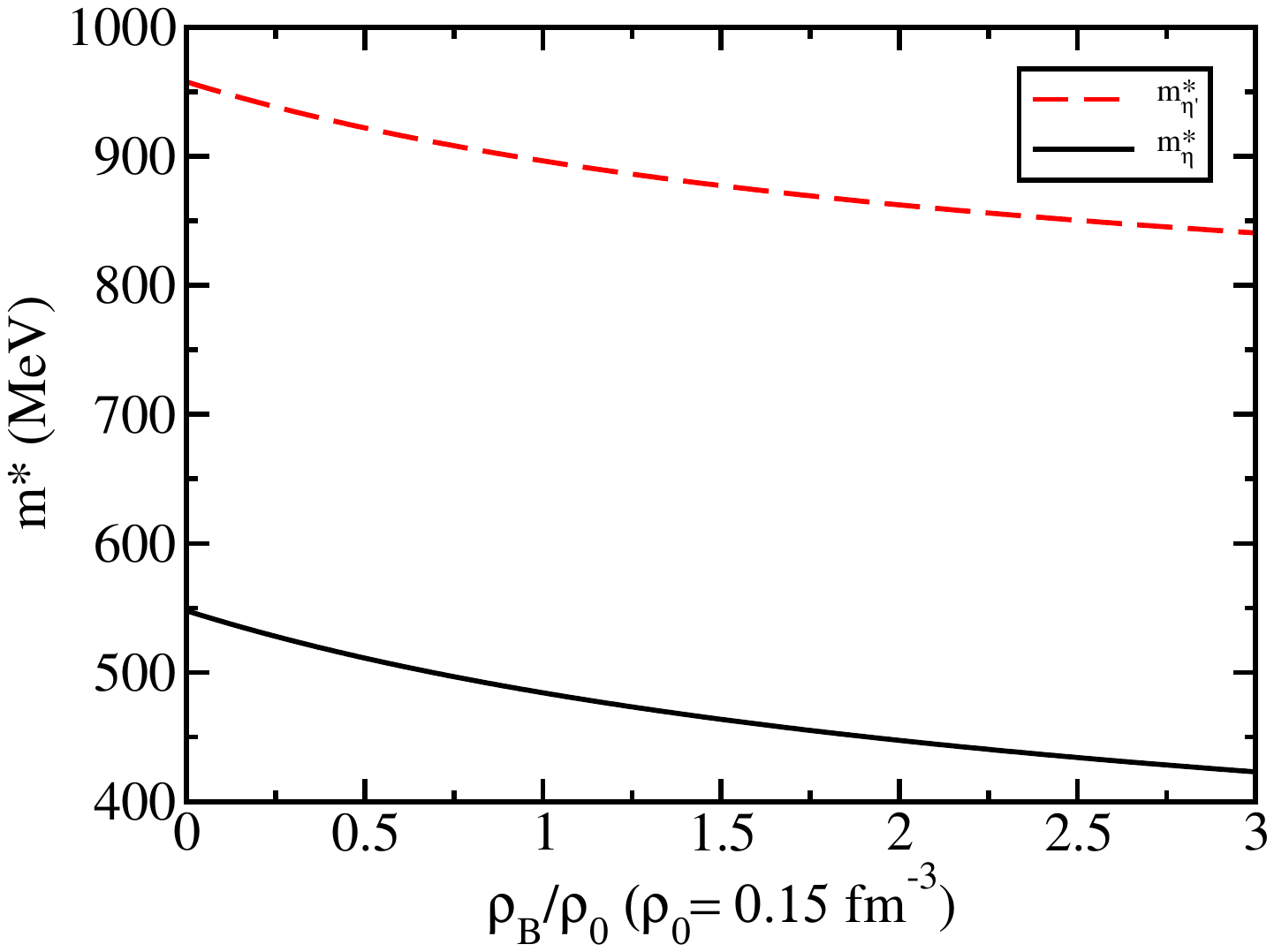} &
 \includegraphics[scale=0.30]{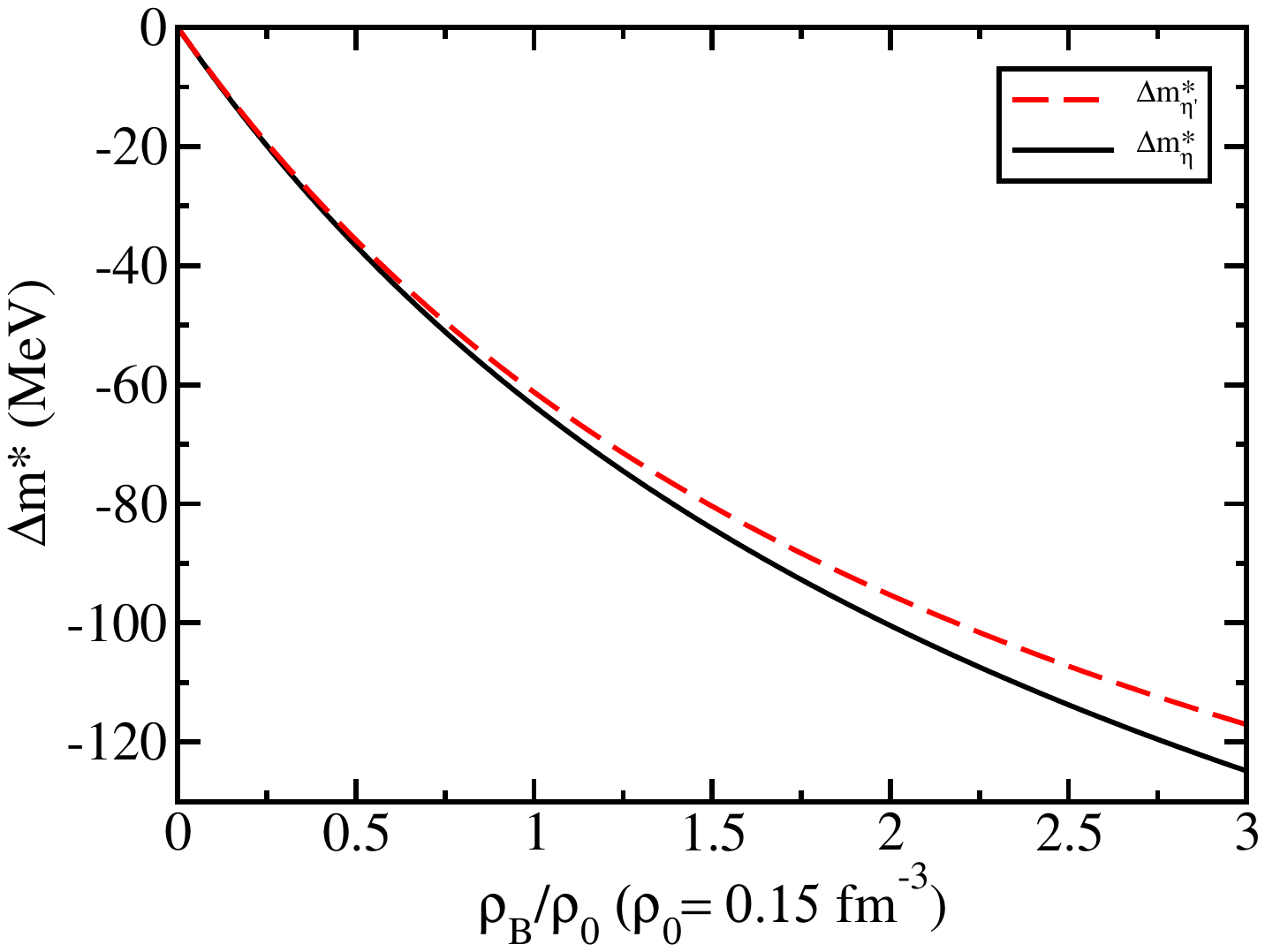}
 \end{tabular}
 }
  \caption{\label{fig:metaetaprime-nm} $\eta$ and $\eta'$ effective masses (left panel)
  and mass shift (right panel) in symmetric nuclear matter
versus baryon density ($\rho_B/\rho_0$), calculated with the QMC model. }
\end{figure}

In Figs.~\ref{fig:mh-nm} and ~\ref{fig:metaetaprime-nm} we
present respectively the QMC model predictions for the effective masses
of $B, B^*, D, D^*, K$ and $K^*$ mesons~\cite{Zeminiani:2023gqc},
and the effective masses and the mass shift $\Delta m_h(\rho_B)\equiv
m_h^*(\rho_B)-m_h$ for $\eta$ and $\eta'$ mesons with $m_h^*$
the in-medium meson mass and $m_h$ the vacuum mass~\cite{Cobos-Martinez:2023hbp},
in symmetric nuclear matter versus nuclear matter density $\rho_0/\rho_B$.
Clearly, the masses of these mesons decrease in the nuclear medium, 
and this fact may be regarded as a signature of partial restoration of chiral symmetry in medium,
although the QMC model does not explicitly have a chiral symmetry mechanism.
Below, we will discuss and use the results shown in
Figs.~\ref{fig:mh-nm} and ~\ref{fig:metaetaprime-nm}.

\section{\label{combinedapproach} Combined the QMC model and effective Lagrangian approach}

Since the Okubo-Zweig-Iizuka rule suppresses the interactions mediated by the exchange of mesons made of light quarks for the case of heavy-heavy mesons, it is therefore necessary to explore other potential sources of attraction which could potentialy lead to the binding of heavy-heavy mesons to atomi nuclei.
Furthermore since the heavy quarks $Q= s,\; c,\; b$ do not directly interact with the 
mean fields in a nuclear medium (see Eqn.~(\ref{DiracQ})), to compute the effective masses 
(Lorentz scalar potentials) for the mesons composed of a (heavy quark)-(heavy antiquark) 
pair, we take a different approach.

This approach consists of the combined treatment with the QMC model and
an effective Lagrangian.
We have already introduced the QMC model above, so we now describe
the effective Lagrangian approach we rely on.


In the effective Lagrangian approach, mesons are treated as the structureless point-like particles,
whose interactions are dictated by a local gauge symmetry principle. In order to be more explicit,
we separate our study according to the different mesons.
Part of the descriptions and treatments reviewed here have already been published
in journals~\cite{Zeminiani:2024rrh,
Cobos-Martinez:2023hbp, Cobos-Martinez:2022fmt, Zeminiani:2020aho, Cobos-Martinez:2020ynh,
Cobos-Martinez:2017woo, Cobos-Martinez:2017vtr},  as well as
presented at various conferences~\cite{CobosMartinez:2021bia,Cobos-Martinez:2021ukw,
Zeminiani:2021xvw, Cobos-Martinez:2019kln, Cobos-Martinez:2017fch,
Cobos-Martinez:2017onm}.

\subsection{The $\phi$ vector meson}

\begin{figure}[ht]
\centering
\includegraphics[scale=0.75]{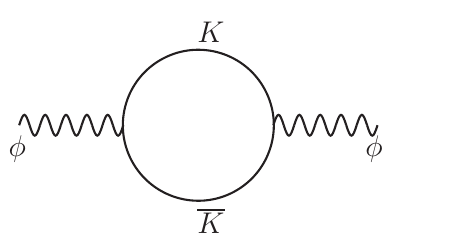}
\caption{\label{fig:phise} $K\Kbar$-loop contribution to the $\phi$ meson 
self-energy.}
\end{figure}
The $\phi$ meson properties in nuclear matter, such as mass
and decay width, are strongly correlated to its coupling to the
$K\Kbar$, which is the dominant decay channel in vacuum. 
Therefore, the density dependence of the $\phi$ meson self-energy
in  nuclear matter arises mainly due to interactions of the kaons and antikaons
with the nuclear medium, and the kaon and antikaon in-medium properties are
calculated in the QMC model~\cite{Tsushima:1997df}
(see also Fig.~\ref{fig:mh-nm} for the effective mass of $K  (=\bar{K})$ meson).
Here we use the effective Lagrangian approach of
Ref.~\cite{Klingl:1996by} to compute the $\phi$ meson self-energy.

The lowest-order interaction Lagrangian providing the coupling of the $\phi$ meson to the $K\Kbar$
pair reads~\cite{Klingl:1996by}
\begin{equation}
\label{eqn:phikklag}
\mathcal{L}_{\phi K\Kbar}=\mi g_{\phi}\phi^{\mu}
\left[\Kbar(\partial_{\mu}K)-(\partial_{\mu}\Kbar)K\right],
\end{equation}
\noindent where $g_{\phi}$ is $\phi K\Kbar$ coupling constant and we use the convention:
\begin{equation}
\label{eqn:isospin}
K=\left(\begin{array}{c} K^{+} \\ K^{0} \end{array} \right),\;
\overline{K}=\left(K^{-}\;\overline{K}^{0}\;\right).
\end{equation}
The scalar self-energy for the $\phi$ meson, $\Pi_{\phi}(p)$, is
determined  from~\eqn{eqn:phikklag}. The Feynman diagram contributing to $\Pi_{\phi}(p)$ at 
$\mathcal{O}(g_{\phi}^{2})$ is depicted in Fig.~\ref{fig:phise}. 
For a $\phi$ meson at rest the scalar self-energy is given by
\begin{equation}
\label{eqn:phise}
\mi\Pi_{\phi}(m_{\phi}^{2})=-\frac{8}{3}g_{\phi}^{2}\int\dfd{3}{3}{q}\textbf{q}^{\,2}
D_{K}(q)D_{K}(q-p),
\end{equation}
\noindent where $D_{K}(q)=\left(q^{2}-m_{K}^{2}+\mi\epsilon\right)^{-1}$ is the
kaon propagator; $p^\mu = (p^{0}=m_{\phi},\textbf{0})$ is the $\phi$
meson four-momentum vector ($\phi$ at rest),  with $m_{\phi}$ the $\phi$ meson
mass; $m_{K} (=m_{\Kbar})$ is  the kaon mass. 
When $m_{\phi}<2m_{K}$ the self-energy $\Pi_{\phi}(p)$ is real. However, when  $m_{\phi}>2m_{K}$,
which is the case here, $\Pi_{\phi}(p)$ acquires an imaginary part.

The mass of the $\phi$ meson is determined from the real part of 
$\Pi_{\phi}(p)$ (see~\eqn{eqn:phimassvacuum}), while its decay width $\Gamma_{\phi}$ to a $K\Kbar$
pair from the imaginary
part of $\Pi_{\phi}(p)$ through the optical theorem (see~\eqn{eqn:optitcalthm}). 
The real and imaginary parts of $\Pi_{\phi}(p)$ can be computed as~\cite{Cobos-Martinez:2017vtr}
\begin{eqnarray}
\label{eqn:repiphi}
\hspace{-1em}
\Re\Pi_{\phi}\hspace{-0.75em}&=&\hspace{-0.75em}-\frac{2}{3}g_{\phi}^{2}
\mathcal{P}\int\dfd{3}{3}{q}\textbf{q}^{\,2}\frac{1}{E_{K}(E_{K}^{2}-m_{\phi}^{2}/4)}, \\
\label{eqn:impiphi}
\hspace{-1em}
\Im\Pi_{\phi}\hspace{-0.75em}&=&\hspace{-0.75em}-\frac{g_{\phi}^{2}}{3\pi}
\frac{1}{m_{\phi}}\left[\frac{m_{\phi}}{2}
\left(1-\frac{4m_{K}^{2}}{m_{\phi}^{2}}\right)^{1/2}\right]^{3},
\end{eqnarray}
\noindent where $\mathcal{P}$ denotes the Principal Value of the integral
and $E_{K}=(\textbf{q}^{\,2}+m_{K}^{2})^{1/2}$.
The integral in \eqn{eqn:repiphi} is divergent but it will be regulated 
using a phenomenological form factor, with cutoff parameter $\Lambda_{K}$, 
as in Ref.~\cite{Krein:2010vp}.


The decay width $\Gamma_{\phi}$ for the process $\phi \to K\overline{K}$ can 
be obtained from the imaginary part of the $\phi$ meson self-energy 
$\Im\Pi_{\phi}$ through the optical theorem
\begin{equation}
\label{eqn:optitcalthm}
\Gamma_{\phi}=-\frac{1}{m_{\phi}}\Im\Pi_{\phi},
\end{equation}
\noindent where $\Im\Pi_{\phi}$ is given by \eqn{eqn:impiphi}. Thus, one gets
\begin{equation}
\label{eqn:phidecaywidth}
\Gamma_{\phi}=\frac{g_{\phi}^{2}}{3\pi}\frac{1}{m_{\phi}^{2}}
\left[\frac{m_{\phi}}{2}\left(1-\frac{4m_{K}^{2}}{m_{\phi}^{2}}\right)^{1/2} 
\right]^{3}.
\end{equation}
The coupling constant $g_{\phi}$ is determined by 
the experimental value for the $\phi \to K\overline{K}$ decay 
width in vacuum, corresponding to a branching ratio 
of $83.1\%$ of the total decay width (4.266 MeV)~\cite{Tanabashi:2018oca}.
For the $\phi$ meson mass $m_{\phi}$ we use its experimental value in vacuum 
$m_{\phi}^{\text{expt}}=1019.461$ MeV~\cite{Tanabashi:2018oca}. 
For the kaon mass $m_{K}$ there is a small ambiguity since 
$m_{K^+}\ne m_{K^0}$ in the real world due to the isospin (or charge) symmetry breaking and
electromagnetic interactions. 
The experimental values for the $K^{+}$ and $K^{0}$ meson masses in vacuum are 
$m_{K^{+}}^{\text{expt}}=493.677$ MeV and $m_{K^{0}}^{\text{expt}}=497.611$ MeV, 
respectively~\cite{Tanabashi:2018oca}. 
For definitiveness we use the average of $m_{K^{+}}^{\text{expt}}$ and 
$m_{K^{0}}^{\text{expt}}$ as the value of $m_{K}$ in vacuum.
(However, the effect of this tiny mass ambiguity on the properties 
of kaon (antikaon) in medium, to be presented in next section, is negligible  
compared with those obtained by using the value 
$m_{K^+} = 493.7$ MeV~\cite{Tsushima:1997df}.)
This gives $g_{\phi}=4.539$~\cite{Cobos-Martinez:2017vtr}.
The mass of the $\phi$ meson will be obtained from the solution of
\begin{equation}
\label{eqn:phimassvacuum}
m_{\phi}^{2}=\left(m_{\phi}^{0}\right)^{2}+\Re\Pi_{\phi}(m_{\phi}^{2})
= \left(m_{\phi}^{0}\right)^{2} - |\Re\Pi_{\phi}(m_{\phi}^{2})|,
\end{equation}
\noindent where $\Re\Pi_{\phi}$ is given by \eqn{eqn:repiphi} and $m_{\phi}^{0}$
is the bare $\phi$ meson mass. In vacuum, \eqn{eqn:phimassvacuum}, 
together with the value obtained for the coupling constant, actually 
fixes the bare $\phi$ meson mass $m_{\phi}^{0}$.

Critical to our results of the in-medium $\phi$ meson mass $m^*_\phi$ and 
decay width $\Gamma_{\phi}^{*}$ at finite baryon density $\rho_B$, 
is the in-medium kaon mass $m_{K}^{*}$.
The nuclear (baryon) density dependence of the $\phi$ meson mass and decay 
width are driven  by the interactions of the kaon with the nuclear medium,
which enter through  $m_{K}^{*}$ in the kaon propagators in~\eqn{eqn:phise}. 
The in-medium kaon mass $m_{K}^{*}$ were calculated previously in the QMC
model, and the results are shown in the right panel of
Fig.~\ref{fig:mh-nm}.
We note that the kaon effective mass at normal nuclear
density  $\rho_0 = 0.15$ fm$^{-3}$ decreases by about $13\%$~\cite{Cobos-Martinez:2017vtr}.
We remind that, to calculate the kaon-antikaon loop contributions to the
$\phi$-meson self-energy in symmetric nuclear matter, the isoscalar-vector 
$\omega$ mean field potentials arise both for the kaon and antikaon. 
However, they have opposite signs and cancel each other, or can be eliminated 
by the variable shift in the loop integral calculation.

\begin{figure}[htb!]%
\centering
\includegraphics[keepaspectratio= true,scale=0.24]{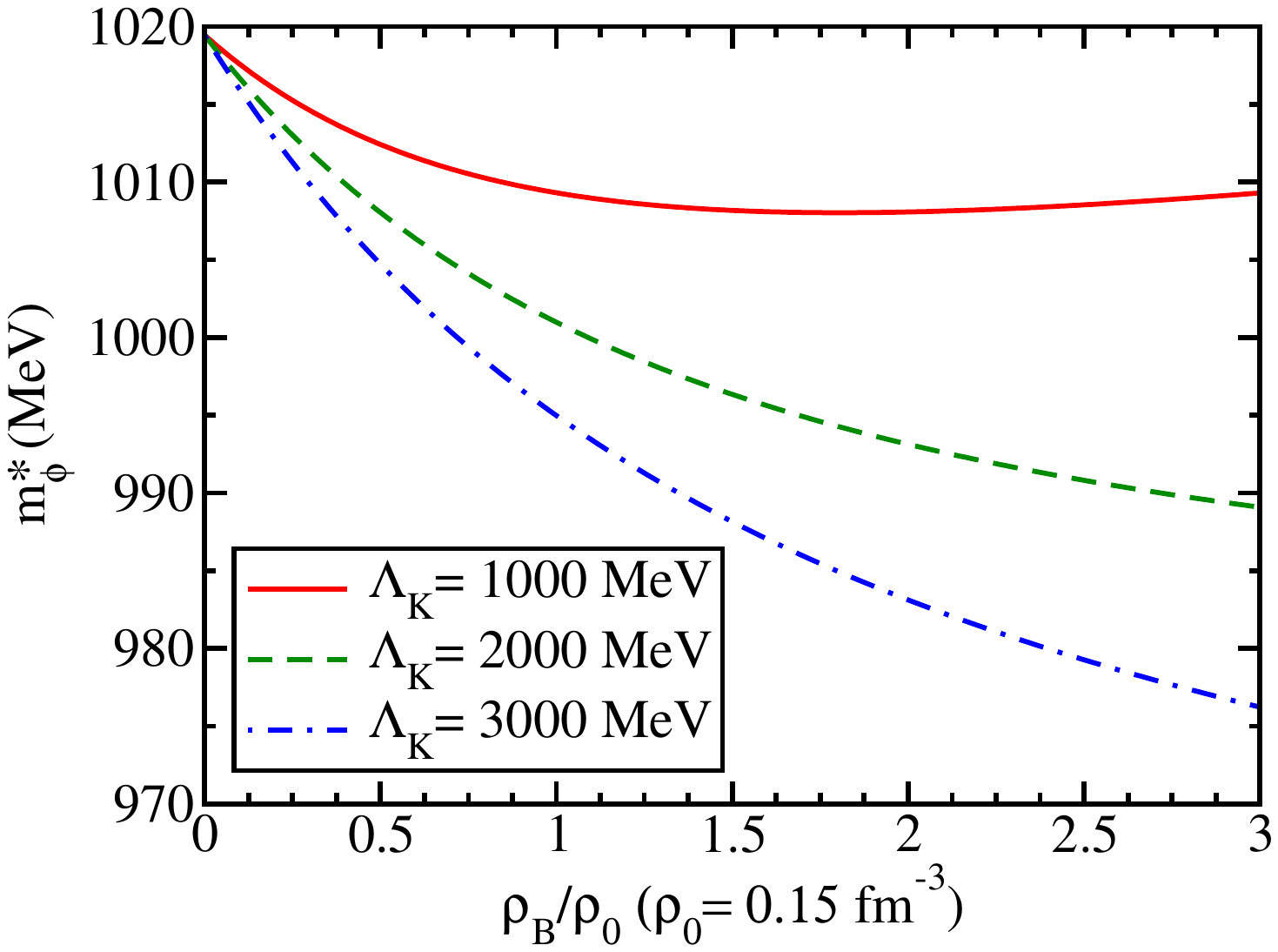}
\includegraphics[keepaspectratio= true,scale=0.24]{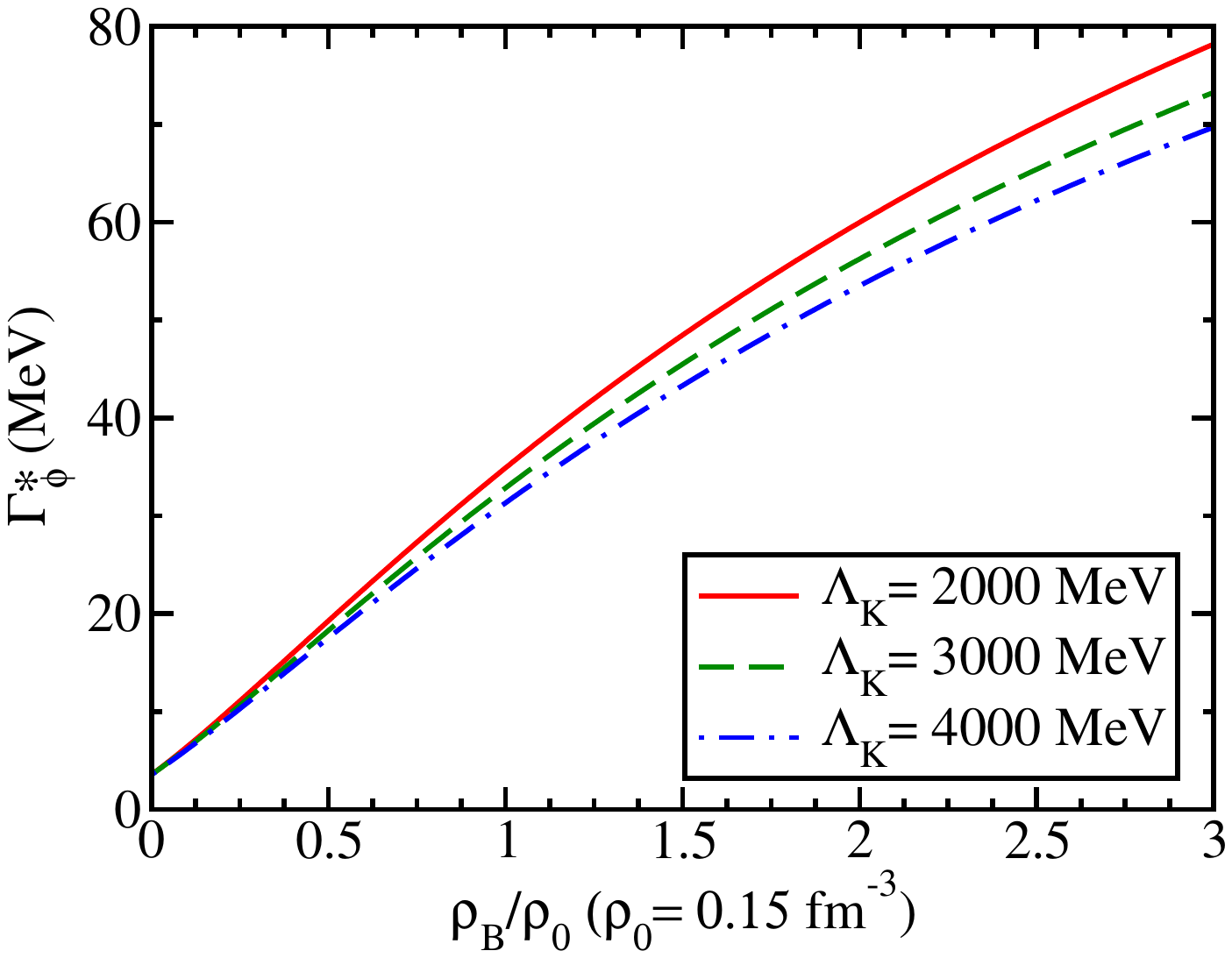}
\caption{\label{fig:phi-nm} In-medium mass (left panel) and decay width (right panel)
of the $\phi$ meson in symmetric nuclear matter versus baryon density $\rho_B/\rho_0$.}%
\label{bksmass}
\end{figure}
%
To calculate the width and mass of the in-medium $\phi$ meson,
$\Gamma_{\phi}^{*}$ and $m_{\phi}^{*}$, respectively, we solve the
corresponding  equations~(\ref{eqn:phidecaywidth}) 
and~(\ref{eqn:phimassvacuum}) in  symmetric nuclear matter by replacing $m_{K}$ by
$m_{K}^{*}$ and $m_{\phi}$ by $m_{\phi}^{*}$ in the self-energy of the
$\phi$ meson. In Fig.~\ref{fig:phi-nm}, we present our results~\cite{Cobos-Martinez:2017vtr}
for the $\phi$ meson mass (left panel) and decay width (right panel) in
nuclear matter up to $\rho_{B}=3\rho_{0}$. 
As can be seen in Fig.~\ref{fig:phi-nm}, the effect of the in-medium
change in kaon mass gives a negative change in $\phi$ meson mass.
However, even for the largest value of density considered in this study,
the downward mass shift is only a few percent for all values of the cutoff
parameter $\Lambda_{K}$.
In Table~\ref{tab:phippties}, we present the values for $m_{\phi^{*}}$ and 
$\Gamma_{\phi}^{*}$ at normal nuclear density $\rho_{0}$.
More quantitatively, from Table~\ref{tab:phippties} we see that the
negative kaon mass shift of 13\% induces only $\approx$ 2\% downward
mass shift of the $\phi$ meson~\cite{Cobos-Martinez:2017vtr}.
On the other hand, from Fig.~(\ref{fig:phi-nm}) we see that
$\Gamma_{\phi}^{*}$ is very sensitive to the change in the kaon mass. 
It increases rapidly with increasing nuclear density, up to a factor of
$\sim 20$ enhancement for the largest nuclear density considered,
$\rho_{B}=3\rho_{0}$~\cite{Cobos-Martinez:2017vtr}. 
As can be seen from Table~\ref{tab:phippties}, the broadening of the $\phi$ meson 
decay width becomes an order of magnitude larger than its vacuum value 
at normal nuclear density.

\begin{table}[h]
\begin{center}
\begin{tabular}{l|rrr}
\hline \hline
& $\Lambda_{K}= 1000$ & $\Lambda_{K}= 2000$  & $\Lambda_{K}= 3000$ \\
\hline
$m_{\phi}^{*}$ & 1009.3 & 1000.9 & 994.9 \\
$\Gamma_{\phi}^{*}$ & 37.7 & 34.8 & 32.8 \\
\hline \hline
\end{tabular}
\caption{\label{tab:phippties} $\phi$ meson mass and width at normal nuclear 
density, $\rho_{0}$. The $\phi$ meson mass decreases by a few percent ($1.8\%$ in 
average), while the decay width increases by an order of magnitude, 
with respect to the corresponding vacuum values. 
All quantities are given in MeV.}
\end{center}
\end{table}

\subsection{$\eta_c$ and $J/\psi$ mesons}
\label{etacpsi}

The study of interactions of charmonium states, such as $\eta_c$ and
$J/\psi$, with atomic nuclei offers the opportunity to gain new insights
into the properties of the strong force and strongly interacting
matter~\cite{Cobos-Martinez:2020ynh}.
Because charmonia and nucleons do not share light quarks
the Okubo-Zweig-Iizuka (OZI) rule~\cite{OZI} suppresses
the interactions mediated by the exchange of mesons composed of light
quarks and/or antiquarks. The situation here is similar to  the $\phi$ meson case
and also generally for quarkonia and two-heavy-flavor mesons).
Thus, it is important and necessary to explore other possible mechanisms which can provide
 attractive (repulsive) interaction that could lead to the binding (unbinding) of charmonia to
  atomic nuclei~\cite{Cobos-Martinez:2020ynh}.
For a review on the subject
see Refs.~\cite{Hosaka:2016ypm,Krein:2016fqh,Metag:2017yuh,Krein:2017usp}.
Here we employ an effective Lagrangian approach and consider
charmed meson loops in the charmonium
self-energy~\cite{Krein:2010vp,Tsushima:2011kh,Tsushima:2011fg,Krein:2013rha,
Cobos-Martinez:2020ynh}, that light quark-antiquark pair is created from the vacuum.

Note that, recent lattice study using the HAL QCD method with nearly realistic
pion mass of $m_\pi = 146$ MeV, which was also able to reproduce well the physical
hadron masses~\cite{Lyu:2024ttm,Lyu:2025jjl}, found that the $N$-$c\bar{c}$ ($N$-$J\psi$ and
$N$-$\eta_c$) interactions to be attractive in all distances.
They predicted mass reduction of $J/\psi$-meson at normal nuclear density of 0.17 fm$^{-3}$
of about 19(3) MeV. This is consistent with our prediction made by the without ''gauge term''
to be shown later.

For the computation of the $\eta_c$ Lorentz scalar potential in nuclear matter,
we use an effective Lagrangian approach at the hadronic level, which is 
an SU(4)-flavor extension of light-flavor chiral-symmetric Lagrangians 
of pseudoscalar and vector mesons~\cite{Klingl:1996by,Lin:1999ve}.
When we treat the mesons contain at least one bottom quark (antiquark), we will use an
SU(5)-flavor Lagrangian~\cite{Lin:2000ke}. However one can expect that the SU(5) 
flavor symmetry breaking is larger than that of SU(4) due to the current quark mass
values of the charm and bottom quarks. Thus for the SU(4) flavor sector,
we use a flavor SU(4) effective Lagrangian, and determine the relevant coupling
constants based the flavor SU(4) symmetry.

We compute the $\eta_c$ self-energy in vacuum and symmetric nuclear matter, following our previous
works~\cite{Krein:2010vp, Tsushima:2011kh,
Tsushima:2011fg, Krein:2013rha,Cobos-Martinez:2017vtr,
Cobos-Martinez:2017woo, Cobos-Martinez:2017onm,Cobos-Martinez:2017fch,
Cobos-Martinez:2019kln}, and consider only $DD^{*}$ loop. 
See Ref.~\cite{Cobos-Martinez:2020ynh} for details.
The interaction Lagrangian density for the $\eta_c D D^{*}$
vertex is given by 
%
%
\begin{equation}
\label{eqn:LetaDDsta}
\mathcal{L}_{\eta_c D D^{*}} = 
\mi g_{\eta_c D D^{*}}(\partial_{\mu} \eta_c)
\left[\Dbar^{*\mu}D - \Dbar D^{*\mu}\right]\, 
-\mi g_{\eta_c D D^{*}}\eta_c
\left[\Dbar^{*\mu}(\partial_\mu D) - (\partial_\mu \Dbar)D^{*\mu}\right]\, , 
\end{equation}
where $D^{(*)}$ represents the $D^{(*)}$-meson field isospin doublet,
and $g_{\eta_c D D^{*}}$ is the coupling constant. 
The $\eta_c$ self-energy
in the rest frame of $\eta_c$ meson, $p^\mu_{\eta_c} = (m_{\eta_c}, \textbf{0})$
is given by~\cite{Cobos-Martinez:2020ynh}
\begin{equation}
    \label{eqn:etac_se}
    \Sigma_{\eta_c}(m_{\eta_c}^{2})= \frac{8g_{\eta_c D D^{*}}^{2}}{\pi^{2}}\int_{0}^{\infty}
    \dx{q}\, q^2 I(q^2),
\end{equation}
where
\begin{align}
I(q^2)&= \left. \frac{m_{\eta_c}^{2}(-1+q_0^2)/m_{D^*}^2)}
{(q_0+\omega_{D^*})(q_0-\omega_{D^*})
(q_0-m_{\eta_c}-\omega_{D})}\right|_{q_0=m_{\eta_c}-\omega_{D^*}}
\nonumber \\
+&\left. \frac{m_{\eta_c}^{2}(-1+q_0^2)/m_{D^*}^2)}
{(q_0-\omega_{D^*})(q_0-m_{\eta_c}+\omega_{D})
(q_0-m_{\eta_c}-\omega_D)}\right|_{q_0=-\omega_{D^*}},
\end{align}
and $\omega_{D^{(*)}}=(q^2+m_{D^{(*)}}^{2})^{1/2}$, with
$q=|\textbf{q}|$.
The integral in Eq.~(\ref{eqn:etac_se}) is divergent, and we regularize it
with a phenomenological vertex form factor
\begin{equation}
    \label{eqn:FF}
    u_{D^{(*)}}(q^{2})=
    \left(\frac{\Lambda_{D^{(*)}}^{2} + m_{\eta_c}^{2}}
    {\Lambda_{D^{(*)}}^{2}+4\omega_{D^{(*)}}^{2}(q^{2})}
    \right)^{2},
\end{equation}
with cutoff parameter $\Lambda_{D^{(*)}}$, as in previous works.
See Ref.~\cite{Cobos-Martinez:2020ynh} and references therein.
Thus, to regularize Eq.~(\ref{eqn:etac_se}) we will introduce the form 
factor $ u_{D}(k^{2})u_{D^{*}}(k^{2})$ into the integrand.
As before, the cutoff parameter $\Lambda_D^{(*)}$ is an unknown input to our calculation (we use $
\Lambda_{D^{*}}=\Lambda_{D}$). However, it may be
fixed phenomenologically, for example, using a quark model. 
In Ref.~\cite{Krein:2010vp} the value of $\Lambda_{D}$  has been estimated
to be $\Lambda\approx 2500$ MeV, and serves as a reasonable guidance 
to quantify the sensitivity of our results to its value. Therefore we vary
it over the the interval 1500-3000 MeV~\cite{Cobos-Martinez:2020ynh}.

Because the flavor $SU(4)$ symmetry is strongly broken (though less than that of SU(5)), we use
the experimental values for the meson masses~\cite{Tanabashi:2018oca} and
known (extracted) empirical values for the coupling constants, as explained
in the following. For the $D$ meson mass, we take the averaged masses of the neutral
and charged states, and similarly for the $D^{*}$. Thus $m_{D}=1867.2$ MeV
and $m_{D^{*}}=2008.6$ MeV.
For the coupling constants, $g_{\eta_c D D^{*}}=0.60\,g_{\psi D D}$ was 
obtained in Ref.~\cite{Lucha:2015dda}, as the residue at the
poles of suitable form factors using a dispersion formulation of
the relativistic constituent quark model, where $g_{\psi D D}=7.64$ 
was estimated in Ref.~\cite{Lin:1999ad} using the vector meson 
dominance (VMD) model and isospin symmetry.
In this study we use the coupling constant, 
$g_{\eta_c D D^{*}}=(0.60/\sqrt{2})\,g_{\psi D D} \simeq 0.424\,g_{\psi D D}$
~\cite{Cobos-Martinez:2020ynh},  where the factor ($1/\sqrt{2}$) is introduced to give
a larger $SU(4)$ symmetry breaking effect than Ref.~\cite{Lucha:2015dda}.

In this subsection. we will show the mass shift of $\eta_c$
with the use of both the SU(4) symmetry coupling constant as well as that with
the broken SU(4) coupling constant. Furthermore,
later we will compare the in-medium masses of $\eta_c$ and $J/\psi$ with
those of the $\eta_b, \Upsilon, B_c$ and $B_c^*$, using the coupling constant
value $g_{\eta_c DD^*} = g_{\psi DD}=7.64 \to 7.7$, without any symmetry breaking factor,
i.e., $g_{\eta_c DD^*}=(0.60/\sqrt{2})\,g_{\psi DD} \simeq 0.424\,g_{\psi DD}$
$\to g_{\eta_c DD} = g_{\psi DD}=7.7$,
where the tiny difference may be ignored.
For the $J/\psi$ mass shift in this subsection, after $\eta_c$ mass shift,
we will use only the SU(4) symmetric coupling constant, $g_{\psi DD}=7.64$.

We are interested in the difference between the in-medium, $m_{\eta_c}^{*}$,
and vacuum, $m_{\eta_c}$, masses of the $\eta_c$,
\begin{equation}
    \Delta m_{\eta_c}= m_{\eta_c}^{*}-m_{\eta_c},
\end{equation}
with the masses obtained self-consistently from
\begin{equation}
    m_{\eta_c}^{2} = (m_{\eta_c}^{0})^{2} + \Sigma_{\eta_c}(m_{\eta_c}^{2})
    = (m_{\eta_c}^{0})^{2} - |\Sigma_{\eta_c}(m_{\eta_c}^{2})|,
\end{equation}
where $m_{\eta_c}^{0}$ is the bare $\eta_c$ mass and the $\eta_c$
self-energy in the rest frame of $\eta_c$ meson, $\Sigma_{\eta_c}(m_{\eta_c}^2)$ is given
by~\eqn{eqn:etac_se}.
The $\Lambda_{D}$-dependent
$\eta_c$-meson bare mass, $m_{\eta_c}^{0}$, is fixed by fitting the
physical $\eta_c$-meson  mass, $m_{\eta_c}=2983.9$ MeV ~\cite{Cobos-Martinez:2020ynh}.

The in-medium $\eta_c$ mass is obtained in a similar way, with the
self-energy calculated with the medium-modified $D$ and $D^{*}$
meson masses.
The nuclear density dependence of the $\eta_c$-meson mass is
influenced and determined by the intermediate-state $D$ and $D^{*}$ meson interactions 
with the nuclear medium through their medium-modified masses.
The in-medium masses $m_{D}^{*}$ and $m_{D^{*}}^{*}$ are 
calculated within the quark-meson coupling (QMC)
model~\cite{Krein:2010vp,Tsushima:2011kh}, in which effective
scalar and vector meson mean fields couple to the light $u$
and $d$ quarks in the charmed
mesons~\cite{Krein:2010vp,Tsushima:2011kh}.

In the middle panel of
Fig.~\ref{fig:mh-nm} we present the resulting 
medium-modified masses for the $D$ and $D^{*}$ mesons, calculated within
the QMC model~{\cite{Krein:2010vp}, as a function of $\rho_{B}/\rho_{0}$,
where $\rho_{B}$ is the baryon density of nuclear matter and
$\rho_{0}=0.15$ fm$^{-3}$ the saturation density of symmetric nuclear
matter. The net reductions in the masses of the $D$ and $D^{*}$ mesons
are nearly the same as a function of density, with each decreasing by
about 60 MeV at $\rho_{0}$.
The behavior of the $D$ meson mass in medium (finite density and/or
temperature) has been studied in a variety of approaches, where
some of these~\cite{Hayashigaki:2000es,Azizi:2014bba,Wang:2015uya}
find a decreasing $D$ meson mass at finite baryon density,
while others~\cite{Suzuki:2015est,Park:2016xrw,Gubler:2020hft,Hilger:2008jg,Carames:2016qhr},
interestingly, find the opposite behavior.
However, it is important to note that none of the studies in nuclear
matter are constrained by the saturation properties of nuclear
matter, despite they are constrained in the present work. Furthermore, some of these works
employ a non relativistic approach, where relativistic effects might be important.

\begin{figure}
\centering
\includegraphics[scale=0.26]{Dmetac_DDsOnly.eps}
\includegraphics[scale=0.26]{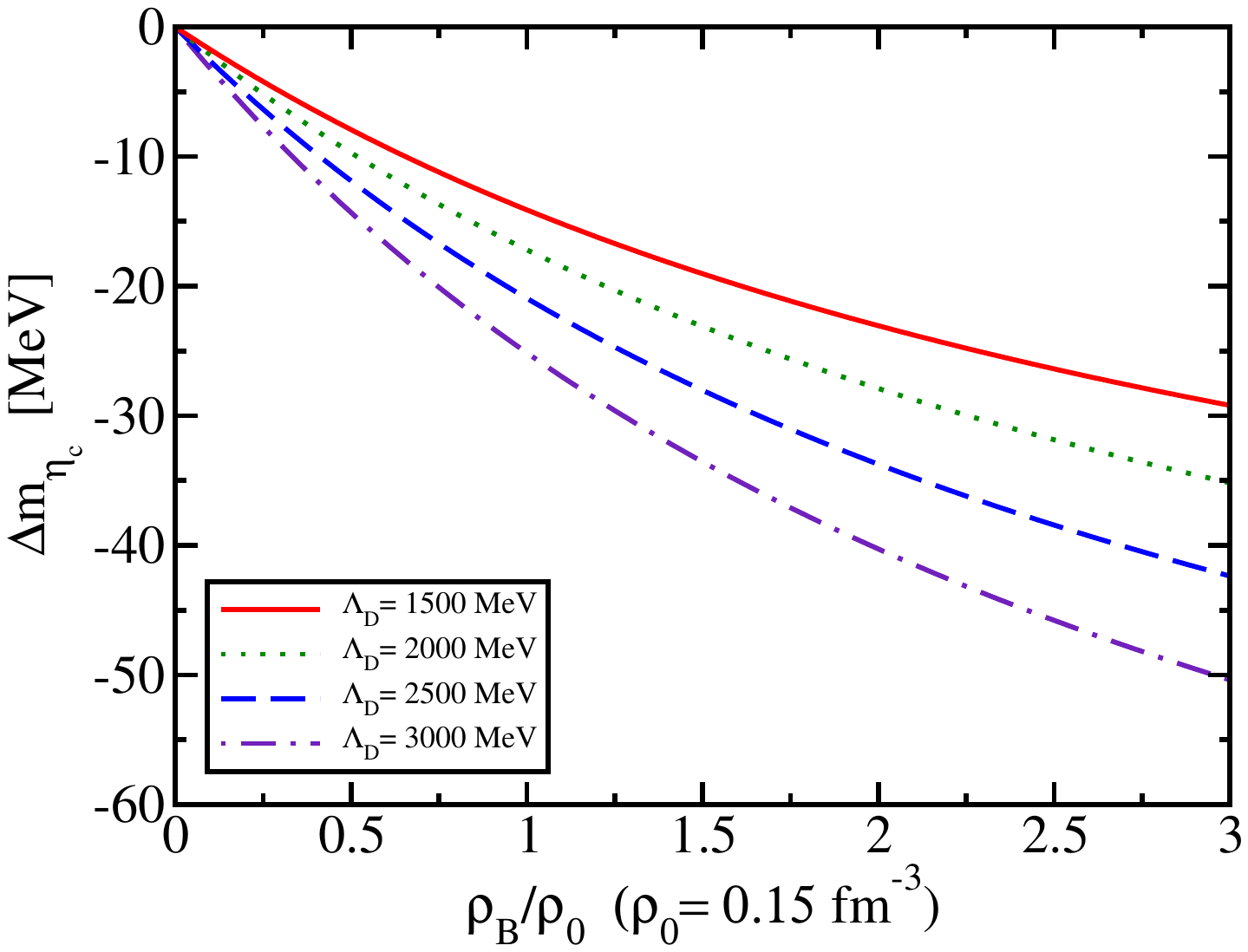}
\caption{\label{fig:Dmetac} $\eta_c$ mass shift (i) with the SU(4) symmetric
coupling~\cite{Zeminiani:2023gqc}, $g_{\eta_c DD}=7.64$ (left-panel),
and (ii) with the broken SU(4) symmetry coupling~\cite{Cobos-Martinez:2020ynh}
$(0.6/\sqrt{2})\times(g_{\eta_c DD^*}=7.64)$
(right panel), versus nuclear matter density for various values of the cutoff
parameter.}
\end{figure}

In Fig.~\ref{fig:Dmetac}, we present the $\eta_c$-meson mass shift,
$\Delta m_{\eta_{c}}$, as a function of the nuclear matter density,
$\rho_{B}$ ($\rho_B/\rho_{0}$), for four values of the cutoff
parameter $\Lambda_{D}$~\cite{Cobos-Martinez:2020ynh}.
As can be seen from the figure, the effect of the in-medium $D$
and $D^{*}$ mass changes is to shift the $\eta_c$ mass downwards.
This is because the reduction in the $D$ and $D^{*}$ masses enhances
the $DD^{*}$-loop contribution in nuclear matter relative to that
in vacuum. This effect increases the larger the cutoff mass
$\Lambda_{D}$ becomes.

The results described above with the two values of the $g_{\eta_c DD^*}$
coupling constants, both support a small downward mass shift for
the $\eta_c$ in nuclear matter, and open the possibility to study the
binding of $\eta_c$ meson to nuclei~\cite{Cobos-Martinez:2020ynh}.

We now turn to the discussion of the $J/\psi$ vector
meson~\cite{Cobos-Martinez:2021ukw,CobosMartinez:2021bia},
following the same procedure as in the $\phi$ meson.
In Refs.~\cite{Krein:2017usp,Krein:2010vp,Tsushima:2011kh}, 
the $J/\psi$ self-energy intermediate states involved the $D$, $\overline{D}$, $D^{*}$,
and $\overline{D^{*}}$ mesons. However, it was
found that the $J/\psi$ self-energy has larger contributions from the loops
involving the $D^{*}$  and $\overline{D^{*}}$ mesons, which is unexpected;
see Ref.~\cite{Krein:2017usp,Krein:2010vp,Tsushima:2011kh} for details on 
the issues and possible explanations. As explained in Ref.~\cite{Krein:2017usp},
this is related to the divergent behavior of the vector meson propagator.
We present results for the $J/\psi$ mass shift in nuclear matter and
nuclei considering only the lightest intermediate state mesons in the $J/\psi$
self-energy, namely the $D\overline{D}$ loop~\cite{Cobos-Martinez:2021ukw,CobosMartinez:2021bia}.

We use the following phenomenological effective Lagrangian densities at the hadronic level, which
are similar to those used above for the $\phi$-meson,
\begin{eqnarray}
{\mathcal L}_{int} &=& {\mathcal L}_{\psi D D} + {\mathcal L}_{\psi\psi D D}, 
\\
{\mathcal L}_{\psi D D} &=& i g_{\psi D D} \, \psi^\mu
\left[\bar D \left(\partial_\mu D\right) -
\left(\partial_\mu \bar D\right) D \right] ,
\label{LpsiDDbar}\\
{\mathcal L}_{\psi\psi D D} &=& g^2_{\psi D D} \psi_\mu \psi^\mu \overline{D} D .
\label{Lpsigauge}
\end{eqnarray}
where $g_{\psi D D}$ is the $J/\psi\,D\overline{D}$ coupling constant and
we use the convention
\begin{eqnarray}
{D = \left(\begin{array}{c}
  D^0 \\
  D^+
\end{array}\right)}, 
\hspace{1.0cm}
\bar D = (\overline{D^0} \;\;\; D^{-} ).
\label{doublets}    
\end{eqnarray}
For notational simplicity we have written $\psi$ the field representing 
the $J/\psi$ vector meson.
We note that the Lagrangians are an $SU$(4) extension of light-flavor
chiral-symmetric Lagrangians of pseudoscalar and vector mesons. In the light
flavor sector, they have been motivated by a local gauge symmetry,
treating vector mesons either as massive gauge bosons or as dynamically
generated gauge bosons. Local gauge symmetry implies in the contact interaction in
Eq.~(\ref{Lpsigauge}) involving two pseudoscalar and two vector mesons.

In view of the fact that $SU$(4) flavor symmetry is strongly broken in nature, and in order to stay
as close as possible to phenomenology, we
use experimental values  for the charmed meson masses and use the
empirically known meson coupling constants. 
For these reasons we  do not use gauged Lagrangians for the study
of $J/\psi$ nuclear bound states -- a similar
attitude was  followed in Ref.~\cite{Lin:1999ve} in a study of hadronic
scattering of charmed mesons. 
However, in order to compare results with Ref.~\cite{Lee:2000csl} and 
assess the impact  of a contact term of the form Eq.~(\ref{Lpsigauge}), 
we also present results for the  $J/\psi$ mass shift including such a term.

We are interested in the difference of the in-medium, $m^*_\psi$, and
vacuum, $m_\psi$,
\begin{equation}
\Delta m_\psi = m^*_\psi - m_\psi,
\label{Delta-m}    
\end{equation}
with the masses obtained from
\begin{equation}
    m^2_\psi = (m^0_\psi)^2 + \Sigma_{D\overline{D}}(m^2_\psi)
    = (m^0_\psi)^2 - |\Sigma_{D\overline{D}}(m^2_\psi)|\, .
\label{m-psi}
\end{equation}
Here $m^0_\psi$ is the bare mass and $\Sigma_{D\overline{D}}(k^2)$ is 
the total $J/\psi$ self-energy obtained from the $D\overline{D}$-loop
contribution only. The in-medium mass, $m^*_\psi$, is obtained likewise,
with the self-energy calculated with medium-modified $D$ meson mass 
calculated by the QMC model (see again the middle panel of Fig.~\ref{fig:mh-nm}).

The scalar self-energy for the $J/\psi$ meson
in the rest frame of $J/\psi$, $\Sigma_{D\overline{D}}(m^2_\psi)$,
is obtained from~\eqn{LpsiDDbar}. The Feynman diagram contributing to
$J/\psi$ self-energy $\mathcal{O}(g_{\psi}^{2})$ is identical to the one
in Fig.~\ref{fig:phise} with the replacements $\phi \to J/\psi$, $K \to D$
and $\overline{K} \to \overline{D}$.

For a $J/\psi$ meson at rest, the self-energy is given by
\begin{equation}
\Sigma_{D\overline{D}}(m^2_\psi) = - \dfrac{ g^2_{\psi \, DD}}{3\pi^2}
\int^\infty_0 dq \, q^{\,2} \, F_{D\overline{D}}(q^{\,2}) \,
K_{D\overline{D}}(q^{\,2}),
\label{Sigma-l}
\end{equation}
where $q = |\textbf{q}|$, and $F_{D\overline{D}}(q^{\,2})=u_{D}(q^2)u_{\overline{D}}
(q^2)$ is the product of vertex form-factors with $u_{D}(q^{2})$ and
$u_{\overline{D}}$ given as in \eqn{eqn:FF} with cutoff parameters
$\Lambda_{D}$ and $\Lambda_{\overline{D}}$,
respectively (we use $\Lambda_{D}=\Lambda_{\overline{D}}$); and
$K_{D\overline{D}}(q^{\,2})$ for the $D\overline{D}$ loop
contribution is given by
\begin{equation}
K_{D\overline{D}}(q^{\,2}) = \dfrac{q^{\,2}}{\omega_D}
\left( \dfrac{q^{\,2}}{\omega^2_D - m^2_\psi/4} - \xi \right) , \label{KDD}
\end{equation}
where $\omega_D = (q^{\,2}+m^2_D)^{1/2}$,
$\xi = 0$ for the non-gauged Lagrangian of Eq.~(\ref{LpsiDDbar}) 
and $\xi = 1$ with Eq.~(\ref{Lpsigauge}), for the gauged Lagrangian of Ref.~\cite{Lee:2000csl}.

As before, the cutoff parameter $\Lambda_D$ is an unknown input to our
calculation.  However, it may be fixed phenomenologically. 
In Ref.~\cite{Krein:2010vp} the value of $\Lambda_{D}$  has been estimated
to be $\Lambda_D \approx 2500$ MeV, and serves as a reasonable guidance
to quantify the sensitivity of our results to its value. 
Since this is a somewhat rough estimate, and it is made solely
to obtain an order of magnitude estimate, we allow the
value of $\Lambda_D$ vary in the range $2000~{\rm MeV} \leq \Lambda_D \leq 6000~{\rm MeV}$; see
Ref.~\cite{Krein:2017usp,Krein:2010vp,Tsushima:2011kh}.

There remain to be fixed the bare $J/\psi$ mass $m^0_\psi$ and the
coupling constants. The bare mass is fixed by fitting the physical mass $m_{J/\psi} = 3096.9$~MeV
using Eq.~(\ref{m-psi}).
As before, because the flavor symmetry $SU(4)$ is strongly broken we
use experimental values for the meson masses and known empirical values for the coupling constants.
For the $D$ meson mass, we take the averaged masses of the neutral
and charged $D$ mesons. Thus $m_{D}=1867.2$ MeV~\cite{Tanabashi:2018oca}.
For the coupling constants we use $g_{\psi DD} = 7.64$, which is obtained by the use of isospin
symmetry~\cite{Lin:1999ad}.
Note that, for $J/\psi$ we use only the SU(4) coupling constant extracted,
different from that of the $\eta_c$ case
(no extra SU(4) breaking effect on the coupling constant).

The nuclear density dependence of the $J/\psi$-meson mass is influenced and determined by the
intermediate-state $D$ and $\overline{D}$ meson
interactions  with the nuclear medium through their medium-modified masses.
The in-medium masses $m_{D}^{*}$ and $m_{\overline{D}^{*}}^{*}=m_{D^*}^*$ are
calculated within the quark-meson coupling (QMC) model~\cite{Krein:2010vp,Tsushima:2011kh},
in which effective
scalar and vector meson mean fields couple to the light $u$
and $d$ quarks in the charmed mesons~\cite{Krein:2010vp,Tsushima:2011kh}.
However in the self-energy of $D\overline{D}$ loop, the vector
potentials cancel out, and no need of considering the effects.

See again the middle panel of Fig.~\ref{fig:mh-nm} we present the
medium-modified masses for the $D$ and $\overline{D}$ mesons
($m_{\overline{D}^{*}}^{*}=m_{D}^{*}$), calculated within the QMC 
model~{\cite{Krein:2010vp}, as a function of $\rho_{B}/\rho_{0}$.
In Fig.~\ref{fig:Dmjpsi} we show the contribution of
the $D\overline{D}$-loop to the $J/\psi$ mass shift for $\xi = 0$.
As the cutoff mass value increases in the form factor, obviously 
the $D\overline{D}$-loop contribution becomes larger.

\begin{figure}[htb]
\centering
\includegraphics[scale=0.27,keepaspectratio=true]{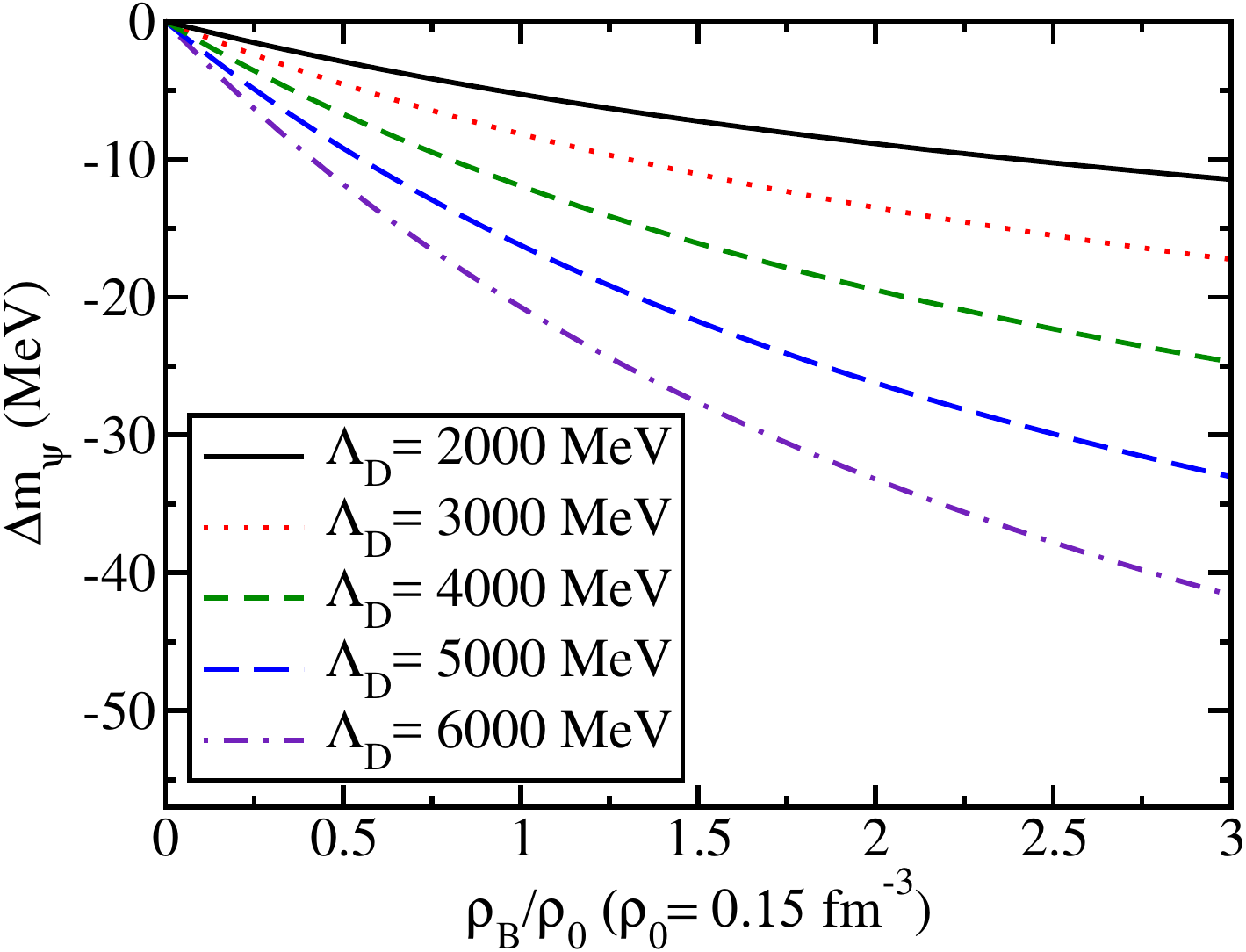}
\includegraphics[scale=0.27,keepaspectratio=true]{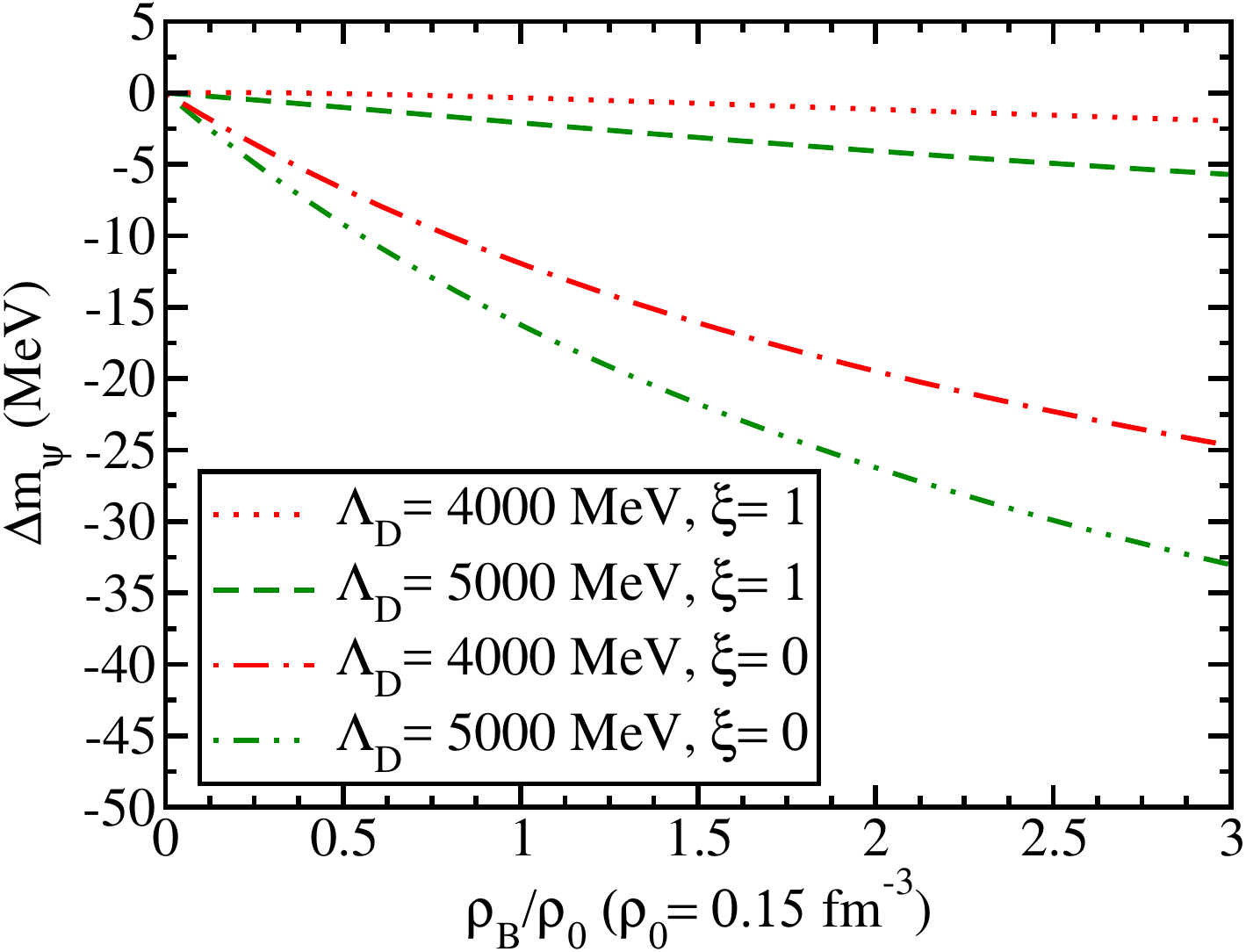}
\caption{Contribution from the $D\overline{D}$-loop to the 
$J/\psi$ mass shift in symmetric nuclear matter without the gauge term
($\xi = 0$) for five different values of the cutoff $\Lambda_D$ 
(left panel), and the comparison with including the gauge term ($\xi = 1$) 
for two values of $\Lambda_D$ (right panel).}
\label{fig:Dmjpsi}
\end{figure}

First, from the result shown in the left panel of Fig.~\ref{fig:Dmjpsi}
without the gauge term ($\xi = 0$), one can see that 
the $J/\psi$ gets the attractive potential for all the values of
the cutoff $\Lambda_D$, $2000 - 6000$ MeV~\cite{Cobos-Martinez:2021ukw,CobosMartinez:2021bia}.
In contrast, one can see from the right panel in Fig.~\ref{fig:Dmjpsi},  
that the effect of the gauge term tends to oppose the effect (repulsion) 
of the contribution of the $D\overline{D}$-loop as noticed in
Ref.~\cite{Krein:2010vp}~\cite{Cobos-Martinez:2021ukw,CobosMartinez:2021bia}.
When the value of $\Lambda_D$ is smaller, the mass shift actually 
becomes positive.
The results shown in Fig.~\ref{fig:Dmjpsi} reveal a negative mass shift (attractive potential)
for the $J/\psi$ meson in symmetric nuclear matter for all
values of the cutoff mass parameter $\Lambda_D$ when $\xi = 0$ and, as
in the $\eta_c$ meson case, open the possibility to study the
binding of $J/\psi$ mesons to nuclei~\cite{Cobos-Martinez:2021ukw,CobosMartinez:2021bia}.

\subsection{\label{upsietab} $\Upsilon$ and $\eta_b$ mesons}

First, we discuss the $\Upsilon$  (vector) meson.
The $\Upsilon$ mass shift in nuclear matter originates from the modifications of the
$BB$, $BB^{*}$, and $B^{*}B^{*}$ meson loops contributions to the $\Upsilon$ self-energy, relative
to those in free space; the lowest order Feynman diagrams associated with these contributions
are similar to Fig.~\ref{fig:phise}. 
The $\Upsilon$ self-energy is calculated using an effective  SU(5)-flavor
symmetric Lagrangian at the hadronic
level~\cite{Zeminiani:2020aho,Lin:2000ke}, where mesons are considered
to be point like, for the interaction vertices $\Upsilon B B$, $\Upsilon 
B^{*}B^{*}$, and $\Upsilon B B^{*}$ neglecting any possible imaginary part.
In  Ref.~\cite{Zeminiani:2020aho} we made an extensive analysis of these
contributions to the $\Upsilon$ self-energy and found that, for example, 
the $B^{*}B^{*}$ loop gives an unexpectedly large contribution, similar to the case of $J/\psi$.
For this reason, and to be consistent with the $\eta_b$ case studied
below, we consider only the $BB$ loop contribution to the $\Upsilon$
self-energy~\cite{Zeminiani:2020aho}, leaving for the future a full study of all three
contributions.
This treatment is also consistent with the $J/\psi$ self-energy calculation with
the lowest $D\overline{D}$ loop contribution, and we can compare the amounts of 
mass shift for the $\Upsilon$ and $J/\psi$ based on a similar footing.
The interaction Lagrangian for the $\Upsilon B B$ vertex is given by~\cite{Zeminiani:2020aho}
\begin{equation}
\label{eqn:LUBB}
{\cal L}_{\Upsilon BB} 
= i g_{\Upsilon BB}\Upsilon^{\mu} 
\left[\overline{B} \partial_{\mu}B - \left(\partial_{\mu}\overline{B}\right)B\right],
\end{equation}
where $g_{\Upsilon BB}$ is the coupling constant for the vertex 
$\Upsilon B B$ vertex, and the following convention is adopted for the isospin doublets of the $B$
mesons
\begin{align*}
B&=\begin{pmatrix}
        B^{+}\\
        B^{0}
       \end{pmatrix}, & \overline{B}=\begin{pmatrix}
       B^{-} & \overline{B^{0}} \end{pmatrix} .  
\end{align*}
The coupling constant $g_{\Upsilon BB}$ is calculated from the 
experimental data for $\Gamma(\Upsilon\to e^{+}e^{-})$ using the vector meson dominance (VMD)
model.
This gives $g_{\Upsilon BB}=13.2$;
see Refs.~\cite{Zeminiani:2020aho,Lin:2000ke} and references therein for details. We note that a
similar approach was taken in  Refs.~\cite{Lin:1999ad,Krein:2010vp} to determine the coupling
constant $g_{J/\psi DD}=7.64$ for the vertex  $J/\psi D D$.

Including only the $BB$ loop, \eqn{eqn:LUBB}, the $\Upsilon$ self-energy $\Sigma_{\Upsilon}$ for an
$\Upsilon$ at rest is given by~\cite{Zeminiani:2020aho}
\begin{equation}
    \label{eqn:upsilon_se}
    \Sigma_{\Upsilon}(m_\Upsilon^{2})= -\frac{g_{\Upsilon B B}^{2}}{3\pi^{2}}\int_{0}^{\infty}
    \dx{q} {q}^2\,I({q}^2)
\end{equation}
where
\begin{equation}
I({q}^2)=\frac{1}{\omega_B}\left(\frac{{q}^2}
{\omega_B-m_{\Upsilon}^2/4} \right),
\end{equation}
with $q = |\textbf{q}|$ and $\omega_B = \left({q}^{2} + m^{2}_{B}\right)^{1/2}$.
As is always the case in an effective Lagrangian approach, when mesons
are treated as point-like particle, the self-energy loop integrals like \eqn{eqn:upsilon_se} are
divergent and therefore need to be regularized. To this end, we  introduce into the integrand of
\eqn{eqn:upsilon_se} a phenomenological vertex form factor
$u_{B}({q}^{2})$ with cutoff parameter $\Lambda_{B}$~\cite{Krein:2010vp, Tsushima:2011kh,
Tsushima:2011fg, Krein:2013rha, Cobos-Martinez:2017vtr, Cobos-Martinez:2017woo,
Cobos-Martinez:2017onm, Cobos-Martinez:2017fch, Cobos-Martinez:2019kln}, for to each $\Upsilon B B$
vertex, as we did in previous cases;  see \eqn{eqn:FF}.
We recall that form factors are necessary to take into account the finite
size of the mesons participating in the vertices, while the cutoff
$\Lambda_B$, which is an unknown input to our calculation, may be
associated with energies needed to probe the internal structure of the
mesons. Thus, in order to reasonably include these effects, and to quantify
the sensitivity of our results to its value, we vary $\Lambda_B$ over the
interval 2000-6000 MeV (roughly up to around the mass of the $B$ meson);
see Ref.~\cite{Zeminiani:2020aho} for a more extensive discussion.
\begin{figure}[htb]%
\centering
\vspace{4ex}
\centering
\includegraphics[scale=0.33]{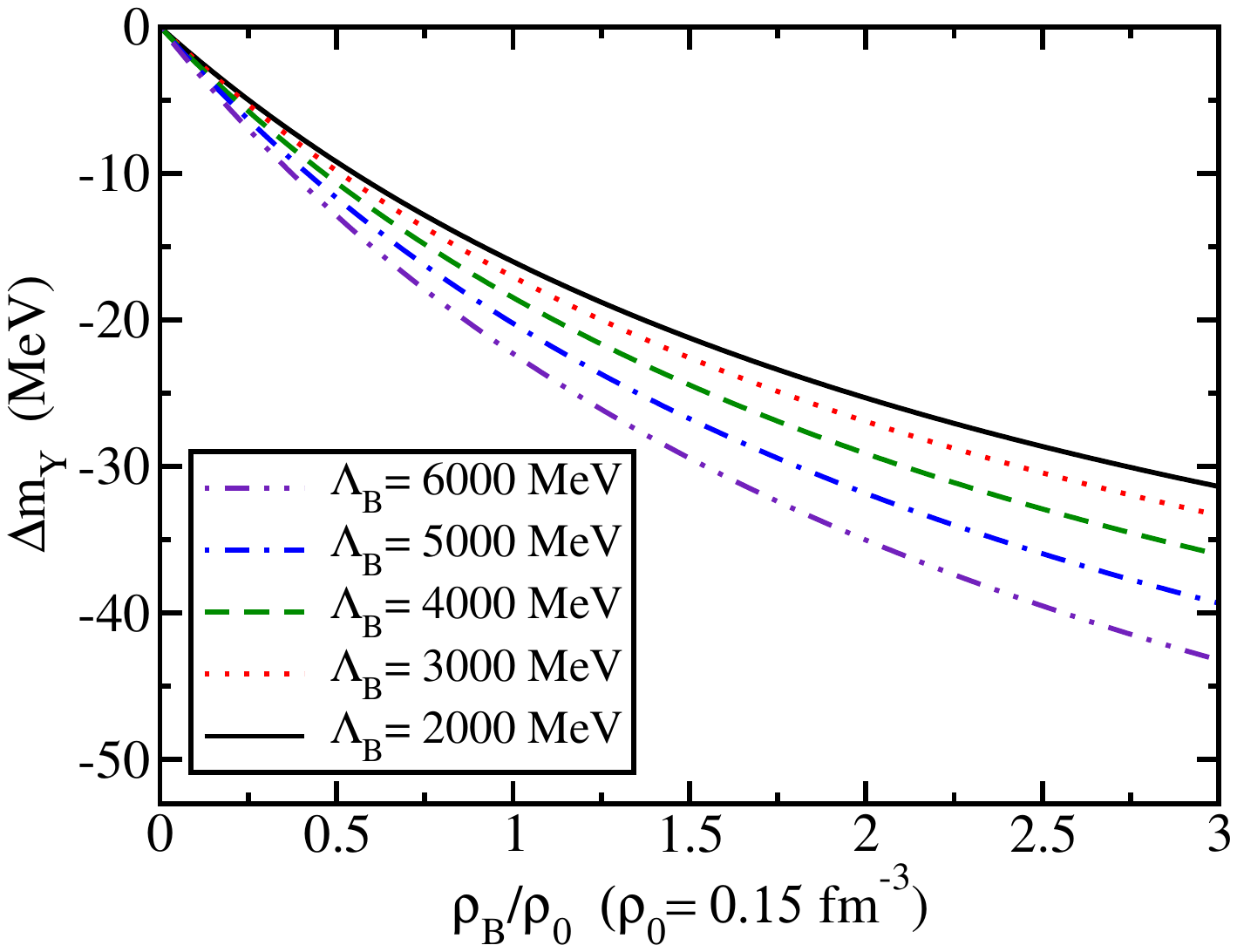}%
\caption{$\Upsilon$ mass shift in symmetric nuclear matter as a function of the nuclear matter
density ($\rho_B/\rho_0$) }%
\label{fig:upsilon_ms}%
\end{figure}

The $\Upsilon$ mass shift in nuclear matter, $\Delta m_{\Upsilon}$, is calculated from the
difference between its mass in the medium, $m_{\Upsilon}^{*}$, and its value in vacuum,
$m_\Upsilon$, in the rest frame of the $\Upsilon$, namely,
\begin{equation}
\label{eqn:upsilon_ms}
\Delta m_{\Upsilon}= m_{\Upsilon}^{*} - m_\Upsilon,
\end{equation}
where these masses are computed self-consistently from
\begin{equation}
\label{eqn:upsilon_mass}
    m_{\Upsilon}^{2} = (m_{\Upsilon}^{0})^{2} + \Sigma_{\Upsilon}(m_{\Upsilon}^{2})
     = (m_{\Upsilon}^{0})^{2} - |\Sigma_{\Upsilon}(m_{\Upsilon}^{2})|,
\end{equation}
with $m_{\Upsilon}^{0}$ the bare $\Upsilon$ mass and the $\Upsilon$
self-energy $\Sigma_{\Upsilon}(m_\Upsilon^{2})$ is given in \eqn{eqn:upsilon_se}.
The $\Lambda_{B}$-dependent $\Upsilon$ bare mass,  $m_{\Upsilon}^{0}$, is
fixed with the physical $\Upsilon$ mass, namely $m_{\Upsilon}=9640$ MeV.

The in-medium $\Upsilon$ mass $m_{\Upsilon}^{*}$ is obtained by 
solving \eqn{eqn:upsilon_mass} with the self-energy calculated with medium-modified
$B$ mass. This medium-modified mass was calculated
using the quark-meson coupling (QMC) model as a 
function of the nuclear matter density $\rho_B$, and the results are
shown in Fig.~\ref{fig:mh-nm} (left panel).
From Fig.~\ref{fig:mh-nm}, it can be seen that the QMC model gives
a similar downward mass shift for the $B$  and $B^{*}$ in symmetric
nuclear matter. For example, at the saturation density 
$\rho_0=0.15\,\text{fm}^{-3}$, the mass shift for the $B$ and $B^*$ mesons
are respectively, $(m^*_B - m_B)=-61$ MeV and  $(m^*_{B^*}-m_{B^*})=-61$
MeV, where the difference in their mass shift values appears in the decimal
place. The values for the masses in vacuum for the $B$ and $B^{*}$ mesons
used are $m_B= 5279$ MeV and $m_{B^{*}}=5325$ MeV, respectively.

The nuclear density dependence of the $\Upsilon$ mass is driven by the
intermediate $B\overline{B}$ state interactions with the nuclear medium, where the
effective scalar and vector meson mean fields couple to the light $u$ 
and $d$ quarks in the bottom mesons, $B$ and $B^{*}$.  
In Fig.~\ref{fig:upsilon_ms} we show the results  for  the $\Upsilon$ mass
shift as a function of the nuclear density, $\rho_B/\rho_0$, for five
values of the cutoff parameter $\Lambda_B$.  As can be seen in
Figs.~\ref{fig:mh-nm} (left panel) and~\ref{fig:upsilon_ms}, a decreasing
of the in-medium $B$ meson mass induces a negative mass shift for $\Upsilon$.
As expected, the mass shift amount of the $\Upsilon$ is dependent on the value
of the cutoff mass $\Lambda_B$ used, being larger for larger $\Lambda_B$;
see  Ref.~\cite{Zeminiani:2020aho} for further details. 
For example, for the values of the cutoff shown in Fig.~\ref{fig:upsilon_ms},
the $\Upsilon$ mass shift amount varies from -16 to -22 MeV, at  $\rho_B=\rho_0$.

For the calculation of the $\eta_b$ mass shift in nuclear matter, we
proceed similarly to the $\Upsilon$ case and take into account only the
$BB^*$ loop (pseudoscalar-pseudoscalar-vector) contribution to the $\eta_b$ self-energy.
In Ref.~\cite{Zeminiani:2020aho} we have also studied the mass shift including the
$\eta_b B^*B^*$ interaction in the $\eta_b$ self-energy, and found that its
contribution to the mass shift amount turned out to be negligible.
Thus, in order to be consistent with the $\Upsilon$ case above, in 
both cases we consider only the minimal contribution,
and here we only give results for the $BB^*$ loop in the $\eta_b$ self-energy.
This is also a consistent treatment with the $\eta_c$ mass shift calculation, 
and later we can compare based on a similar footing of the self-energy calculation.

\begin{figure}[htb]%
\centering
\vspace{4ex}
\centering
\includegraphics[scale=0.33]{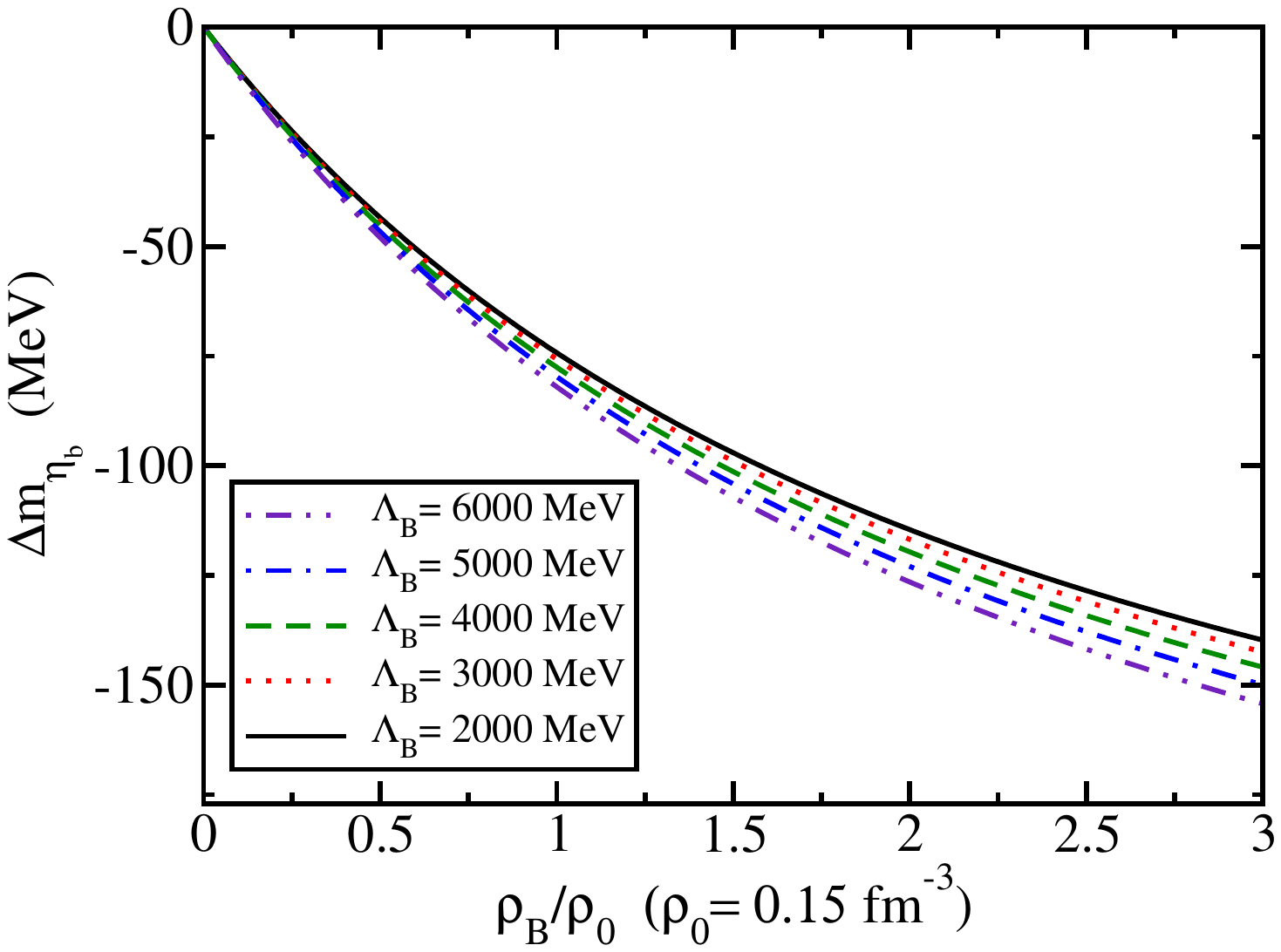}%
\caption{$\eta_b$ mass shift in nuclear matter as a function of the nuclear density
$\rho_B/\rho_0$. }%
\label{fig:etab_ms}%
\end{figure}

For the calculation of the $\eta_b$ mass shift in symmetric nuclear matter, we
proceed similarly to the $\Upsilon$ and $\eta_c$ cases, and take into
account only the $BB^*$ loop contribution to the $\eta_b$ self-energy.
As already mentioned, in Ref.~\cite{Zeminiani:2020aho}, we have also studied including
the $\eta_b B^*B^*$ interaction in the $\eta_b$ self-energy and found that
its contribution to the mass shift amount is negligible.

The effective Lagrangian for the $\eta_b BB^*$ interaction is~\cite{Zeminiani:2020aho}
\begin{equation}
\label{eqn:LetabBBast}
{\cal L}_{\eta_b BB^*} 
= i g_{\eta_b BB^*}
\left[ (\partial^\mu \eta_b) 
\left( \overline{B}^*_\mu B - \overline{B} B^*_\mu \right)
- \eta_b 
\left( \overline{B}^*_\mu (\partial^\mu B) - (\partial^\mu \overline{B}) B^*_\mu \right)
\right ], 
\end{equation}
where $g_{\eta_b BB^*}$ is the coupling constant for the $\eta_b BB^*$
vertex. We will use its value in the  SU(5) 
scheme~\cite{Zeminiani:2020aho}, namely $g_{\eta_b BB^*} = g_{\Upsilon BB} = g_{\Upsilon B^*B^*} =
\frac{5g}{4\sqrt{10}}$. Using \eqn{eqn:LetabBBast},
the $\eta_b$ self-energy for an $\eta_b$ at rest  is given 
by~\cite{Cobos-Martinez:2020ynh}
\begin{equation}
\label{eqn:etab_se}
\Sigma_{\eta_b} 
= \frac{8 g_{\eta_b B B^*}^{2}}{\pi^{2}}\int_{0}^{\infty}
    \dx{q} {q}^{2} I({q}^{2}),
\end{equation}
where 
\begin{eqnarray}
\hspace{-5mm}I({q}^{2})
&=& \left. \frac{m_{\eta_b}^{2}(-1+q_0^2/m_{B^{*}}^{2})}
{(q_0^2-\omega_{B^{*}}^2) 
(q_{0}-m_{\eta_b}-\omega_{B})}\right|_{q_{0}=m_{\eta_b}- \omega_B} 
\nonumber \\
&+& \left. \frac{m_{\eta_b}^{2}(-1+q_0^2/m_{B^{*}}^{2})}
{(q_{0}-\omega_{B^{*}})\left((q_{0}-m_{\eta_b})^2-\omega_{B}^2\right) 
}\right|_{q_{0}=-\omega_{B^*}},  \nonumber \\
\label{IBBs}
\end{eqnarray}
with $q = |\textbf{q}|$ and $\omega_{B^*}=\sqrt{{q}^{2}+m_{B^*}^{2}}$.
The mass of the $\eta_b$ meson, in vacuum and in nuclear matter, is 
computed similarly to the $\Upsilon$ case~\cite{Zeminiani:2020aho}.  First, we introduce form
factors, as in \eqn{eqn:FF}, into each $\eta_b BB^*$ vertex, with $\Lambda_B=\Lambda_{B^{*}}$, to
regularize the
divergent integral in the self-energy \eqn{eqn:etab_se}. Second, we fix 
the value of the $\eta_b$  bare mass using the physical (vacuum) mass of the $\eta_b$,  namely
$m_{\eta_b}=9399$ MeV, using
\eqn{eqn:upsilon_mass} appropriately written  for the $\eta_b$ case. 
 Then, for the calculation of the $\eta_b$ mass shift in nuclear matter,
 the self-energy $\Sigma_{\eta_b} $ is computed using the medium-modified
 $B$ and $B^{*}$ masses calculated with the QMC model and shown in
 Fig.~\ref{fig:mh-nm} (left panel). 
 The results for the $\eta_b$ mass shift behavior in nuclear matter are shown in
 Fig.~\ref{fig:etab_ms} as a function of the nuclear matter density $\rho_B/\rho_0$,
 for the same range of values for the cutoff mass $\Lambda_B$ as for
 the $\Upsilon$~\cite{Zeminiani:2020aho}.
 As can be seen from Fig.~\ref{fig:etab_ms}, the mass of the $\eta_b$ is shifted
 downwards in nuclear matter for all values of the cutoff $\Lambda_B$,
 similarly to the $\Upsilon$. For example, at  the normal density of
 symmetric nuclear matter $\rho_0$, the mass shift value varies from -75 MeV to -82 MeV
 when the cutoff varies from $\Lambda_B=2000$ MeV to $\Lambda_B=6000$ MeV.
 Similarly to the $\Upsilon$ mass shift, the dependence of the $\eta_b$ mass shift amount
 on the values of the cutoff is small, for example, just -7 MeV
 when the cutoff is increased by a factor of 3 at $\rho_B=\rho_0$~\cite{Zeminiani:2020aho}.

\subsection{$B_c$ and $B_c^*$ mesons}

The $B_c$ ($B^*_c$) in-medium downwards mass shift comes from
the enhanced $B^*D + BD^*$ ($BD$) loop contribution to the self-energy,
relative to those in free space.
See Fig.~\ref{fig:phise} but replacing the $K\overline{K}$ loop  
properly by $B^*D$ and $BD^*$ loops ($BD$ loop). 
See Refs.~\cite{Zeminiani:2023gqc} for details.
By expanding the SU(5) flavor symmetric effective meson Lagrangian~\cite{Lin:2000ke} 
in terms of the components of pseudoscalar ($P$) and vector ($V$) $5 \times 5$ matrices, 
we obtain the following Lagrangians for the interactions $B_c B^*D$, $B_c BD^*$ and
$B^*_c BD$~\cite{Zeminiani:2023gqc}:
\begin{eqnarray}
    \mathcal{L}_{B_{c}B^{*}D} &=& ig_{B_{c}B^{*}D}
    [(\partial_{\mu}B^{-}_{c}){D}
    - B^{-}_{c}(\partial_{\mu}{D})] B^{* \mu} + h.c.,
    \nonumber \\
    \mathcal{L}_{B_{c}BD^{*}} &=& ig_{B_{c}BD^{*}}
    [(\partial_{\mu}B^{+}_{c})\overline{B}
    - B_c^{+}(\partial_{\mu}\overline{B})] \overline{D^{*}}^\mu + h.c.,
   \nonumber \\
   \mathcal{L}_{B^{*}_{c}BD} &=& -ig_{B^{*}_{c}BD}
     B^{*+{\mu}}_{c} [\overline{B}(\partial_{\mu} \overline{D}) -
(\partial_{\mu} \overline{B}) \overline{D}] + h.c.,
\end{eqnarray}
where the conventions for the $B, D$ and  $B^*$ mesons have been already given.

The SU(5) symmetric universal coupling $g$ yields the relations,
$g_{B_c B^* D} = g_{B_c B D^*} = g_{B^*_c B D}$.
The value of $g$ is fixed by
$g_{\Upsilon BB} = \frac{5g}{4\sqrt{10}} \approx 13.2$
by the $\Upsilon$ decay data $\Gamma(\Upsilon \to e^+ e^-)$
with the vector meson dominance (VMD) model~\cite{Lin:2000ke,Zeminiani:2020aho},
and thus we get,
\begin{equation}
g_{B_c B^* D} = \frac{2}{\sqrt{5}}g_{\Upsilon BB}, \hspace{3ex}
g_{B_c B^* D} = g_{B_c B D^*} = g_{B^*_c B D} = \frac{g}{2\sqrt{2}}
\approx 11.9.
\end{equation}

The in-medium mass shift of the $B_c$ meson, $\Delta m_{B_c}$, is computed by the difference of the
in-medium $m^*_{B_c}$ and the free space $m_{B_c}$ masses
\begin{equation}
\Delta m_{B_c} = m^*_{B_c} - m_{B_c},
\end{equation}
where, the free space mass $m_{B_c}$ (input) is used to determine the bare mass $m^0_{B_c}$ by
\begin{equation}
m^2_{B_c} = \left( m^0_{B_c} \right)^2 + \Sigma_{B_c}(m^2_{B_c})
          = \left( m^0_{B_c} \right)^2 - |\Sigma_{B_c}(m^2_{B_c})|.
\label{m0}
\end{equation}
Note that, the total self-energy $\Sigma_{B_c}$ is calculated by the sum of the $B^*D$ and $BD^*$
meson loop contributions in free space
ignoring the possible $B_c$ meson as well as all the other meson widths (or imaginary part)
in the self-energy.
The in-medium $B_c$ mass $m^{* 2}_{B_c}$
is similarly calculated, with the same bare mass
value $m^0_{B_c}$ determined in free space,
and the in-medium masses of the
($B,B^*,D,D^*$) mesons ($m^*_{B},m^*_{B^*},m^*_{D},m^*_{D^*}$), namely,
\begin{eqnarray}
m^2_{B_c}
&=& \left[ m^0_{B_c}(B^*D+BD^*) \right]^2
- \left|\Sigma_{B_c}(B^*D) + \Sigma_{B_c}(BD^*) \right|(m^2_{B_c}),
\label{m02}\\
m^{* 2}_{B_c}
&=& \left[ m^0_{B_c}(B^*D+BD^*) \right]^2
- \left|\Sigma^*_{B_c}(B^*D) + \Sigma^*_{B_c}(BD^*) \right|(m^{* 2}_{B_c}).
\label{m03}
\end{eqnarray}

We note that, when the self-energy graphs contain different contributions,
as $\Sigma_{B_c} ({\rm total}) = \Sigma(B^*D) + \Sigma(BD^*)$,
$m^0$ depends on both $\Sigma(B^*D)$ and $\Sigma(BD^*)$ to reproduce the physical mass
$m_{B_c}$. Thus, one must be careful when discussing the $B_c$
in-medium mass and mass shift of each loop contribution
$\Sigma(B^*D)$ and $\Sigma(BD^*)$,
since $m^0(B^*D+BD^*) \ne m^0(B^*D) \ne m^0(BD^*)$,
and $m^0(B^*D + BD^*) \ne m^0(B^*D) + m^0(BD^*)$.
The dominant loop contribution can be known by the decomposition
of the self-energy
$\Sigma^{(*)}_{B_c} (B^*D + BD^*) = \Sigma^{(*)}_{B_c}(B^*D) + \Sigma^{(*)}_{B_c}(BD^*)$.
It turned out that the dominant contribution is from the $BD^*$ 
loop~\cite{Zeminiani:2023gqc}.
This is due to the dominant contribution from the lighter vector meson $D^*$
due to the vector meson propagator Lorentz structure.

As an example, in the case considering solely the $B^*D$ loop without the $BD^*$ loop,
the ''in-medium'' $B_c$ self-energy in the rest frame of $B_c$ is given by
\begin{equation}
    \Sigma^{B^{*}D}_{B_{c}}(m^{*}_{B_{c}}) = \frac {-4g^{2}_{B_{c}B^{*}D}}{\pi^{2}}
    \int dq
    q^{2} I_{B_c}^{B^{*}D}(q^2)
    F_{B_c B^* D} (q^2),
\label{SigBsD}
\end{equation}
with $q = |\textbf{q}|$, and $I_{B_c}^{B^{*}D} (q^2)$ is expressed,
after the Cauchy integral with respect to $q^0$ complex plane shifting $q^0$
variable for the vector potentials as,
\begin{eqnarray}
    I_{B_c}^{B^{*}D} (q^2) &=&
    \left. \frac{m^{*2}_{B_c} \left(-1 + q^2_0 / m^{*2}_{B^*} \right)}
    {(q_0 - \omega^*_{B^*}) (k_0 - m^{*}_{B_c} + \omega^*_D)
    (q_0 - m^{*}_{B_c} -\omega^*_D)}
    \right|_{q_{0} = -\omega^*_{B^*} }
    \nonumber \\
    && \left. \hspace{5ex} + \frac{m^{*2}_{B_c} \left( -1 + q^2_0 / m^{*2}_{B^*} \right) }
    {(q_0 + \omega^*_{B^*}) (q_0 - \omega^*_{B^*})
    (q_0 - m^{*}_{B_c} -\omega^*_D)}
    \right|_{q_{0} = m^{*}_{B_c} - \omega^*_D},
\label{IBsD}
\end{eqnarray}
where, $F_{B_c B^* D}$ in Eq.~(\ref{SigBsD}) is the product of vertex form factors
in medium to regularize the divergence in the loop integral,
$F_{B_c B^* D} (q^2) = u_{B_cB^*}(q^2)u_{B_c D}(q^2)$.
They are given by using the corresponding meson in-medium masses,
$u_{B_c B^*} = \left( \frac{\Lambda^2_{B^*} + m^{* 2}_{B_c}}
{\Lambda^2_{B^*} + 4\omega^{* 2}_{B^*} (q^2)} \right)^2$ and
$u_{B_c D} = \left( \frac{\Lambda^2_{D} + m^{* 2}_{B_c}}
{\Lambda^2_{D} + 4\omega^{* 2}_{D} (q^2)} \right)^2$
with $\Lambda_{B^*}$ and $\Lambda_{D}$ being the cutoff masses associated with the $B^*$ 
and $D$ mesons, respectively.
We use the common value $\Lambda = \Lambda_{B^*} = \Lambda_{D}$. 
A similar calculation is performed to obtain the $BD^*$ loop contribution,
namely, in Eqs.~(\ref{SigBsD}) and~(\ref{IBsD}), as well as in the form factors,
by replacing $(B^*,D) \to (B,D^*)$.

The choice of cutoff value has nonegligible impacts on the results.
We use the common cutoff $\Lambda \equiv \Lambda_{B,B^*,D,D^*,K,K^*}$
by varying the $\Lambda$ value. The $\Lambda$ value may be associated
with the energies to probe the internal structure of the mesons.
In the previous study~\cite{Zeminiani:2020aho} it was observed that when the values
of the cutoff becomes close to the masses of the mesons in calculating
the self-energies, a certain larger cutoff mass value range did
not make sense to serve as the form factors. This is because the Compton wavelengths of the
corresponding cutoff mass values reach near and/or smaller than those of the meson sizes.
Therefore, we need to constrain the cutoff $\Lambda$ value in such
a way that the form factors reflect properly the finite size of the mesons.
Based on the heavy quark and heavy meson symmetry,
we use the same range of values for $\Lambda$ as it was practiced for
the quarkonia~\cite{Zeminiani:2020aho}.
Thus, we use the values, $\Lambda$ = 2000, 3000, 4000, 5000 and 6000 MeV.

\begin{figure}[htb]
\centering
 \includegraphics[width=8.0cm]{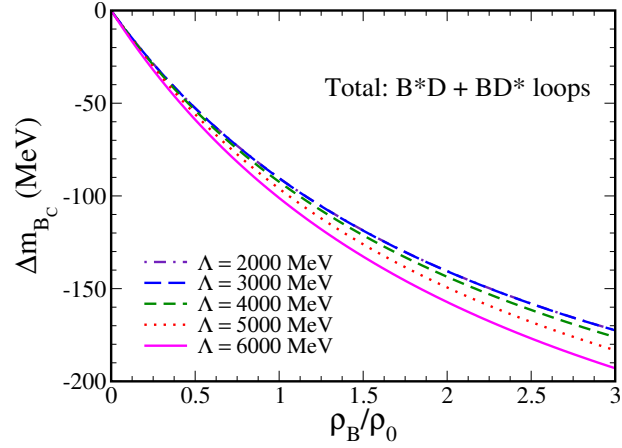}
 \caption{Total ($B^*D$ + $BD^*$) loop contribution for the
in-medium $B_c$ mass shift 
versus baryon density ($\rho_B/\rho_0$) for five different
values of the cutoff mass $\Lambda$.}
\label{totbc}
\end{figure}

Including the total ($BD^* + B^*D$) loop contributions,
the $B_c$ mass shift amount $\Delta m_{B_c}(BD^*+B^*D)$
at $\rho_0$ ranges from -90.4 to -101.1 MeV
($m^*_{B_c}(BD^* + B^*D) =$ 6184.1 to 6173.4 MeV).
Later, we will compare the $B_c$ mass shift and those of the $\eta_b$ and $\eta_c$.

Next, we study the in-medium mass shift of the $B^*_c$ meson calculated in the
rest frame of $B_c^*$.
For the $B^*_c$ self-energy, we include only the $BD$ loop contribution,
as already commented based on the $\Upsilon$ and $J/\psi$
self-energies~\cite{Zeminiani:2020aho},
\begin{equation}
\Sigma^{BD}_{B^{*}_{c}}(m^*_{B_c^*}) = \frac{-4g^{2}_{B^{*}_{c}BD}}{3\pi^{2}} \int dq
q^{4} I^{BD}_{B^{*}_{c}}(q^2)
F_{B^*_c BD} (q^2),
\label{IBcs}
\end{equation}
where $I^{BD}_{B^{*}_{c}}(q^2)$ is expressed by,
\begin{eqnarray}
    I^{BD}_{B^{*}_{c}}(q^2) &=&
    \left. \frac{1}{(q_0 - \omega^*_B) (q_0 - m^*_{B^*_c} + \omega^*_D)
    (q_0 - m^*_{B^*_c} - \omega^*_D)} \right|_{q_0 = - \omega^*_B}
    \nonumber \\
    && \left. \hspace{5ex} + \frac{1}
    {(q_0 + \omega^*_B) (q_0 - \omega^*_B)
    (q_0 -m^*_{B^*_c} - \omega^*_D)} \right|_{q_0 = m^*_{B^*_c} - \omega^*_D},
\end{eqnarray}
with $q = |\textbf{q}|$. In Eq.~(\ref{IBcs}), $F_{B^*_c BD} (q^2)$ is given by
the product of the form factors,
$F_{B^*_c BD} (q^2) = u_{B^*_c B}(q^2)
u_{B^*_c D}(q^2)$,
with $u_{B^*_c B}$ and $u_{B^*_c B}$ being
\newline
$u_{B^*_c B} = \left( \frac{\Lambda^2_{B} + m^{* 2}_{B^*_c}}
{\Lambda^2_{B} + 4\omega^{* 2}_{B} (q^2)} \right)^2$ and
$u_{B^*_c B} = \left( \frac{\Lambda^2_{D} + m^{* 2}_{B^*_c}}
{\Lambda^2_{D} + 4\omega^{* 2}_{D} (q^2)} \right)^2$.
Again we use $\Lambda = \Lambda_B = \Lambda_D$ ranging 2000 to 6000 MeV.

\begin{figure}[htb!]
\centering
\includegraphics[width=8.0cm]{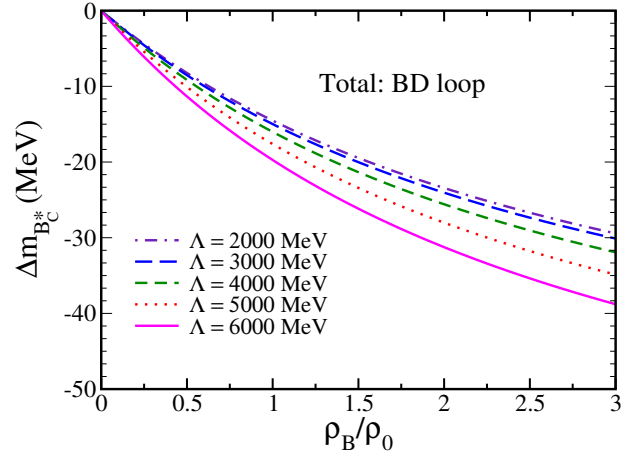}
 \caption{$BD$ loop (total) contribution for the
in-medium $B^*_c$ mass shift versus baryon density
($\rho_B/\rho_0$) for five different values of the cutoff mass $\Lambda$.}
\label{totbcs}
\end{figure}

\subsection{Comparison with heavy quarkonia}
\label{hvqrkcomp}

We now compare in Fig.~\ref{comp} the results of $B_c$ and $B^*_c$~\cite{Zeminiani:2023gqc},
with those of the heavy
quarkonia~\cite{Krein:2010vp,Krein:2017usp,Zeminiani:2020aho,Cobos-Martinez:2020ynh}.
Since the $B_c$ meson is a pseudoscalar meson, we compare with the bottomonium $\eta_b$
and charmonium $\eta_c$ (upper panel), while for the $B_c^*$ meson, we compare with
those of the $\Upsilon$ and $J/\psi$ (lower panel).

For the comparison, we would like to emphasize that
we use the empirically extracted SU(4) sector coupling constants
for the charm sector ($\eta_c$ and $J/\psi$),
which would be more reasonable than using the empirically
extracted SU(5) sector coupling constant from the $\Gamma (\Upsilon \rightarrow e^+ e^-)$,
since the SU(5) flavor symmetry breaking is expected to be much larger than
that of the SU(4) based on the quark masses.

The value for the coupling constant of the vertex $J/\psi DD$
used in the calculation of $J/\psi$ mass shift, was obtained from the
experimental data for $\Gamma (J/\psi \rightarrow e^+ e^-)$ by the VMD
hypothesis
(note that the slight difference, 7.64 $\to$ 7.7 below, but the difference is negligible)

\begin{equation}
\label{gJPsi}
    g_{J/\psi DD} = \frac{g}{\sqrt{6}} \approx 7.7,
\end{equation}
where $g$ is the universal SU(4) coupling constant.

For the coupling constant $g_{\eta_{c} DD^{*}}$ used in the calculation of the $\eta_c$ mass shift,
we also adopt the SU(4) symmetry for the charm sector, which gives the relation

\begin{equation}
 g_{\eta_{c} DD^{*}} = g_{J/\psi DD} = \frac{g}{\sqrt{6}} \approx 7.7.
\end{equation}

A comprehensive list of the values used for the coupling constants is presented in
Table.~\ref{tblcpl45}.

\begin{table}
\caption{\label{tblcpl45} Coupling constant values in SU(4) and SU(5) symmetries.}
\begin{center}
\begin{tabular}{ll|r}
  \hline \hline
  & & \multicolumn{1}{c}{SU(4)} \\
\hline
&$g$ & 18.9 \\
& $g_{J/\psi DD}$ & 7.7 \\
& $g_{\eta_{c} DD^{*}}$ & 7.7 \\
  \hline \hline
  & & \multicolumn{1}{c}{SU(5)} \\
\hline
&$g$ & 33.4 \\
& $g_{\Upsilon BB}$ & 13.2 \\
& $g_{\eta_{b} BB^{*}}$ & 13.2 \\
& $g_{B_{c}B^{*}D}$ & 11.9 \\
& $g_{B^{*}_{s}BD}$ & 11.9 \\
\hline \hline
\end{tabular}
\end{center}
\end{table}
%

Although we make this comparison, we repeat 
that this is not done based on a rigorous SU(5) symmetry of the same footing.
Namely, the coupling constant $g$ is calculated
for the charm sector ($J/\psi$, $\eta_c$) based on the SU(4) symmetry, and
for the bottom sector ($\Upsilon$, $\eta_b$) and ($B_c$, $B^*_c$) based on the SU(5) symmetry.
This comparison would make a sense based on
the fact that SU(5) symmetry is much more broken
by the quark masses than that of SU(4).

Note that, although for the mass shift amount
$\Delta m_{\eta_c}$~\cite{Cobos-Martinez:2020ynh},
the cutoff mass values $\Lambda =\Lambda_D=\Lambda_{D^*} = 3000$ and $5000$ MeV
are missing, it is irrelevant
to see the mass shift range for the cutoff range between the 2000 MeV and 6000 MeV.
\begin{figure}[htb!]
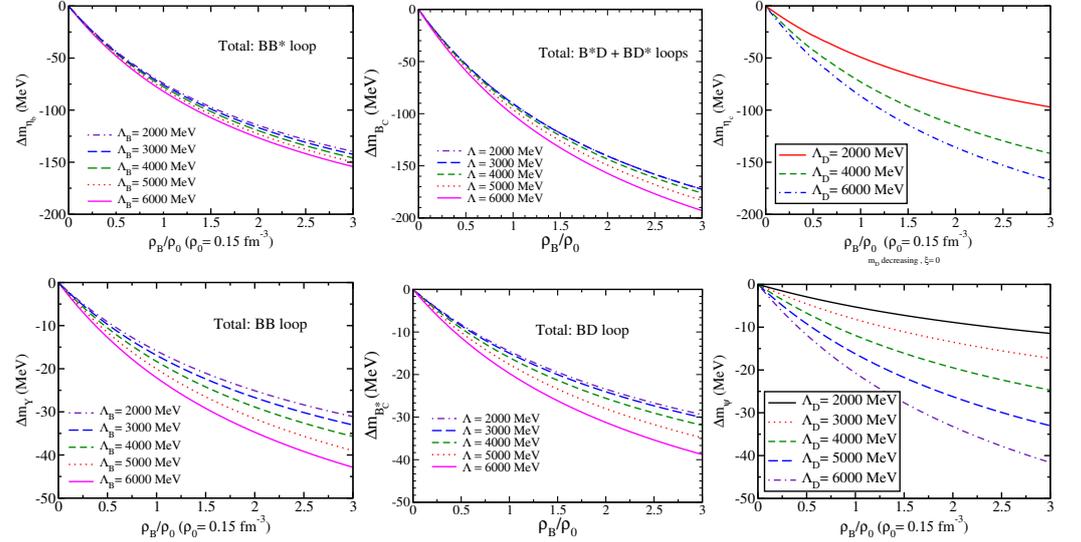
%
\centering
\includegraphics[width=4.5cm]{etab_BBs.eps}
\includegraphics[width=4.5cm]{Bc_totalpot.eps}
\includegraphics[width=4.5cm]{Dmetac_DDsOnly.eps}
\\
\includegraphics[width=4.5cm]{Upsilon_BB.eps}
\includegraphics[width=4.5cm]{Bcspot.eps}
\includegraphics[width=4.5cm]{DJPsi_in_medium_XiOff_OnlyDDbar.eps}
\caption{Comparison of the mass shift of $B_c$ with $\eta_b$ and $\eta_c$ (upper panel)
as well as of $B^*_c$ with $\Upsilon$ and $J/\psi$ (lower panel).}%
\label{comp}%
\end{figure}

In the study of the $\eta_c$ mass shift, only the $DD^*$ loop contribution was included,
and it corresponds to the mass shift value
$\Delta m_{\eta_c} (DD^*)$
at $\rho_0$ ranges  -49.2 to -86.5,
for the cutoff mass values $\Lambda_D = \Lambda_{D^*}$ of 2000, 4000 and 6000 MeV.
The estimated values for the $\eta_b$ mass shift 
$\Delta m_{\eta_b} (BB^*)$
at $\rho_0$ including only the $BB^*$ loop, ranges from -74.2 to -82.0 MeV,
where the same range of the cutoff mass value is applied for the present study.
The total $B^*D + BD^*$ loop
contributions for the  $B_c$ mass shift give more negative mass shift
than those of the $\eta_b$ and $\eta_c$.
This fact indicates that the $B_c$ mass shift value does not show the middle
range mass shift value between those of the $\eta_c$ and $\eta_b$,
may be different from one's naive expectation.

Next, we compare the mass shift behaviors of $\Upsilon, B_c^*$ and $J/\psi$ in Fig~\ref{comp}
(lower panel).
The $\Upsilon$ and $J/\psi$ mass shift values are calculated by taking respectively only the
(minimal) $BB$, and $DD$ loop contributions
corresponding to the present $B^*_c$ meson treatment with only the $BD$ loop.
The mass shift value $\Delta m_{\Upsilon} (BB)$ at $\rho_0$ ranges from -15.9 to -22.1
MeV, while $\Delta m_{J/\psi} (DD)$ at $\rho_0$ ranges from
-5.3 to -20.7, when the common range
of the $\Lambda$ (2000 to 6000 MeV) is used.
The corresponding $B^*_c$ mass shift value $\Delta m_{B_c^*} (BD)$
at $\rho_0$ ranges from -14.5 to - 19.7 MeV.
The $B^*_c$ meson in-medium mass shift value is less dependent on the cutoff
mass value than that of the $J/\psi$.
Although the mass shift behavior is depend on the cutoff mass value,
the global trend shown in the lower panel of Fig.~\ref{comp}
indicates that the $\Delta m_{B^*_c}$ is more or less
the middle of the corresponding $\Delta m_\Upsilon$ and $\Delta m_{J/\psi}$.

\section{\label{nuclearpotentials} Meson-nucleus potential}

\noindent

The baryon density dependence of the mass shift behaviors of the $\eta$, $\eta'$, $\phi$,
$\eta_c$, $J/\psi$, $\eta_b$, $\Upsilon, B_c$, and $B_c^*$ mesons in nuclear matter, shown in Figs.
\ref{fig:metaetaprime-nm}, \ref{fig:phi-nm} (left panel), \ref{fig:Dmetac},
\ref{fig:Dmjpsi}, \ref{fig:upsilon_ms}, \ref{fig:etab_ms}, \ref{totbc}, and \ref{totbcs}
indicate that the nuclear medium provides attraction to these mesons, and opens
the possibility for their binding to nuclei.

Therefore, we now consider the nuclear bound states for several of these
mesons, which we generally denote the meson as $h$, when the mesons
have been produced nearly at rest inside nucleus $A$, and study the
following nuclei in a wide range of masses, namely $^{4}$He, $^{12}$C,
$^{16}$O, $^{40}$Ca, $^{48}$Ca, $^{90}$Zr,  $^{197}$Au, and $^{208}$Pb. 

In a local density approximation, the meson $h$ potential within a nucleus
$A$ is given by
\begin{equation}
\label{eqn:VhA}
V_{hA}(r)= U_{hA}(r)-\frac{\mi}{2}W_{hA}(r),
\end{equation}
\noindent where $r$ is the distance from the center of the nucleus;
$U_{hA}(r)=\Delta m_{h}(\rho_{B}(r))$, with $\Delta m_{h}(\rho_B)$ 
the value of mass shift computed previously for meson $h$ as a function of nuclear
density $\rho_B$; and $\rho_{B}^{A}(r)$ is  the baryon density distribution in the nucleus $A$.
The imaginary part of the potential $W_{hA}(r)$, is related to  the absorption
of the meson $h$ in
the nuclear medium, is included only for the $\phi$, $\eta$, and $\eta'$ mesons in the present
 study. For the $\phi$ meson it is given by
$W_{\phi A}(r)=\Gamma_{\phi}(\rho_{B}^{A}(r))$ where $\Gamma_{\phi}(\rho_{B})$
is the $\phi$ decay width in a nucleus $A$, \eqn{eqn:phidecaywidth}.
For the $\eta$ and $\eta'$ mesons $W_{hA}(r)=-\gamma \Delta m_{h}(\rho_{B}
(r)) + \gamma_h^{\text{vac}}$. Here $\Gamma_{h}^{\text{vac}}$ the meson
decay width in vacuum ($\Gamma_{\eta}^{\text{vac}}=1.31$ keV and
$\Gamma_{\eta'}^{\text{vac}}=0.188$ MeV~\cite{Workman:2022ynf}), and
$\gamma$ is a phenomenological parameter used to simulate the strength of
the absorption of the meson in the nuclear medium. The values of the $\gamma$ parameter used below cover the estimated widths of the $\eta$ and $\eta'$ mesons in the nuclear medium~\cite{Cobos-Martinez:2023hbp}.
The nuclear density distributions  $\rho_{B}^{A}(r)$ for the nuclei listed
above are calculated using the QMC model~\cite{Saito:1996sf}, except for
$^{4}$He, which we take from Ref.~\cite{Saito:1997ae}.
Before proceeding, a comment on the use of the local density approximation might be useful, in
 particular for $^{4}$He nuclei. At the position $r$ inside nucleus $A$, the nuclear density is 
 $\rho_A(r)$, and the potentials (effective masses) are taken from the  uniform (constant) nuclear
  density calculation in nuclear matter. For small nuclei, such as $^{4}$He, this might appear
   problematic at first sight since for such nucleus th $r$-dependence of the nuclear density is
    expected to be relatively rapid (strong). Because the $r$-dependence of the nuclear density can 
    be faster than for larger nuclei, depending on the interval value $\Delta r$, to use the local 
    density approximation, it might not be good enough to assume the uniform nuclear density
     between the interval $\Delta r$. However, our calculation  uses $\Delta r=0.04$ fm with the
      interpolation, and we expect the local density approximation even for the  $^{4}$He nucleus is
       sufficiently good. 
     For the $^{4}$He nucleus, the nuclear density change within the interval 0.04 fm  is very small
 and thus can be regarded as a constant density. 

In the following figures we present the meson-nucleus potentials for some
selected nuclei computed using~\eqn{eqn:VhA}.

\begin{figure}[ht]
\centering
\scalebox{1.0}{
\begin{tabular}{cc}
    \includegraphics[scale=0.27,angle=-90]{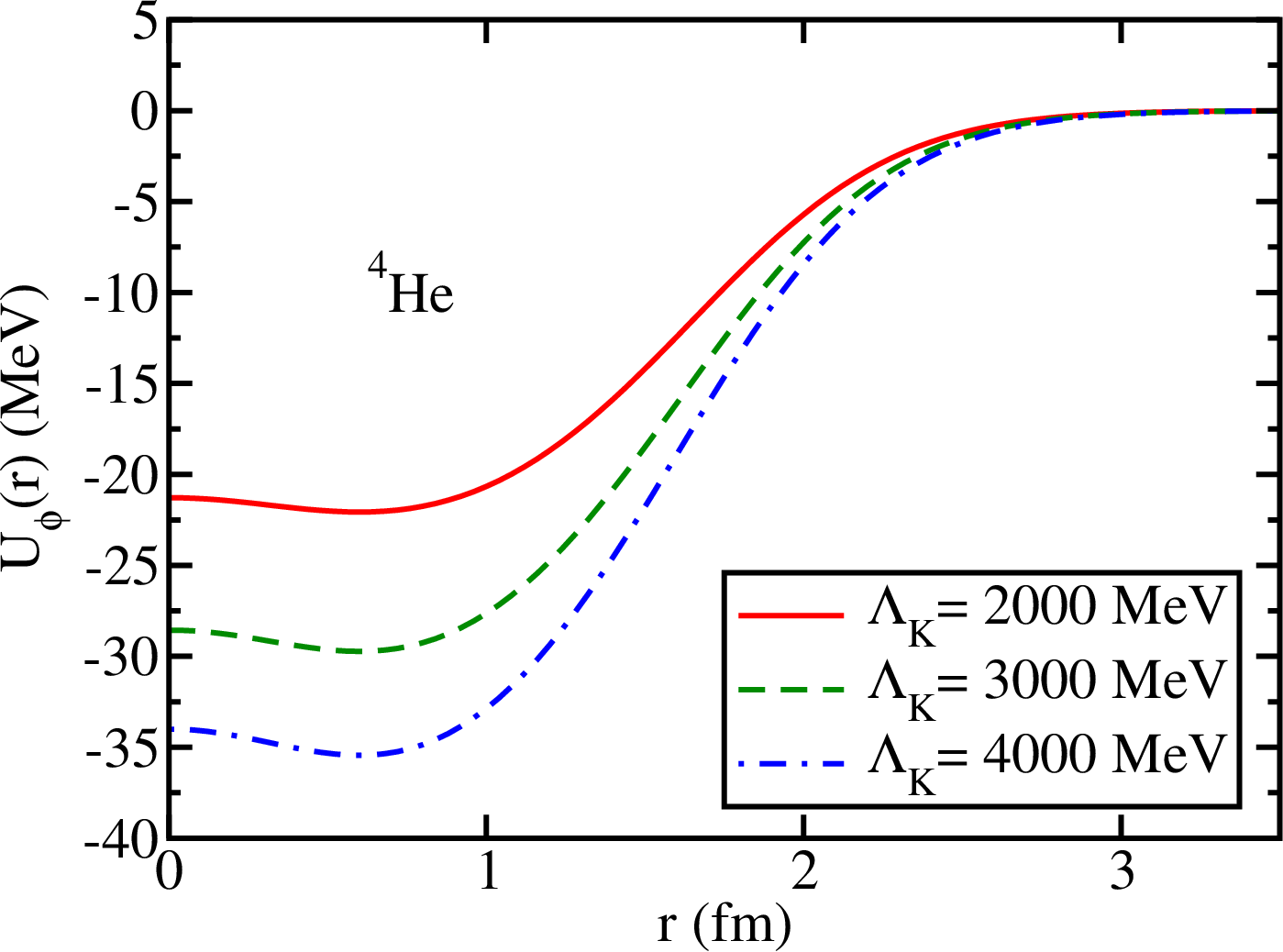} &  
    \includegraphics[scale=0.27,angle=-90]{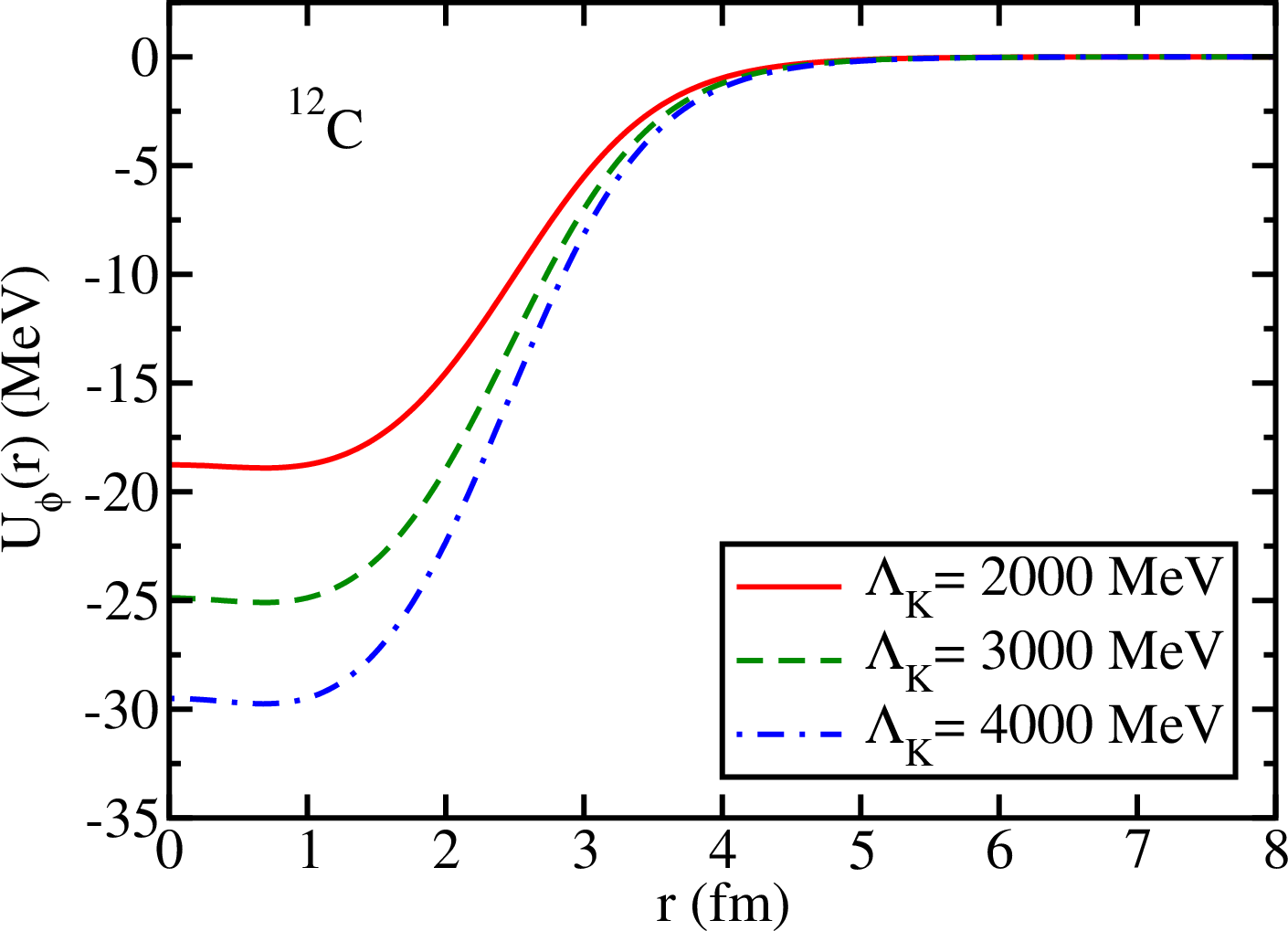} \\
    \includegraphics[scale=0.27,angle=-90]{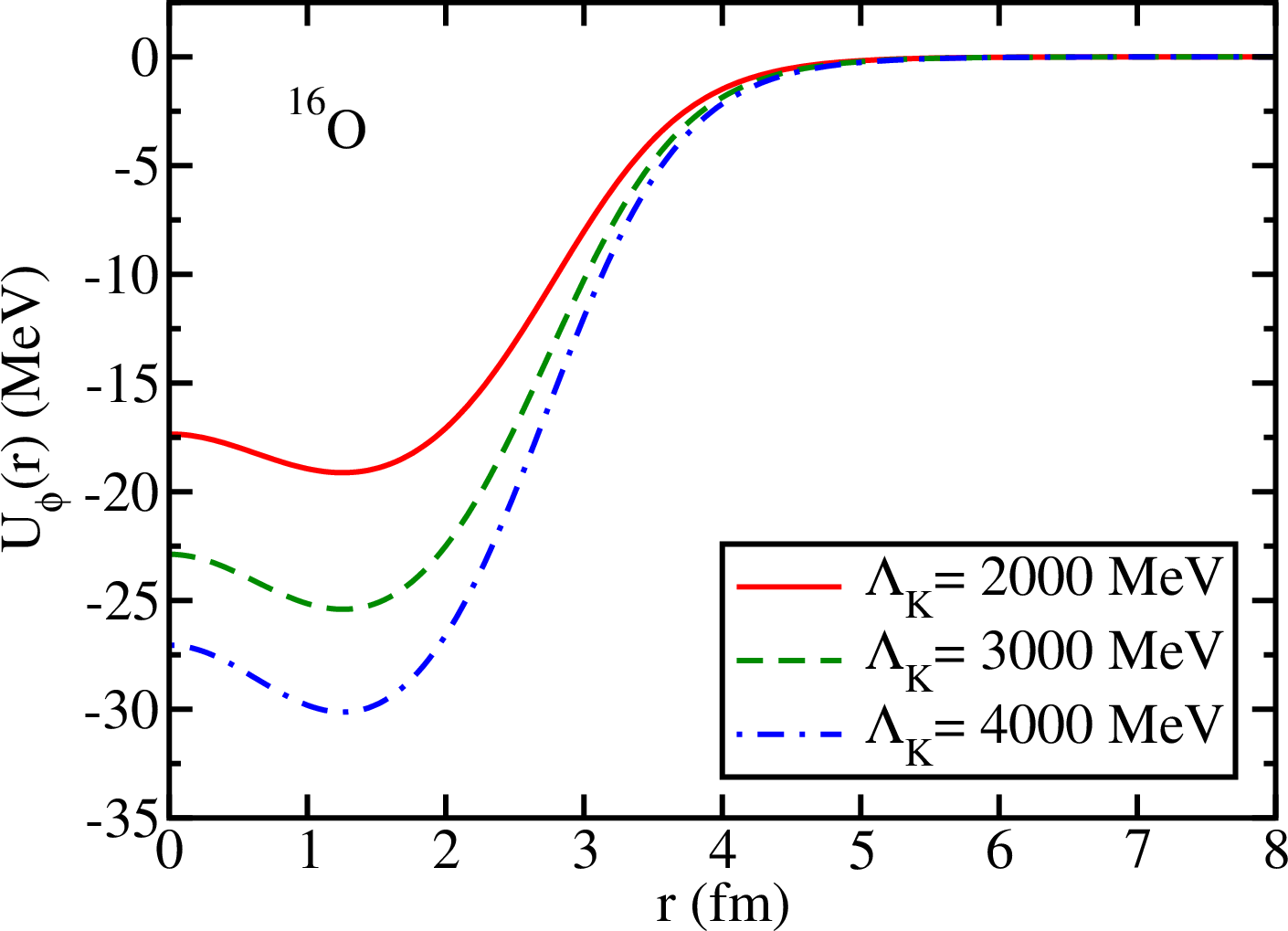}  &
    \includegraphics[scale=0.27,angle=-90]{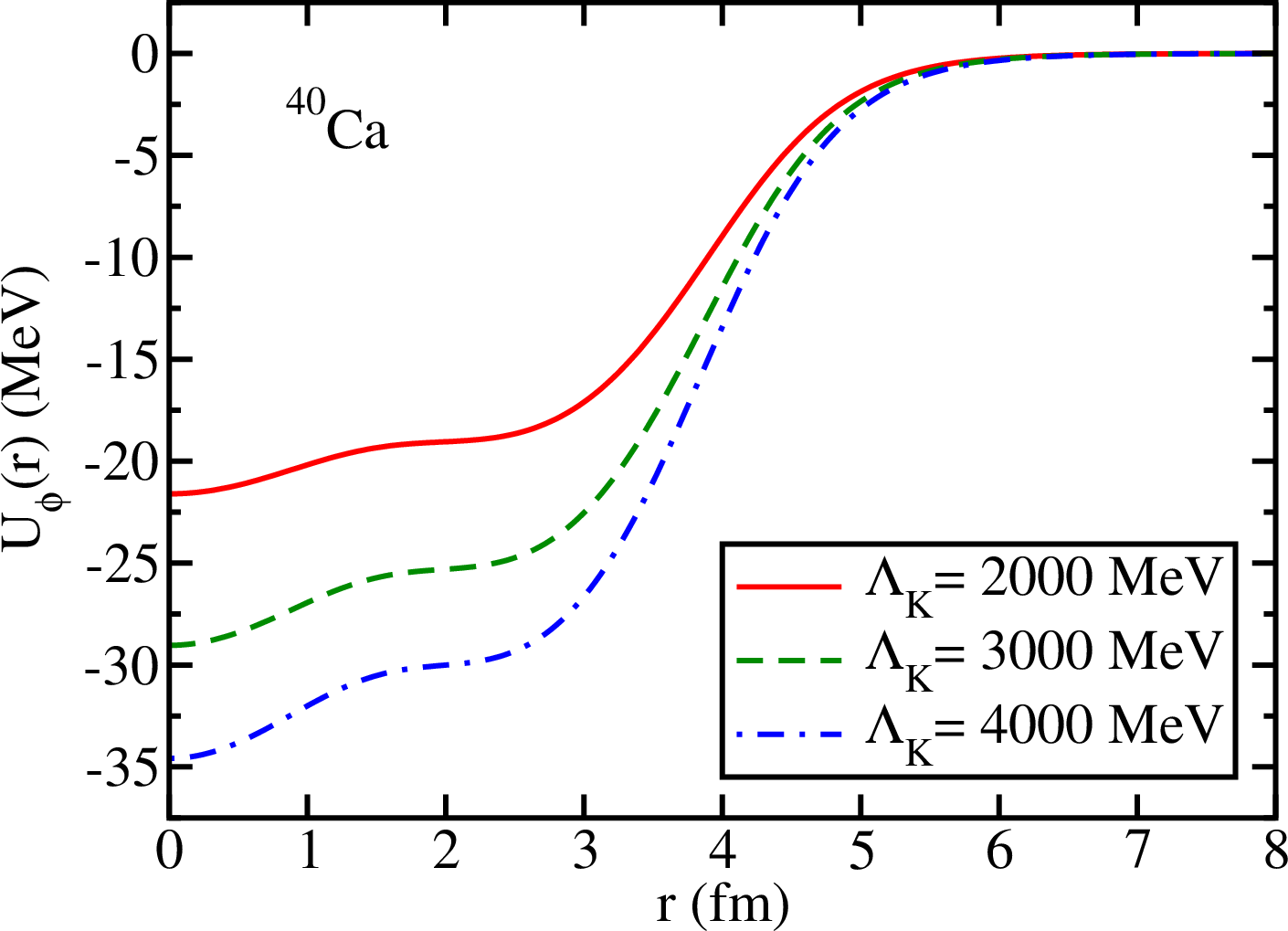} \\
    \includegraphics[scale=0.27,angle=-90]{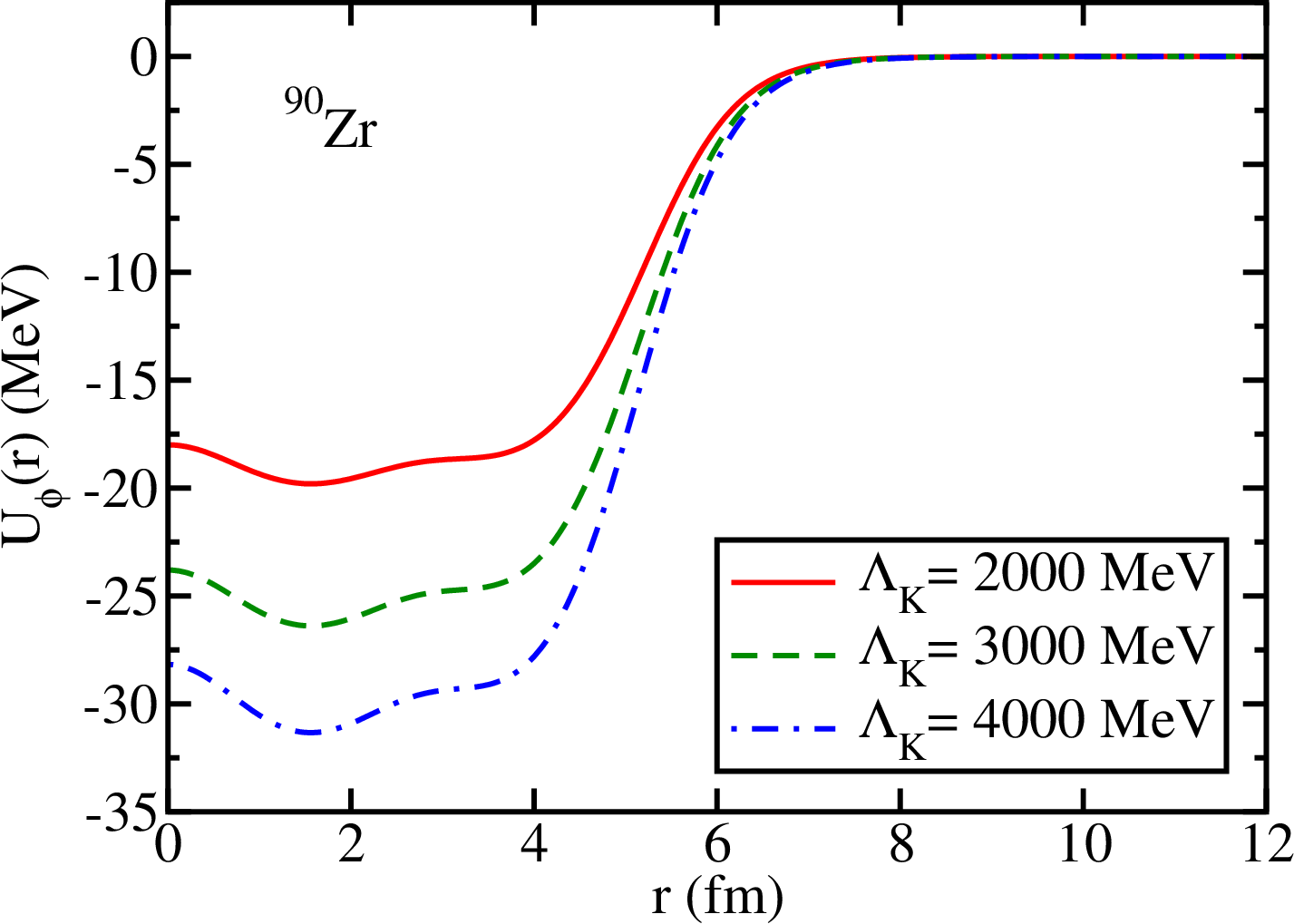} &  
    \includegraphics[scale=0.27,angle=-90]{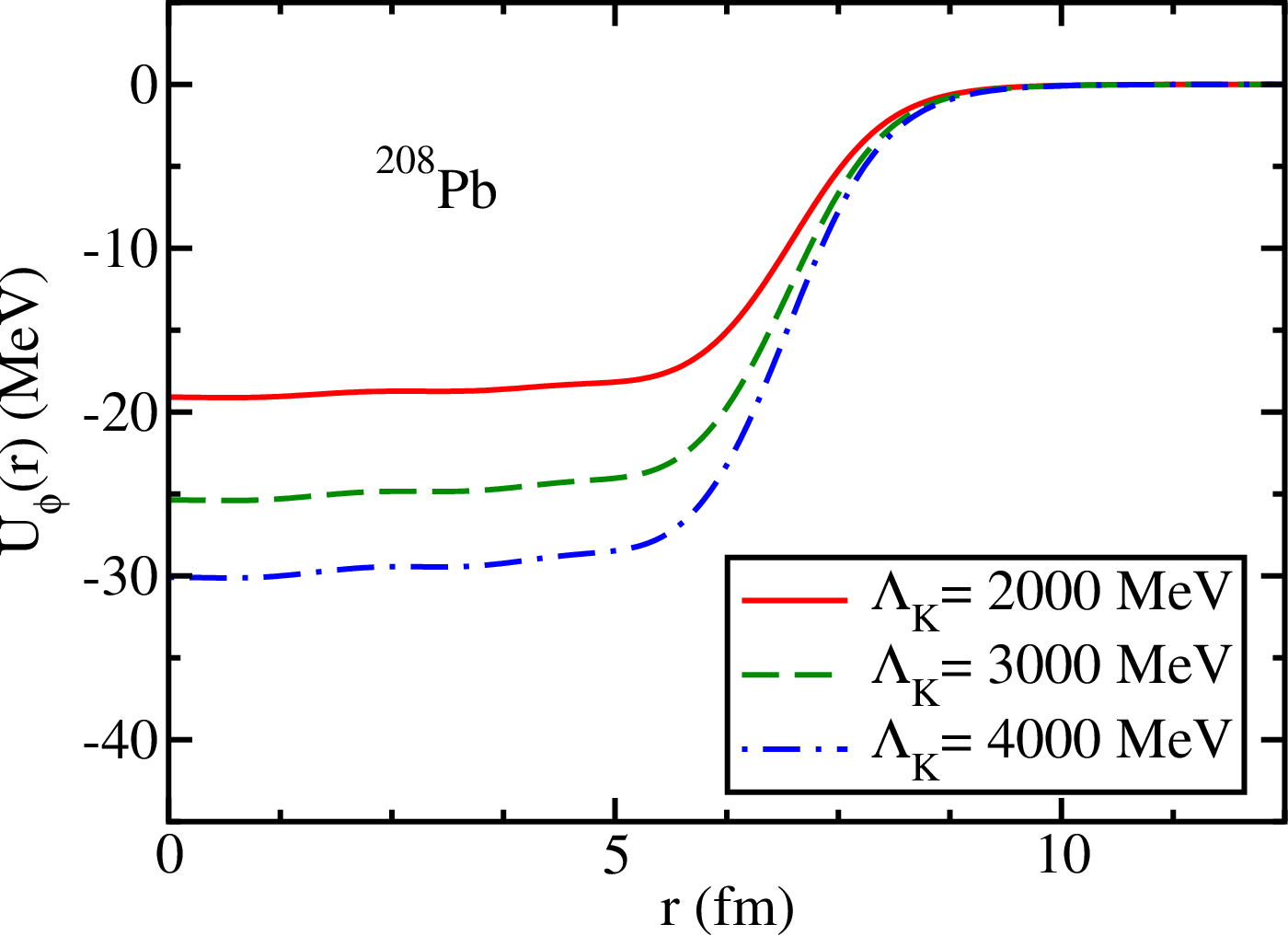}
  \end{tabular}
}
  \caption{\label{fig:ReVphiA} Real [$U_{\phi}(r)](r)$] part of the
    $\phi$-meson-nucleus potentials in some nuclei selected,  for three values of the cutoff
parameter $\Lambda_{K}$.}
\end{figure}
\begin{figure}[ht]
\centering
\scalebox{1.0}{
\begin{tabular}{cc}
    \includegraphics[scale=0.27,angle=-90]{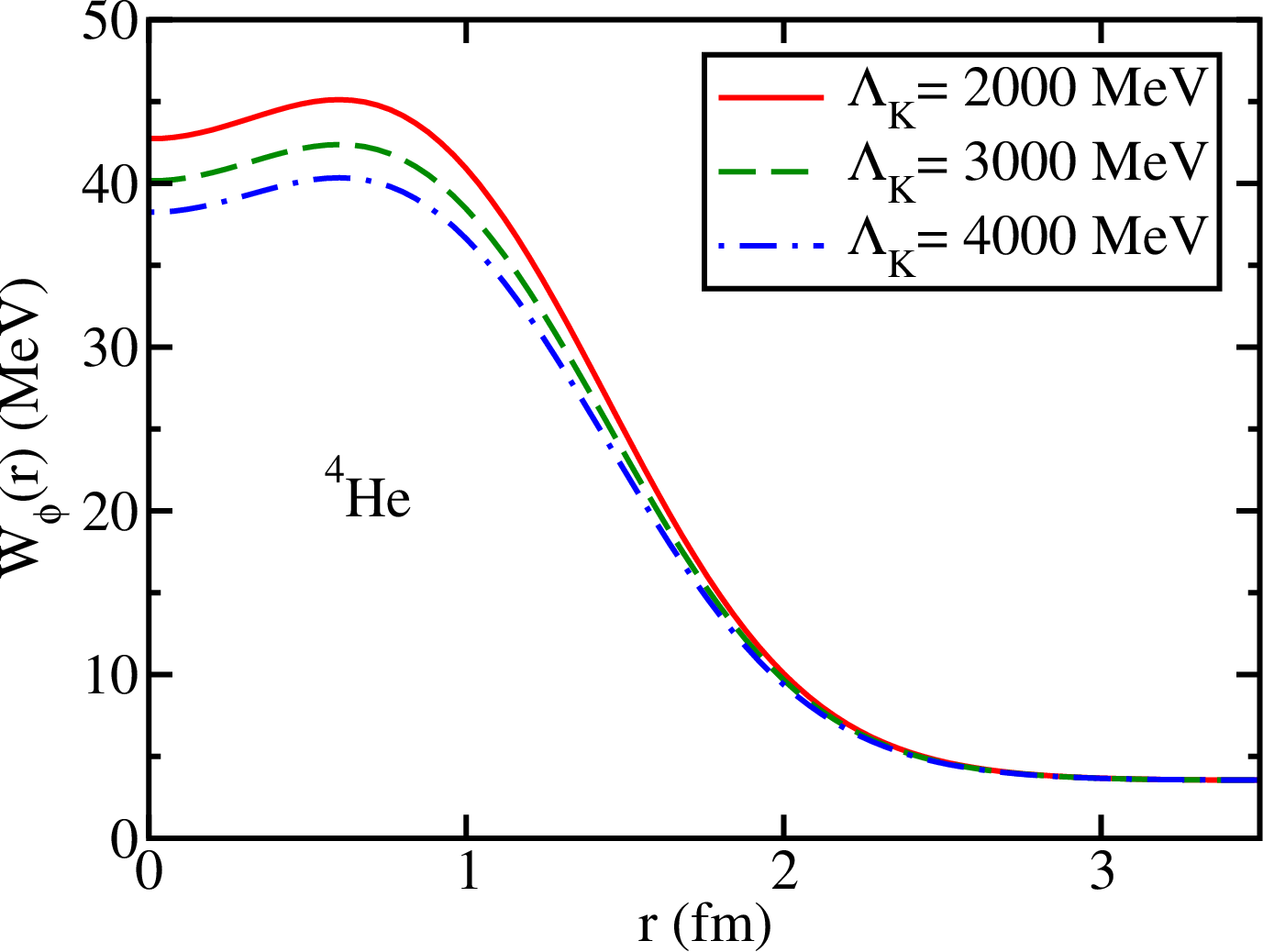} &  
    \includegraphics[scale=0.27,angle=-90]{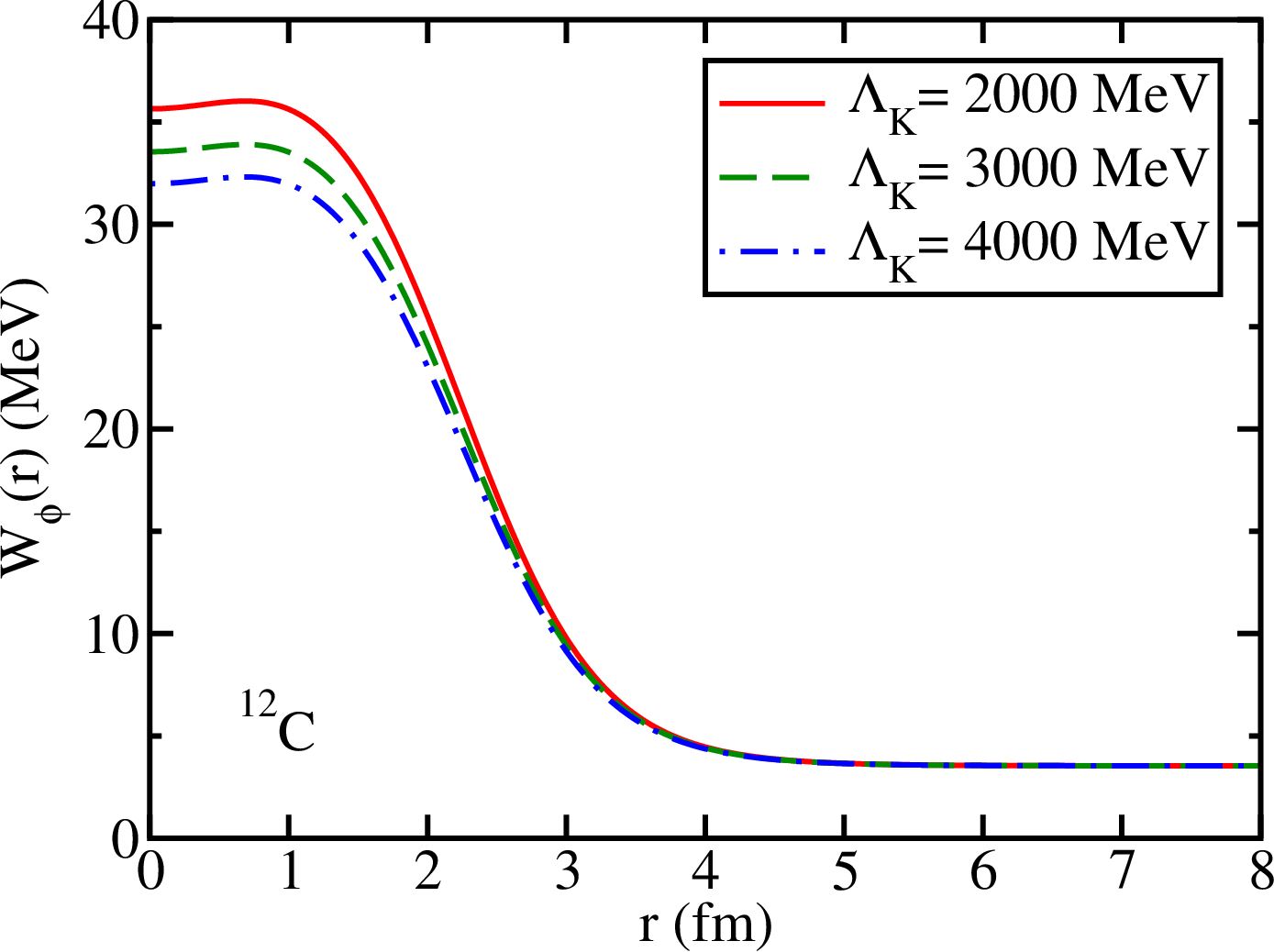} \\
    \includegraphics[scale=0.27,angle=-90]{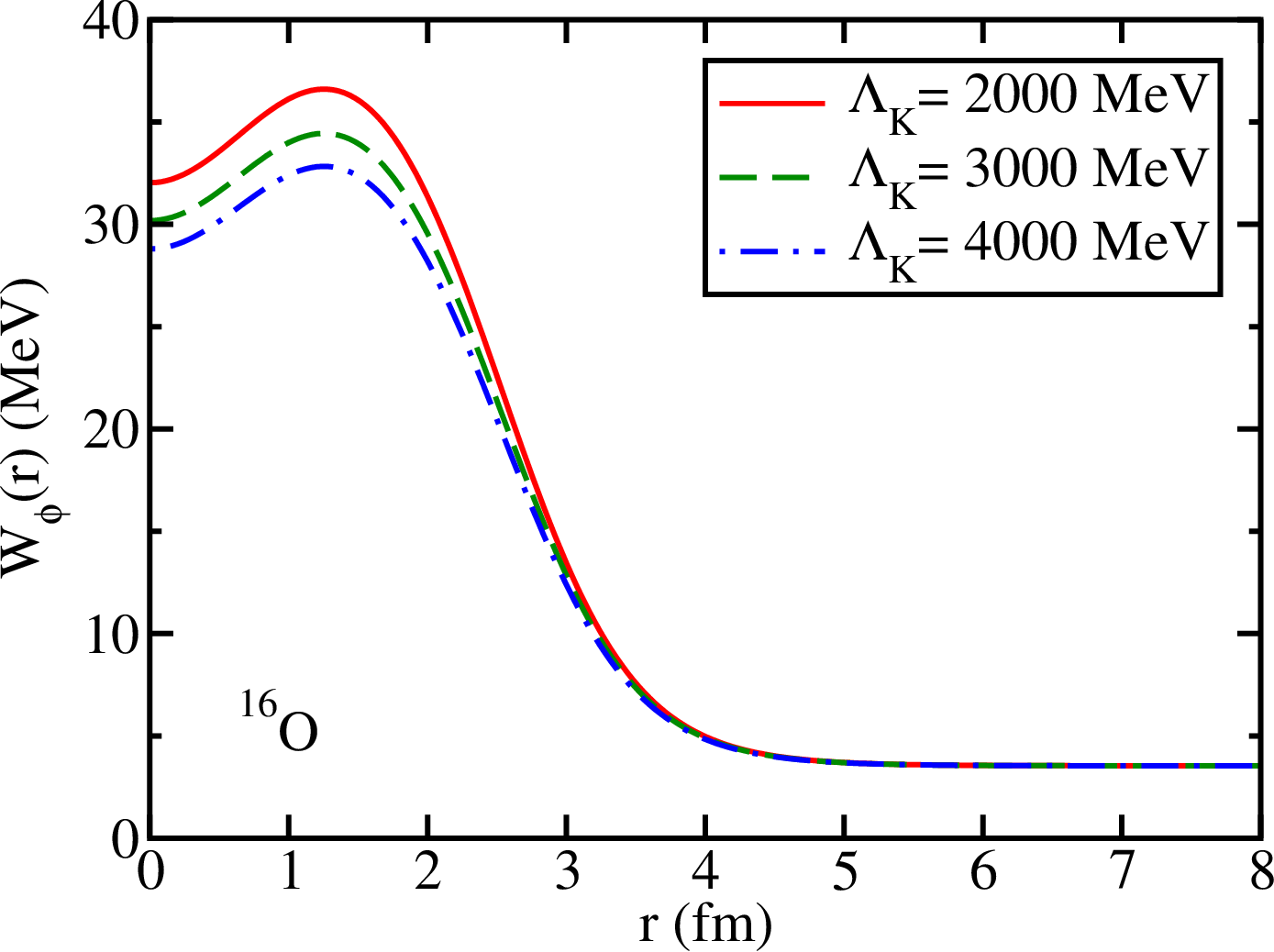}  &
    \includegraphics[scale=0.27,angle=-90]{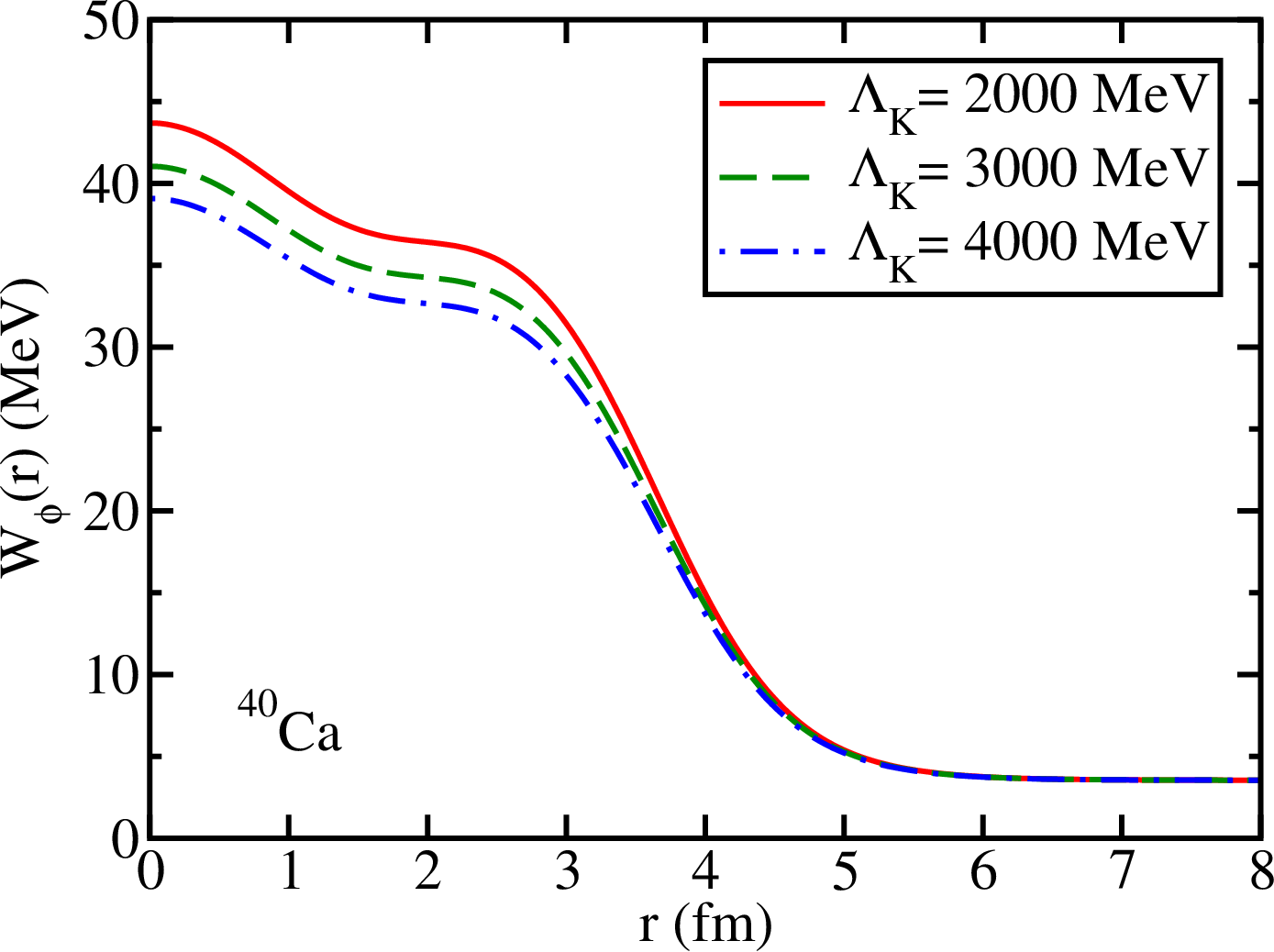} \\
    \includegraphics[scale=0.27,angle=-90]{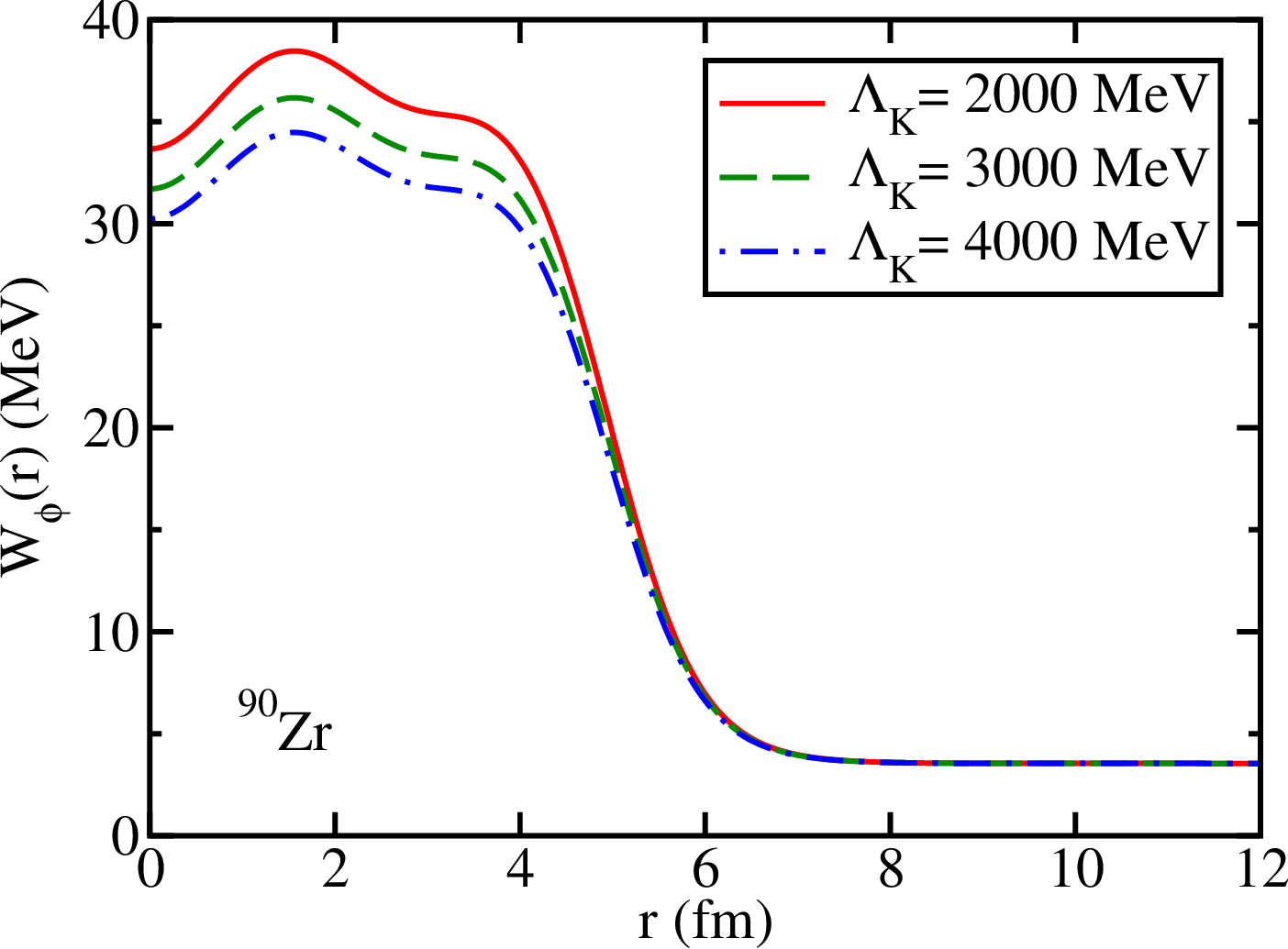} &  
    \includegraphics[scale=0.27,angle=-90]{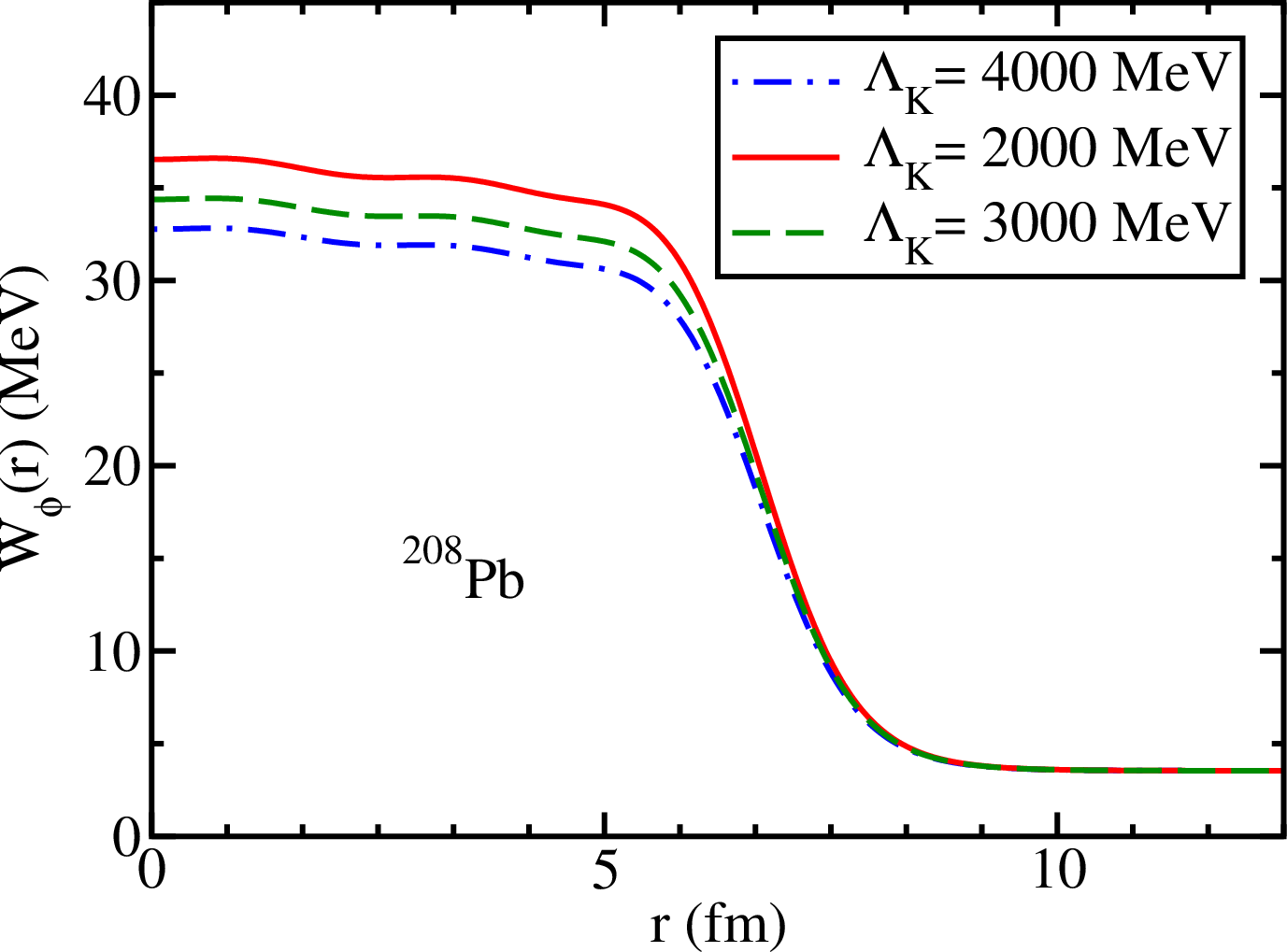}
  \end{tabular}
}
  \caption{\label{fig:ImVphiA} Imaginary  [$W_{\phi}(r)$] part of the
    $\phi$-meson-nucleus  potentials in some nuclei,  for three values of the cutoff parameter
    $\Lambda_{K}$.}
\end{figure}

In Figs.~\ref{fig:ReVphiA} and~\ref{fig:ImVphiA} we present the
$\phi$-meson potentials calculated for some nuclei, for three values
of the cutoff parameter $\Lambda_{K}$, $2000, 4000$ and $6000$ MeV. 
One can see that the depth of the real part of the potential, $U_\phi(r)$,
is sensitive to the cutoff parameter, from -20 MeV to -35
MeV for $^{4}$He and from -20 MeV to -30 MeV for $^{208}$Pb~\cite{Cobos-Martinez:2017woo}.
In addition, one can see that the imaginary part does not vary much with
$\Lambda_{K}$. Furthermore, note the imaginary part of the potential
is repulsive. This observation may well have consequences for the 
feasibility of experimental observation of the expected bound states~\cite{Cobos-Martinez:2017woo}.

\begin{figure}[ht]
\centering
\scalebox{0.9}{
  \begin{tabular}{cc}
\includegraphics[scale=0.25]{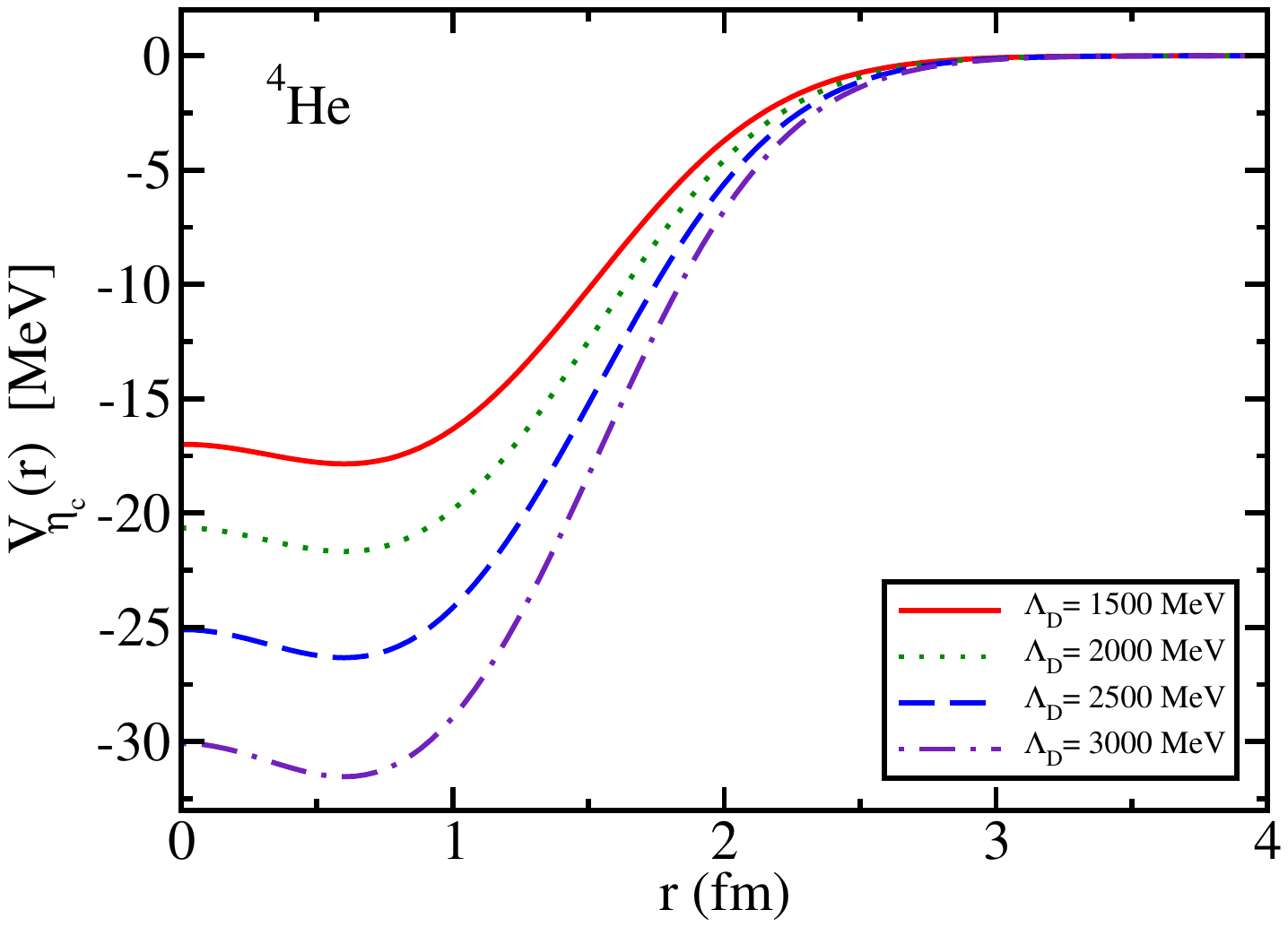} & 
\includegraphics[scale=0.25]{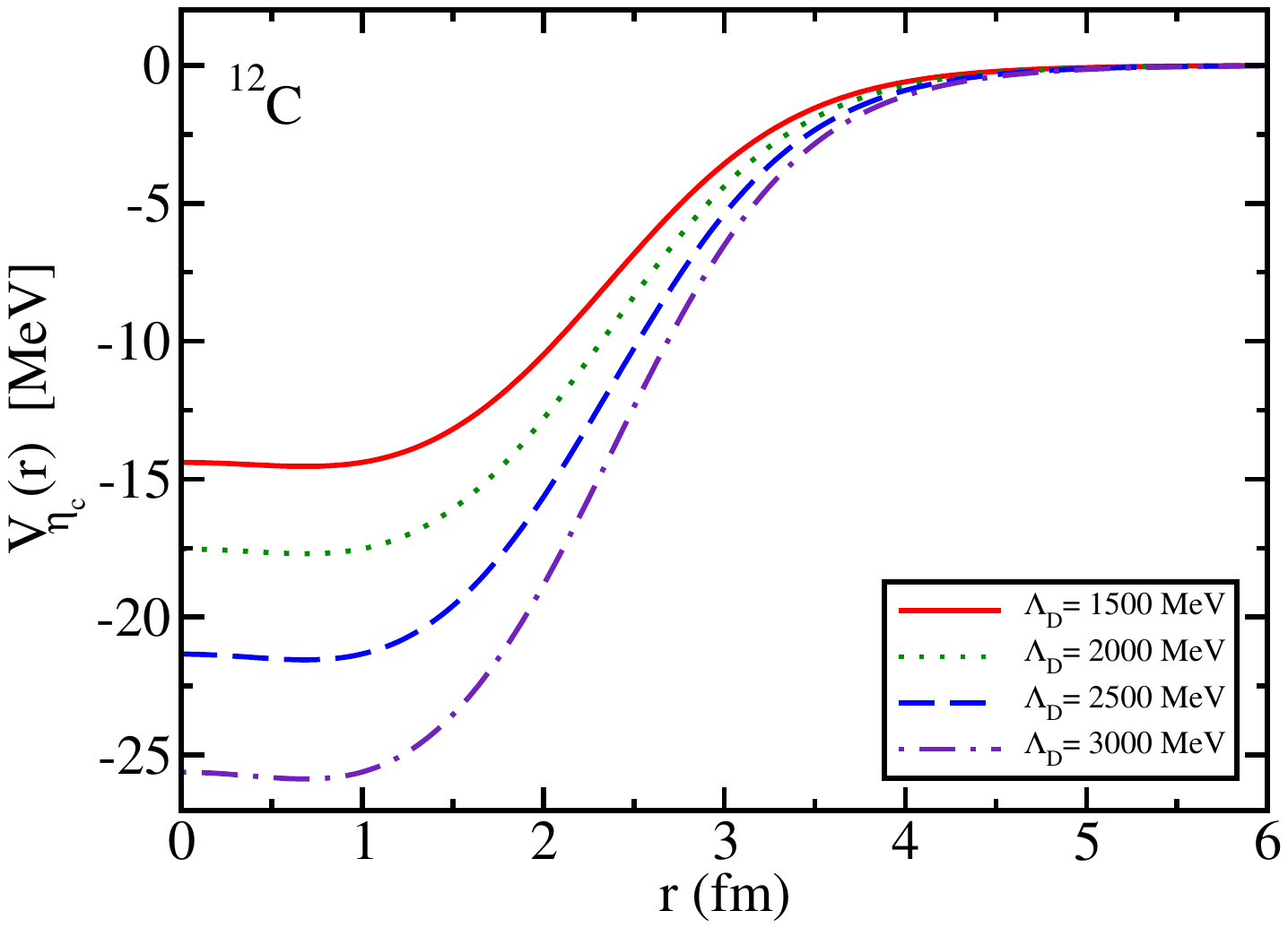} \\
\includegraphics[scale=0.25]{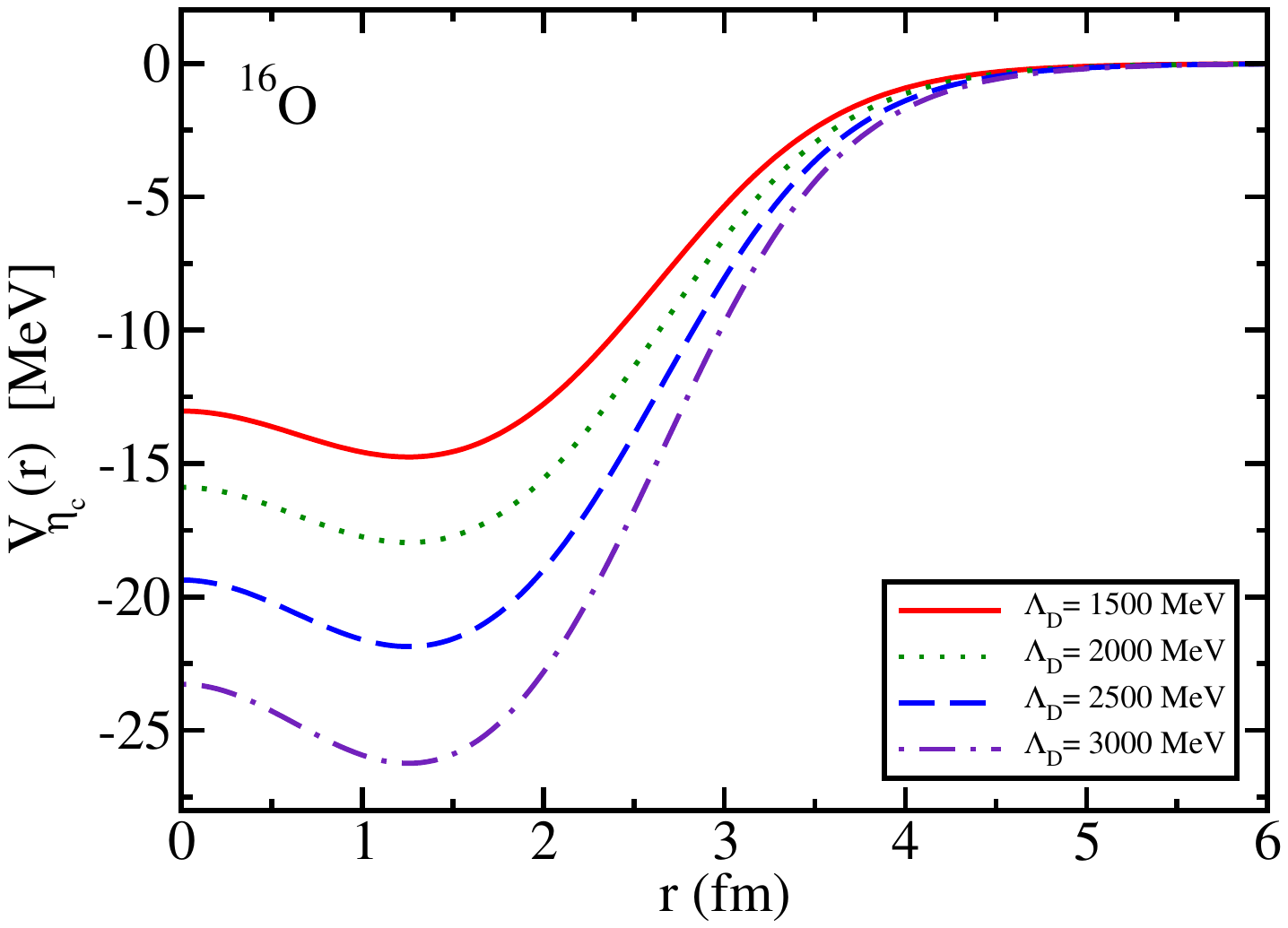} &
\includegraphics[scale=0.25]{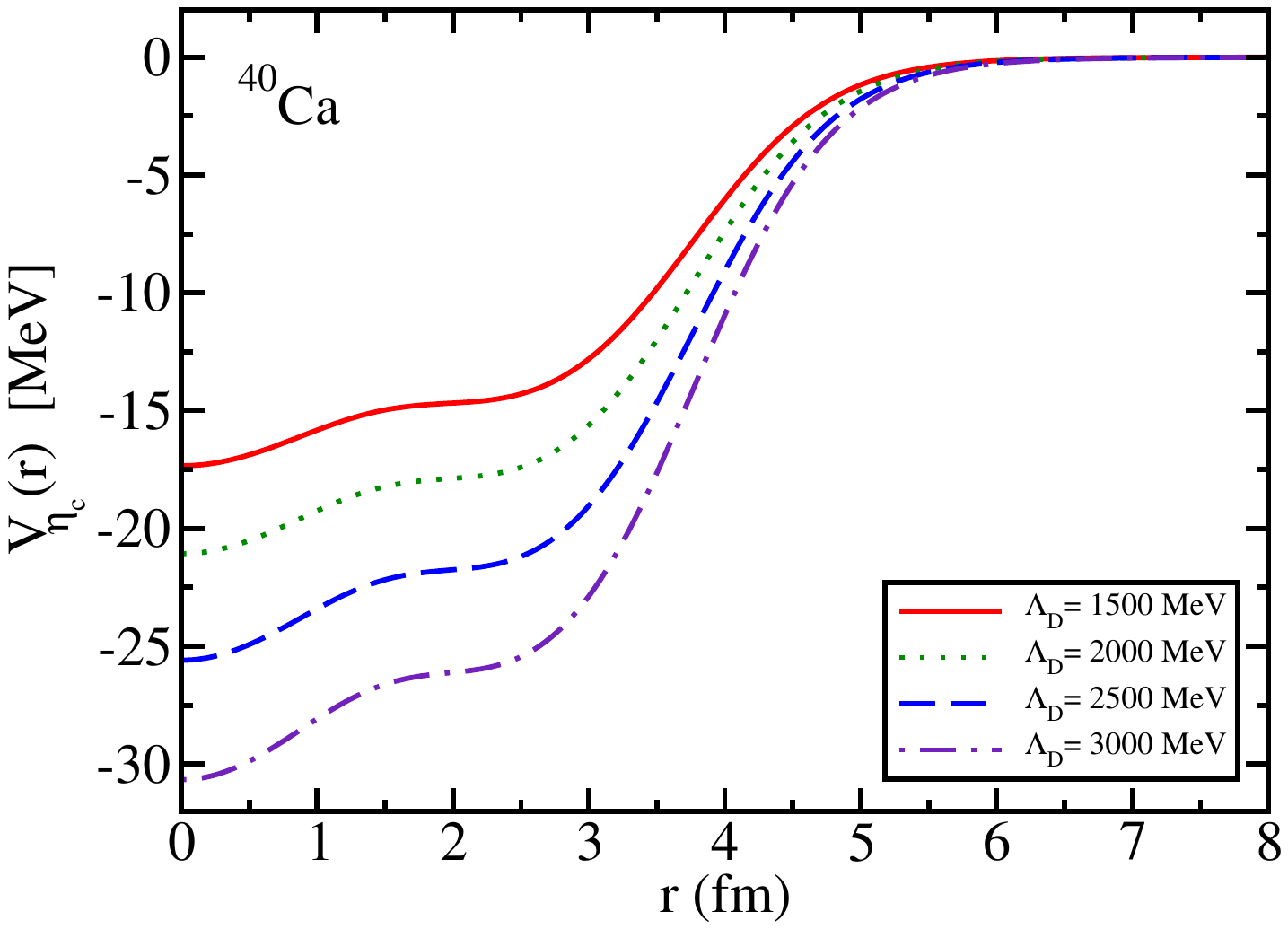} \\ 
\includegraphics[scale=0.25]{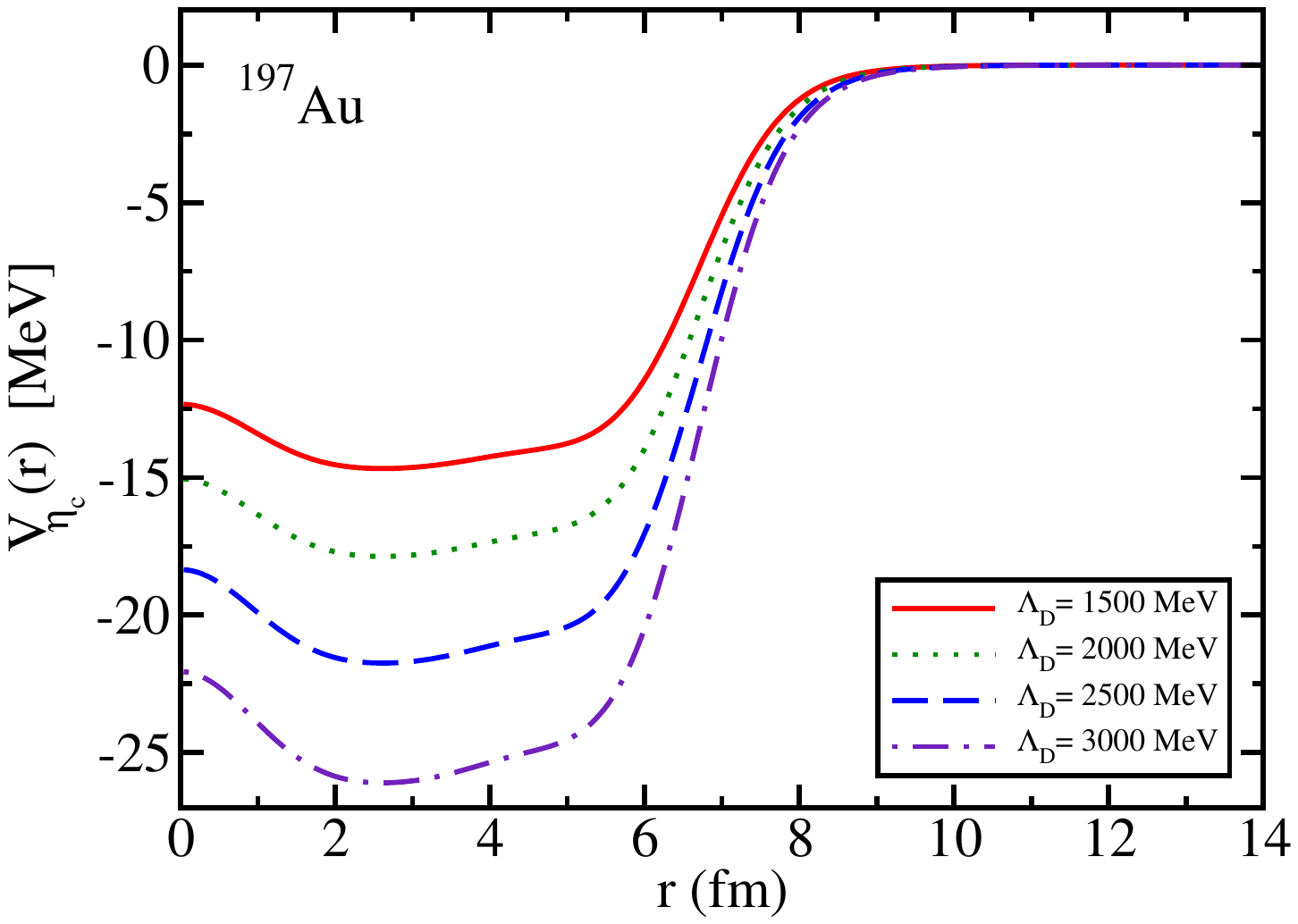} & 
\includegraphics[scale=0.25]{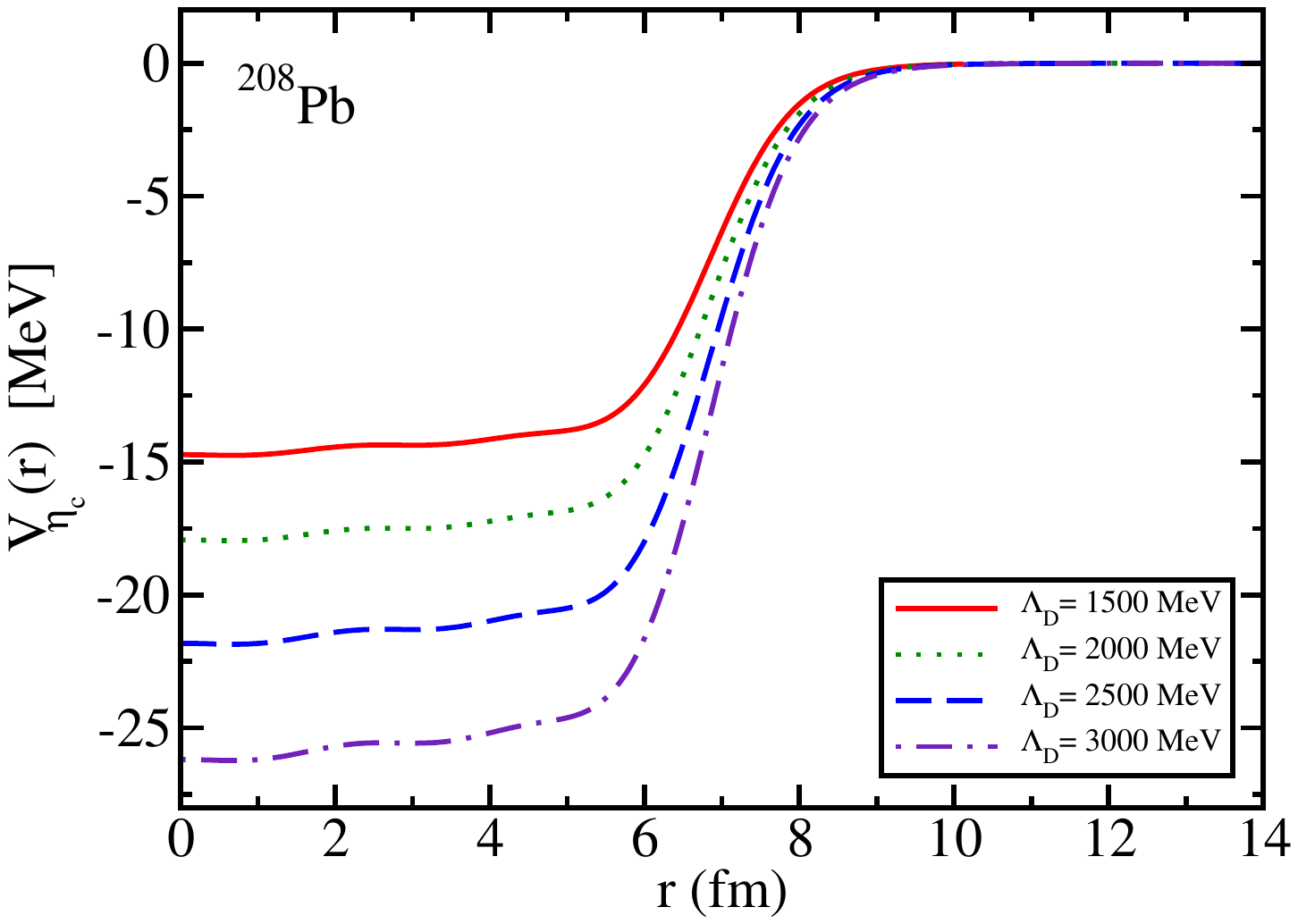} 
  \end{tabular}
  }
  \caption{\label{fig:VetacA}
    $\eta_c$-nucleus potentials for various nuclei and values of the
    cutoff parameter $\Lambda_{D}$~\cite{Cobos-Martinez:2020ynh}.
    Note that the potentials are calculated with the SU(4) breaking parameter,
    $0.6/\sqrt{2}$ for the coupling constant, as explained in Sec.~\ref{etacpsi}.
    }
\end{figure}

\begin{figure}[ht]
\centering
\scalebox{0.9}{
  \begin{tabular}{cc}
\includegraphics[scale=0.25]{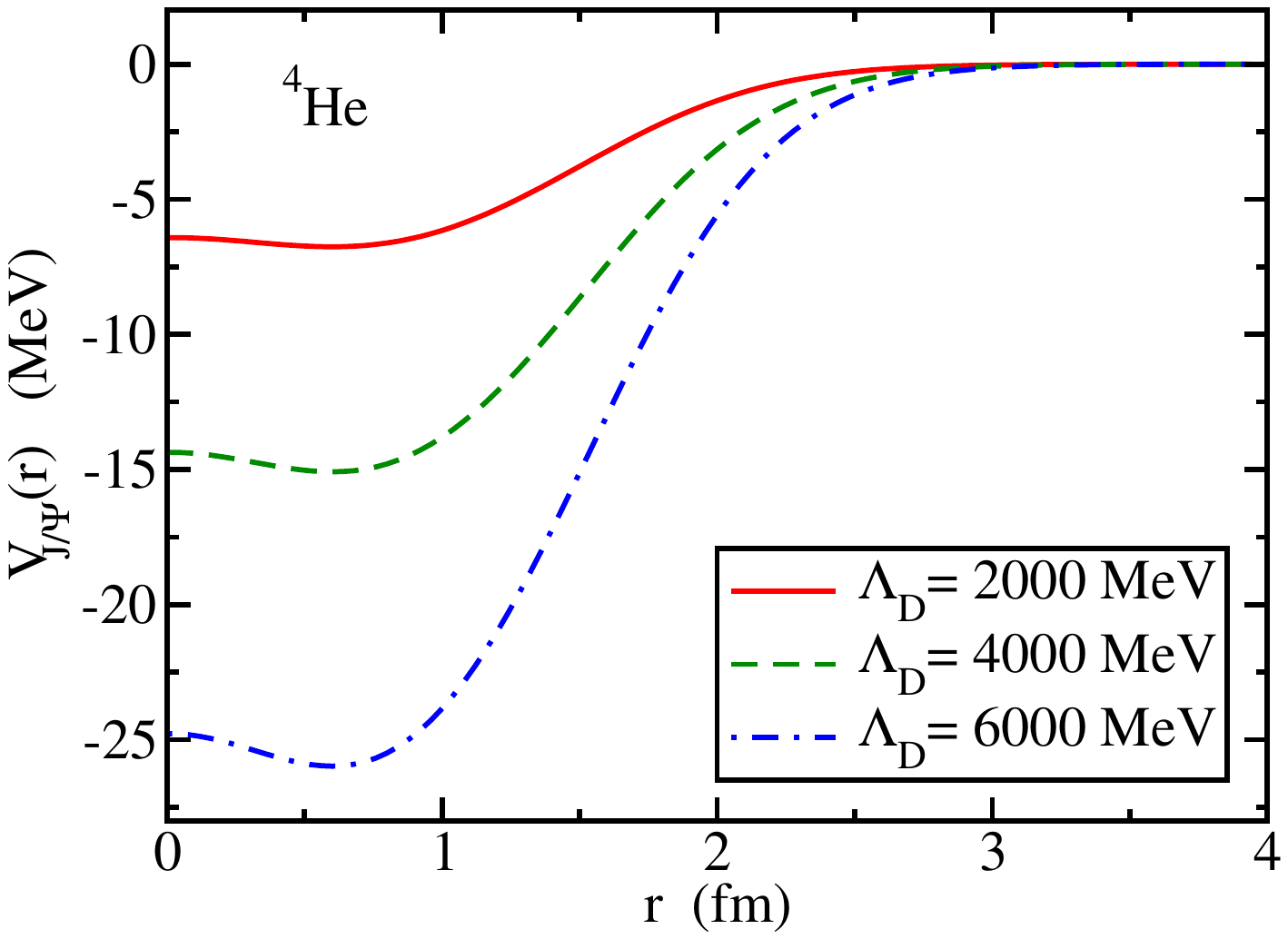} &  
\includegraphics[scale=0.25]{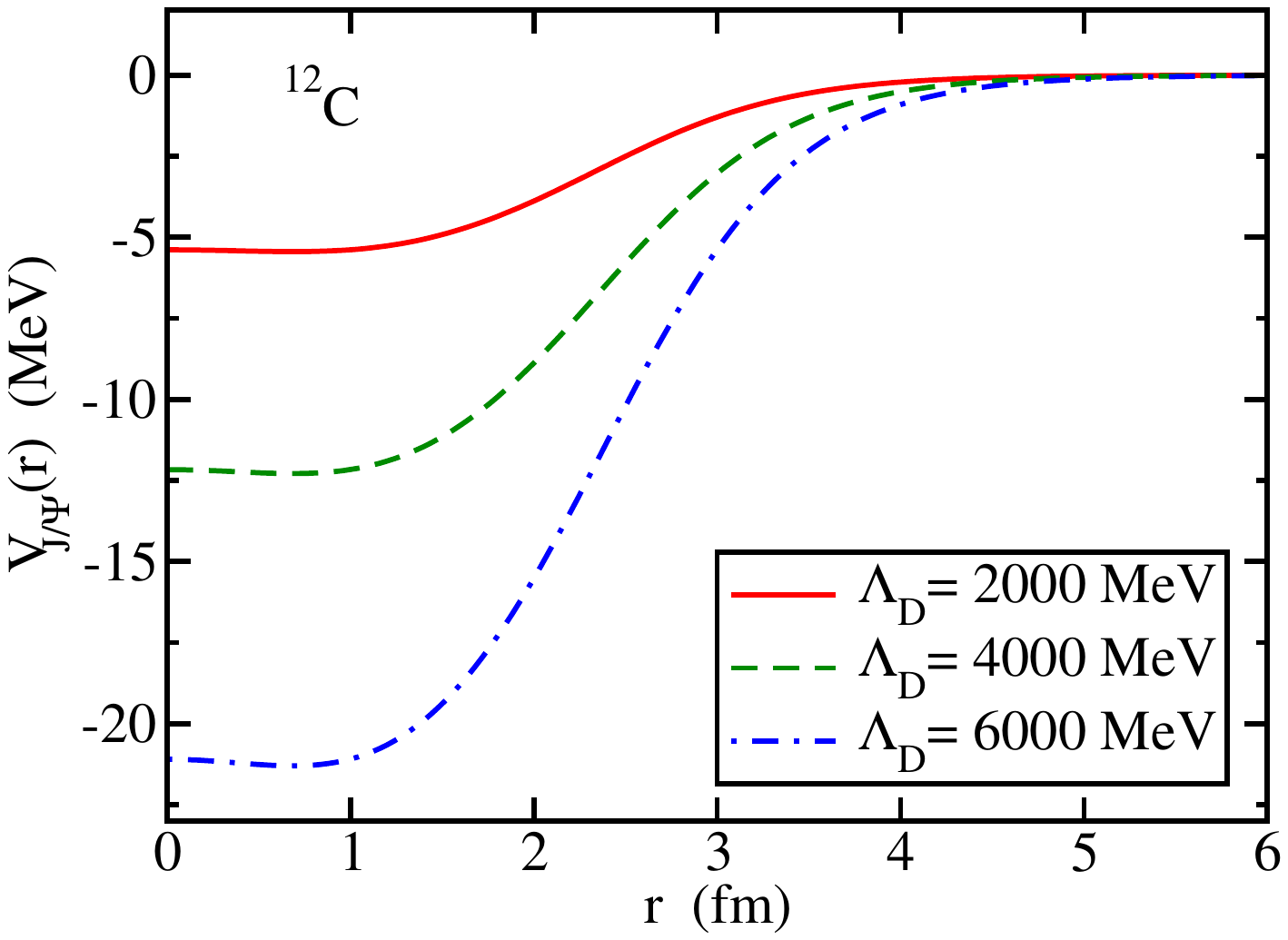} \\
\includegraphics[scale=0.25]{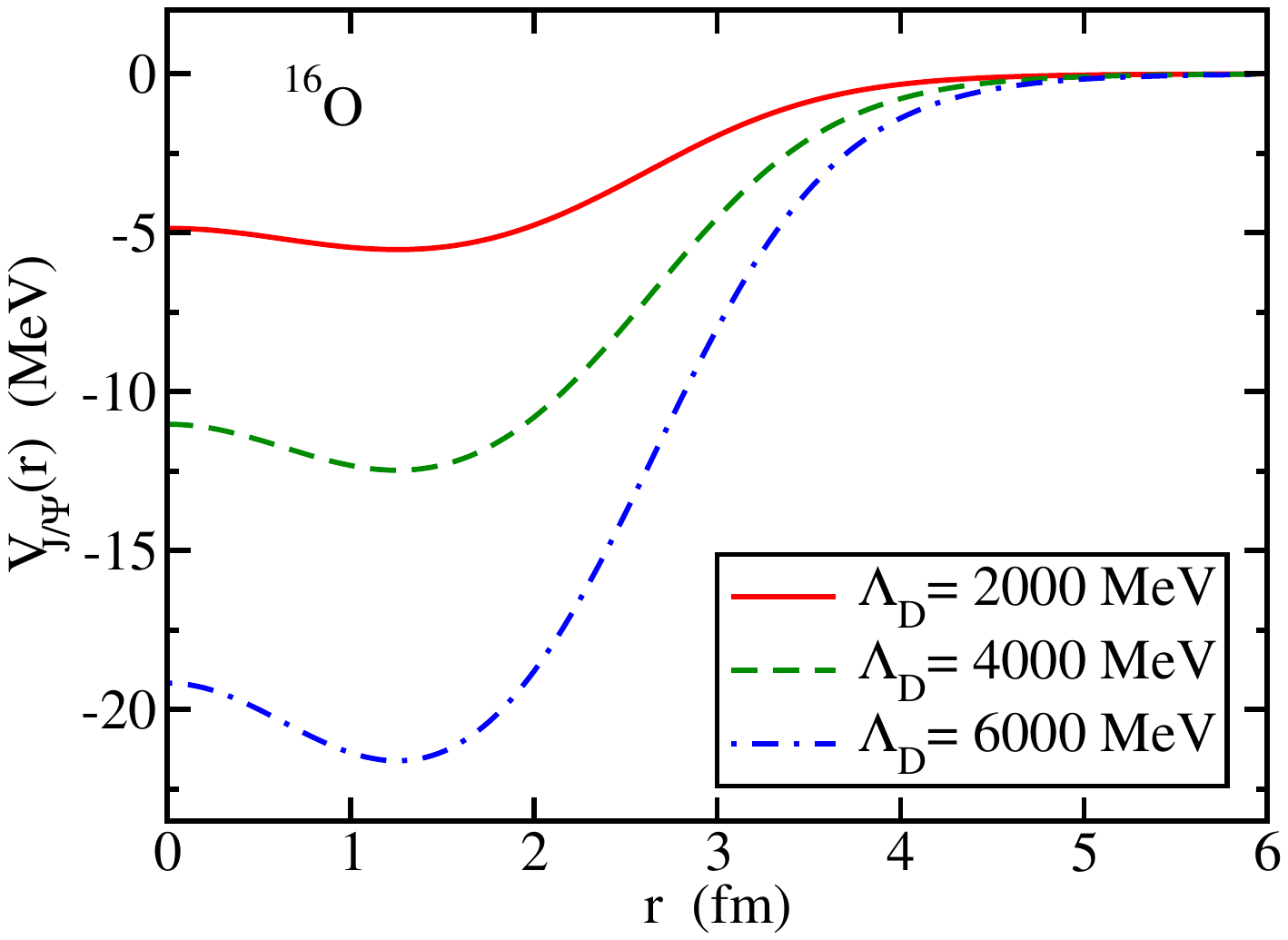} & 
\includegraphics[scale=0.25]{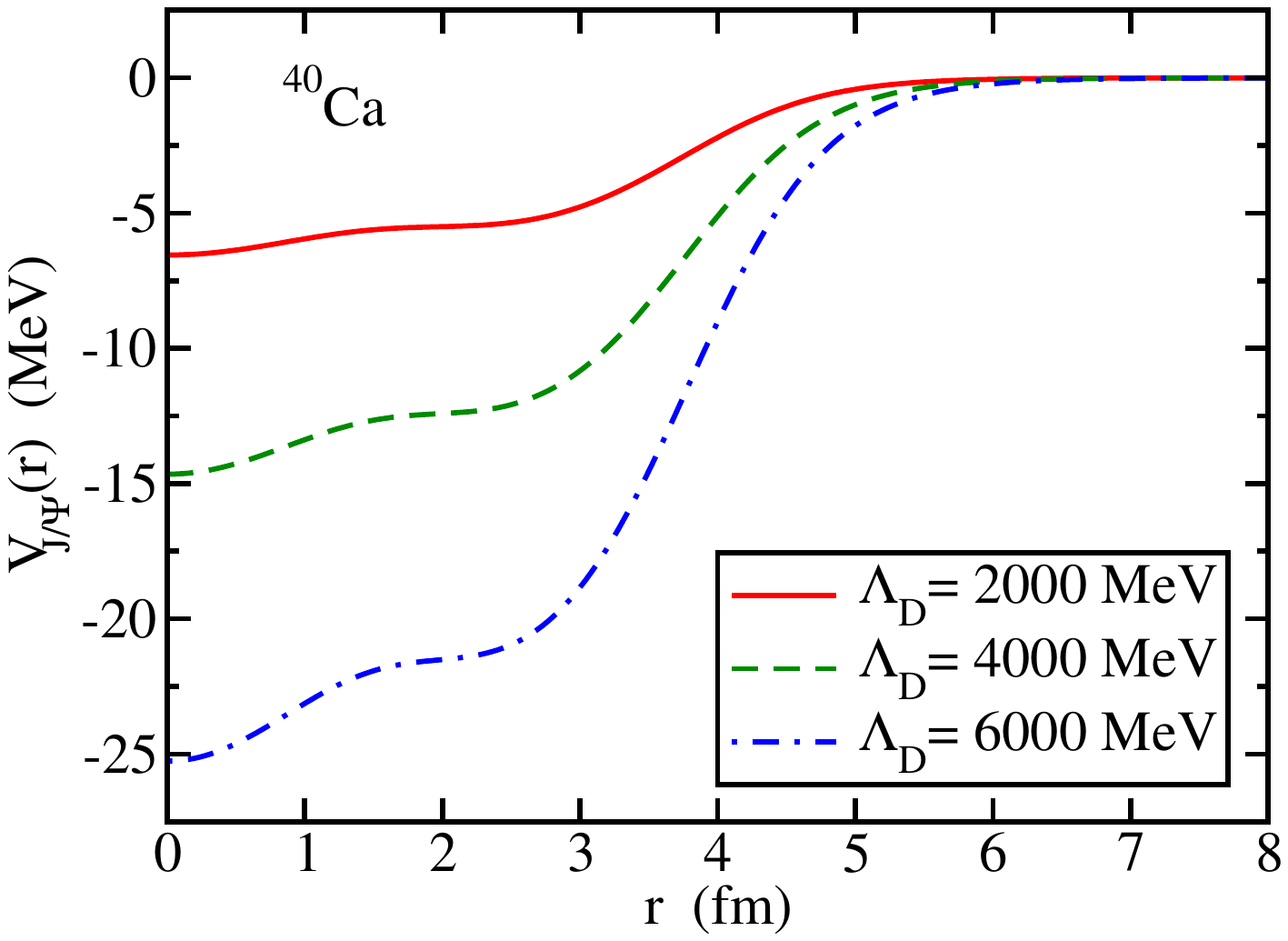} \\
\includegraphics[scale=0.25]{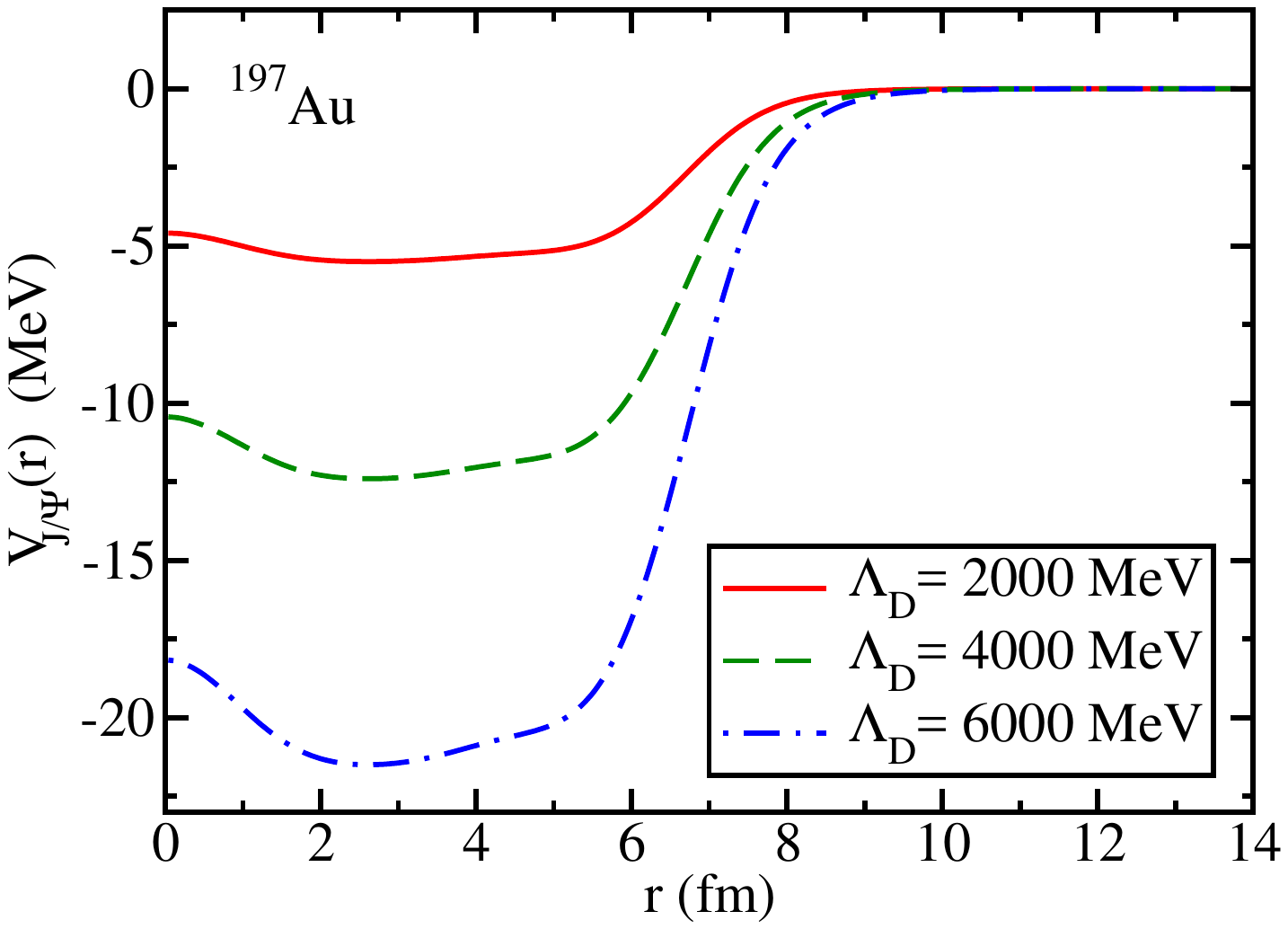} & 
\includegraphics[scale=0.25]{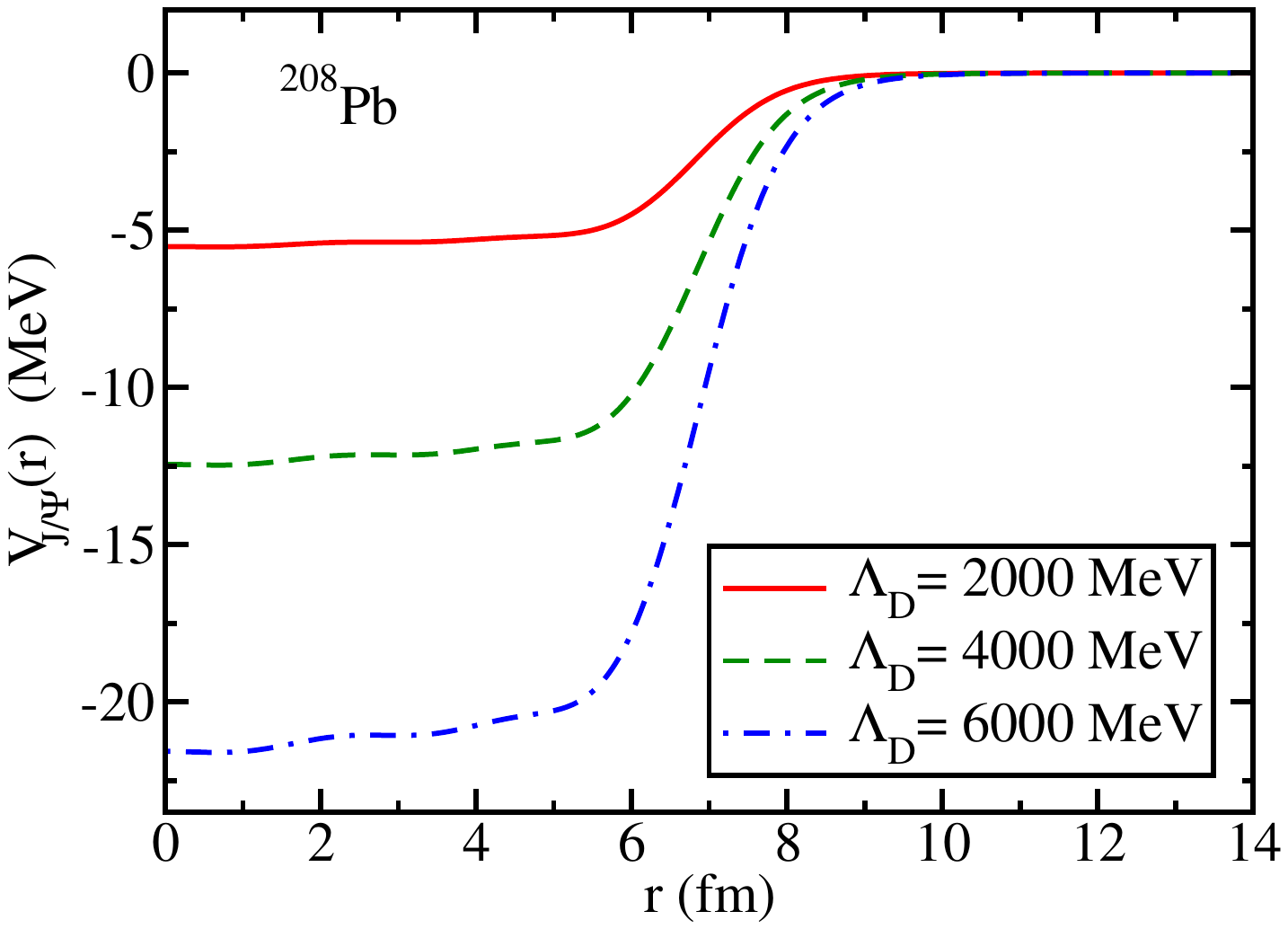} \\
  \end{tabular}
  }
  \caption{\label{fig:VJPsiA}
    $J/\psi$-nucleus  potentials for various nuclei and values of the
    cutoff parameter $\Lambda_{D}$.}
\end{figure}
%
In Figs.~\ref{fig:VetacA} and~\ref{fig:VJPsiA} we present, respectively,
the $\eta_c$-meson potentials for selected nuclei listed above
and various values of the  cutoff parameter $\Lambda_{D}$~\cite{Cobos-Martinez:2020ynh},
with the SU(4) breaking parameter of $0.6/\sqrt{2}$ for the coupling constant as
explained in Sec.~\ref{etacpsi}.
From the figures one can see that the $\eta_c$ and $J/\psi$ potentials in
nuclei are attractive in all cases but its depth depends on the value of
the cutoff parameter, being deeper the larger $\Lambda_{D}$ becomes.
This dependence is, indeed, an uncertainty in the results obtained in our
approach, when using an effective Lagrangian approach. Note that this is
the same conclusion we reached from the mass shift computed in the previous
section~\ref{etacpsi}~\cite{Cobos-Martinez:2020ynh}.

\begin{figure}[ht]
\centering
\scalebox{0.9}{
  \begin{tabular}{cc}
\includegraphics[scale=0.25]{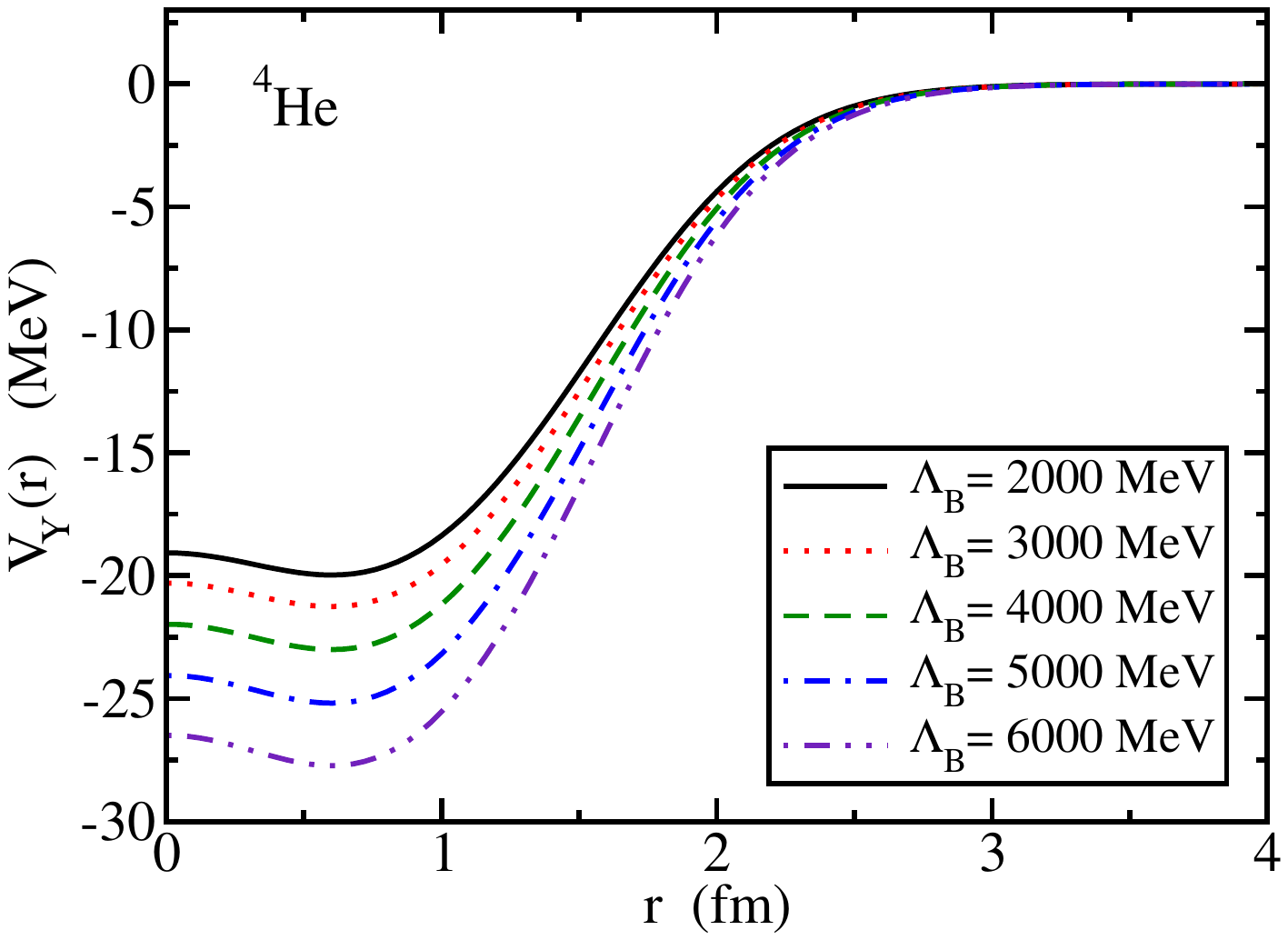} &  
\includegraphics[scale=0.25]{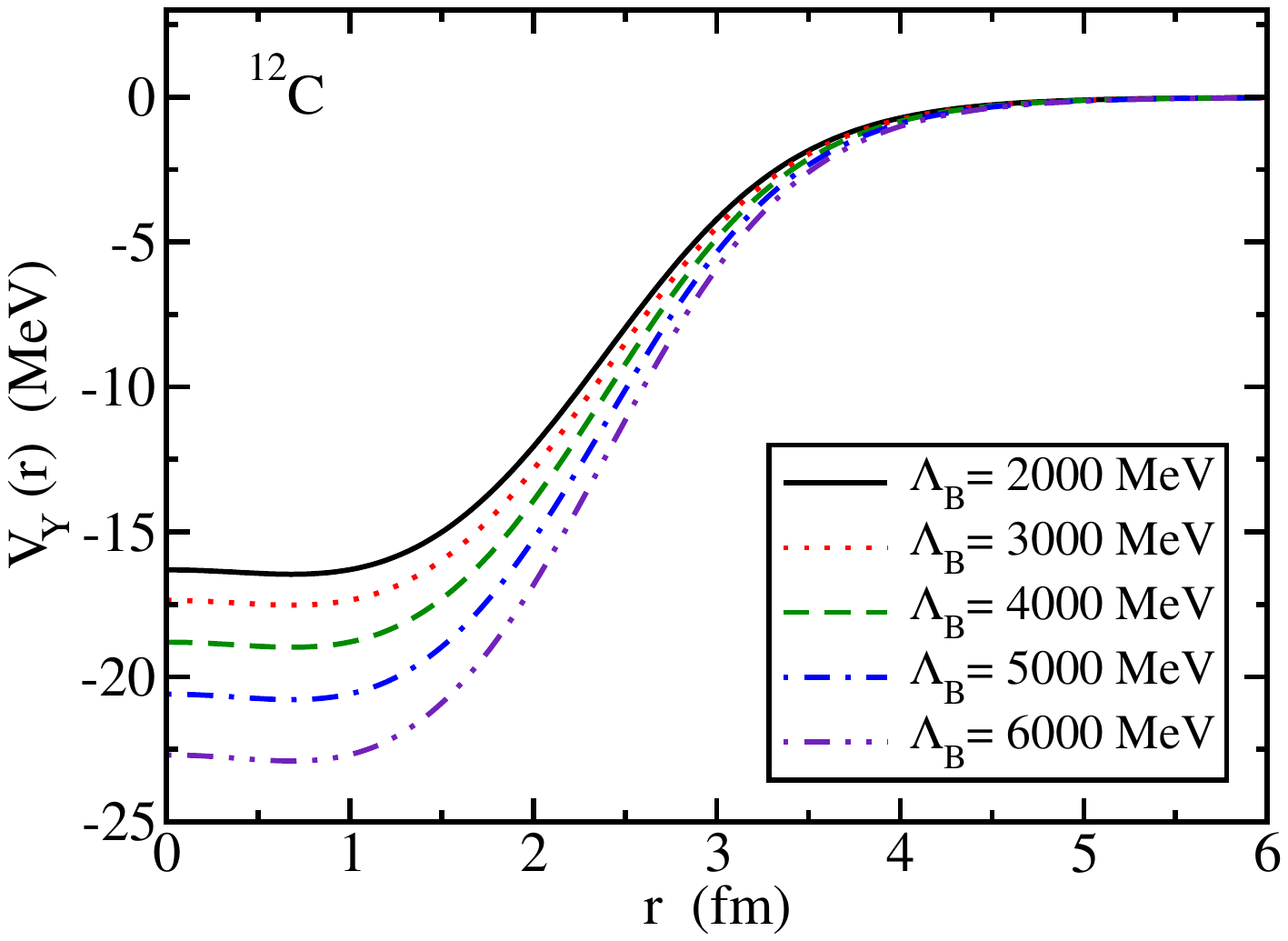} \\
\includegraphics[scale=0.25]{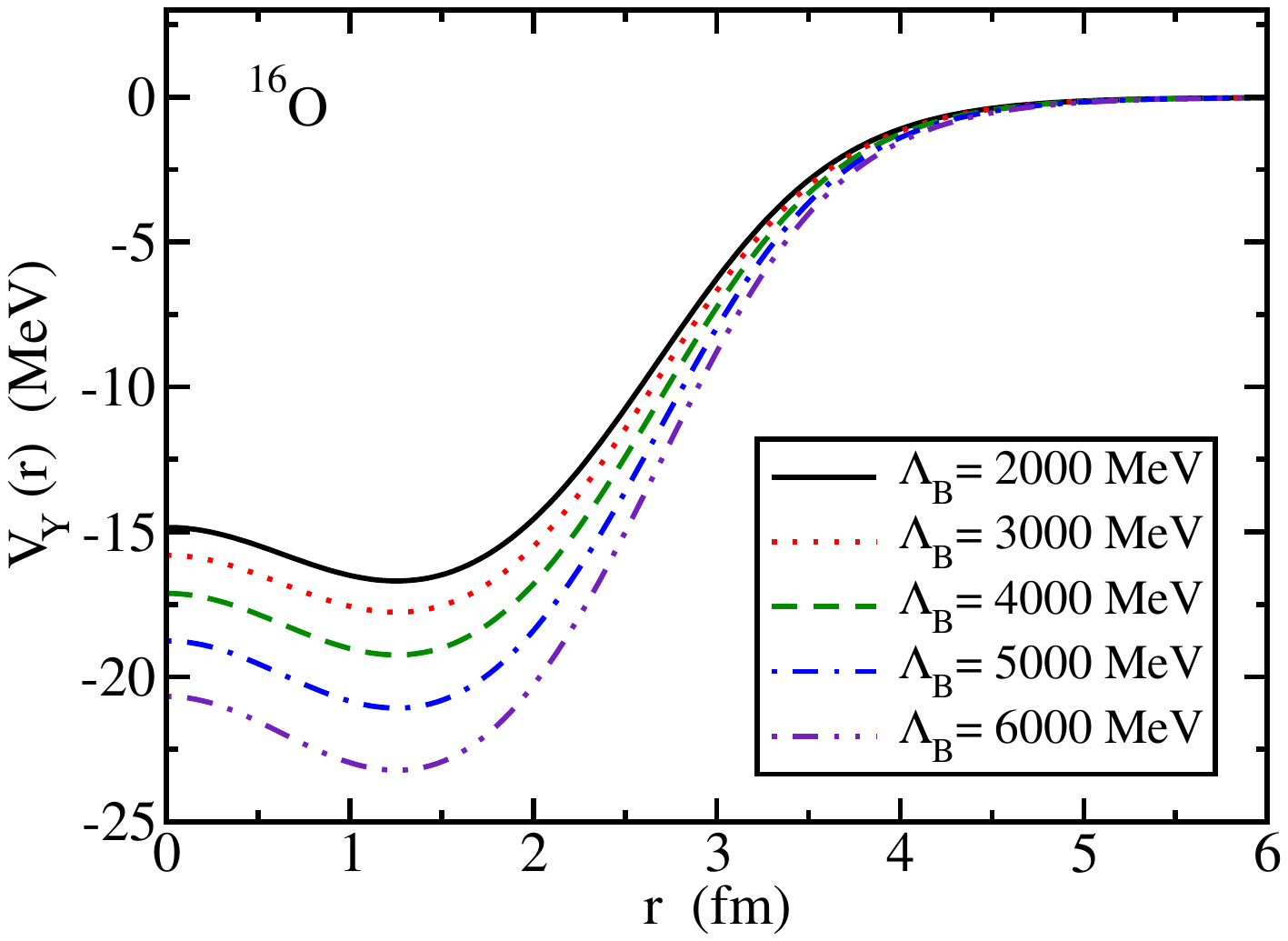} & 
\includegraphics[scale=0.25]{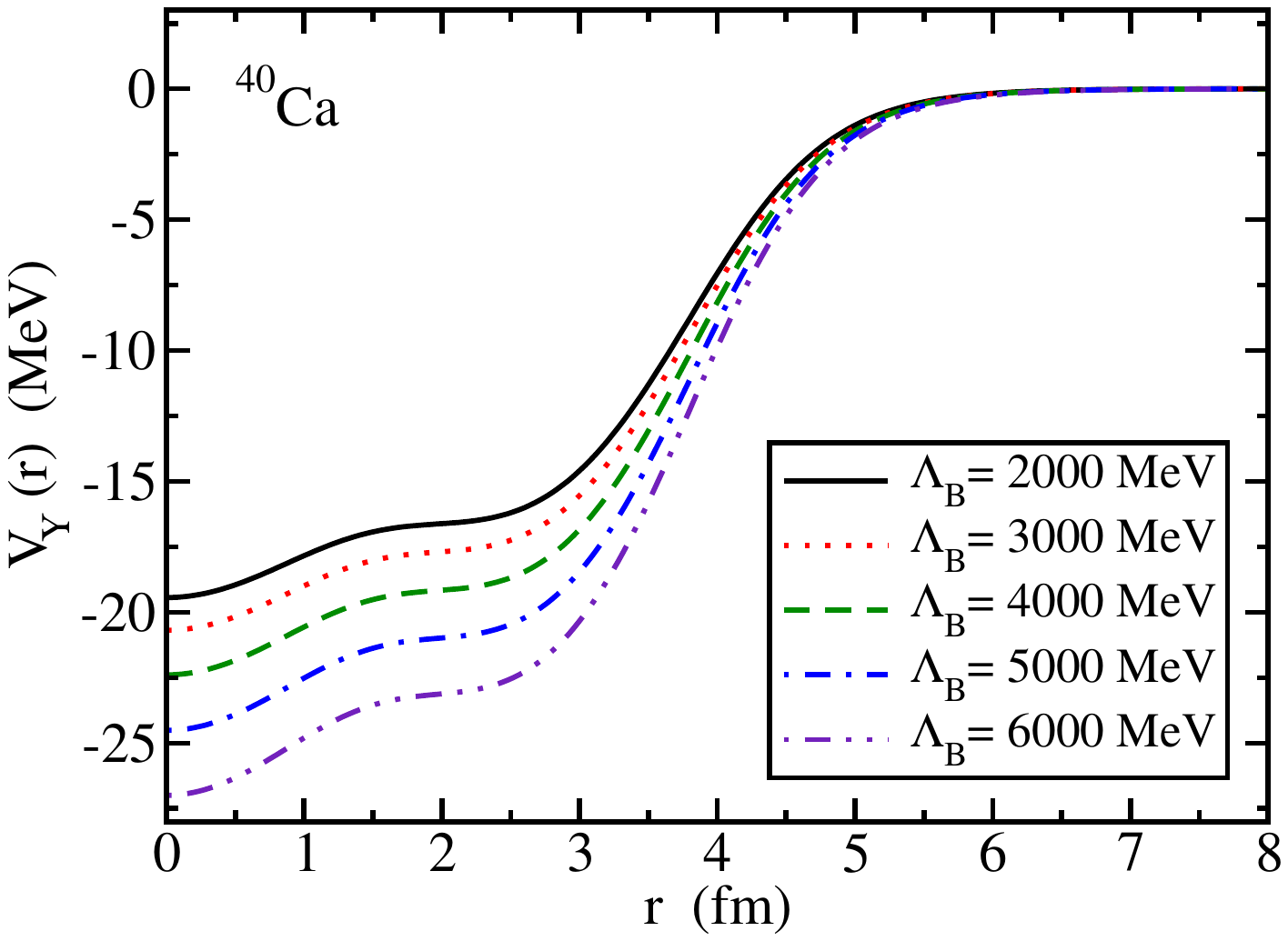} \\
\includegraphics[scale=0.25]{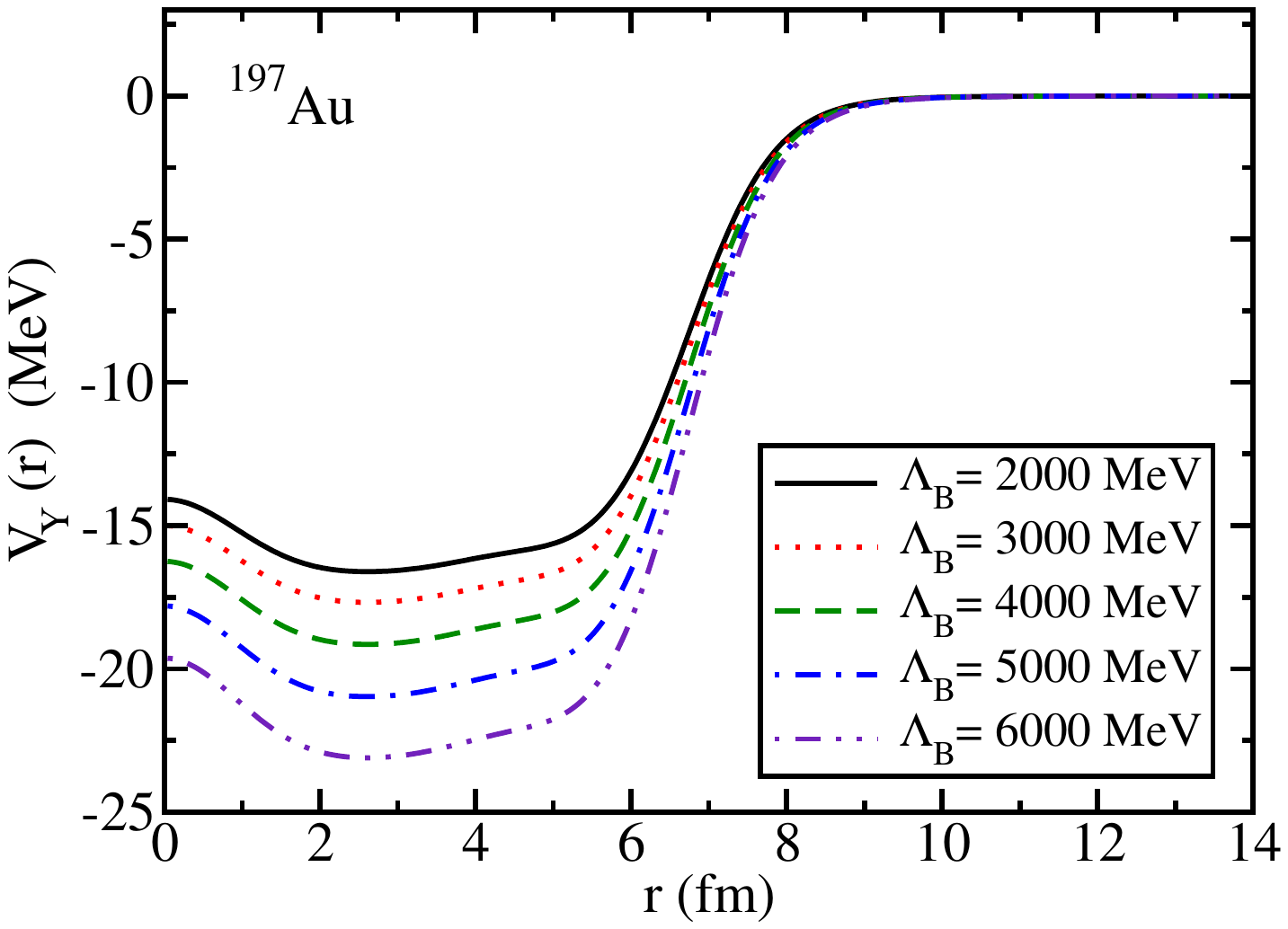} & 
\includegraphics[scale=0.25]{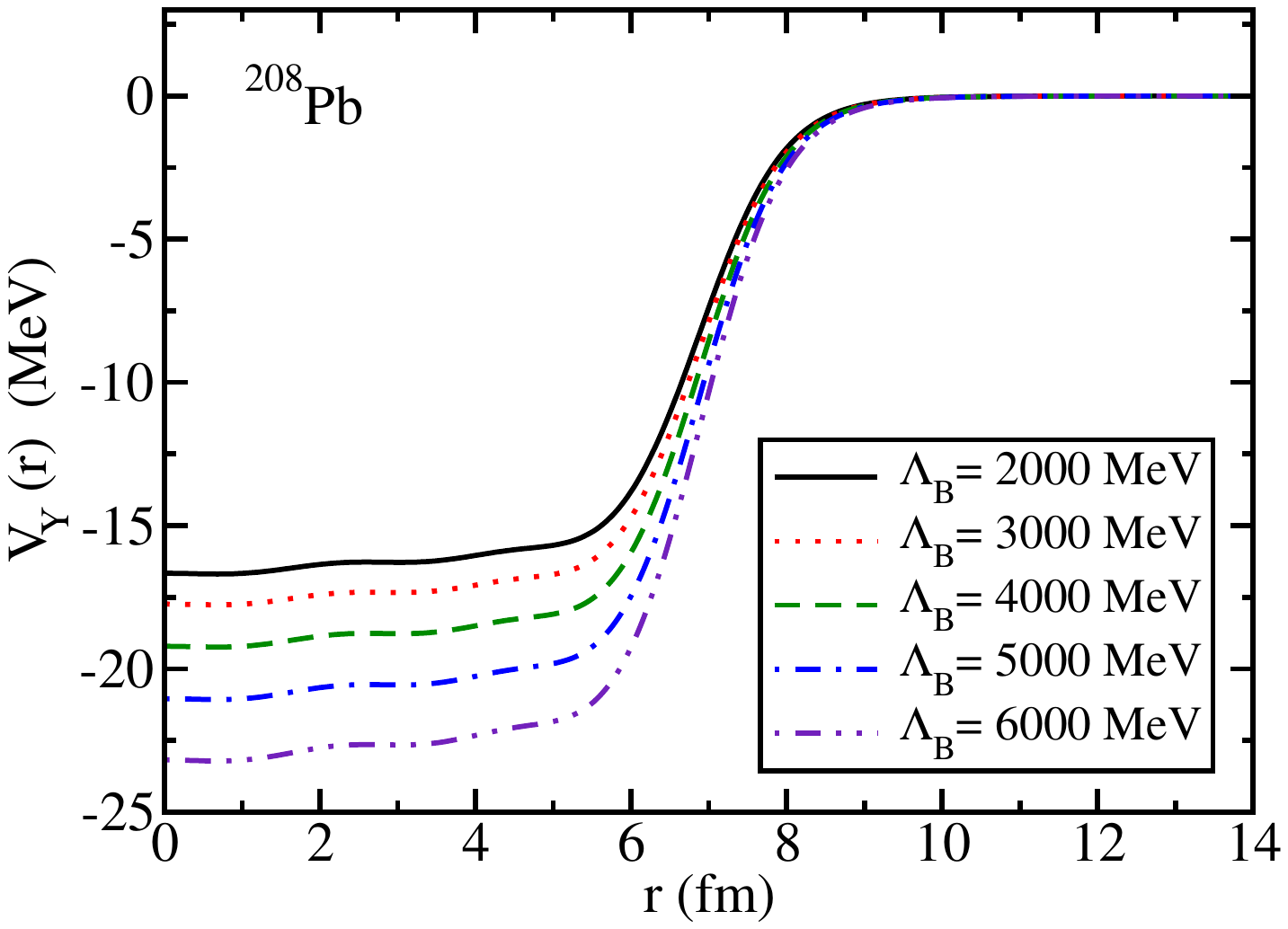} \\
  \end{tabular}
  }
  \caption{\label{fig:VUpsilonA}
    $\Upsilon$-nucleus  potentials for various nuclei with several values of the
    cutoff parameter $\Lambda_{B}$.}
\end{figure}
%
\begin{figure}[ht]
\centering
\scalebox{0.9}{
\begin{tabular}{cc}
\includegraphics[scale=0.25]{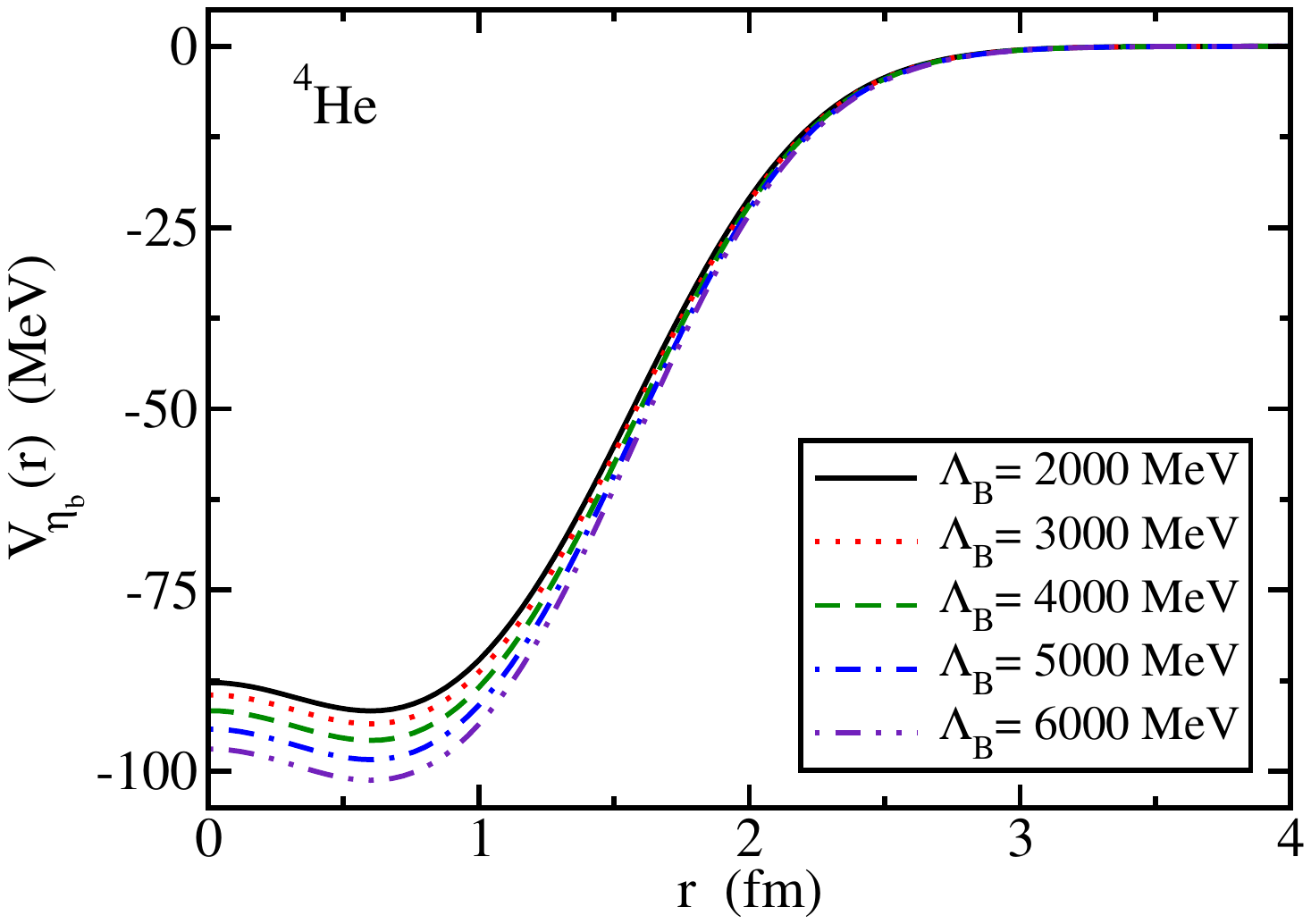} &  
\includegraphics[scale=0.25]{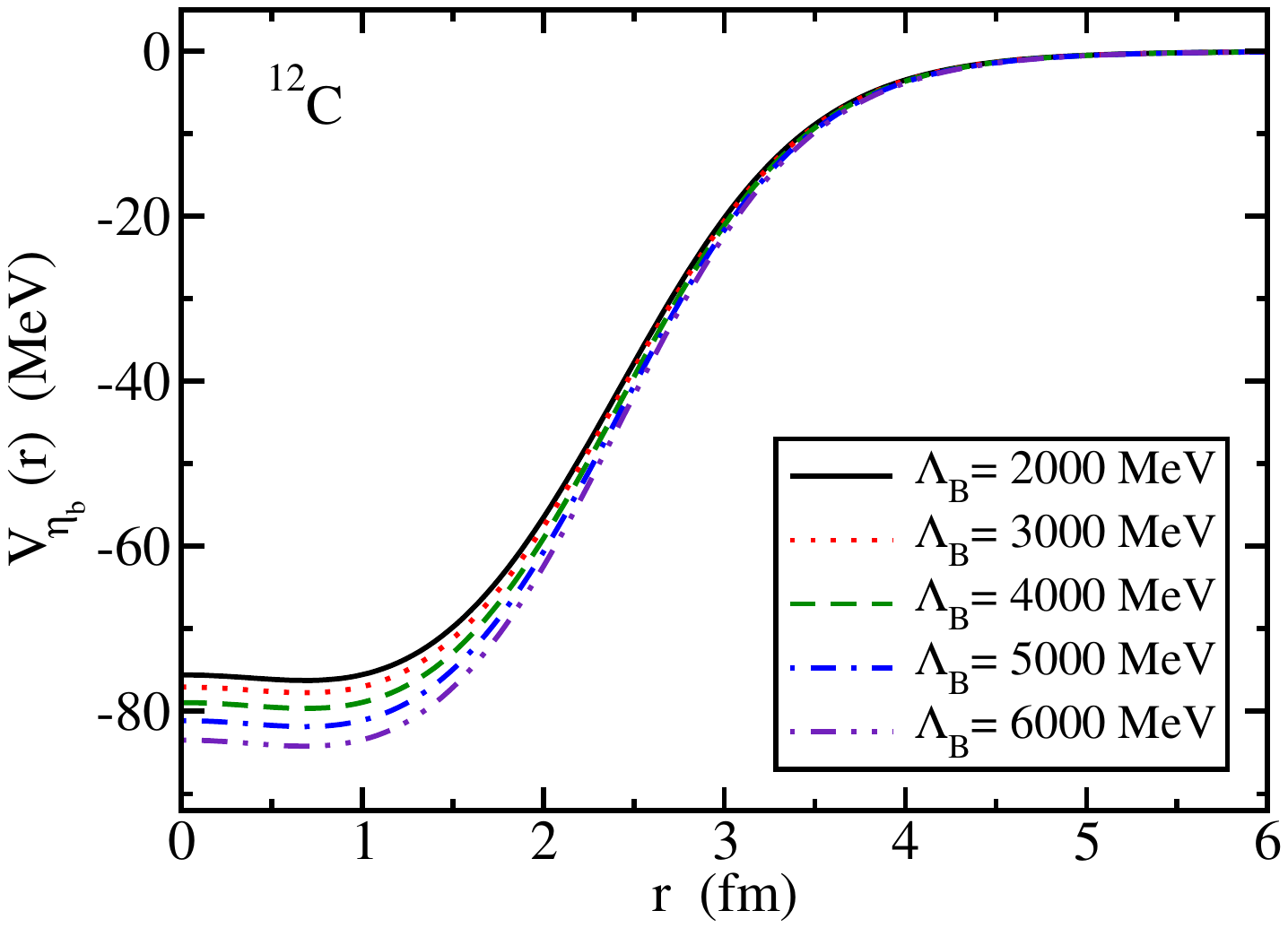} \\
\includegraphics[scale=0.25]{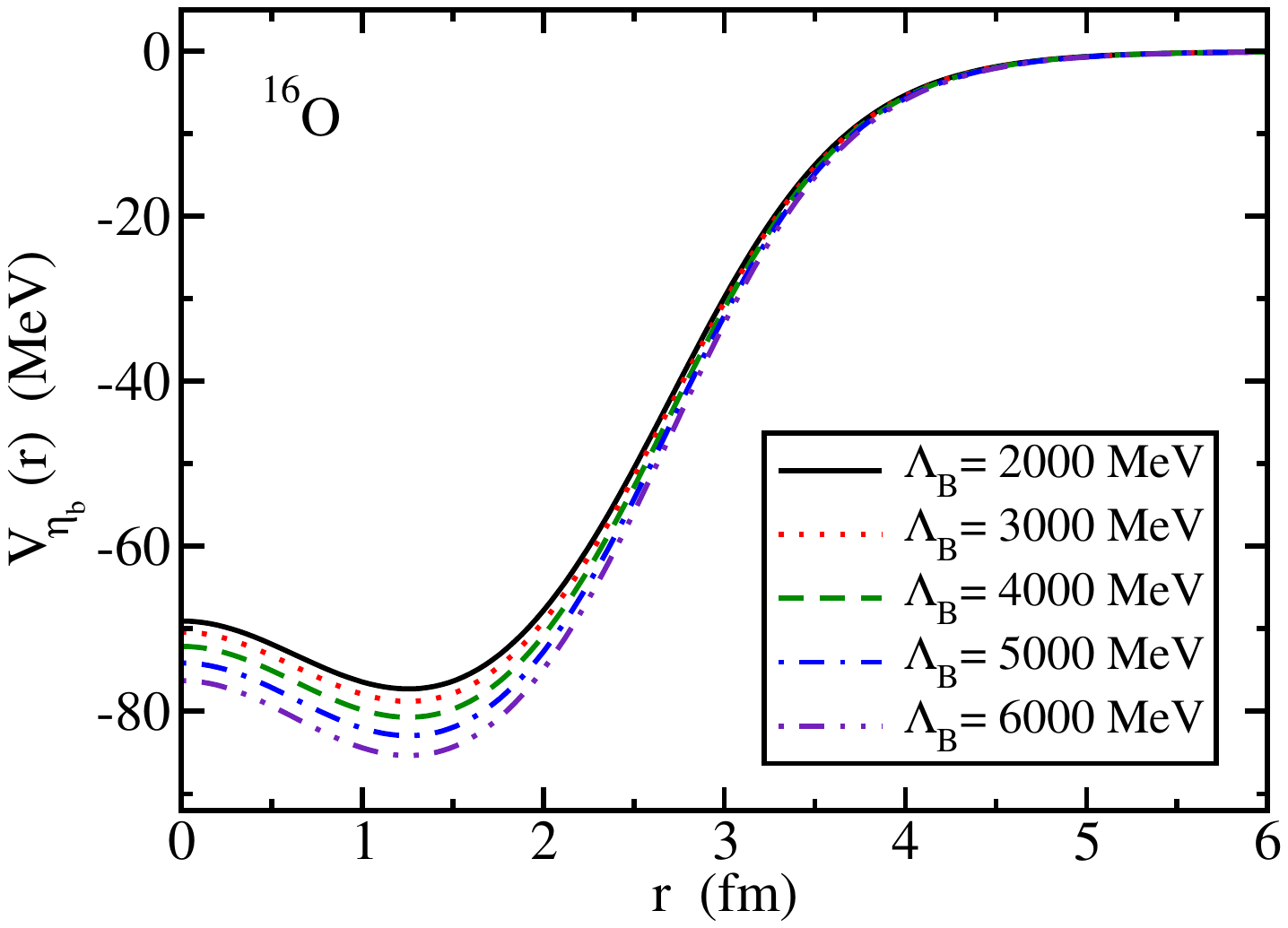} & 
\includegraphics[scale=0.25]{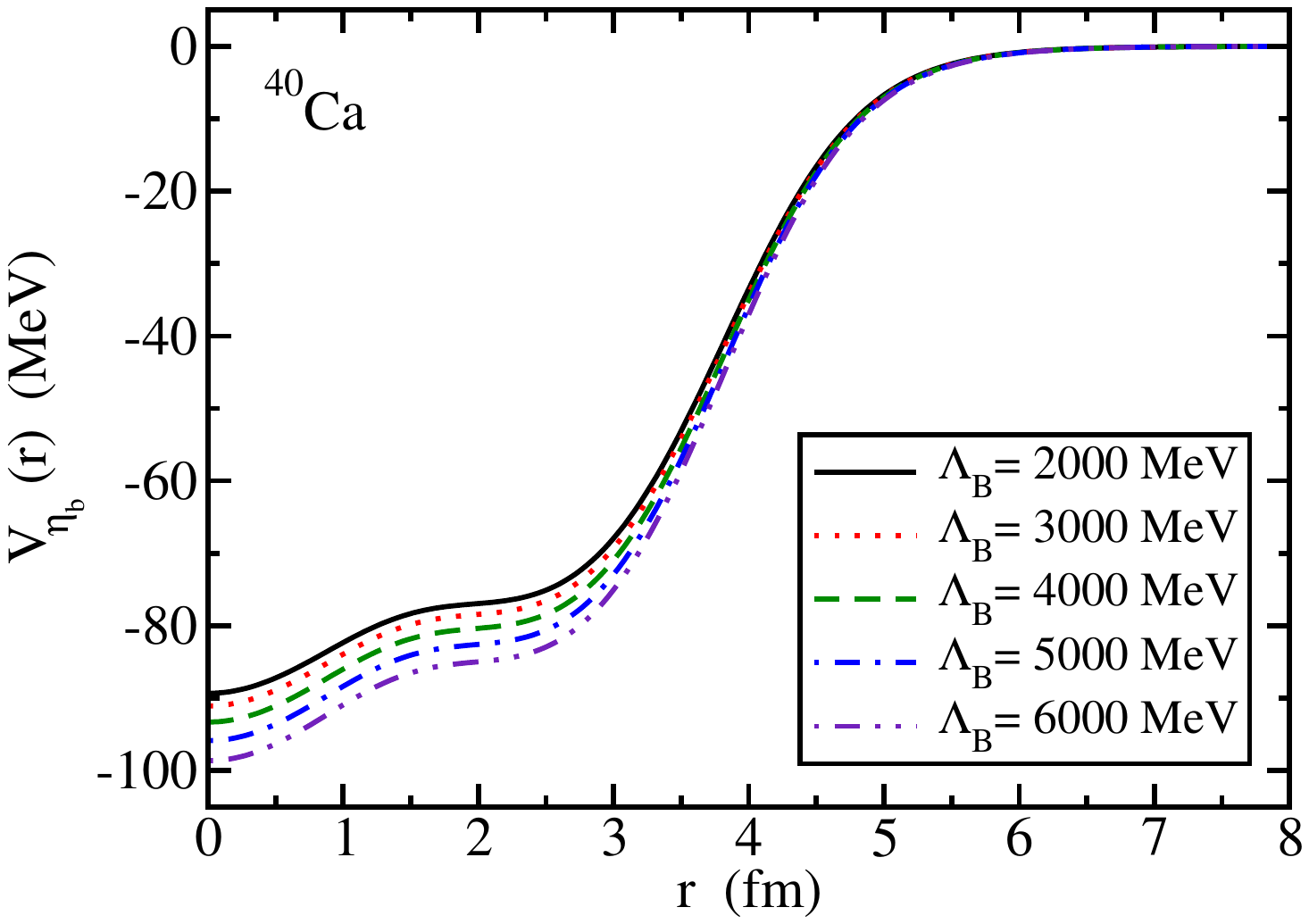} \\
\includegraphics[scale=0.25]{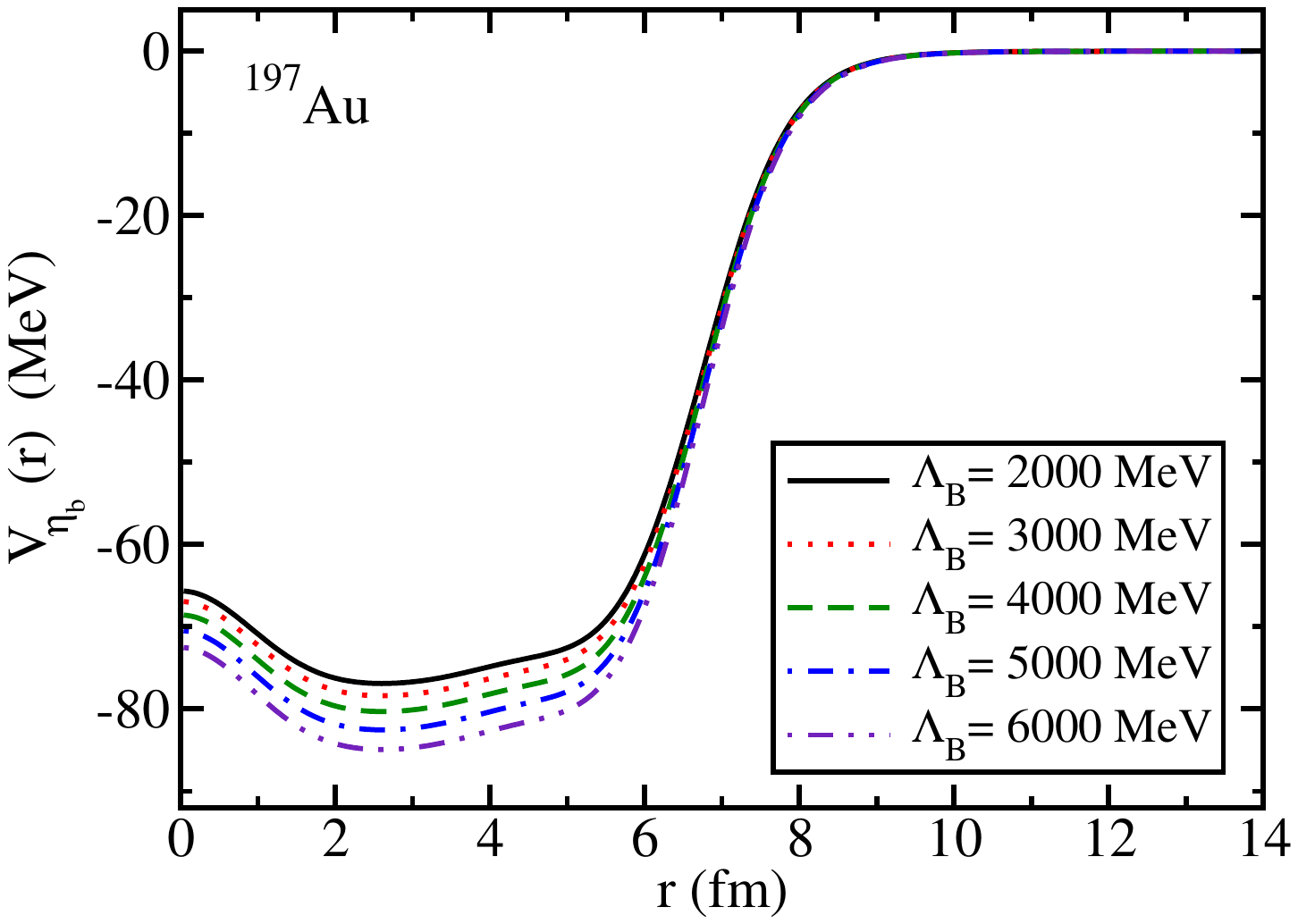} & 
\includegraphics[scale=0.25]{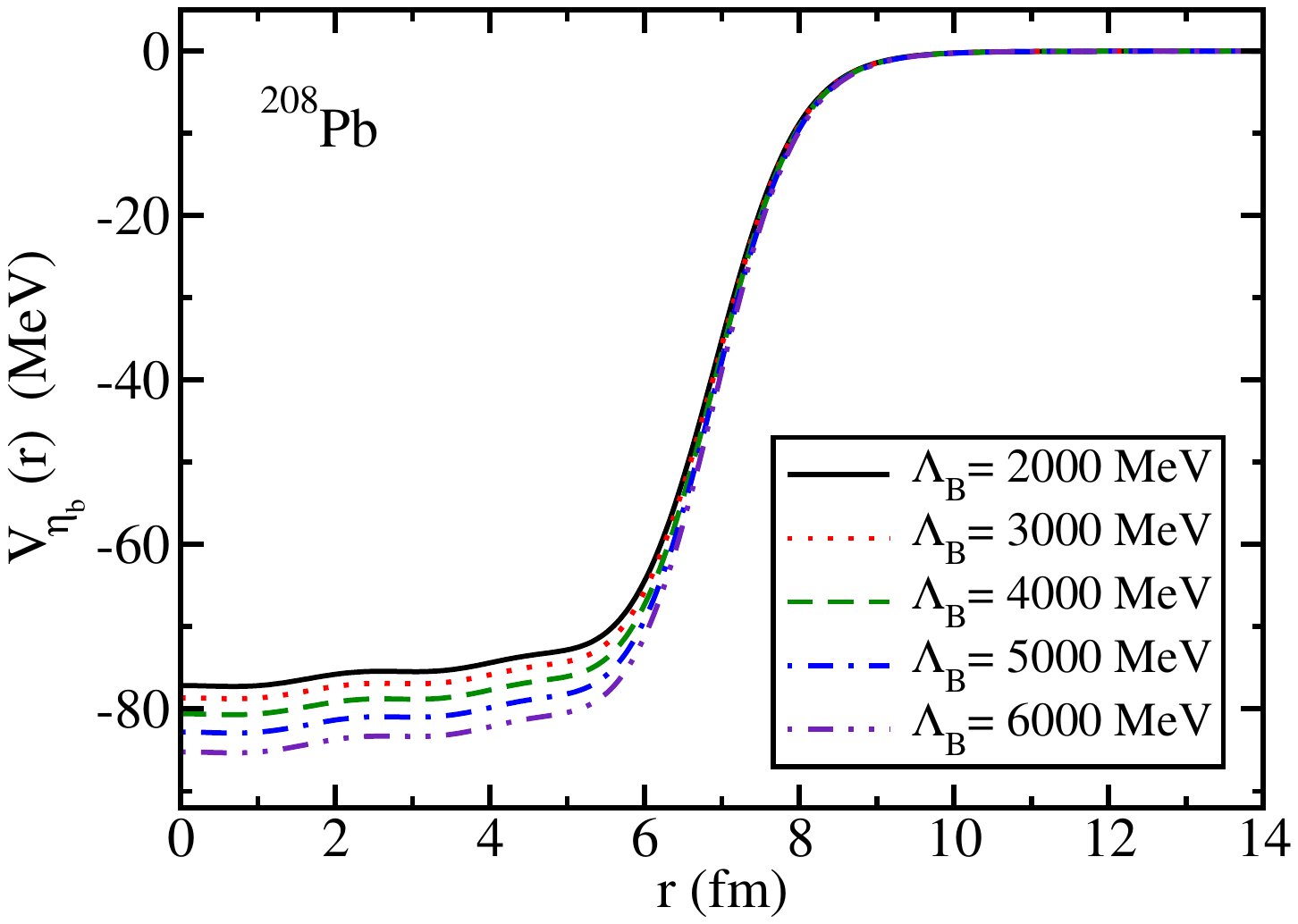} \\
 \end{tabular}
 }
  \caption{\label{fig:VetabA}
    $\eta_b$-nucleus potentials for various nuclei with several values of the
    cutoff parameter $\Lambda_{B}$.}
\end{figure}

In Figs.~\ref{fig:VUpsilonA} and~\ref{fig:VetabA} we present
the bottomonium-nucleus potentials for some of the nuclei listed above and the same
values of the cutoff parameter $\Lambda_{B}$ that were used in the
computation of the mass shift in the previous
section~\ref{upsietab}~\cite{Cobos-Martinez:2022fmt}.
We can see from Figs.~\ref{fig:VUpsilonA} and~\ref{fig:VetabA} that the
$V_{hA}$ potentials, for $h=\Upsilon$ and $\eta_b$, respectively, are
attractive for all nuclei and all values of the cutoff mass
parameter used~\cite{Cobos-Martinez:2022fmt}.
However, for each nuclei, the depth of the potential depends on the value
of the  cutoff parameter,  being more attractive the larger $\Lambda_{B}$
becomes. This dependence is expected and is, indeed, an uncertainty in the
results obtained in our approach~\cite{Cobos-Martinez:2022fmt}.
\begin{figure}[ht]
\centering
\scalebox{0.9}{
\begin{tabular}{cc}
 \includegraphics[scale=0.25]{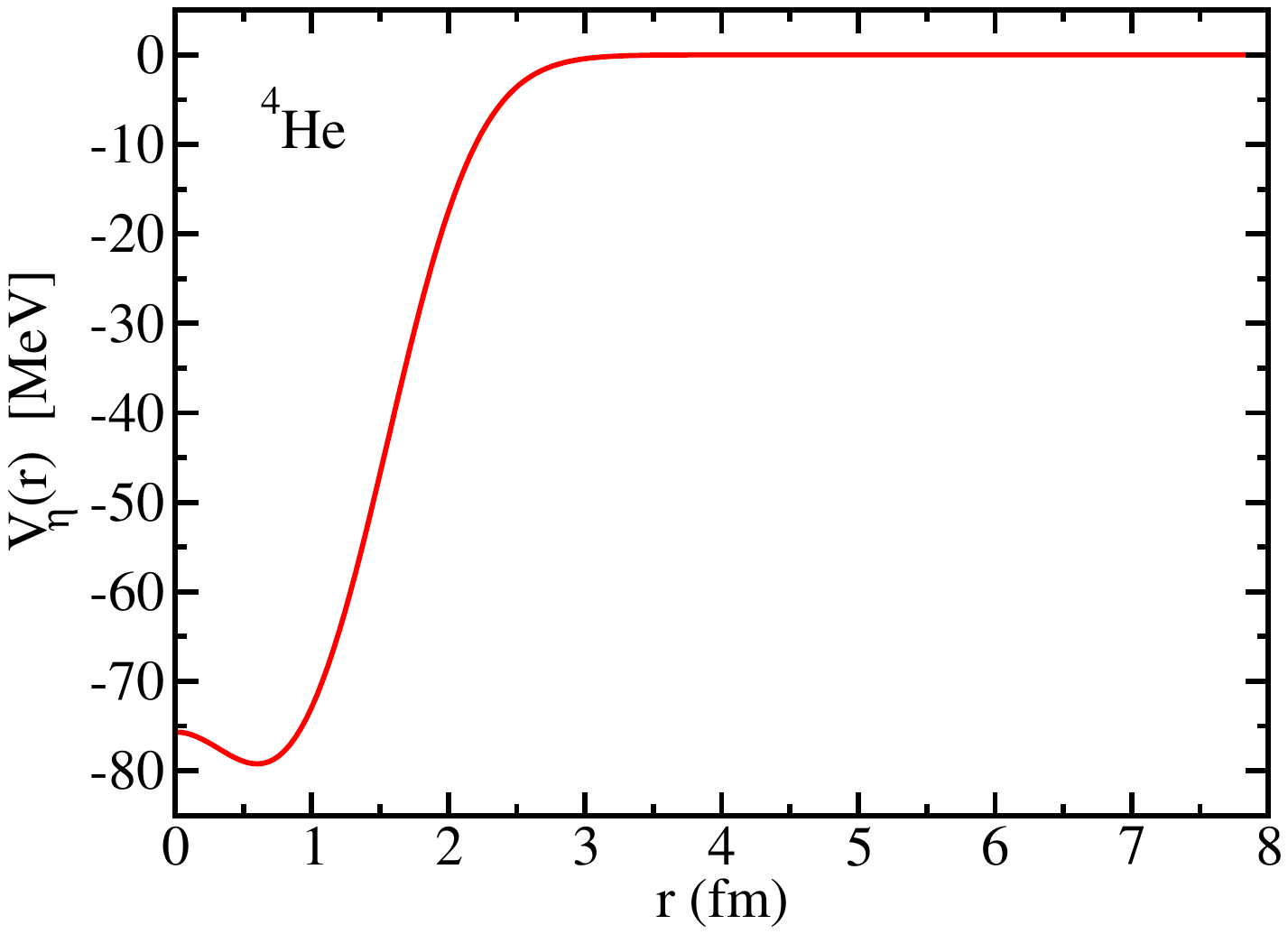} &
 \includegraphics[scale=0.25]{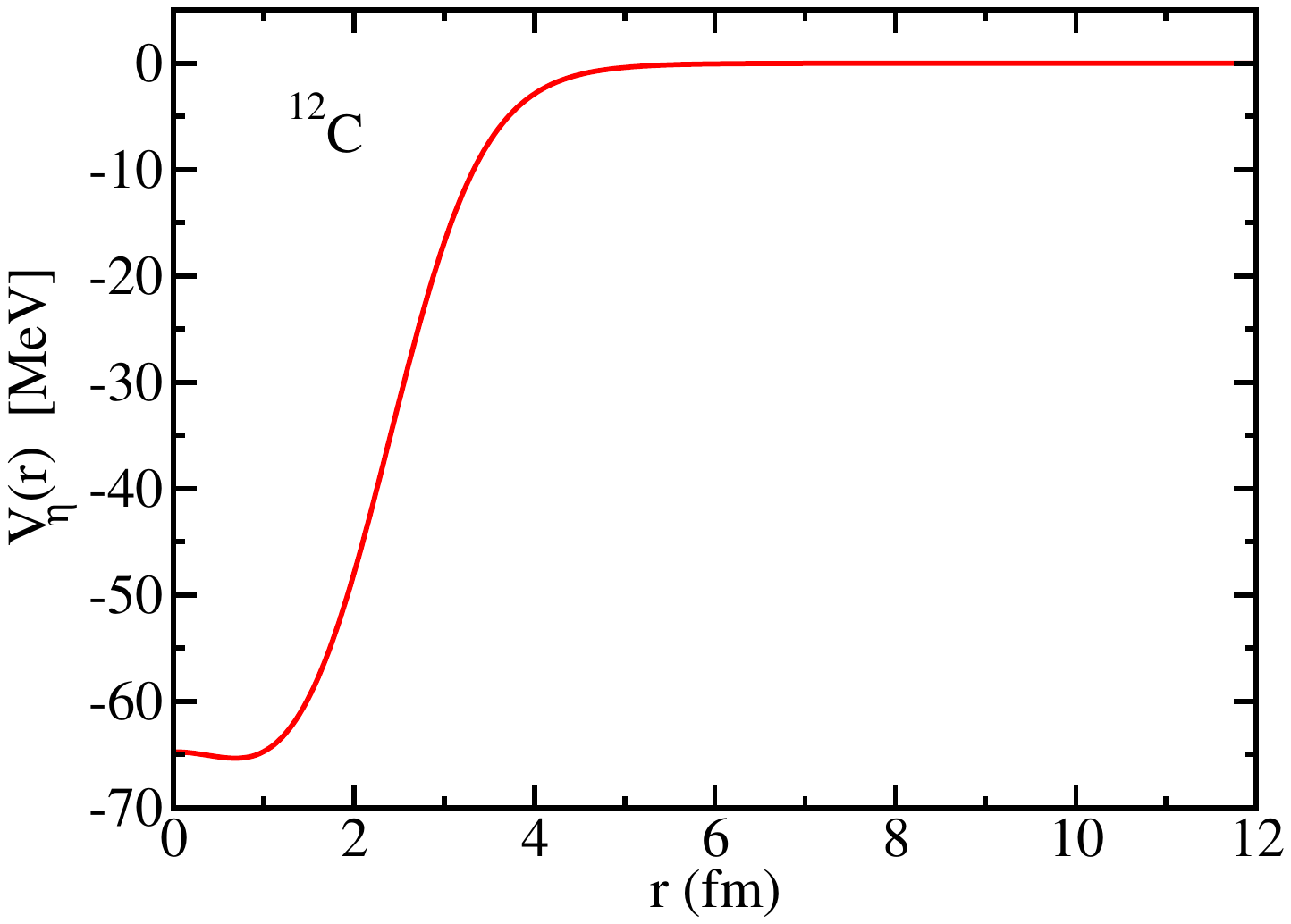} \\
 \includegraphics[scale=0.25]{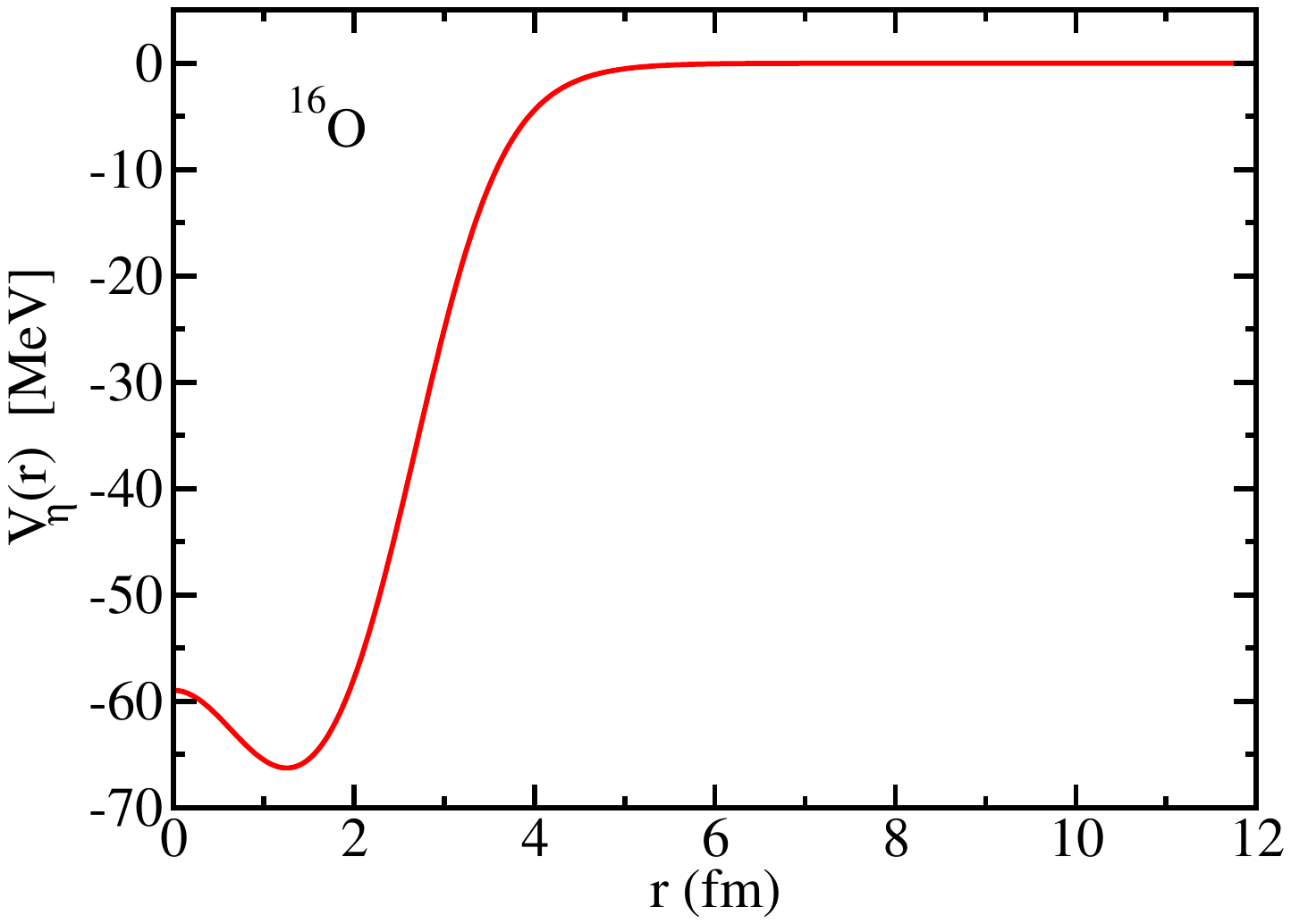} &
 \includegraphics[scale=0.25]{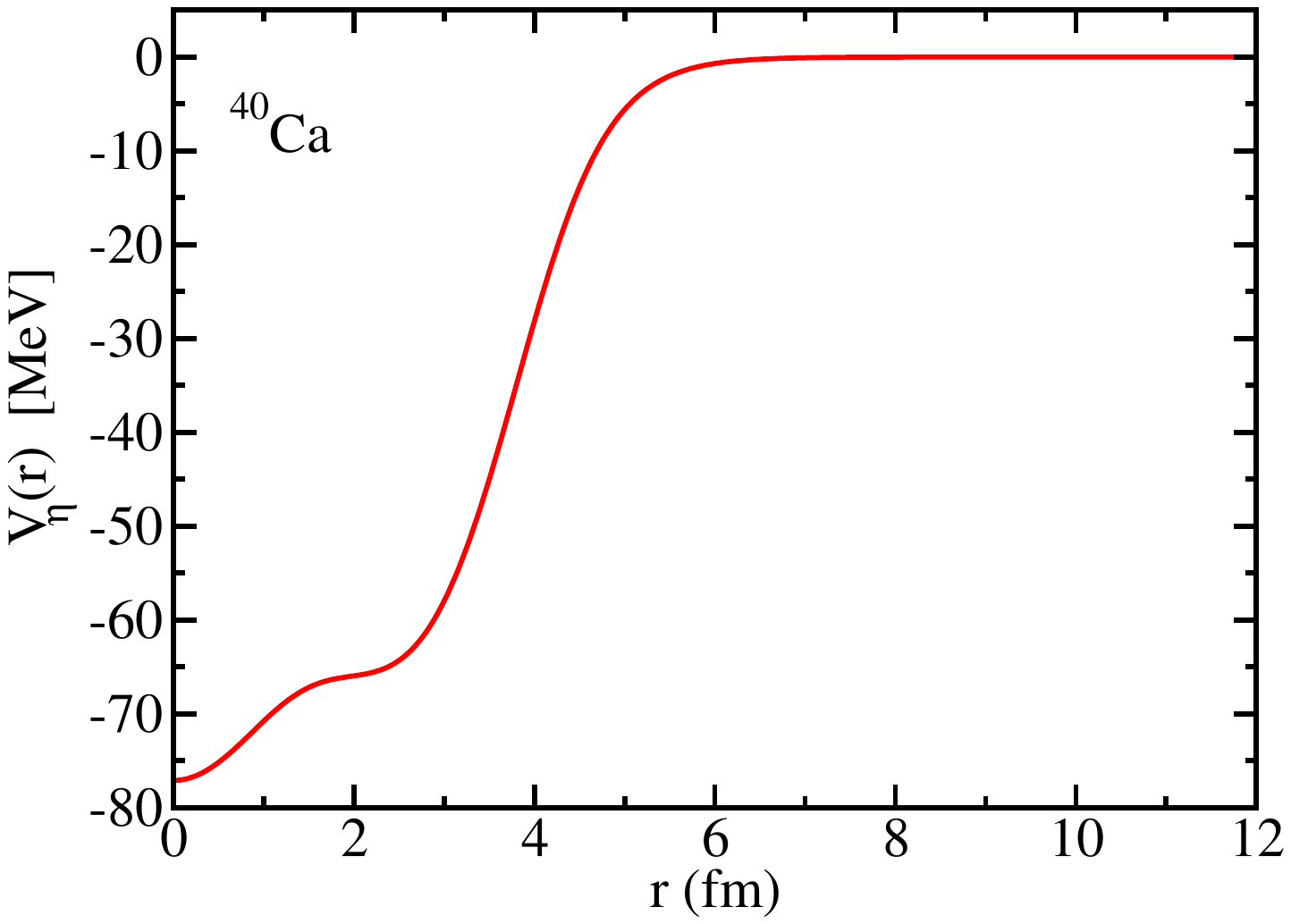} \\
 \includegraphics[scale=0.25]{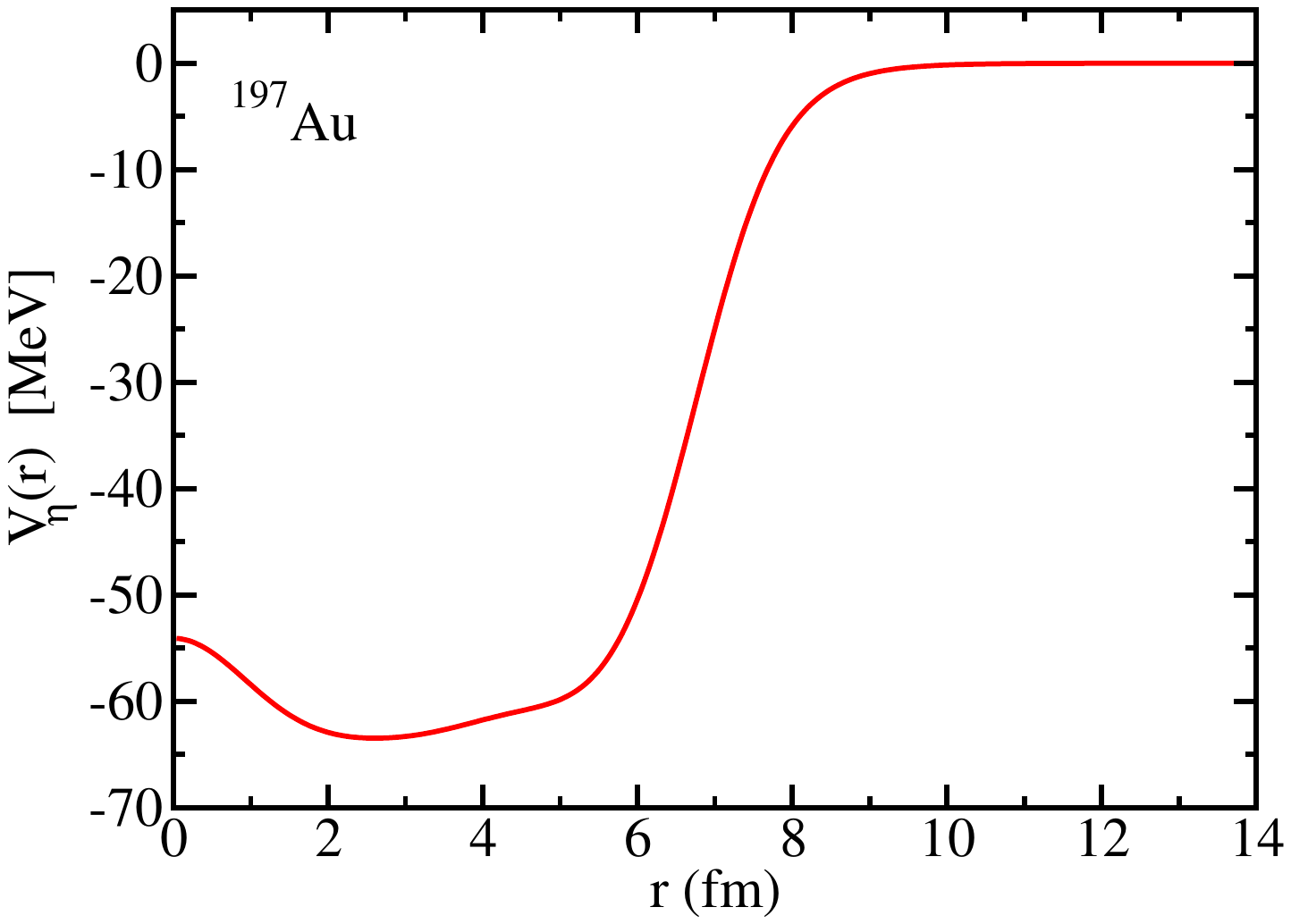}  &
 \includegraphics[scale=0.25]{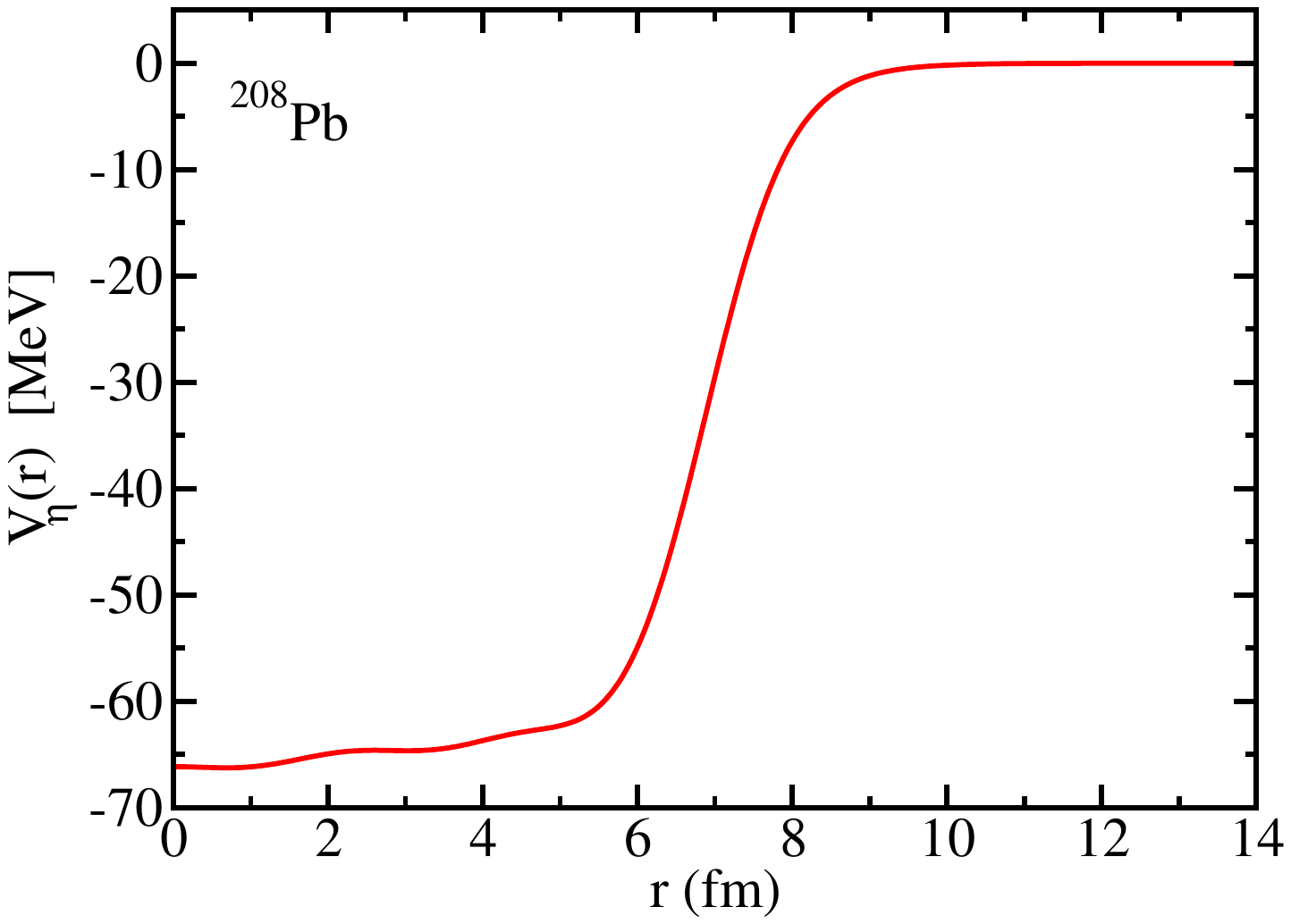} 
 \end{tabular}
 }
\caption{\label{fig:v_eta_nucl} $\eta$-nucleus potentials for several nuclei.}
\end{figure}
\begin{figure}[ht]
\centering
\scalebox{0.9}{
\begin{tabular}{cc}
 \includegraphics[scale=0.25]{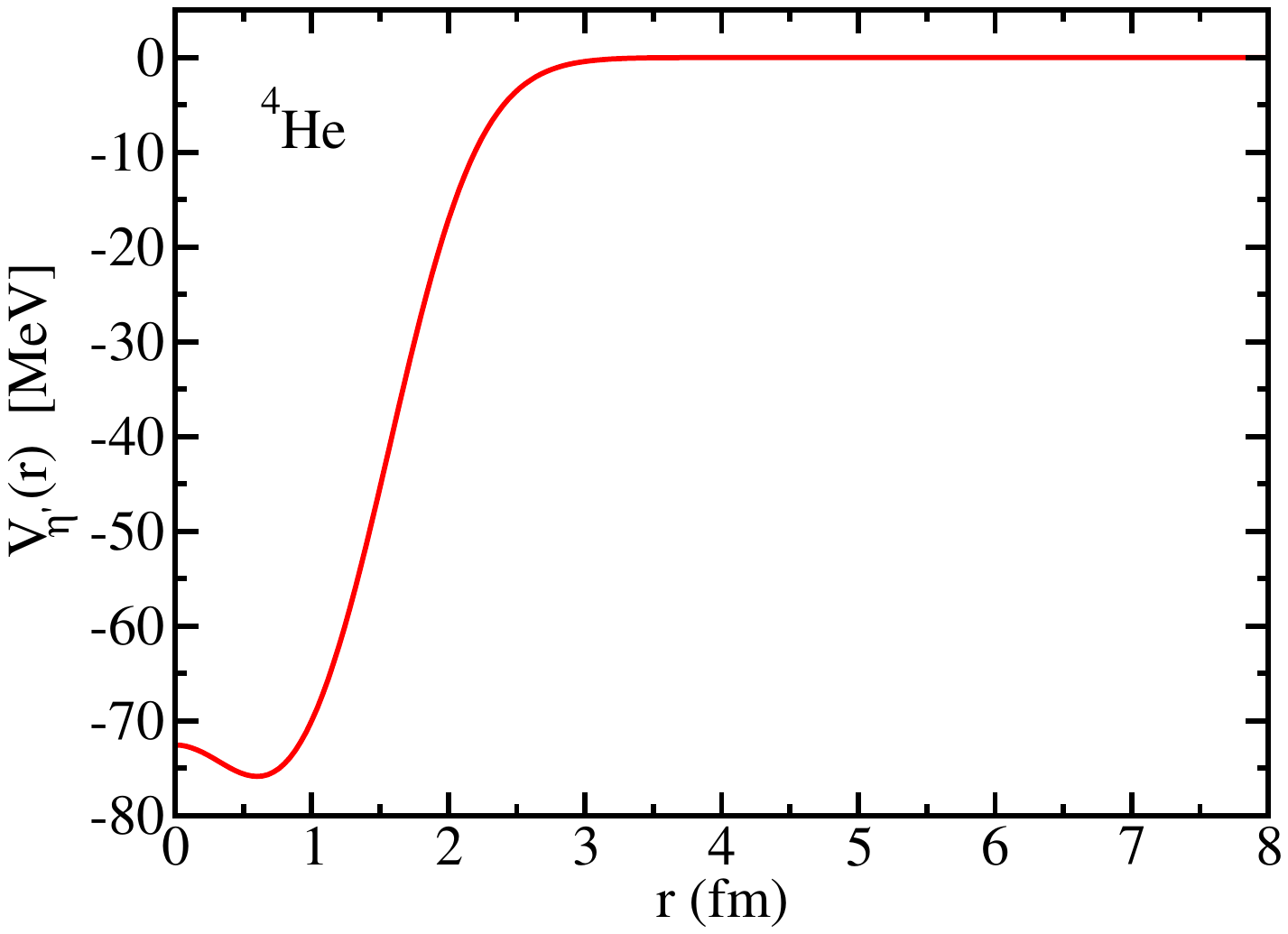} &
 \includegraphics[scale=0.25]{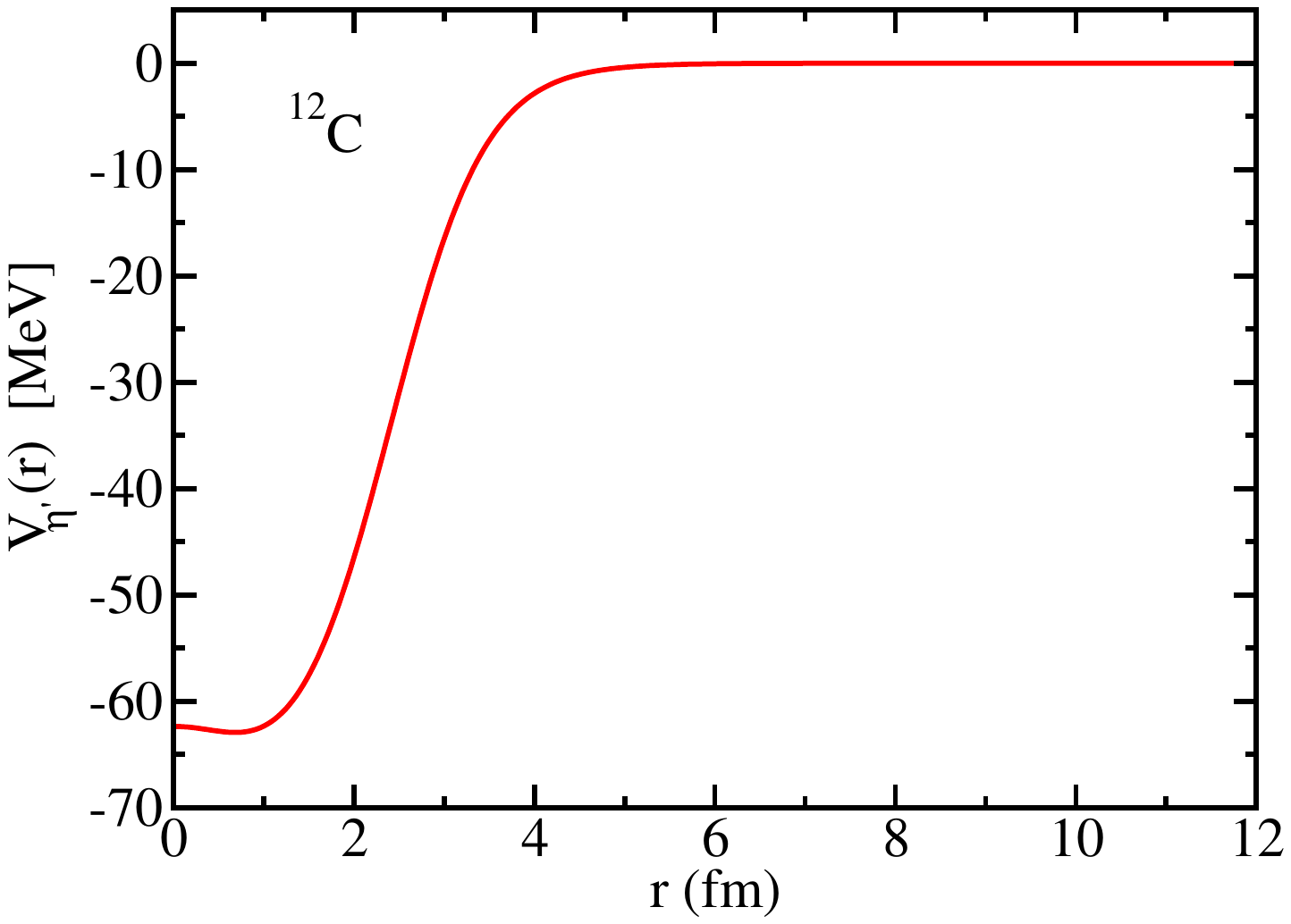} \\
 \includegraphics[scale=0.25]{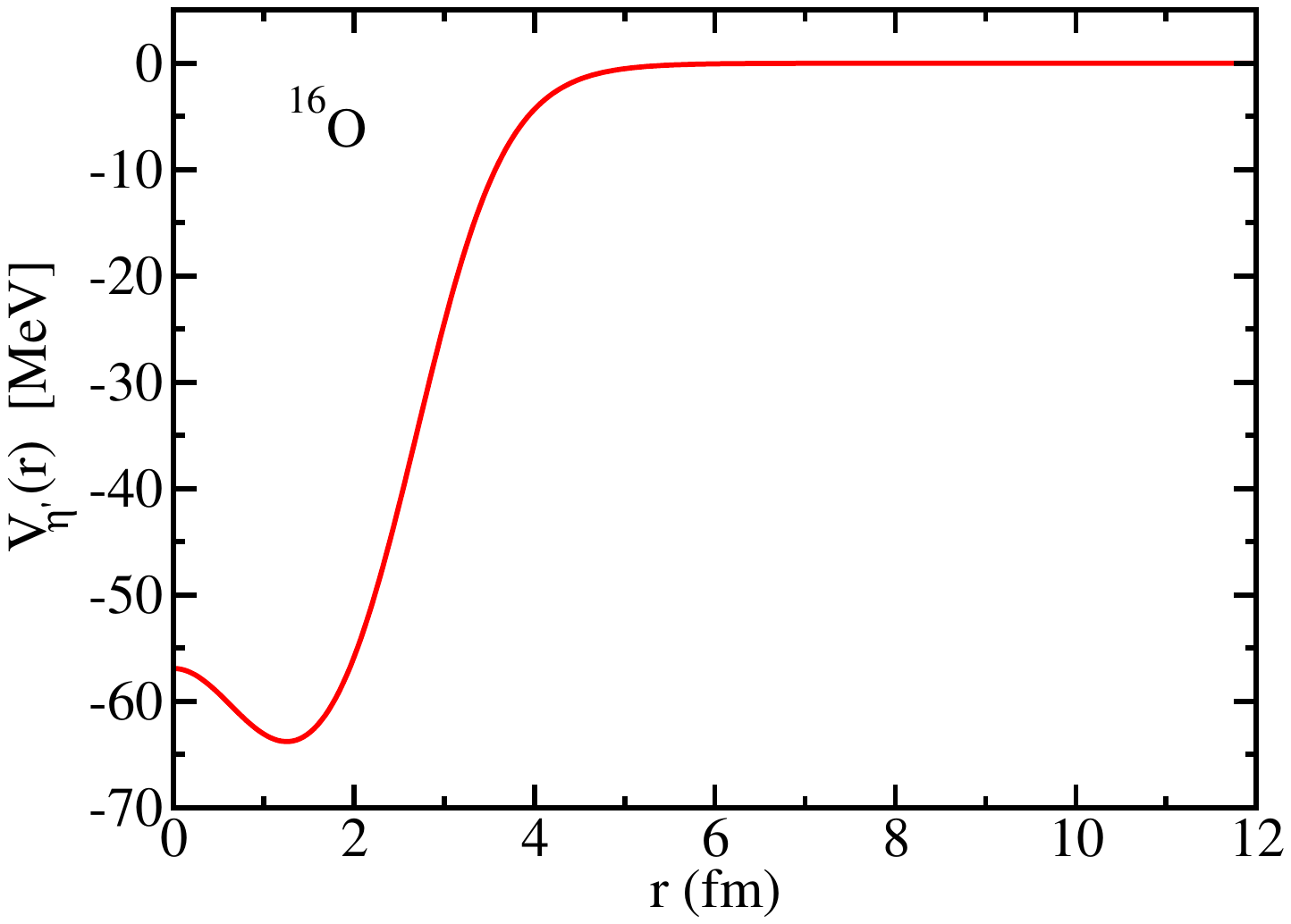} &
 \includegraphics[scale=0.25]{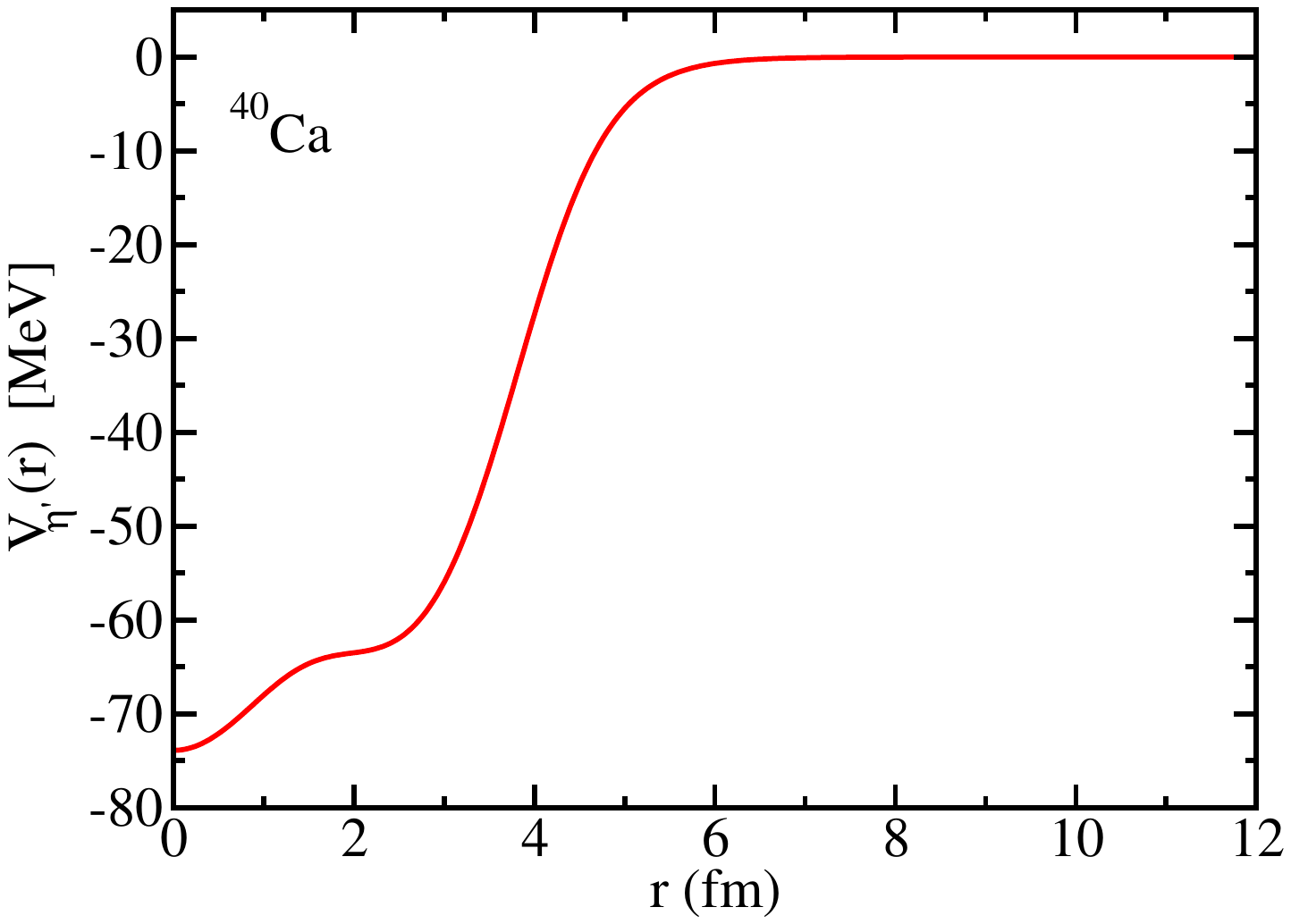} \\
 \includegraphics[scale=0.25]{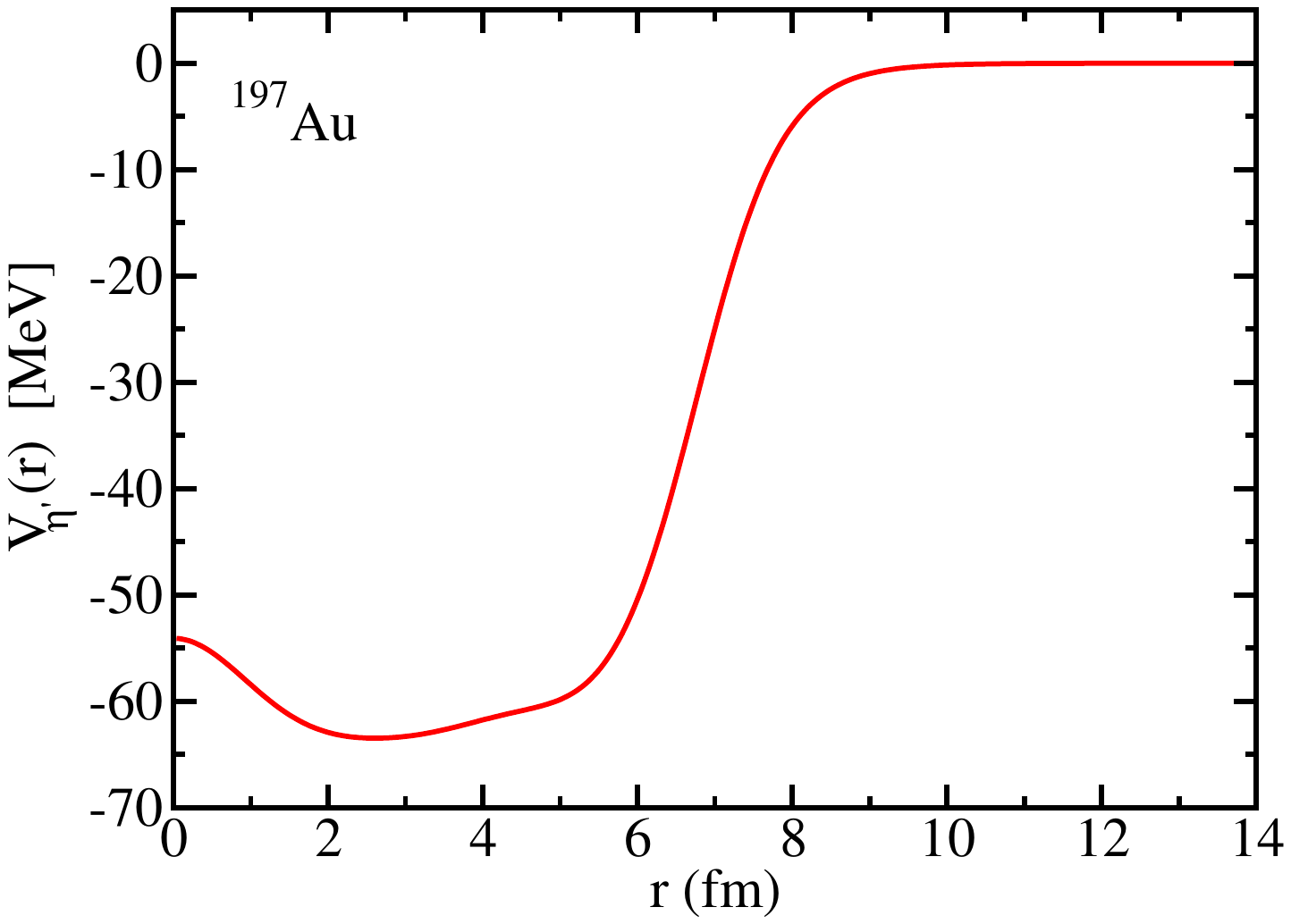}  &
 \includegraphics[scale=0.25]{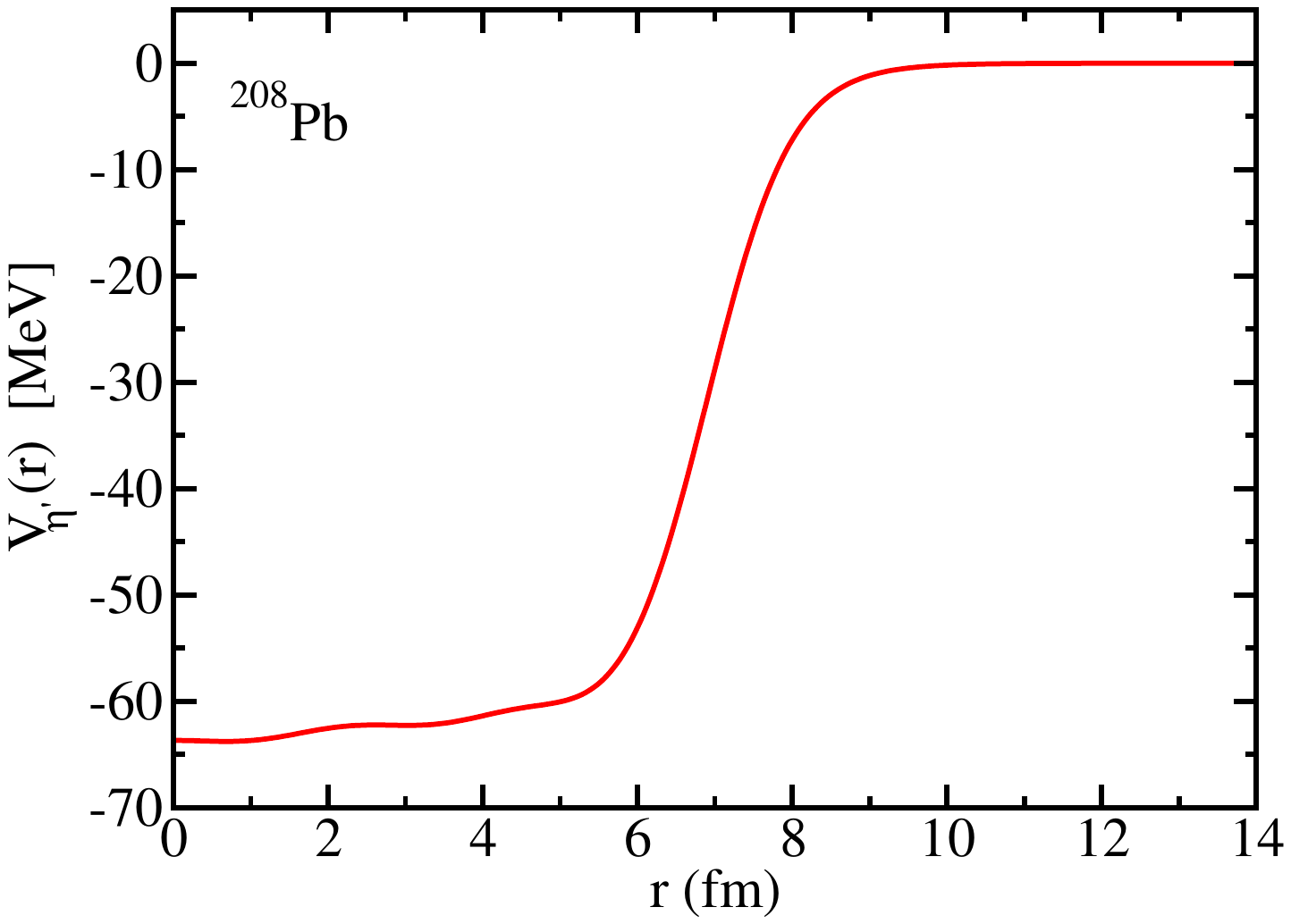} 
 \end{tabular}
 }
\caption{\label{fig:v_etaprime_nucl} $\eta'$-nucleus potentials for
several nuclei.}
\end{figure}

Next, the calculated potentials for the $\eta$ and $\eta'$ mesons in nuclei are shown in
Figs.~\ref{fig:v_eta_nucl} and~\ref{fig:v_etaprime_nucl}~\cite{Cobos-Martinez:2023hbp}.
These figures show that all potentials for the  $\eta$ and $\eta'$ in nuclei are attractive.
This is so because the corresponding value of the mass shift
(in nuclear matter) is negative for both mesons (see Sec.~\ref{qmcmodel}). The differences in the
potentials, for a given meson, reflect the differences in the baryon
density distributions for the nuclei studied~\cite{Cobos-Martinez:2023hbp}.
Furthermore, note that for a given nucleus, the potentials for
the $\eta$ and $\eta'$ are very similar, the reason for this
is that values of the mass shift for the are very similar, as shown in
Fig.~\ref{fig:metaetaprime-nm}~\cite{Cobos-Martinez:2023hbp}.

Finally, the nuclear strong interaction potentials for the $B^{\pm}_c$-A systems
are presented in Fig.~\ref{fig:bcnuccoul},
together with the attractive and repulsive Coulomb potentials,
where the Coulomb potentials are not added, but they will be included in
calculating the bound state energies in next section.

\begin{figure}[ht]
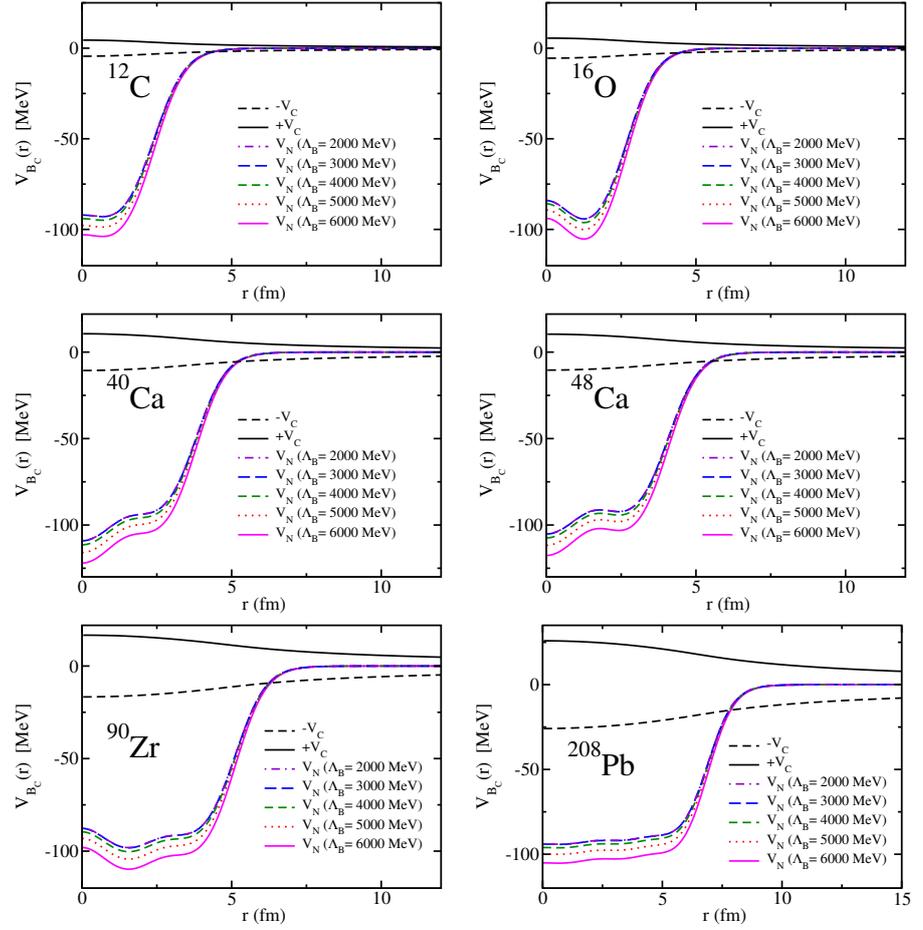

\centering
\scalebox{1.0}{
\begin{tabular}{cc}
    \includegraphics[scale=0.23]{Bc_nucl_Coul_pot_C12} &  
    \includegraphics[scale=0.23]{Bc_nucl_Coul_pot_O16} \\
    \includegraphics[scale=0.23]{Bc_nucl_Coul_pot_Ca40}  &
    \includegraphics[scale=0.23]{Bc_nucl_Coul_pot_Ca48} \\
    \includegraphics[scale=0.23]{Bc_nucl_Coul_pot_Zr90} &  
    \includegraphics[scale=0.23]{Bc_nucl_Coul_pot_Pb208}
  \end{tabular}
}
  \caption{\label{fig:bcnuccoul} Attractive and repulsive Coulomb potentials, together with the
strong nuclear potentials for the
$B^{\pm}_c$-A systems.}
\end{figure}

\section{\label{bsenergies} Numerical results for the meson-nucleus bound state energies}

We now compute the meson $h$-nucleus $A$ bound state energies, for
$h= \eta$, $\eta'$, $\phi$, $\eta_c$, $J/\psi$, $\eta_b$, $\Upsilon$ and $B^{\pm}_c$,
in a wide range of nuclear masses $A= ^{4}\hspace{-1mm}\text{He}$, $^{12}$C,
$^{16}$O, $^{40}$Ca, $^{48}$Ca, $^{90}$Zr, $^{197}$Au, and $^{208}$Pb
by solving the Klein-Gordon equation (KGE)
\begin{equation}
    \label{eqn:kge}
    \left(-\nabla^2 + (m + V(\textbf{r}))^2\right)\psi(\textbf{r})=\mathcal{E}^2\psi(\textbf{r})
\end{equation}
where $V(\textbf{r})= V(r)$ is the scalar nuclear potential
associated with mass shift, given
by~\eqn{eqn:VhA}, $r=|\textbf{r}|$ is the distance from the nucleus, and
$m$ is the reduced mass of the meson $h$-nucleus $A$ system 
$m_h m_A/(m_h + m_A)$, in vacuum.
The bound state energies $E$ and widths $\Gamma$ are given by
are given by $E=\mathcal{E}-m$ and $\Gamma=-2\mathrm{Im}\,\mathcal{E}$,
respectively, where $\mathcal{E}$ is the energy eigenvalue in \eqn{eqn:kge}.
Note that, when the Coulomb or vector potential is relevant, the right hand side of
Eq.~(\ref{eqn:kge}) must be modified properly as
$\mathcal{E}^2 \to (\mathcal{E} - V_\text{V,Coul})^2$ with $V_\text{V,Coul}$ being the
vector and Coulomb potentials, respectively. See Ref.~\cite{Zeminiani:2024rrh} for details.

Before proceeding to solve \eqn{eqn:kge} we note that we have also solved
to approximations of~\eqn{eqn:kge}, namely the Schr\"{o}dinger equation and
also the KGE dropping the $V^2(\textbf{r})$ term in~\eqn{eqn:kge}
for some meson-nucleus systems.
In all cases we obtain essentially the same results,
which do not change the conclusions about the existence
of the bound states.

We solve the Klein-Gordon equation using the momentum space
methods~\cite{Kwan:1978zh}. Here, \eqn{eqn:kge} is first converted to
momentum space representation via a Fourier transform, followed by a 
partial wave-decomposition of the Fourier-transformed potential, 
or we obtain directly the partial wave decomposition in momentum space
by a double Spherical Bessel transform.
For $\eta$, $\eta'$, $\phi$, $\eta_c$ and $J/\psi$, the method used is the
partial wave-decomposition of the Fourier-transformed potential.
For the $B^{\pm}_c$ we employ the direct double Spherical Bessel transform, and
for $\Upsilon$ and $\eta_b$ we use both methods. 
Then, for a given value of angular momentum $\ell$,
the eigenvalues of the resulting equation
are found by the inverse iteration eigenvalue algorithm.
The detailed comparison and discussions were made
in Ref.~\cite{Zeminiani:2024rrh}, and it turned out that the main conclusions
remain valid in both methods. The calculated bound state energies
are similar, with, at most,  few MeV difference, which we would think within the desired 
experimental accuracy for the strong interaction bound state energy measurement.

\begin{table}[ht]
\begin{center}
\caption{\label{tab:phibse} $\phi$-nucleus single-particle energies, $E$,
and half widths, $\Gamma/2$, obtained with and without the imaginary part
of the potential, for three values of the cutoff parameter $\Lambda_K$.
When only the real part is included, where the corresponding single-particle energy
$E$ is given in parenthesis and $\Gamma=0$ for all nuclei.``n'' indicates that no bound state is found. All quantities are given in MeV.}
\scalebox{0.85}{
  \begin{tabular}{ll|rr|rr|rr} 
\hline \hline
& & \multicolumn{2}{c|}{$\Lambda_{K}=2000$} &
\multicolumn{2}{c|}{$\Lambda_{K}=3000$} & 
\multicolumn{2}{c}{$\Lambda_{K}=4000$}  \\
\hline
 & & $E$ & $\Gamma/2$ & $E$ & $\Gamma/2$ & $E$ & $\Gamma/2$ \\
\hline
$^{4}_{\phi}\text{He}$ & 1s & n (-0.8) & n & n (-1.4) & n & -1.0 (-3.2) & 8.3 \\
\hline
$^{12}_{\phi}\text{C}$ & 1s & -2.1 (-4.2) & 10.6 & -6.4 (-7.7) & 11.1 & -9.8 (-10.7) & 11.2 \\
\hline
$^{16}_{\phi}\text{O}$ & 1s & -4.0 (-5.9) & 12.3 & -8.9 (-10.0) & 12.5 & -12.6 (-13.4) & 12.4 \\
& 1p & n (n) & n & n (n) & n & n (-1.5) & n \\
\hline
$^{40}_{\phi}\text{Ca}$ & 1s & -9.7 (-11.1) & 16.5 & -15.9 (-16.7) & 16.2 & -20.5 (-21.2) & 15.8 \\
& 1p & -1.0 (-3.5) & 12.9 & -6.3 (-7.8) & 13.3 & -10.4 (-11.4) & 13.3 \\
& 1d & n (n) & n & n (n) & n & n (-1.4) & n \\
\hline
$^{48}_{\phi}\text{Ca}$ & 1s & -10.5 (-11.6) & 16.5 & -16.5 (-17.2) & 16.0 & -21.1 (-21.6) & 15.6 \\
& 1p & -2.5 (-4.6) & 13.6 & -7.9 (-9.2) & 13.7 & -12.0 (-12.9) & 13.6 \\
& 1d & n (n) & n & n (-0.8) & n & -2.1 (-3.6) & 11.1 \\
\hline
$^{90}_{\phi}\text{Zr}$ & 1s & -12.9 (-13.6) & 17.1 & -19.0 (-19.5) & 16.4 & -23.6 (-24.0) & 15.8 \\
& 1p & -7.1 (-8.4) & 15.5 & -12.8 (-13.6) & 15.2 & -17.2 (-17.8) & 14.8 \\
& 1d & -0.2 (-2.5) & 13.4 & -5.6 (-6.9) & 13.5 & -9.7 (-10.6) & 13.4 \\
& 2s & n (-1.4) & n & -3.4 (-5.1) & 12.6 & -7.4 (-8.5) & 12.7 \\
& 2p & n (n) & n & n (n) & n & n (-1.1) & n \\
\hline
$^{208}_{\phi}\text{Pb}$ & 1s & -15.0 (-15.5) & 17.4 & -21.1 (-21.4) & 16.6 & -25.8 (-26.0) & 16.0
\\
& 1p & -11.4 (-12.1) & 16.7 & -17.4 (-17.8) & 16.0 & -21.9  (-22.2) & 15.5 \\
& 1d & -6.9 (-8.1) & 15.7 & -12.7 (-13.4) & 15.2 & -17.1 (-17.6) & 14.8 \\
& 2s & -5.2 (-6.6) & 15.1 & -10.9 (-11.7) & 14.8 & -15.2 (-15.8) & 14.5 \\
& 2p & n (-1.9) & n & -4.8 (-6.1) & 13.5 & -8.9 (-9.8) & 13.4 \\
& 2d & n (n) & n & n (-0.7) & n & -2.2 (-3.7) & 11.9 \\
\hline \hline
\end{tabular}}
\end{center}
\end{table}

In Table~\ref{tab:phibse} we show our results for the  $\phi$-nucleus bound
state energies and half widths, obtained with and without the imaginary
part of the potential, for three values of the cutoff parameter~\cite{Cobos-Martinez:2017woo}.

We first analyse the case in which the imaginary part of the $\phi$-nucleus
potential, $W_{\phi}(r)$, is set to zero. These results are shown in
parenthesis in Table~\ref{tab:phibse}. From the values shown in parenthesis,
we see that the $\phi$-meson is expected to form bound states with all the
seven nuclei selected, for all values of the cutoff parameter $\Lambda_{K}$
studied.  However, the bound state energy is obviously dependent on
$\Lambda_{K}$, increasing in magnitude with $\Lambda_K$~\cite{Cobos-Martinez:2017woo}.

Next, we discuss the results obtained when the imaginary part of the potential is retained. 
Adding the absorptive part of the potential, 
the situation changes considerably. From the results presented in
Table~\ref{tab:phibse} we note that for the largest value of the cutoff parameter,
which yields the deepest attractive potentials, the $\phi$-meson is
expected to form bound states in all the nuclei selected, including the
lightest $^4$He nucleus. However, in this case, whether or not the bound
states can be observed experimentally, is sensitive to the value of the cutoff parameter
$\Lambda_K$~\cite{Cobos-Martinez:2017woo}. Given that the widths are large,
the signal for the formation of the $\phi$-nucleus bound states may be difficult to identify
experimentally.

We also observe that the width of the bound state is insensitive to the
values of $\Lambda_{K}$ for all nuclei. Furthermore, since the so-called
dispersive effect of the absorptive potentials are repulsive,
the bound states disappear completely in some cases, even though they were
found when the absorptive part was set to zero~\cite{Cobos-Martinez:2017woo}. 
This feature is obvious for the  $^4$He nucleus, making it especially relevant to the
future experiments, planned at J-PARC and JLab using light and medium-heavy
nuclei~\cite{Buhler:2010zz, Ohnishi:2014xla,Csorgo:2014sat,JLabphi}.

%
\begin{table}[ht!]
\begin{center}
\caption{\label{tab:etac-A-kg-be} $\eta_c$-nucleus bound
state energies for different values of the cutoff parameter $\Lambda_{D}$.
All dimensional quantities are in MeV.}
\scalebox{0.75}{
\begin{tabular}{ll|r|r|r|r}
  \hline \hline
  & & \multicolumn{4}{c}{Bound state energies} \\
  \hline
& $n\ell$ & $\Lambda_{D}=1500$ & $\Lambda_{D}=2000$ & $\Lambda_{D}=2500$ & 
$\Lambda_{D}= 3000$ \\
\hline
$^{4}_{\eta_{c}}\text{He}$
    & 1s & -1.49 & -3.11 & -5.49 & -8.55 \\
\hline
$^{12}_{\eta_{c}}\text{C}$
    & 1s & -5.91 & -8.27 & -11.28 & -14.79 \\
    & 1p & -0.28 & -1.63 & -3.69  & -6.33 \\
\hline
$^{16}_{\eta_{c}}\text{O}$
    & 1s & -7.35 & -9.92 & -13.15 & -16.87 \\
    & 1p & -1.94 & -3.87 & -6.48  & -9.63 \\
\hline
$^{40}_{\eta_{c}}\text{Ca}$
    & 1s & -11.26 & -14.42 & -18.31 & -22.73 \\
    & 1p & -7.19  & -10.02 & -13.59 & -17.70 \\
    & 1d & -2.82  &  -5.22 &  -8.36 & -12.09 \\
    & 2s & -2.36  &  -4.51 &  -7.44 & -10.98 \\
\hline
$^{48}_{\eta_{c}}\text{Ca}$
    & 1s & -11.37 & -14.46 & -18.26 & -22.58 \\
    & 1p & -7.83  & -10.68 & -14.23 & -18.32 \\
    & 1d & -3.88  & -6.40  & -9.63  & -13.41 \\
    & 2s & -3.15  & -5.47  & -8.54  & -12.17 \\
\hline
$^{90}_{\eta_{c}}\text{Zr}$
    & 1s & -12.26 & -15.35 & -19.14 & -23.43 \\
    & 1p & -9.88  & -12.86 & -16.53 & -20.70 \\
    & 1d & -7.05  & -9.87  & -13.38 & -17.40 \\
    & 2s & -6.14  & -8.87  & -12.29 & -16.24 \\
    & 1f & -3.90  & -6.50  & -9.81  & -13.65 \\
\hline
$^{197}_{\eta_{c}}\text{Au}$
      & 1s & -12.57 & -15.59 & -19.26 & -23.41 \\
      & 1p & -11.17 & -14.14 & -17.77 & -21.87 \\
      & 1d & -9.42  & -12.31 & -15.87 & -19.90 \\
      & 2s & -8.69  & -11.53 & -15.04 & -19.02 \\
      & 1f & -7.39  & -10.19 & -13.70 & -17.61 \\
\hline
$^{208}_{\eta_{c}}\text{Pb}$
    & 1s & -12.99 & -16.09 & -19.82 & -24.12 \\
    & 1p & -11.60 & -14.64 & -18.37 & -22.59 \\
    & 1d & -9.86  & -12.83 & -16.49 & -20.63 \\
    & 2s & -9.16  & -12.09 & -15.70 & -19.80 \\
    & 1f & -7.85  & -10.74 & -14.30 & -18.37 \\
\hline 
\hline
\end{tabular}
}
\end{center}
\end{table}

The bound state energies $E$ of the $\eta_c$-nucleus system were
calculated for four values of the cutoff parameter $\Lambda_{D}$ and are
listed in Table~\ref{tab:etac-A-kg-be}~\cite{Cobos-Martinez:2020ynh}.
Note that, the $\eta_c$ bound state energies are calculated with
the SU(4) broken coupling constant by $(0.6/\sqrt{2}) g_{eta_c DD}$, thus the values shown
below are expected to be smaller in magnitude than those calculated with
the SU(4) symmetric coupling constant, $g_{eta_c DD}$. (See Sec.~\ref{etacpsi}.)
These results show that the $\eta_c$-meson is expected to form bound
states with all the nuclei studied, and this prediction is independent
of the value of the cutoff parameter $\Lambda_{D}$~\cite{Cobos-Martinez:2020ynh}.
However, the particular values for the bound state energies are
clearly dependent on $\Lambda_{D}$, namely, each of them increases  in
absolute value  as $\Lambda_D$ increases. This was expected from the
behavior of the $\eta_c$ potentials, since these are deeper for larger
values of the cutoff parameter. Note also that the $\eta_c$ bounds more
strongly to heavier nuclei~\cite{Cobos-Martinez:2020ynh}..

We remind that we have ignored the natural width o $\sim 31$ MeV~\cite{PDG2024},
in free space of the $\eta_c$ and this could be an issue related to the observability of the
predicted bound states. Furthermore, we have no reason to believe the width
will be suppressed in medium. Thus, even though it could be difficult to
resolve the  individual states, it should be possible to see that there
are bound states which is the main point here. It remains to be seen by 
how much the inclusion of a repulsive imaginary part will affect the predicted bound states. We
believe this can be done in future work.

\begin{table}[htb]
\begin{center}
\caption{\label{tab:jpsi-nucleus-be} $J/\psi-$nucleus bound state
  energies taking into account the change in the self-energy in medium,  calculated 
  with the Schr\"{o}dinger equation. All dimensioned quantities are given in MeV.
  \vspace{1ex}
  }
\begin{tabular}{ll|r|r|r|r|r}
  \hline 
  & & \multicolumn{5}{c}{Bound state energies} \\
  \hline
  & & $\Lambda_{D}=2000$ & $\Lambda_{D}=3000$ & $\Lambda_{D}= 4000$ &
$\Lambda_{D}= 5000$ & $\Lambda_{D}= 6000$ \\
\hline
$^{4}_{J/\psi}{\rm He}$ & 1s & n & n & -0.70 & -2.70 & -5.51 \\
\hline
$^{12}_{J/\psi}{\rm C}$ & 1s & -0.52 & -1.98 & -4.47 & -7.67 & -11.26 \\
                      & 1p & n & n & n & -1.38 & -3.84 \\
\hline
$^{16}_{J/\psi}{\rm O}$ & 1s & -1.03 & -2.87 & -5.72 & -9.24 & -13.09 \\
                      & 1p & n & n & -0.94 & -3.48 & -6.60 \\
\hline
$^{40}_{J/\psi}{\rm Ca}$ & 1s & -2.78 & -5.44 & -9.14 & -13.50 & -18.12 \\
                       & 1p & -0.38 & -2.32 & -5.43 & -9.32 & -13.56 \\
                       & 1d & n & n & -1.52 & -4.74 & -8.49 \\
                       & 2s & n & n & -1.27 & -4.09 & -7.60 \\
\hline
$^{48}_{J/\psi}{\rm Ca}$ & 1s & -2.96 & -5.62 & -9.28 & -13.55 & -18.08 \\
                       & 1p & -0.73 & -2.83 & -6.03 & -9.95 & -14.18 \\
                       & 1d & n & n & -2.46 & -5.87 & -9.73 \\
                       & 2s & n & -0.07 & -1.90 & -5.00 & -8.65 \\
\hline
$^{90}_{J/\psi}{\rm Zr}$ & 1s & -3.64 & -6.40 & -10.12 & -14.41 & -18.92 \\
                       & 1p & -1.93 & -4.42 & -7.92 & -12.03 & -16.40 \\
                       & 1d & -0.03 & -2.13 & -5.31 & -9.18 & -13.37 \\
                       & 2s & -0.02 & -1.56 & -4.51 & -8.26 & -12.37 \\
                       & 2p & n & n & -1.52 & -4.71 & -8.45 \\
\hline
$^{208}_{J/\psi}{\rm Pb}$ & 1s & -4.25 & -7.08 & -10.82 & -15.11 & -19.60 \\
                      & 1p & -3.16 & -5.86 & -9.52 & -13.74 & -18.18 \\
                      & 1d & -1.84 & -4.38 & -7.90 & -12.01 & -16.37 \\
                      & 2s & -1.41 & -3.81 & -7.25 & -11.30 & -15.61 \\
                      & 2p & -0.07 & -1.95 & -5.10 & -8.97 & -13.14 \\
\hline 
\end{tabular}
\end{center}
\end{table}

The results for the $J/\psi$-nucleus bound states are presented in
Table~\ref{tab:jpsi-nucleus-be}. These results show that the $J/\psi$ is
expected to form $J/\psi$-nuclear bound states for nearly all the nuclei
considered, except some cases for $^4$He, for all values of the cutoff
parameter $\Lambda_D$~\cite{Cobos-Martinez:2021ukw,CobosMartinez:2021bia}. 
Therefore, it will be possible to search for the
bound states, for example, in a $^{208}$Pb nucleus at JLab the 12 GeV upgraded facility.
In addition, one can expect quite rich spectra for medium
and heavy mass nuclei. Of course, the main issue is to produce the $J/\psi$
meson with nearly stopped kinematics, or nearly zero momentum relative to
the nucleus. Since the present results imply that many nuclei should form 
$J/\psi$-nuclear bound states, it may be possible to find such kinematics
by careful selection of the beam and target
nucleus~\cite{Cobos-Martinez:2021ukw,CobosMartinez:2021bia}.

The bound state energies $E$ of the $\Upsilon$-nucleus and $\eta_b$-nucleus
systems are listed in Tables~\ref{tab:upsilon-A-BSE} to~\ref{tab:etab-A-BSE-bssl}, respectively,
for all nuclei listed at the beginning of this section and the same
range of values for the cutoff mass
parameter as used in the mass shift calculation (see
Sec.~\ref{upsietab})~\cite{Cobos-Martinez:2022fmt}.
We note that for the $\Upsilon$-nucleus systems we have only listed a few bound
states for each nucleus, since that number increases with the mass of the
nucleus and for the heaviest of these, $^{208}$Pb, the number of bound states quiet
large. For the  $^{208}$Pb} nucleus we have found $\sim$ 70
states~\cite{Cobos-Martinez:2022fmt}.
%
\begin{table}[h]
  \caption{\label{tab:upsilon-A-BSE} $\Upsilon$-nucleus bound state
  energies obtained by the Woods-Saxon Fourier transform for several nuclei $A$. All dimensioned
quantities are in MeV.}
\begin{center}
\scalebox{0.8}{
\begin{tabular}{ll|r|r|r|r|r}
  \hline \hline
  & & \multicolumn{5}{c}{Bound state energies} \\
  \hline
& $n\ell$ & $\Lambda_{B}=2000$ & $\Lambda_{B}=3000$ & $\Lambda_{B}= 4000$ &
$\Lambda_{B}= 5000$ & $\Lambda_{B}= 6000$ \\
\hline
$^{4}_{\Upsilon}\text{He}$
& 1s &  -5.6 &  -6.4 & -7.5 & -9.0 & -10.8 \\
\hline
$^{12}_{\Upsilon}\text{C}$
& 1s & -10.6 & -11.6 & -12.8 & -14.4 & -16.3 \\
& 1p & -6.1 & -6.8 & -7.9 & -9.3 & -10.9 \\
& 1d & -1.5 & -2.1 & -2.9 & -4.0 & -5.4 \\
& 2s & -1.6 & -2.1 & -2.8 & -3.8 & -5.1 \\
\hline
$^{16}_{\Upsilon}\text{O}$
& 1s & -11.9 & -12.9 & -14.2 & -15.8 & -17.8 \\
& 1p & -8.3 & -9.2 & -10.4 & -11.9 & -13.7 \\
& 1d & -4.4 & -5.1 & -6.2 & -7.5 & -9.2 \\
& 2s & -3.7 & -4.4 & -5.4 & -6.7 & -8.3 \\
& 1f & n & -0.9 & -1.8 & -2.9 & -4.3 \\
\hline
$^{40}_{\Upsilon}\text{Ca}$
& 1s & -15.5 & -16.6 & -18.2 & -20.0 & -22.3 \\
& 1p & -13.3 & -14.4 & -15.9 & -17.7 & -19.8 \\
& 1d & -10.8 & -11.9 & -13.3 & -15.0 & -17.1 \\
& 2s & -10.3 & -11.3 & -12.7 & -14.4 & -16.4 \\
& 1f & -8.1 & -9.1 & -10.4 & -12.1 & -14.0 \\
\hline
$^{48}_{\Upsilon}\text{Ca}$
& 1s & -15.3 & -16.4 & -17.9 & -19.7 & -21.8 \\
& 1p & -13.5 & -14.6 & -16.0 & -17.8 & -19.9 \\
& 1d & -11.4 & -12.4 & -13.8 & -15.6 & -17.6 \\
& 2s & -10.8 & -11.8 & -13.2 & -14.9 & -16.9 \\
& 1f & -9.1 & -10.1 & -11.4 & -13.1 & -15.0 \\
\hline
$^{90}_{\Upsilon}\text{Zr}$
& 1s & -15.5 & -16.6 & -18.1 & -19.9 & -22.0 \\
& 1p & -14.5 & -15.5 & -17.0 & -18.8 & -20.9 \\
& 1d & -13.2 & -14.2 & -15.7 & -17.4 & -19.5 \\
& 2s & -12.7 & -13.8 & -15.2 & -16.9 & -19.0 \\
& 1f & -11.7 & -12.7 & -14.1 & -15.9 & -17.9 \\
\hline
$^{197}_{\Upsilon}\text{Au}$
& 1s & -15.3 & -16.3 & -17.7 & -19.4 & -21.5 \\
& 1p & -14.7 & -15.8 & -17.2 & -18.9 & -20.9 \\
& 1d & -14.0 & -15.0 & -16.4 & -18.1 & -20.1 \\
& 2s & -13.7 & -14.7 & -16.0 & -17.8 & -19.8 \\
& 1f & -13.2 & -14.2 & -15.6 & -17.3 & -19.3 \\
\hline
$^{208}_{\Upsilon}\text{Pb}$
& 1s & -15.7 & -16.8 & -18.2 & -20.0 & -22.1 \\
& 1p & -15.2 & -16.2 & -17.7 & -19.4 & -21.5 \\
& 1d & -14.5 & -15.5 & -16.9 & -18.7 & -20.8 \\
& 2s & -14.1 & -15.2 & -16.6 & -18.3 & -20.4 \\
& 1f & -13.6 & -14.7 & -16.1 & -17.8 & -19.9 \\
\hline
\end{tabular}
}
\end{center}
\end{table}

\begin{table}[h]
  \caption{\label{tab:upsilon-A-BSE-bssl} $\Upsilon$-nucleus bound state
  energies obtained by the Direct Bessel transform for several nuclei $A$. 
All dimensioned quantities are in MeV.}
\begin{center}
\scalebox{0.8}{
\begin{tabular}{ll|r|r|r}
  \hline \hline
  & & \multicolumn{3}{c}{Bound state energies (MeV)} \\
  \hline
& & \multicolumn{3}{c}{Direct Bessel transform} \\
\hline
& $n\ell$ & $\Lambda_{B}=2000$ & $\Lambda_{B}= 4000$ &
$\Lambda_{B}= 6000$ \\
\hline
$^{4}_{\Upsilon}\text{He}$
& 1s & -5.93 & -6.25 & -6.56 \\
\hline
$^{12}_{\Upsilon}\text{C}$
& 1s & -13.22 & -15.26 & -18.41 \\
& 1p & -8.30 & -9.57 & -11.51 \\
\hline
$^{16}_{\Upsilon}\text{O}$
& 1s & -14.30 & -16.57 & -20.06 \\
& 1p & -10.81 & -12.37 & -14.73 \\
\hline
$^{40}_{\Upsilon}\text{Ca}$
& 1s & -18.17 & -21.63 & -23.16 \\
& 1p & -15.22 & -18.11 & -19.58 \\
\hline
$^{48}_{\Upsilon}\text{Ca}$
& 1s & -16.74 & -19.33 & -23.20 \\
& 1p & -15.36 & -17.76 & -21.53 \\
\hline
$^{90}_{\Upsilon}\text{Zr}$
& 1s & -15.87 & -18.24 & -21.89 \\
& 1p & -12.52 & -14.78 & -18.32 \\
\hline
$^{208}_{\Upsilon}\text{Pb}$
& 1s & -15.95 & -18.41 & -22.23 \\
& 1p & -13.23 & -15.49 & -19.91 \\
\hline
\end{tabular}
}
\end{center}
\end{table}
%
In Table~\ref{tab:etab-A-BSE} we show the $\eta_b$-nucleus bound state energies
for the same nuclei and range of values of the cutoff mass parameter as in
Table~\ref{tab:upsilon-A-BSE}~\cite{Cobos-Martinez:2022fmt}.
Furthermore, as in the case of the  $\Upsilon$-nucleus bound
state energies, for each nucleus we have listed only a few bound states.
For the $^{208}$Pb nucleus we have $\sim$ 200 states and clearly is not
practical to show them all~\cite{Cobos-Martinez:2022fmt}.

%
\begin{table}[h]
  \caption{\label{tab:etab-A-BSE} $\eta_b$-nucleus bound state
  energies obtained by the Woods-Saxon Fourier transform for several nuclei $A$. All dimensioned
quantities are in MeV.}
\begin{center}
\scalebox{0.8}{
\begin{tabular}{ll|r|r|r|r|r}
  \hline \hline
  & & \multicolumn{5}{c}{Bound state energies} \\
  \hline
& $n\ell$ & $\Lambda_{B}=2000$ & $\Lambda_{B}=3000$ & $\Lambda_{B}= 4000$ &
$\Lambda_{B}= 5000$ & $\Lambda_{B}= 6000$ \\
\hline
$^{4}_{\eta_b}\text{He}$
& 1s & -63.1 & -64.7 & -66.7 & -69.0 & -71.5 \\ 
& 1p & -40.6 & -42.0 & -43.7 & -45.8 & -48.0 \\ 
& 1d & -17.2 & -18.3 & -19.7 & -21.4 & -23.2 \\ 
& 2s & -15.6 & -16.6 & -17.9 & -19.4 & -21.1 \\ 
\hline
$^{12}_{\eta_b}\text{C}$
& 1s & -65.8 & -67.2 & -69.0 & -71.1 & -73.4 \\
& 1p & -57.0 & -58.4 & -60.1 & -62.1 & -64.3 \\
& 1d & -47.5 & -48.8 & -50.4 & -52.3 & -54.4 \\
& 2s & -46.3 & -47.5 & -49.1 & -51.0 & -53.0 \\
& 1f & -37.5 & -38.7 & -40.2 & -42.0 & -43.9 \\
\hline
$^{16}_{\eta_b}\text{O}$
& 1s & -67.8 & -69.2 & -71.0 & -73.1 & -75.4 \\
& 1p & -61.8 & -63.2 & -64.9 & -67.0 & -69.2 \\
& 1d & -54.9 & -56.2 & -57.9 & -59.9 & -62.0 \\
& 2s & -53.2 & -54.6 & -56.3 & -58.2 & -60.3 \\
& 1f  & -47.3 & -48.6 & -50.2 & -52.1 & -54.2 \\
\hline
$^{40}_{\eta_b}\text{Ca}$
& 1s & -79.0 & -80.6 & -82.6 & -85.0 & -87.5 \\
& 1p & -75.4 & -77.0 & -79.0 & -81.4 & -83.9 \\
& 1d & -71.4 & -73.0 & -74.9 & -77.2 & -79.7 \\
& 2s & -70.5 & -72.0 & -74.0 & -76.3 & -78.8 \\
& 1f & -67.0 & -68.5 & -70.4 & -72.7 & -75.1 \\
\hline
$^{48}_{\eta_b}\text{Ca}$
& 1s & -76.7 & -78.2 & -80.2 & -82.5 & -85.0 \\
& 1p & -74.0 & -75.5 & -77.4 & -79.7 & -82.1 \\
& 1d & -70.8 & -72.3 & -74.2 & -76.4 & -78.8 \\
& 2s & -69.9 & -71.4 & -73.3 & -75.5 & -77.9 \\
& 1f & -67.2 & -68.6 & -70.6 & -72.8 & -75.1 \\
\hline
$^{90}_{\eta_b}\text{Zr}$
& 1s & -75.5 & -77.0 & -78.9 & -81.1 & -83.5 \\
& 1p & -74.1 & -75.6 & -77.5 & -79.7 & -82.1 \\ 
& 1d & -72.3 & -73.8 & -75.7 & -77.9 & -80.2 \\ 
& 2s & -71.6 & -73.0 & -74.9 & -77.1 & -79.5 \\ 
& 1f & -70.2 & -71.7 & -73.6 & -75.8 & -78.1 \\ 
\hline
$^{197}_{\eta_b}\text{Au}$
& 1s & -72.8 & -74.2 & -76.1 & -78.2 & -80.5 \\ 
& 1p & -72.3 & -73.7 & -75.6 & -77.7 & -80.0 \\ 
& 1d & -71.3 & -72.8 & -74.6 & -76.7 & -79.0 \\ 
& 2s & -70.7 & -72.1 & -74.0 & -76.1 & -78.4 \\ 
& 1f & -70.2 & -71.7 & -73.5 & -75.6 & -77.9 \\ 
\hline
$^{208}_{\eta_b}\text{Pb}$
& 1s & -74.7 & -76.2 & -78.1 & -80.3 & -82.6 \\ 
& 1p & -74.2 & -75.7 & -77.5 & -79.7 & -82.1 \\ 
& 1d & -73.2 & -74.7 & -76.6 & -78.8 & -81.1 \\ 
& 2s & -72.7 & -74.1 & -76.0 & -78.2 & -80.5 \\ 
& 1f & -72.1 & -73.6 & -75.5 & -77.6 & -80.0 \\ 
\hline
\end{tabular}
}
\end{center}
\end{table}

\begin{table}[h]
  \caption{\label{tab:etab-A-BSE-bssl} $^{4}_{\eta_b}\text{A}$ bound state
  energies obtained by the Direct Bessel transform for several nuclei $A$. 
All dimensioned quantities are in MeV.}
\begin{center}
\scalebox{0.8}{
\begin{tabular}{ll|r|r|r}
  \hline \hline
  & & \multicolumn{3}{c}{Bound state energies (MeV)} \\
  \hline
& & \multicolumn{3}{c}{Direct Bessel transform} \\
\hline
& $n\ell$ & $\Lambda_{B}=2000$ & $\Lambda_{B}= 4000$ &
$\Lambda_{B}= 6000$ \\
\hline
$^{4}_{\eta_b}\text{He}$
& 1s & -68.71 & -71.59 & -75.44 \\
& 1p & -39.97 & -41.50 & -43.54 \\
& 1d & -37.73 & -39.56 & -42.03 \\
& 2s & -29.14 & -30.09 & -31.38 \\
\hline
$^{12}_{\eta_b}\text{C}$
& 1s & -63.70 & -66.93 & -70.27 \\
& 1p & -53.17 & -55.13 & -59.38 \\
& 1d & -46.47 & -48.50 & -51.17 \\
& 2s & -34.53 & -36.30 & -39.43 \\
\hline
$^{16}_{\eta_b}\text{O}$
& 1s & -68.37 & -71.25 & -75.14 \\
& 1p & -57.02 & -59.58 & -63.02 \\
& 1d & -47.05 & -49.37 & -52.50 \\
& 2s & -23.18 & -25.50 & -28.69 \\
\hline
$^{40}_{\eta_b}\text{Ca}$
& 1s & -79.11 & -82.59 & -87.27 \\
& 1p & -70.60 & -73.86 & -78.26 \\
& 1d & -53.31 & -55.99 & -59.61 \\
& 2s & -48.35 & -51.31 & -55.32 \\
\hline
$^{48}_{\eta_b}\text{Ca}$
& 1s & -63.94 & -66.88 & -70.83 \\
& 1p & -58.60 & -61.10 & -64.43 \\
& 1d & -34.04 & -36.40 & -39.60 \\
& 2s & -26.35 & -28.30 & -30.95 \\
\hline
$^{90}_{\eta_b}\text{Zr}$
& 1s & -71.32 & -74.52 & -78.85 \\
& 1p & -63.78 & -67.03 & -71.42 \\
& 1d & -57.78 & -60.82 & -64.93 \\
& 2s & -51.53 & -54.07 & -57.46 \\
\hline
$^{208}_{\eta_b}\text{Pb}$
& 1s & -61.44 & -64.25 & -68.02 \\
& 1p & -59.82 & -62.95 & -67.18 \\
& 1d & -51.36 & -54.05 & -57.65 \\
& 2s & -48.71 & -51.25 & -54.66 \\
\hline
\end{tabular}
}
\end{center}
\end{table}
These results given in Tables~\ref{tab:upsilon-A-BSE} to 
\ref{tab:etab-A-BSE-bssl} show that the $\Upsilon$ and $\eta_b$ mesons are
expected to form bound states with all the nuclei studied, independent of
the value  of the cutoff parameter $\Lambda_{B}$. However, the particular
values  for the bound state energies are dependent on the cutoff parameter  values,
increasing in absolute value as the cutoff parameter increases.
This dependence was  expected from the behavior of the
bottomonium-nucleus potentials, since these are more attractive for larger values of the cutoff parameter.
Note also that bottomonium ($\eta_b$ or $\Upsilon$) bounds more
strongly to heavier nuclei and
therefore a richer spectrum is expected for these nuclei~\cite{Cobos-Martinez:2022fmt}.

However, from Tables ~\ref{tab:upsilon-A-BSE}  to~\ref{tab:etab-A-BSE-bssl},
we see that the  bound state energies for the $\eta_b$ are larger than
those of the $\Upsilon$ for the same nuclei and range of cutoff values
explored. These differences are probably due to two reasons: ({\bf a}) 
the couplings $g_{\eta_b BB^*}$ and $g_{\Upsilon BB}$ are very different.
Indeed, the results obtained in Ref.~\cite{Cobos-Martinez:2020ynh} on the
$\eta_c$ nuclear bound state energies are closer to those of the $J/\psi$
when the $SU(4)$ flavor symmetry is broken, such that 
$g_{\eta_c DD^*}=(0.6/\sqrt{2})\,g_{J/\psi DD} \simeq 0.424\,g_{J/\psi
DD}$~\cite{Cobos-Martinez:2020ynh,Lucha:2015dda}.
Thus, a reduced coupling $g_{\eta_b BB^*}$ can bring the $\eta_b$ nuclear
bound state energies closer to those the $\Upsilon$,
since the $\eta_b$ self-energy is proportional to $g_{\eta_b BB^*}^2$.
({\bf b}) the form factors are not equal for the vertices
$\Upsilon B B$ and $\eta_b B B^*$, and we have to
readjust the cutoff values, which means $\Lambda_B\ne \Lambda_{B^*}$, and
the comparisons for the mass shift values and bound state energies have to be
made for different values of the cutoff parameters.

The bound-state energies associated with each energy level can be
confirmed by analyzing the number of nodes in the
corresponding coordinate space wave function as described in Ref.~\cite{Zeminiani:2024rrh}.
We present the wave functions of some meson-nucleus systems and cutoff values in
Appendix~\ref{ApxA},
which will help to understand better the meson-nucleus bound systems.

\begin{table}[t]
\begin{center}
  \caption{\label{tab:eta-A-kge-complex} Bound state energies ($E$)
  and full widths ($\Gamma$) of $\eta$ meson in nucleus of mass
  number $A$ obtained by solving the Klein-Gordon equation for various
  values of the parameter $\gamma$.
}
\scalebox{0.85}{
\begin{tabular}{ll|rr|rr|rr|rr}
  \hline \hline
  & &  \multicolumn{2}{c}{$\gamma=0$}   &
  \multicolumn{2}{c}{$\gamma=0.25$}   &
  \multicolumn{2}{c}{$\gamma=0.5$} & \multicolumn{2}{c}{$\gamma=1.0$}  \\
  \hline
& $n\ell$ & $E$ & $\Gamma$ & $E$ & $\Gamma$ & $E$ & $\Gamma$ & $E$ & 
$\Gamma$ \\
  \hline
  $^{4}_{\eta}\text{He}$
 & $1s$ &  -10.99 & 0 & -10.79 &  8.21 & -10.20 & 16.65 & -8.13 & 34.94 \\
  \hline
$^{12}_{\eta}\text{C}$
& $1s$ &  -25.25 & 0 & -25.16 & 10.86 & -24.91 & 21.82 & -24.02 &  44.29 \\
& $1p$ &   -0.87 & 0 & -0.43 & 4.97  & N & N & N & N \\
  \hline
$^{16}_{\eta}\text{O}$
& $1s$ &  -30.78 & 0 & -30.72 & 12.00 & -30.53 & 24.07 & -29.86 & 48.67 \\
& $1p$ &   -6.47 & 0 & -6.26 & 7.84 & -5.67 & 15.99  & -3.77 & 33.80 \\
  \hline
$^{40}_{\eta}\text{Ca}$
 & $1s$ &  -46.93 & 0 & -46.89 & 15.12 & -46.79 & 30.28 & -46.43 & 60.87 \\
& $1p$ &  -26.93 & 0 & -26.85 & 12.67 & -26.61 & 25.44 & -25.77 & 51.59 \\
& $1d$ &   -6.67 & 0 & -6.47 & 9.48 & -5.91 & 19.27  & -4.15 & 40.31 \\
& $2s$ &   -5.43 & 0 & -5.09 & 7.51 & -4.18 & 15.59 & N & N \\
  \hline
$^{48}_{\eta}\text{Ca}$
& $1s$ &  -47.78 & 0 & -47.75 & 14.98 & -47.66 & 30.00  & -47.38 & 60.25 \\
& $1p$ &  -29.97 & 0 & -29.90 & 12.99 & -29.71 & 26.06 & -29.04 & 52.71 \\
& $1d$ &  -11.08 & 0 &  -10.93 & 10.45  & -10.51 & 21.10 & -9.15 & 43.52 \\
& $2s$ &   -8.7 & 0 &  -8.11 & 8.83   & -7.42 & 18.06  & N & N \\
  \hline
  $^{90}_{\eta}\text{Zr}$
& $1s$ &  -52.56 & 0 & -52.54 & 15.34 & -52.50 & 30.71 & -52.34 & 61.56 \\
& $1p$ &  -39.85 & 0 & -39.81 & 14.17 & -39.71 & 28.40 & -39.36 & 57.11 \\
& $1d$ &  -25.32 & 0 & -25.25 & 12.74 & -25.06 & 25.57 & -24.40 & 51.75 \\
& $2s$ &  -21.04 & 0 & -20.94 & 11.95 & -20.65 & 24.04 & -19.70 & 49.03 \\
  \hline
$^{197}_{\eta}\text{Au}$
& $1s$ &  -55.12 & 0 & -55.11 & 15.20 & -55.09 & 30.41  & -55.01 & 60.89 \\
& $1p$ &  -47.13 & 0 & -47.11 & 14.58 & -47.06 & 29.19 & -46.90 & 58.53 \\
& $1d$ &  -37.60 & 0 & -37.58 & 13.83 & -37.49 & 27.69 & -37.20 & 55.67 \\
& $2s$ &  -34.01 & 0 & -33.97 & 13.45  & -33.86 & 26.96 & -33.47 & 54.31 \\
  \hline
$^{208}_{\eta}\text{Pb}$
& $1s$ &  -56.85 & 0 & -56.84 & 15.61 & -56.82 & 31.24 & -56.75 & 62.55 \\
& $1p$ &  -48.92 & 0 & -48.90 & 14.99 & -48.86 & 30.01  &  -48.70 &  60.17  \\
& $1d$ &  -39.81 & 0 & -39.45 &  14.24 & -39.37 & 28.51 & -39.09 & 57.29 \\
& $2s$ &  -35.95 & 0 & -35.91 & 13.87 & -35.80 & 27.80  & -35.43 & 55.96  \\
 \hline \hline
\end{tabular}
}
\end{center}
\end{table}

\begin{table}[t]
\begin{center}
 \caption{\label{tab:etaprime-A-kge-complex} Bound state energies ($E$)
  and full widths ($\Gamma$) of $\eta'$ meson in nucleus of mass
  number $A$ obtained by solving the Klein-Gordon equation for various values
  of the parameter $\gamma$.
}
\scalebox{0.85}{
\begin{tabular}{ll|rr|rr|rr|rr}
  \hline \hline
  & &  \multicolumn{2}{c}{$\gamma=0$}   &  \multicolumn{2}{c}{$\gamma=0.25$} &
  \multicolumn{2}{c}{$\gamma=0.5$} & \multicolumn{2}{c}{$\gamma=1.0$}  \\
  \hline
  & $n\ell$ & $E$ & $\Gamma$ & $E$ & $\Gamma$ & $E$ & $\Gamma$ &
  $E$ & $\Gamma$ \\
  \hline
  $^{4}_{\eta'}\text{He}$
 & $1s$ &  -22.11 & 0 & -21.96 & 11.37 & -21.55 & 22.89  & -20.06 &  46.83  \\
  \hline
$^{12}_{\eta'}\text{C}$
& $1s$ &  -33.88 & 0 & -33.82 & 12.30 &  -33.64 & 24.66 & -33.00 & 49.73  \\
& $1p$ &  -12.72 & 0 & -12.57 & 9.06 &  -12.15 & 18.29  & -10.67 & 37.68  \\
  \hline
$^{16}_{\eta'}\text{O}$
& $1s$ &  -38.64 & 0 & -38.59 & 13.06 &  -38.46 & 26.17 & -38.00 & 52.65  \\
& $1p$ &  -19.75 & 0 & -19.65 & 10.76 &  -19.34 & 21.64  & -18.28 & 44.07 \\
& $2s$ &   -1.39 & 0 & -0.84 & 4.48 &  N & N & N & N \\
& $1d$ &   -0.33 & 0 & -0.69 & 7.20 &  N & N & N & N \\
  \hline
$^{40}_{\eta'}\text{Ca}$
 & $1s$ &  -52.38 & 0 & -52.35 & 15.59 &  -52.28 & 31.22 & -52.00 & 62.61 \\
 & $1p$ &  -38.41 & 0 & -38.35 & 14.18 &  -38.19 & 28.41 & -37.63 & 57.22 \\
 & $1d$ &  -23.12 & 0 & -23.02 & 12.46 &  -22.74 & 25.03 & -21.75 & 50.81  \\
& $2s$ &  -20.38 & 0 & -20.25 & 11.72 &  -19.87 & 23.60 & -18.58 & 48.25  \\
  \hline
$^{48}_{\eta'}\text{Ca}$
& $1s$ &  -52.40 & 0 & -52.38 & 15.29 &  -52.32 & 30.60  & -52.11 & 61.35  \\
& $1p$ &  -40.30 & 0 & -40.26 & 14.18 &  -40.13 & 28.40 & -39.68 & 57.12  \\
& $1d$ &  -26.68 & 0 & -26.59 & 12.82 &  -26.37 & 25.72 & -25.58 & 52.02  \\
& $2s$ &  -23.45 & 0 & -23.34 & 12.19 &  -23.04 & 24.51  & -22.01 & 49.85  \\
  \hline
  $^{90}_{\eta'}\text{Zr}$ 
& $1s$ &  -55.20 & 0 & -55.19 & 15.31 &  -55.16 & 30.63  & -55.04 & 61.35  \\
& $1p$ &  -47.05 & 0 & -47.02 & 14.70  &  -46.96 & 29.43  & -46.72 & 59.04  \\
& $1d$ &  -37.42 & 0 & -37.38 & 13.96  &  -37.27 & 27.96  & -36.86 & 56.22 \\
& $2s$ &  -34.19 & 0 & -34.14 & 13.61 &  -33.99 & 27.29  & -33.47 & 54.98  \\
  \hline
$^{197}_{\eta'}\text{Au}$
& $1s$ &  -56.03 & 0 & -56.03 & 14.94  &  -56.01 & 29.89  & -55.96 & 59.83  \\
& $1p$ &  -51.12 & 0 & -51.10 & 14.64 &  -51.07 & 29.30  & -50.96 & 58.67 \\
& $1d$ &  -45.15 & 0 & -45.14 & 14.27 &  -45.08 & 28.56  & -44.89 & 57.26  \\
& $2s$ &  -42.80 & 0 & -42.78 & 14.10  &  -42.71 & 28.22  & -42.47 & 56.63  \\
  \hline
$^{208}_{\eta'}\text{Pb}$
& $1s$ &  -57.65 & 0 & -57.64 & 15.34 &  -57.63 & 30.68  & -57.57& 61.40 \\
& $1p$ &  -52.77 & 0 & -52.76 & 15.03 &  -52.73 & 30.07  & -52.62 & 60.23  \\
& $1d$ &  -46.87 & 0 & -46.85 & 14.66 &  -46.80 & 29.33  & -46.61 & 58.80  \\
& $2s$ &  -44.56 & 0 & -44.54 & 14.49 &  -44.47 & 29.00  & -44.24 & 58.19 \\
 \hline \hline
\end{tabular}
}
\end{center}
\end{table}

In Tables~\ref{tab:eta-A-kge-complex} and~\ref{tab:etaprime-A-kge-complex},
we show, respectively, the results for the bound state energies ($E$) 
and full widths ($\Gamma$) of the $\eta$- and $\eta'$-mesic nuclei
of mass number $A$, obtained by solving the Klein-Gordon equation,
for various values of the strength of the imaginary part of the potential
$\gamma= 0.0,\, 0.25,\,0.5,\, 1.0$. (See below Eq.~(\ref{eqn:VhA}) about the $\gamma$.)
The results for $\gamma=0$, for both the $\eta$ and $\eta'$ mesons,
correspond to the case where the imaginary part of the potential
has been ignored. 
The bound state energies and full widths are obtained from the complex
energy eigenvalue $\mathcal{E}$ as $\mathcal{E}= E  + m - i\Gamma/2$.
We also note that for each nucleus, we have computed all bound states but
have only listed up to four. In fact, the number of bound states increases
with the mass of the nucleus
such that for the heavier nuclei we have a richer structure of bound
states. Furthermore, we note that the relativistic corrections
shallower the
bound state energies for the $\eta$ and $\eta'$ by approximately 2 MeV and
1 MeV, respectively.

From Tables~\ref{tab:eta-A-kge-complex} 
and~\ref{tab:etaprime-A-kge-complex}, (column with $\gamma= 0$) we
conclude that the $\eta$ and $\eta'$ are expected to form bound states
with all the nuclei considered. 

However, the situation changes appreciably once we take into account the
absorption effects of these mesons by nuclei, which we simulate with nonzero
phenomenological parameter $\gamma$. We study the values
$\gamma=0.25,\,0.5,\, 1.0$, where a larger values means a stronger
absorption of the meson by the nuclear medium. When $\gamma \ne 0$,
some of the bound states that present when $\gamma=0$ disappear.
The columns with $\gamma=0.25$, $\gamma=0.5$, and $\gamma=1.0$ in 
Tables~\ref{tab:eta-A-kge-complex} and~\ref{tab:etaprime-A-kge-complex}
show the results for the bound state energies $E$ and full widths 
$\Gamma$ of the $\eta$- and $\eta'$-mesic nuclei of mass number $A$,
obtained by solving the  Klein-Gordon equation, for some values of
the strength of the imaginary part of the potential
$\gamma= 0.25,\,0.5,\, 1.0$. 

Considering only the ground states, adding and absorptive part of the
potential changes the situation appreciably, where the effects are larger 
the larger $\gamma$ is. 
Clearly, the imaginary part of the potential is repulsive, being more
repulsive for $\gamma=1$.
Whether or not  the bound states can be observed experimentally is sensitive
to the value of the parameter $\gamma$, since $\Gamma$ increases with
increasing $\gamma$.
Furthermore, because the so-called dispersive effect of the absorptive
potential is repulsive, the binding energies for all nuclei decrease with
$\gamma$. However, they decrease very little. 
Even for the largest value of $\gamma$, there is at least one bound state. 
We have found similar results for the $\phi$ meson in our past
work~\cite{Cobos-Martinez:2017woo}.
Note that, the width of the ground state increases with $\gamma$ for all
nuclei, as expected, since a larger $\gamma$ means that the strength of the
imaginary part of the potential is larger. 

\begin{table}[ht!]
\caption{\label{tblplus} Bound state energies of $B^{-}_c$ in nucleus of mass number $A$, 
obtained by the Direct Bessel transform method. All dimensioned quantities are in MeV.}
\begin{center}
\begin{tabular}{ll|r|r|r}
  \hline \hline
  & & \multicolumn{3}{c}{Bound state energies (MeV)} \\
  \hline
& & \multicolumn{3}{c}{Direct Bessel transform} \\
\hline
& $n\ell$ & $\Lambda_{B}=2000$ & $\Lambda_{B}= 4000$ &
$\Lambda_{B}= 6000$ \\
\hline
$^{12}_{B^{-}_c}\text{C}$
& 1s & -79.12 & -80.63 & -87.03 \\
& 1p & -56.15 & -57.53 & -63.38 \\
\hline
$^{16}_{B^{-}_c}\text{O}$
& 1s & -75.00 & -76.16 & -80.94 \\
& 1p & -54.86 & -56.13 & -61.55 \\
\hline
$^{40}_{B^{-}_c}\text{Ca}$
& 1s & -104.27 & -105.69 & -111.87 \\
& 1p & -81.71 & -83.51 & -91.34 \\
\hline
$^{48}_{B^{-}_c}\text{Ca}$
& 1s & -96.63 & -98.37 & -105.81 \\
& 1p & -72.02 & -73.56 & -80.24 \\
\hline
$^{90}_{B^{-}_c}\text{Zr}$
& 1s & -96.34 & -98.32 & -106.82 \\
& 1p & -83.82 & -85.44 & -92.35 \\
\hline
$^{208}_{B^{-}_c}\text{Pb}$
& 1s & -95.88 & -97.39 & -103.79 \\
& 1p & -70.46 & -71.76 & -77.34 \\
\hline
\end{tabular}
\end{center}
\end{table}
\begin{table}[htb!]
\caption{\label{tblminus} Bound state energies of $B^{+}_c$ in nucleus of mass number $A$, 
obtained by the Direct Bessel transform method. All dimensioned quantities are in MeV.}
\begin{center}
\begin{tabular}{ll|r|r|r}
  \hline \hline
  & & \multicolumn{3}{c}{Bound state energies (MeV)} \\
  \hline
& & \multicolumn{3}{c}{Direct Bessel transform} \\
\hline
& $n\ell$ & $\Lambda_{B}=2000$ & $\Lambda_{B}= 4000$ &
$\Lambda_{B}= 6000$ \\
\hline
$^{12}_{B^{+}_c}\text{C}$
& 1s & -71.01 & -72.53 & -78.94 \\
& 1p & -49.11 & -50.49 & -56.32 \\
\hline
$^{16}_{B^{+}_c}\text{O}$
& 1s & -64.64 & -65.80 & -70.59 \\
& 1p & -45.94 & -47.22 & -52.64 \\
\hline
$^{40}_{B^{+}_c}\text{Ca}$
& 1s & -84.89 & -86.31 & -92.49 \\
& 1p & -62.80 & -64.57 & -72.23 \\
\hline
$^{48}_{B^{+}_c}\text{Ca}$
& 1s & -77.09 & -78.83 & -86.26 \\
& 1p & -53.64 & -55.13 & -61.60 \\
\hline
$^{90}_{B^{+}_c}\text{Zr}$
& 1s & -65.51 & -67.49 & -75.99 \\
& 1p & -55.11 & -56.75 & -63.75 \\
\hline
$^{208}_{B^{+}_c}\text{Pb}$
& 1s & -48.61 & -50.13 & -56.53 \\
& 1p & -29.27 & -30.58 & -36.22 \\
\hline
\end{tabular}
\end{center}
\end{table}

Finally, in Tables~\ref{tblplus} and~\ref{tblminus} we present the $B_c^{\pm}$-nucleus bound state energies for several nuclei restricting to the 1s and 1p states,
where we certainly expect more shallower baond states.
(More detailed results will be presented elsewhere in the near future.)
For details of the momentum space and the Coulomb potential treatment
focusing on the $^{12}_{B_c}$C case, see Ref.~\cite{Zeminiani:2024rrh}.
From Tables~\ref{tblplus} and~\ref{tblminus} we conclude that the
$B_c^{\pm}$ are expected to form bound states with all the nuclei studied.

\section{\label{conclusions} Summary and Conclusions}

We have computed the mass shift amount $\Delta m_h \equiv m_h^{*}-m_h$
with being $m_h$ the meson mass in vacuum and $m_h^{*}$ that in nuclear medium,
of the mesons $h = \eta$, $\eta'$, $\phi$, $\eta_c$, $J/\psi$,
$\eta_b$, $\Upsilon$, and $B^{\pm}_c$}, in symmetric nuclear matter and nuclei.
For this, we have used two approaches, namely the quark-meson coupling
model, and an hybrid approach that combines the quark-meson coupling model
with an effective Lagrangian, for some meson in this study.

We found in all cases that the mass shift amount (Lorentz scalar potential)
is negative, which means that
the nuclear medium provides attraction to these mesons (these mesons do not aquire any repulsive vector potentials)
and opens the possibility of their binding to nuclei.
 Even though the precise values for the negative mass shifts reported in this work are 
based on the quark-meson coupling model and effective lagrangians approach, negative
mass shifts have also been observed in other approaches and experimental results. Thus, we believe this is a roubus prediction of our approach. 
Using the baryon density distributions of several nuclei calculated
in the quark-meson coupling model except for $^{4}$He nucleus
(taken from Ref.~\cite{Saito:1997ae}), and the
mass shift amount computed previously, we have calculated
the meson-nucleus potentials
in a local density approximation for these mesons in nuclei in
a wide range of nuclear masses, namely $A= ^{4}\hspace{-1mm}\text{He}$,
$^{12}$C, $^{16}$O, $^{40}$Ca, $^{48}$Ca, $^{90}$Zr, $^{197}$Au, and
$^{208}$Pb. In all the nucleus cases selected for each meson,
the resulting nuclear potentials have turned out to be attractive,
reflecting the characteristics of the mass shift in the nuclear medium.

Finally, we have solved the Schr\"{o}dinger or Klein-Gordon equation with
the calculated nuclear potentials to obtain the meson-nucleus bound
state energies, and widths when the nuclear potential is complex.
Although the details differ for each meson, we have found that all
the mesons studied are expected to form bound states with nuclei.
For the nuclear potential is complex, the signal for the formation of the
meson-nucleus bound state might be difficult to identify experimentally,
depending on the obtained bound state energy imaginary part.

\authorcontributions{
All authors have contributed equally, read the manuscript, and
agreed to the published version of the manuscript. 
}

\dataavailability{

The data are available upon the agreement between the requested party and the authors.

}

\acknowledgments{J.J.C.M acknowlledges financial support from Universidad de Sonora under grant
USO315009105.
G.N.Z.~was supported by the Coordena\c{c}\~ao de Aperfei\c{c}oamento de Pessoal
de N\'ivel Superior- Brazil (CAPES), FAPESP Process No.~2023/073-3-6,
and Instituto Nacional de Ci\^{e}ncia e Tecnologia - Nuclear Physics
and Applications (INCT-FNA), Brazil, Process No.~464898/2014-5.
The work of K.T. was
supported by Conselho Nacional de Desenvolvimento
Cient\'{i}ıfico e Tecnol\'{o}gico (CNPq, Brazil), Processes
No. 304199/2022-2, and FAPESP No. 2023/07313-6, and his
work was also part of the projects, Instituto Nacional de
Ci\^{e}ncia e Tecnologia - Nuclear Physics and Applications
(INCT-FNA), Brazil, Process No. 464898/2014-5.
}

\conflictsofinterest{The authors declare no conflicts of interest.} 



\abbreviations{Abbreviations}{
The following abbreviations are used in this manuscript:\\

\noindent 
\begin{tabular}{@{}ll}
MDPI & Multidisciplinary Digital Publishing Institute\\
DOAJ & Directory of open access journals\\
TLA & Three letter acronym\\
LD & Linear dichroism
\end{tabular}
}

\appendixtitles{no} 
\appendixstart
\appendix
\section[\appendixname~\thesection]{Wave functions}
\label{ApxA}
The results for the $B^{\pm}_c$-nucleus system wave functions, for different
cutoff values, are presented in Figs.~\ref{fig:psirbcc12} to~\ref{fig:psirbcpb208}.
The wave functions for the $\Upsilon$- and $\eta_b$-nucleus systems,
for different methods of partial wave decomposition and
cutoff values, are also presented in Figs.~\ref{fig:psiryhe4} to~\ref{fig:psiretabpb208}.

The wave functions obtained when using the Bessel transform of the original
potential and the decomposition of the Fourier transform of the fitted Woods-Saxon potential
produce different shapes of wave function distributions at various energy levels.
These treatments give slight dependence of the bound-state energies
on the method used to obtain the partial-wave decomposition
of the momentum space potential. For the future possible improvements of the treatments,
and so that one can compare with different treatments, we present
all the wave functions obtained for the two different methods.
These will be very useful in the future.
However, we believe that the difference originated from the numerical
procedure and treatments, the difference will not change our main conclusions,
especially in connection with the accuracy required and achieved
associated with the strong interaction experimental measurement.

\begin{figure}[htb]
\centering
\scalebox{1.0}{
\begin{tabular}{cc}
    \includegraphics[scale=0.23]{Psir_Bc_AttCoul_C12_2000_1s1p.eps} &  
    \includegraphics[scale=0.23]{Psir_Bc_AttCoul_C12_4000_1s1p.eps} \\
    \includegraphics[scale=0.23]{Psir_Bc_AttCoul_C12_6000_1s1p.eps}  &
    \includegraphics[scale=0.23]{Psir_Bc_RepCoul_C12_2000_1s1p.eps} \\
    \includegraphics[scale=0.23]{Psir_Bc_RepCoul_C12_4000_1s1p.eps} &  
    \includegraphics[scale=0.23]{Psir_Bc_RepCoul_C12_6000_1s1p.eps}
  \end{tabular}
}
  \caption{\label{fig:psirbcc12}Coordinate space wave functions for the 1s and 1p states of the $B^{\pm}_c$-$^{12}$C
systems with the Coulomb potentials for
different values of $\Lambda$.}
\end{figure}

\begin{figure}[ht]
\centering
\scalebox{1.0}{
\begin{tabular}{cc}
    \includegraphics[scale=0.23]{Psir_Bc_AttCoul_O16_2000_1s1p.eps} &  
    \includegraphics[scale=0.23]{Psir_Bc_AttCoul_O16_4000_1s1p.eps} \\
    \includegraphics[scale=0.23]{Psir_Bc_AttCoul_O16_6000_1s1p.eps}  &
    \includegraphics[scale=0.23]{Psir_Bc_RepCoul_O16_2000_1s1p.eps} \\
    \includegraphics[scale=0.23]{Psir_Bc_RepCoul_O16_4000_1s1p.eps} &  
    \includegraphics[scale=0.23]{Psir_Bc_RepCoul_O16_6000_1s1p.eps}
  \end{tabular}
}
  \caption{\label{fig:psirbco16}Coordinate space wave functions for the 1s and 1p states of the $B^{\pm}_c$-$^{16}$O
systems with the Coulomb potentials for
different values of $\Lambda$.}
\end{figure}

\begin{figure}[ht]
\centering
\scalebox{1.0}{
\begin{tabular}{cc}
    \includegraphics[scale=0.23]{Psir_Bc_AttCoul_Ca40_2000_1s1p.eps} &  
    \includegraphics[scale=0.23]{Psir_Bc_AttCoul_Ca40_4000_1s1p.eps} \\
    \includegraphics[scale=0.23]{Psir_Bc_AttCoul_Ca40_6000_1s1p.eps}  &
    \includegraphics[scale=0.23]{Psir_Bc_RepCoul_Ca40_2000_1s1p.eps} \\
    \includegraphics[scale=0.23]{Psir_Bc_RepCoul_Ca40_4000_1s1p.eps} &  
    \includegraphics[scale=0.23]{Psir_Bc_RepCoul_Ca40_6000_1s1p.eps}
  \end{tabular}
}
  \caption{\label{fig:psirbcca40}Coordinate space wave functions for the 1s and 1p states of the $B^{\pm}_c$-$^{40}$Ca
systems with the Coulomb potentials for
different values of $\Lambda$.}
\end{figure}

\begin{figure}[ht]
\centering
\scalebox{1.0}{
\begin{tabular}{cc}
    \includegraphics[scale=0.23]{Psir_Bc_AttCoul_Ca48_2000_1s1p.eps} &  
    \includegraphics[scale=0.23]{Psir_Bc_AttCoul_Ca48_4000_1s1p.eps} \\
    \includegraphics[scale=0.23]{Psir_Bc_AttCoul_Ca48_6000_1s1p.eps}  &
    \includegraphics[scale=0.23]{Psir_Bc_RepCoul_Ca48_2000_1s1p.eps} \\
    \includegraphics[scale=0.23]{Psir_Bc_RepCoul_Ca48_4000_1s1p.eps} &  
    \includegraphics[scale=0.23]{Psir_Bc_RepCoul_Ca48_6000_1s1p.eps}
  \end{tabular}
}
  \caption{\label{fig:psirbcca48}Coordinate space wave functions for the 1s and 1p states of the $B^{\pm}_c$-$^{48}$Ca
systems with the Coulomb potentials for
different values of $\Lambda$.}
\end{figure}

\begin{figure}[ht]
\centering
\scalebox{1.0}{
\begin{tabular}{cc}
    \includegraphics[scale=0.23]{Psir_Bc_AttCoul_Zr90_2000_1s1p.eps} &  
    \includegraphics[scale=0.23]{Psir_Bc_AttCoul_Zr90_4000_1s1p.eps} \\
    \includegraphics[scale=0.23]{Psir_Bc_AttCoul_Zr90_6000_1s1p.eps}  &
    \includegraphics[scale=0.23]{Psir_Bc_RepCoul_Zr90_2000_1s1p.eps} \\
    \includegraphics[scale=0.23]{Psir_Bc_RepCoul_Zr90_4000_1s1p.eps} &  
    \includegraphics[scale=0.23]{Psir_Bc_RepCoul_Zr90_6000_1s1p.eps}
  \end{tabular}
}
  \caption{\label{fig:psirbczr90}Coordinate space wave functions for the 1s and 1p states of the $B^{\pm}_c$-$^{90}$Zr
systems with the Coulomb potentials for
different values of $\Lambda$.}
\end{figure}

\begin{figure}[ht]
\centering
\scalebox{1.0}{
\begin{tabular}{cc}
    \includegraphics[scale=0.23]{Psir_Bc_AttCoul_Pb208_2000_1s1p.eps} &  
    \includegraphics[scale=0.23]{Psir_Bc_AttCoul_Pb208_4000_1s1p.eps} \\
    \includegraphics[scale=0.23]{Psir_Bc_AttCoul_Pb208_6000_1s1p.eps}  &
    \includegraphics[scale=0.23]{Psir_Bc_RepCoul_Pb208_2000_1s1p.eps} \\
    \includegraphics[scale=0.23]{Psir_Bc_RepCoul_Pb208_4000_1s1p.eps} &  
    \includegraphics[scale=0.23]{Psir_Bc_RepCoul_Pb208_6000_1s1p.eps}
  \end{tabular}
}
  \caption{\label{fig:psirbcpb208}Coordinate space wave functions for the 1s and 1p states of the $B^{\pm}_c$-$^{208}$Pb
systems with the Coulomb potentials for
different values of $\Lambda$.}
\end{figure}


\begin{figure}[ht]
\centering
\scalebox{1.0}{
\begin{tabular}{cc}
    \includegraphics[scale=0.23]{Psir_Y_He4_2000_1s.eps} &  
    \includegraphics[scale=0.23]{Psir_Y_He4_4000_1s.eps} \\
    \includegraphics[scale=0.23]{Psir_Y_He4_6000_1s.eps}  &
    \includegraphics[scale=0.23]{Psir_WS_Y_He4_2000_1s.eps} \\
    \includegraphics[scale=0.23]{Psir_WS_Y_He4_4000_1s.eps} &  
    \includegraphics[scale=0.23]{Psir_WS_Y_He4_6000_1s.eps}
  \end{tabular}
}
  \caption{\label{fig:psiryhe4}Coordinate space 1s state wave functions of the $\Upsilon$-$^4$He system
for different values of cutoff $\Lambda$, obtained by the direct Bessel transform and by the Fourier transform of
the fitted Woods-Saxon form potential (FWS).}
\end{figure}

\begin{figure}[ht]
\centering
\scalebox{1.0}{
\begin{tabular}{cc}
    \includegraphics[scale=0.23]{Psir_Y_C12_2000_total.eps} &  
    \includegraphics[scale=0.23]{Psir_Y_C12_4000_total.eps} \\
    \includegraphics[scale=0.23]{Psir_Y_C12_6000_total.eps}  &
    \includegraphics[scale=0.23]{Psir_WS_Y_C12_2000_total.eps} \\
    \includegraphics[scale=0.23]{Psir_WS_Y_C12_4000_total.eps} &  
    \includegraphics[scale=0.23]{Psir_WS_Y_C12_6000_total.eps}
  \end{tabular}
}
  \caption{\label{fig:psiryc12}Coordinate space 1s and 1p state wave functions of
the $\Upsilon$-$^{12}$C system for different values of cutoff $\Lambda$, obtained by the direct Bessel transform and by the Fourier
transform of the fitted Woods-Saxon form potential (FWS).}
\end{figure}

\begin{figure}[ht]
\centering
\scalebox{1.0}{
\begin{tabular}{cc}
    \includegraphics[scale=0.23]{Psir_Y_O16_2000_total.eps} &  
    \includegraphics[scale=0.23]{Psir_Y_O16_4000_total.eps} \\
    \includegraphics[scale=0.23]{Psir_Y_O16_6000_total.eps}  &
    \includegraphics[scale=0.23]{Psir_WS_Y_O16_2000_total.eps} \\
    \includegraphics[scale=0.23]{Psir_WS_Y_O16_4000_total.eps} &  
    \includegraphics[scale=0.23]{Psir_WS_Y_O16_6000_total.eps}
  \end{tabular}
}
  \caption{\label{fig:psiryo16}Coordinate space 1s and 1p state wave functions of
the $\Upsilon$-$^{16}$O system for different values of cutoff $\Lambda$, obtained by the direct Bessel transform and by the Fourier transform of
the fitted Woods-Saxon form potential (FWS).}
\end{figure}

\begin{figure}[ht]
\centering
\scalebox{1.0}{
\begin{tabular}{cc}
    \includegraphics[scale=0.23]{Psir_Y_Ca40_2000_total.eps} &  
    \includegraphics[scale=0.23]{Psir_Y_Ca40_4000_total.eps} \\
    \includegraphics[scale=0.23]{Psir_Y_Ca40_6000_total.eps}  &
    \includegraphics[scale=0.23]{Psir_WS_Y_Ca40_2000_total.eps} \\
    \includegraphics[scale=0.23]{Psir_WS_Y_Ca40_4000_total.eps} &  
    \includegraphics[scale=0.23]{Psir_WS_Y_Ca40_6000_total.eps}
  \end{tabular}
}
  \caption{\label{fig:psiryca40}Coordinate space 1s and 1p state wave functions of
the $\Upsilon$-$^{40}$Ca system for different values of cutoff $\Lambda$, obtained by the direct Bessel transform and by the Fourier transform of
the fitted Woods-Saxon form potential (FWS).}
\end{figure}

\begin{figure}[ht]
\centering
\scalebox{1.0}{
\begin{tabular}{cc}
    \includegraphics[scale=0.23]{Psir_Y_Ca48_2000_total.eps} &  
    \includegraphics[scale=0.23]{Psir_Y_Ca48_4000_total.eps} \\
    \includegraphics[scale=0.23]{Psir_Y_Ca48_6000_total.eps}  &
    \includegraphics[scale=0.23]{Psir_WS_Y_Ca48_2000_total.eps} \\
    \includegraphics[scale=0.23]{Psir_WS_Y_Ca48_4000_total.eps} &  
    \includegraphics[scale=0.23]{Psir_WS_Y_Ca48_6000_total.eps}
  \end{tabular}
}
  \caption{\label{fig:psiryca48}Coordinate space 1s and 1p state wave functions of
the $\Upsilon$-$^{48}$Ca system for different values of cutoff $\Lambda$, obtained by the direct Bessel transform and by the Fourier transform of
the fitted Woods-Saxon form potential (FWS).}
\end{figure}

\begin{figure}[ht]
\centering
\scalebox{1.0}{
\begin{tabular}{cc}
    \includegraphics[scale=0.23]{Psir_Y_Zr90_2000_total.eps} &  
    \includegraphics[scale=0.23]{Psir_Y_Zr90_4000_total.eps} \\
    \includegraphics[scale=0.23]{Psir_Y_Zr90_6000_total.eps}  &
    \includegraphics[scale=0.23]{Psir_WS_Y_Zr90_2000_total.eps} \\
    \includegraphics[scale=0.23]{Psir_WS_Y_Zr90_4000_total.eps} &  
    \includegraphics[scale=0.23]{Psir_WS_Y_Zr90_6000_total.eps}
  \end{tabular}
}
  \caption{\label{fig:psiryzr90}Coordinate space 1s and 1p state wave functions of
the $\Upsilon$-$^{90}$Zr system for different values of cutoff $\Lambda$, obtained by the direct Bessel transform and by the Fourier transform of
the fitted Woods-Saxon form potential (FWS).}
\end{figure}

\begin{figure}[ht]
\centering
\scalebox{1.0}{
\begin{tabular}{cc}
    \includegraphics[scale=0.23]{Psir_Y_Pb208_2000_total.eps} &  
    \includegraphics[scale=0.23]{Psir_Y_Pb208_4000_total.eps} \\
    \includegraphics[scale=0.23]{Psir_Y_Pb208_6000_total.eps}  &
    \includegraphics[scale=0.23]{Psir_WS_Y_Pb208_2000_total.eps} \\
    \includegraphics[scale=0.23]{Psir_WS_Y_Pb208_4000_total.eps} &  
    \includegraphics[scale=0.23]{Psir_WS_Y_Pb208_6000_total.eps}
  \end{tabular}
}
  \caption{\label{fig:psirypb208}Coordinate space 1s and 1p state wave functions of
the $\Upsilon$-$^{208}$Pb system for different values of cutoff $\Lambda$, obtained by the direct Bessel transform and by the Fourier transform of
the fitted Woods-Saxon form potential (FWS).}
\end{figure}



\begin{figure}[ht]
\centering
\scalebox{1.0}{
\begin{tabular}{cc}
    \includegraphics[scale=0.23]{Psir_etab_He4_2000_total.eps} &  
    \includegraphics[scale=0.23]{Psir_etab_He4_4000_total.eps} \\
    \includegraphics[scale=0.23]{Psir_etab_He4_6000_total.eps}  &
    \includegraphics[scale=0.23]{Psir_WS_etab_He4_2000_total.eps} \\
    \includegraphics[scale=0.23]{Psir_WS_etab_He4_4000_total.eps} &  
    \includegraphics[scale=0.23]{Psir_WS_etab_He4_6000_total.eps}
  \end{tabular}
}
  \caption{\label{fig:psiretabhe4}Coordinate space wave functions for the 1s to 2s states of the $\eta_b$-$^{4}$He system for different values of cutoff $\Lambda$, obtained by the direct Bessel transform and by the Fourier transform of
the fitted Woods-Saxon form potential (FWS).}
\end{figure}

\begin{figure}[ht]
\centering
\scalebox{1.0}{
\begin{tabular}{cc}
    \includegraphics[scale=0.23]{Psir_etab_C12_2000_total.eps} &  
    \includegraphics[scale=0.23]{Psir_etab_C12_4000_total.eps} \\
    \includegraphics[scale=0.23]{Psir_etab_C12_6000_total.eps}  &
    \includegraphics[scale=0.23]{Psir_WS_etab_C12_2000_total.eps} \\
    \includegraphics[scale=0.23]{Psir_WS_etab_C12_4000_total.eps} &  
    \includegraphics[scale=0.23]{Psir_WS_etab_C12_6000_total.eps}
  \end{tabular}
}
  \caption{\label{fig:psiretabc12}Coordinate space wave functions for the 1s to 2p states of the $\eta_b$-$^{12}$C system for different values of cutoff $\Lambda$, obtained by the direct Bessel transform and by the Fourier transform of
the fitted Woods-Saxon form potential (FWS).}
\end{figure}

\begin{figure}[ht]
\centering
\scalebox{1.0}{
\begin{tabular}{cc}
    \includegraphics[scale=0.23]{Psir_etab_O16_2000_total.eps} &  
    \includegraphics[scale=0.23]{Psir_etab_O16_4000_total.eps} \\
    \includegraphics[scale=0.23]{Psir_etab_O16_6000_total.eps}  &
    \includegraphics[scale=0.23]{Psir_WS_etab_O16_2000_total.eps} \\
    \includegraphics[scale=0.23]{Psir_WS_etab_O16_4000_total.eps} &  
    \includegraphics[scale=0.23]{Psir_WS_etab_O16_6000_total.eps}
  \end{tabular}
}
  \caption{\label{fig:psiretabo16}Coordinate space wave functions for the 1s to 2s states of the $\eta_b$-$^{16}$O system for different values of cutoff $\Lambda$, obtained by the direct Bessel transform and by the Fourier transform of
the fitted Woods-Saxon form potential (FWS).}
\end{figure}

\begin{figure}[ht]
\centering
\scalebox{1.0}{
\begin{tabular}{cc}
    \includegraphics[scale=0.23]{Psir_etab_Ca40_2000_total.eps} &  
    \includegraphics[scale=0.23]{Psir_etab_Ca40_4000_total.eps} \\
    \includegraphics[scale=0.23]{Psir_etab_Ca40_6000_total.eps}  &
    \includegraphics[scale=0.23]{Psir_WS_etab_Ca40_2000_total.eps} \\
    \includegraphics[scale=0.23]{Psir_WS_etab_Ca40_4000_total.eps} &  
    \includegraphics[scale=0.23]{Psir_WS_etab_Ca40_6000_total.eps}
  \end{tabular}
}
  \caption{\label{fig:psiretabca40}Coordinate space wave functions for the 1s to 2s states of the $\eta_b$-$^{40}$Ca system for different values of cutoff $\Lambda$, obtained by the direct Bessel transform and by the Fourier transform of
the fitted Woods-Saxon form potential (FWS).}
\end{figure}

\begin{figure}[ht]
\centering
\scalebox{1.0}{
\begin{tabular}{cc}
    \includegraphics[scale=0.23]{Psir_etab_Ca48_2000_total.eps} &  
    \includegraphics[scale=0.23]{Psir_etab_Ca48_4000_total.eps} \\
    \includegraphics[scale=0.23]{Psir_etab_Ca48_6000_total.eps}  &
    \includegraphics[scale=0.23]{Psir_WS_etab_Ca48_2000_total.eps} \\
    \includegraphics[scale=0.23]{Psir_WS_etab_Ca48_4000_total.eps} &  
    \includegraphics[scale=0.23]{Psir_WS_etab_Ca48_6000_total.eps}
  \end{tabular}
}
  \caption{\label{fig:psiretabca48}Coordinate space wave functions for the 1s to 2s states of the $\eta_b$-$^{48}$Ca system for different values of cutoff $\Lambda$, obtained by the direct Bessel transform and by the Fourier transform of
the fitted Woods-Saxon form potential (FWS).}
\end{figure}

\begin{figure}[ht]
\centering
\scalebox{1.0}{
\begin{tabular}{cc}
    \includegraphics[scale=0.23]{Psir_etab_Zr90_2000_total.eps} &  
    \includegraphics[scale=0.23]{Psir_etab_Zr90_4000_total.eps} \\
    \includegraphics[scale=0.23]{Psir_etab_Zr90_6000_total.eps}  &
    \includegraphics[scale=0.23]{Psir_WS_etab_Zr90_2000_total.eps} \\
    \includegraphics[scale=0.23]{Psir_WS_etab_Zr90_4000_total.eps} &  
    \includegraphics[scale=0.23]{Psir_WS_etab_Zr90_6000_total.eps}
  \end{tabular}
}
  \caption{\label{fig:psiretabzr90}Coordinate space wave functions for the 1s to 2s states of the $\eta_b$-$^{90}$Zr system for different values of cutoff $\Lambda$, obtained by the direct Bessel transform and by the Fourier transform of
the fitted Woods-Saxon form potential (FWS).}
\end{figure}

\begin{figure}[ht]
\centering
\scalebox{1.0}{
\begin{tabular}{cc}
    \includegraphics[scale=0.23]{Psir_etab_Pb208_2000_total.eps} &  
    \includegraphics[scale=0.23]{Psir_etab_Pb208_4000_total.eps} \\
    \includegraphics[scale=0.23]{Psir_etab_Pb208_6000_total.eps}  &
    \includegraphics[scale=0.23]{Psir_WS_etab_Pb208_2000_total.eps} \\
    \includegraphics[scale=0.23]{Psir_WS_etab_Pb208_4000_total.eps} &  
    \includegraphics[scale=0.23]{Psir_WS_etab_Pb208_6000_total.eps}
  \end{tabular}
}
  \caption{\label{fig:psiretabpb208}Coordinate space wave functions for the 1s to 2s states of the $\eta_b$-$^{208}$Pb system for different values of cutoff $\Lambda$, obtained by the direct Bessel transform and by the Fourier transform of
the fitted Woods-Saxon form potential (FWS).}
\end{figure}


\newpage
\begin{adjustwidth}{-\extralength}{0cm}

\reftitle{References}

\reftitle{References}
\PublishersNote{}
\end{adjustwidth}

\newpage
\begin{thebibliography}{999}

\bibitem{Accardi:2023chb}
A.~Accardi, P.~Achenbach, D.~Adhikari, A.~Afanasev, C.~S.~Akondi, N.~Akopov, M.~Albaladejo, H.~Albataineh, M.~Albrecht and B.~Almeida-Zamora, \textit{et al.}
Eur. Phys. J. A \textbf{60}, 173 (2024) 
[arXiv:2306.09360 [nucl-ex]].

\bibitem{Brodsky:2015aia}
S.~J.~Brodsky, A.~L.~Deshpande, H.~Gao, R.~D.~McKeown, C.~A.~Meyer, Z.~E.~Meziani, R.~G.~Milner, J.~Qiu, D.~G.~Richards and C.~D.~Roberts,
[arXiv:1502.05728 [hep-ph]].

\bibitem{Heinz:2015tua}
U.~Heinz, P.~Sorensen, A.~Deshpande, C.~Gagliardi, F.~Karsch, T.~Lappi, Z.~E.~Meziani, R.~Milner, B.~Muller and J.~Nagle, \textit{et al.}
[arXiv:1501.06477 [nucl-th]].

\bibitem{Alkofer:2000wg}
R.~Alkofer and L.~von Smekal,
Phys. Rept. \textbf{353}, 281 (2001),
[arXiv:hep-ph/0007355 [hep-ph]].

\bibitem{Brambilla:2014jmp}
N.~Brambilla, S.~Eidelman, P.~Foka, S.~Gardner, A.~S.~Kronfeld, M.~G.~Alford, R.~Alkofer, M.~Butenschoen, T.~D.~Cohen and J.~Erdmenger, \textit{et al.}
Eur. Phys. J. C \textbf{74}, 2981 (2014), 
[arXiv:1404.3723 [hep-ph]].

\bibitem{Bashir:2012fs}
A.~Bashir, L.~Chang, I.~C.~Cloet, B.~El-Bennich, Y.~X.~Liu, C.~D.~Roberts and P.~C.~Tandy,
Commun. Theor. Phys. \textbf{58}, 79 (2012), 
[arXiv:1201.3366 [nucl-th]].

\bibitem{Cloet:2013jya}
I.~C.~Cloet and C.~D.~Roberts,
Prog. Part. Nucl. Phys. \textbf{77}, 1 (2014), 
[arXiv:1310.2651 [nucl-th]].


\bibitem{Hosaka:2016ypm}
A.~Hosaka, T.~Hyodo, K.~Sudoh, Y.~Yamaguchi and S.~Yasui,
Prog. Part. Nucl. Phys. \textbf{96}, 88 (2017),
[arXiv:1606.08685 [hep-ph]].

\bibitem{Krein:2016fqh}
G.~Krein,
AIP Conf. Proc. \textbf{1701}, 
020012 (2016).

\bibitem{Metag:2017yuh}
V.~Metag, M.~Nanova and E.~Y.~Paryev,
Prog. Part. Nucl. Phys. \textbf{97}, 199 (2017),
[arXiv:1706.09654 [nucl-ex]].

\bibitem{Krein:2017usp}
G.~Krein, A.~W.~Thomas and K.~Tsushima,
Prog. Part. Nucl. Phys. \textbf{100}, 161 (2018),
[arXiv:1706.02688 [hep-ph]].

\bibitem{Hatsuda:1994pi}
T.~Hatsuda and T.~Kunihiro,
Phys. Rept. \textbf{247}, 221-367 (1994),
[arXiv:hep-ph/9401310 [hep-ph]].

\bibitem{Leupold:2009kz}
S.~Leupold, V.~Metag and U.~Mosel,
Int. J. Mod. Phys. E \textbf{19}, 147-224 (2010),
[arXiv:0907.2388 [nucl-th]].

\bibitem{Hayano:2008vn}
R.~S.~Hayano and T.~Hatsuda,
Rev. Mod. Phys. \textbf{82}, 2949 (2010),
[arXiv:0812.1702 [nucl-ex]].

\bibitem{piAF:2022gvw}
T.~Nishi \textit{et al.} [piAF],
Nature Phys. \textbf{19}, 788 (2023), 
[arXiv:2204.05568 [nucl-ex]].

%
\bibitem{Schroedter:2000ek}
M.~Schroedter, R.~L.~Thews, and J.~Rafelski,
Phys. Rev. C \textbf{62}, 024905 (2000),
[arXiv:hep-ph/0004041 [hep-ph]].

\bibitem{Andronic:2015wma}
A.~Andronic, F.~Arleo, R.~Arnaldi, A.~Beraudo, E.~Bruna, D.~Caffarri, Z.~C.~del Valle,
J.~G.~Contreras, T.~Dahms, A.~Dainese \textit{et al.,}
Eur. Phys. J. C \textbf{76}, 107 (2016),
[arXiv:1506.03981 [nucl-ex]].

\bibitem{Li:2023mrj}
Y.~S.~Li and X.~Liu,
Phys. Rev. D \textbf{108}, 093005 (2023),
[arXiv:2309.08191 [hep-ph]].

\bibitem{Bird:2004ts}
C.~Bird, P.~Jackson, R.~V.~Kowalewski and M.~Pospelov,
Phys. Rev. Lett. \textbf{93}, 201803 (2004),
[arXiv:hep-ph/0401195 [hep-ph]].

\bibitem{Altmannshofer:2014rta}
W.~Altmannshofer and D.~M.~Straub,
Eur. Phys. J. C \textbf{75}, 382 (2015),
[arXiv:1411.3161 [hep-ph]].


\bibitem{Buras:2014fpa}
A.~J.~Buras, J.~Girrbach-Noe, C.~Niehoff and D.~M.~Straub,
JHEP \textbf{02}, 184 (2015),
[arXiv:1409.4557 [hep-ph]].


\bibitem{Descotes-Genon:2015uva}
S.~Descotes-Genon, L.~Hofer, J.~Matias and J.~Virto,
JHEP \textbf{06}, 092 (2016),
[arXiv:1510.04239 [hep-ph]].

\bibitem{Dutta:2017xmj}
R.~Dutta and A.~Bhol,
Phys. Rev. D \textbf{96}, 076001 (2017),
[arXiv:1701.08598 [hep-ph]].

\bibitem{CMS:2021hug}
A.~Tumasyan \textit{et al.} [CMS],
Phys. Rev. Lett. \textbf{129}, 032001 (2022),
[arXiv:2111.02219 [hep-ex]].

\bibitem{Akram:2013dhd}
F.~Akram,
``Hadronic Cross Sections of $B_c$ Mesons,''
PhD Punjab U. (2013).

\bibitem{Lodhi:2007zz}
M.~A.~K.~Lodhi and R.~Marshall,
Nucl. Phys. A \textbf{790}, 323 (2007),

\bibitem{Lodhi:2011zz}
M.~A.~K.~Lodhi, F.~Akram and S.~Irfan,
Phys. Rev. C \textbf{84}, 034901 (2011),
[arXiv:1309.2912 [nucl-th]].

\bibitem{Wu:2023djn}
B.~Wu, Z.~Tang, M.~He and R.~Rapp,
Phys. Rev. C \textbf{109}, 014906 (2024),
[arXiv:2302.11511 [nucl-th]].

\bibitem{CMS:2022sxl}
A.~Tumasyan \textit{et al.} [CMS],
Phys. Rev. Lett. \textbf{128}, 252301 (2022),
[arXiv:2201.02659 [hep-ex]].

\bibitem{Harris:2023tti}
J.~W.~Harris and B.~M\"uller,
[arXiv:2308.05743 [hep-ph]].

\bibitem{Lin:2000ke}
Z.~w.~Lin and C.~M.~Ko,
Phys. Lett. B \textbf{503}, 104 (2001),
[arXiv:nucl-th/0007027 [nucl-th]].

\bibitem{CMS:2018zza}
A.~M.~Sirunyan \textit{et al.} [CMS],
Phys. Lett. B \textbf{790}, 270 (2019),
[arXiv:1805.09215 [hep-ex]].

\bibitem{ATLAS:2022exb}
G.~Aad \textit{et al.} [ATLAS],
Phys. Rev. C \textbf{107}, 054912 (2023),
[arXiv:2205.03042 [nucl-ex]].


\bibitem{STAR:2022rpk}
B.~Aboona \textit{et al.} [STAR],
Phys. Rev. Lett. \textbf{130}, 112301 (2023),
[arXiv:2207.06568 [nucl-ex]].

\bibitem{Guichon:1987jp}
P.~A.~M.~Guichon,
Phys. Lett. B \textbf{200}, 235 (1988).

\bibitem{Brodsky:1997gh}
S.~J.~Brodsky and G.~A.~Miller,
Phys. Lett. B \textbf{412}, 125 (1997),
[arXiv:hep-ph/9707382 [hep-ph]].

\bibitem{Krein:2010vp}
G.~Krein, A.~W.~Thomas and K.~Tsushima,
Phys. Lett. B \textbf{697}, 136 (2011),
[arXiv:1007.2220 [nucl-th]].

\bibitem{Tsushima:2011kh}
K.~Tsushima, D.~H.~Lu, G.~Krein and A.~W.~Thomas,
Phys. Rev. C \textbf{83}, 065208 (2011),
[arXiv:1103.5516 [nucl-th]].

\bibitem{Ko:1992tp}
C.~M.~Ko, P.~Levai, X.~J.~Qiu and C.~T.~Li,
Phys. Rev. C \textbf{45}, 1400 (1992).

\bibitem{Asakawa:1992ht}
M.~Asakawa, C.~M.~Ko, P.~Levai and X.~J.~Qiu,
Phys. Rev. C \textbf{46}, R1159 (1992).

\bibitem{Friedman:1984yg}
E.~Friedman and G.~Soff,
J. Phys. G \textbf{11}, L37 (1985).

\bibitem{Yamazaki:1996ch}
Yamazaki, T., Hayano, R.S., Itahashi, K. et al.,
Z. Physik A
355, 219 (1996).


\bibitem{Davies:1979aj}
J.~D.~Davies, G.~J.~Pyle, G.~T.~A.~Squier, C.~J.~Batty, S.~F.~Biagi, S.~D.~Hoath, P.~Sharman and
A.~S.~Clough,
Phys. Lett. B \textbf{83}, 55 (1979).

\bibitem{Lee:1994jj}
C.~H.~Lee, G.~E.~Brown, D.~P.~Min and M.~Rho,
Nucl. Phys. A \textbf{585}, 401 (1995),
[arXiv:hep-ph/9406311 [hep-ph]].

\bibitem{Ito:1998yi}
T.~M.~Ito, R.~S.~Hayano, S.~N.~Nakamura, T.~P.~Terada, M.~Iwasaki, D.~R.~Gill, L.~Lee, A.~Olin,
M.~Salomon and S.~Yen, \textit{et al.}
Phys. Rev. C \textbf{58}, 2366 (1998).

\bibitem{Curceanu:2019uph}
C.~Curceanu, C.~Guaraldo, M.~Iliescu, M.~Cargnelli, R.~Hayano, J.~Marton, J.~Zmeskal, T.~Ishiwatari,
M.~Iwasaki and S.~Okada, \textit{et al.}
Rev. Mod. Phys. \textbf{91}, 025006 (2019).


\bibitem{Hirenzaki:2000da}
S.~Hirenzaki, Y.~Okumura, H.~Toki, E.~Oset and A.~Ramos,
Phys. Rev. C \textbf{61}, 055205 (2000).


\bibitem{Hayano:1998sy}
R.~S.~Hayano, S.~Hirenzaki and A.~Gillitzer,
Eur. Phys. J. A \textbf{6}, 99 (1999),
[arXiv:nucl-th/9806012 [nucl-th]].

\bibitem{Tsushima:1998qw}
K.~Tsushima, D.~H.~Lu, A.~W.~Thomas and K.~Saito,
Phys. Lett. B \textbf{443}, 26 (1998),
[arXiv:nucl-th/9806043 [nucl-th]].

\bibitem{Tsushima:1998ru}
K.~Tsushima, D.~H.~Lu, A.~W.~Thomas, K.~Saito and R.~H.~Landau,
Phys. Rev. C \textbf{59}, 2824 (1999),
[arXiv:nucl-th/9810016 [nucl-th]].

\bibitem{Brodsky:1989jd}
S.~J.~Brodsky, I.~A.~Schmidt and G.~F.~de Teramond,
Phys. Rev. Lett. \textbf{64}, 1011 (1990).


\bibitem{Lee:2000csl}
S.~H.~Lee and C.~M.~Ko,
Phys. Rev. C \textbf{67}, 038202 (2003),
[arXiv:nucl-th/0208003 [nucl-th]].

\bibitem{Krein:2013rha}
G.~Krein,
J. Phys. Conf. Ser. \textbf{422}, 012012 (2013).

\bibitem{Klingl:1998sr}
F.~Klingl, S.~s.~Kim, S.~H.~Lee, P.~Morath and W.~Weise,
Phys. Rev. Lett. \textbf{82} (1999), 3396; \textbf{83}, 4224 (1999),
[arXiv:nucl-th/9811070 [nucl-th]].

\bibitem{Hayashigaki:1998ey}
A.~Hayashigaki,
Prog. Theor. Phys. \textbf{101} (1999), 923.
[arXiv:nucl-th/9811092 [nucl-th]].

\bibitem{Kumar:2010hs}
A.~Kumar and A.~Mishra,
Phys. Rev. C \textbf{82}, 045207 (2010),
[arXiv:1005.2748 [nucl-th]].

\bibitem{Belyaev:2006vn}
V.~B.~Belyaev, N.~V.~Shevchenko, A.~I.~Fix and W.~Sandhas,
Nucl. Phys. A \textbf{780}, 100 (2006),
[arXiv:nucl-th/0601058 [nucl-th]].

\bibitem{Yokota:2013sfa}
A.~Yokota, E.~Hiyama and M.~Oka,
PTEP \textbf{2013}, 113D01 (2013),
[arXiv:1308.6102 [nucl-th]].

\bibitem{Peskin:1979va}
M.~E.~Peskin,
Nucl. Phys. B \textbf{156}, 365 (1979).

\bibitem{Kharzeev:1995ij}
D.~Kharzeev,
Proc. Int. Sch. Phys. Fermi \textbf{130}, 105 (1996),
[arXiv:nucl-th/9601029 [nucl-th]].

\bibitem{Kaidalov:1992hd}
A.~B.~Kaidalov and P.~E.~Volkovitsky,
Phys. Rev. Lett. \textbf{69}, 3155 (1992).

\bibitem{Luke:1992tm}
M.~E.~Luke, A.~V.~Manohar and M.~J.~Savage,
Phys. Lett. B \textbf{288} (1992), 355.
[arXiv:hep-ph/9204219 [hep-ph]].

\bibitem{deTeramond:1997ny}
G.~F.~de Teramond, R.~Espinoza and M.~Ortega-Rodriguez,
Phys. Rev. D \textbf{58}, 034012 (1998),
[arXiv:hep-ph/9708202 [hep-ph]].

\bibitem{Sibirtsev:2005ex}
A.~Sibirtsev and M.~B.~Voloshin,
Phys. Rev. D \textbf{71}, 076005 (2005),
[arXiv:hep-ph/0502068 [hep-ph]].

\bibitem{Voloshin:2007dx}
M.~B.~Voloshin,
Prog. Part. Nucl. Phys. \textbf{61}, 455 (2008),
[arXiv:0711.4556 [hep-ph]].

\bibitem{TarrusCastella:2018php}
J.~Tarr\'us Castell\`a and G.~Krein,
Phys. Rev. D \textbf{98}, 014029 (2018),
[arXiv:1803.05412 [hep-ph]].

\bibitem{Cobos-Martinez:2020ynh}
J.~J.~Cobos-Mart\'\i{}nez, K.~Tsushima, G.~Krein and A.~W.~Thomas,
Phys. Lett. B \textbf{811}, 135882 (2020),
[arXiv:2007.04476 [hep-ph]].



\bibitem{Yokokawa:2006td}
K.~Yokokawa, S.~Sasaki, T.~Hatsuda and A.~Hayashigaki,
Phys. Rev. D \textbf{74}, 034504 (2006),
[arXiv:hep-lat/0605009 [hep-lat]].

\bibitem{Kawanai:2010ev}
T.~Kawanai and S.~Sasaki,
Phys. Rev. D \textbf{82}, 091501 (2010),
[arXiv:1009.3332 [hep-lat]].

\bibitem{Skerbis:2018lew}
U.~Skerbis and S.~Prelovsek,
Phys. Rev. D \textbf{99}, 094505 (2019),
[arXiv:1811.02285 [hep-lat]].

\bibitem{Chizzali:2022pjd}
E.~Chizzali, Y.~Kamiya, R.~Del Grande, T.~Doi, L.~Fabbietti, T.~Hatsuda and Y.~Lyu,
Phys. Lett. B \textbf{848}, 138358 (2024),
[arXiv:2212.12690 [nucl-ex]].

\bibitem{Zeminiani:2020aho}
G.~N.~Zeminiani, J.~J.~Cobos-Martinez and K.~Tsushima,
Eur. Phys. J. A \textbf{57}, 259 (2021),
[arXiv:2012.11381 [hep-ph]].

\bibitem{Zeminiani:2021vaq}
G.~N.~Zeminiani,
[arXiv:2201.09158 [nucl-th]].

\bibitem{Zeminiani:2021xvw}
G.~Zeminiani, J.~J.~Cobos-Mart\'\i{}nez and K.~Tsushima,
PoS \textbf{PANIC2021}, 208 (2022),
[arXiv:2109.08636 [hep-ph]].

\bibitem{Cobos-Martinez:2022fmt}
J.~J.~Cobos-Mart\'\i{}nez, G.~N.~Zeminiani and K.~Tsushima,
Phys. Rev. C \textbf{105}, 025204 (2022),
[arXiv:2201.05696 [nucl-th]].


\bibitem{EuropeanMuon:1983wih}
J.~J.~Aubert \textit{et al.} [European Muon],
Phys. Lett. B \textbf{123}, 275 (1983).


\bibitem{Geesaman:1995yd}
For a review, D.~F.~Geesaman, K.~Saito and A.~W.~Thomas,
Ann. Rev. Nucl. Part. Sci. \textbf{45}, 337 (1995).  



\bibitem{Dieterich:2000mu}
S.~Dieterich, P.~Bartsch, D.~Baumann, J.~Bermuth, K.~Bohinc, R.~Bohm, D.~Bosnar, S.~Derber, M.~Ding
and M.~Distler, \textit{et al.}
Phys. Lett. B \textbf{500}, 47 (2001),
[arXiv:nucl-ex/0011008 [nucl-ex]].


\bibitem{POLE} S. Strauch (Jefferson Lab E93-049 Collaboration), 
Eur. Phys. J. A \textbf{19}, 153 (2004), 
S. Strauch {\it et al.},
Phys. Rev. Lett. \textbf{91}, 052301 (2003).
%

\bibitem{Saito:2005rv}
K.~Saito, K.~Tsushima and A.~W.~Thomas,
Prog. Part. Nucl. Phys. \textbf{58}, 1 (2007),
[arXiv:hep-ph/0506314 [hep-ph]].

\bibitem{Guichon:2008zz}
P.~A.~M.~Guichon, A.~W.~Thomas and K.~Tsushima,
Nucl. Phys. A \textbf{814}, 66 (2008),
[arXiv:0712.1925 [nucl-th]].

\bibitem{Shyam:2019laf}
R.~Shyam and K.~Tsushima,
[arXiv:1901.06090 [nucl-th]].

\bibitem{Guichon:2018uew}
P.~A.~M.~Guichon, J.~R.~Stone and A.~W.~Thomas,
Prog. Part. Nucl. Phys. \textbf{100}, 262 (2018).



\bibitem{Zeminiani:2023gqc}
G.~N.~Zeminiani, S.~L.~P.~G.~Beres and K.~Tsushima,
Phys. Rev. D \textbf{110}, 094045 (2024),
[arXiv:2401.00250 [hep-ph]].


\bibitem{Tsushima:1997df}
K.~Tsushima, K.~Saito, A.~W.~Thomas and S.~V.~Wright,
Phys. Lett. B \textbf{429}, 239 (1998),
[erratum: Phys. Lett. B \textbf{436}, 453 (1998)],
[arXiv:nucl-th/9712044 [nucl-th]].

\bibitem{Sibirtsev:1999js}
  A.~Sibirtsev, K.~Tsushima and A.~W.~Thomas,
  Eur.\ Phys.\ J.\ A {\bf 6}, 351 (1999),
  [nucl-th/9904016].


  
\bibitem{Sibirtsev:1999jr}
  A.~Sibirtsev, K.~Tsushima, K.~Saito and A.~W.~Thomas,
  Phys.\ Lett.\ B {\bf 484}, 23 (2000),
  [nucl-th/9904015].

\bibitem{Tsushima:2002cc}
  K.~Tsushima and F.~C.~Khanna,
  Phys.\ Lett.\ B {\bf 552}, 138 (2003),
  [nucl-th/0207036].


\bibitem{Guichon:1995ue}
P.~A.~M.~Guichon, K.~Saito, E.~N.~Rodionov and A.~W.~Thomas,
Nucl. Phys. A \textbf{601}, 349 (1996),
[arXiv:nucl-th/9509034 [nucl-th]].

\bibitem{Tsushima:2020gun}
K.~Tsushima,
PTEP \textbf{2022}, 043D02 (2022),
[arXiv:2008.03724 [hep-ph]].

\bibitem{Workman:2022ynf}
R.~L.~Workman \textit{et al.} [Particle Data Group],
``Review of Particle Physics,''
PTEP \textbf{2022}, 083C01 (2022).

\bibitem{Klingl:1996by} 
  F.~Klingl, N.~Kaiser and W.~Weise,
  Z.\ Phys.\ A {\bf 356}, 193 (1996),
  [hep-ph/9607431].

\bibitem{Tanabashi:2018oca} 
  M.~Tanabashi {\it et al.} [Particle Data Group],
  Phys.\ Rev.\ D {\bf 98}, 030001 (2018).


\bibitem{OZI}
S. Okubo, Phys. Lett. \textbf{5}, 165 (1963), 
Phys. Rev. D \textbf{16}, 2336 (1977);
G. Zweig, CERN Report No. 8419 TH 412, 1964, (unpublished); 
J. Iizuka, K. Okada, and O. Shito, Prog. Theor. Phys. \textbf{35}, 1061 (1966),
J. Iizuka, Prog. Theor. Phys. Suppl. \textbf{37}, 21 (1966).




\bibitem{Tsushima:2011fg}
K.~Tsushima, D.~Lu, G.~Krein and A.~W.~Thomas,
AIP Conf. Proc. \textbf{1354}, 39 (2011),  
[arXiv:1101.3389 [nucl-th]].


\bibitem{Lyu:2024ttm}
Y.~Lyu, T.~Doi, T.~Hatsuda and T.~Sugiura,
Phys. Lett. B \textbf{860}, 139178 (2025),
[arXiv:2410.22755 [hep-lat]].

\bibitem{Lyu:2025jjl}
Y.~Lyu, T.~Doi, T.~Hatsuda and T.~Sugiura,
[arXiv:2502.00054 [hep-lat]].




\bibitem{Klingl:1996by} 
  F.~Klingl, N.~Kaiser and W.~Weise,
  Z.\ Phys.\ A {\bf 356}, 193 (1996),
  [hep-ph/9607431].


\bibitem{Lin:1999ve} 
  Z.~w.~Lin, C.~M.~Ko and B.~Zhang,
  Phys.\ Rev.\ C {\bf 61}, 024904 (2000),
  [nucl-th/9905003].


\bibitem{Cobos-Martinez:2017vtr}
J. J.~Cobos-Mart\'inez, K.~Tsushima, G.~Krein and A.~W.~Thomas,
Phys. Lett. B \textbf{771}, 113 (2017),
[arXiv:1703.05367 [nucl-th]].


\bibitem{Cobos-Martinez:2017woo}
J. J.~Cobos-Mart\'inez, K.~Tsushima, G.~Krein and A.~W.~Thomas,
Phys. Rev. C \textbf{96}, 035201 (2017), 
[arXiv:1705.06653 [nucl-th]].


\bibitem{Cobos-Martinez:2017onm}
J. J.~Cobos-Mart\'inez, K.~Tsushima, G.~Krein and A.~W.~Thomas,
J. Phys. Conf. Ser. \textbf{912}, 012009 (2017),
[arXiv:1711.06358 [nucl-th]].


\bibitem{Cobos-Martinez:2017fch}
J. J.~Cobos-Mart\'inez, K.~Tsushima, G.~Krein and A.~W.~Thomas,
PoS \textbf{Hadron2017}, 209 (2018),
[arXiv:1711.09895 [nucl-th]].


\bibitem{Cobos-Martinez:2019kln}
J. J.~Cobos-Mart\'inez, K.~Tsushima, G.~Krein and A.~W.~Thomas,
JPS Conf. Proc. \textbf{26}, 024033 (2019),
[arXiv:1901.07404 [hep-ph]].


\bibitem{Lucha:2015dda} 
  W.~Lucha, D.~Melikhov, H.~Sazdjian and S.~Simula,
  Phys.\ Rev.\ D {\bf 93}, no. 1, 016004 (2016);
  Addendum: [Phys.\ Rev.\ D {\bf 93}, 019902 (2016)], 
  [arXiv:1506.09213 [hep-ph]].

\bibitem{Lin:1999ad} 
  Z.~w.~Lin and C.~M.~Ko,
  Phys.\ Rev.\ C {\bf 62}, 034903 (2000),
  [nucl-th/9912046].


\bibitem{Hayashigaki:2000es}
A.~Hayashigaki,
Phys. Lett. B \textbf{487} 96 (2000),
[arXiv:nucl-th/0001051 [nucl-th]].


\bibitem{Azizi:2014bba}
K.~Azizi, N.~Er and H.~Sundu,
Eur. Phys. J. C \textbf{74}, 3012 (2014),
[arXiv:1405.3058 [hep-ph]].


\bibitem{Wang:2015uya}
Z.~G.~Wang,
Phys. Rev. C \textbf{92}, 065205 (2015),
[arXiv:1501.05093 [hep-ph]].

\bibitem{Suzuki:2015est}
K.~Suzuki, P.~Gubler and M.~Oka,
Phys. Rev. C \textbf{93}, 045209 (2016), 
[arXiv:1511.04513 [hep-ph]].


\bibitem{Park:2016xrw}
A.~Park, P.~Gubler, M.~Harada, S.~H.~Lee, C.~Nonaka and W.~Park,
Phys. Rev. D \textbf{93}, 054035 (2016), 
[arXiv:1601.01250 [nucl-th]].


\bibitem{Gubler:2020hft}
P.~Gubler, T.~Song and S.~H.~Lee,
Phys. Rev. D \textbf{101}, 114029 (2020), 
[arXiv:2003.09073 [hep-ph]].


\bibitem{Hilger:2008jg}
T.~Hilger, R.~Thomas and B.~Kampfer,
Phys. Rev. C \textbf{79}, 025202 (2009), 
[arXiv:0809.4996 [nucl-th]].


\bibitem{Carames:2016qhr}
T.~F.~Caram\'es, C.~E.~Fontoura, G.~Krein, K.~Tsushima, J.~Vijande and A.~Valcarce,
Phys. Rev. D \textbf{94}, 034009 (2016), 
[arXiv:1608.04040 [hep-ph]].

\bibitem{Zeminiani:2024rrh}
G.~N.~Zeminiani, J.~J.~Cobos-Mart\'\i{}nez and K.~Tsushima,
[arXiv:2406.11114 [nucl-th]].

\bibitem{Saito:1996sf}
K.~Saito, K.~Tsushima and A.~W.~Thomas,
Nucl. Phys. A \textbf{609}, 339 (1996),
[arXiv:nucl-th/9606020 [nucl-th]].

\bibitem{Saito:1997ae} 
  K.~Saito, K.~Tsushima and A.~W.~Thomas,
  Phys.\ Rev.\ C {\bf 56}, 566 (1997),
 [nucl-th/9703011].


\bibitem{Kwan:1978zh} 
  Y.~R.~Kwan and F.~Tabakin,
  Phys.\ Rev.\ C {\bf 18}, 932 (1978).

\bibitem{Buhler:2010zz} 
  P.~Buhler {\it et al.},
  Prog.\ Theor.\ Phys.\ Suppl.\  {\bf 186}, 337 (2010).
  
\bibitem{Ohnishi:2014xla} 
  H.~Ohnishi {\it et al.},
  Acta Phys.\ Polon.\ B {\bf 45}, 819 (2014).

\bibitem{Csorgo:2014sat}
  T.~Cs\"{o}rg\H{o}, M.~Csan\'ad and T.~Nov\'ak,
  Proceedings, 10th Workshop on Particle Correlations and Femtoscopy (WPCF 2014) : Gy\"{o}ngy\"{o}s,
  Hungary, August 25-29, 2014, SLAC-econf-C140825.8.

  \bibitem{JLabphi}
  \url{https://www.jlab.org/exp_prog/PACpage/PAC42/PAC42_FINAL_Report.pdf}

\bibitem{Cobos-Martinez:2023hbp}
J.~J.~Cobos-Martinez and K.~Tsushima,
Phys. Rev. C \textbf{109}, 2 (2024),
[arXiv:2308.07836 [nucl-th]].

\bibitem{Cobos-Martinez:2021ukw}
J.~J.~Cobos-Martinez, K.~Tsushima, G.~Krein and A.~W.~Thomas,
PoS \textbf{CHARM2020}, 041 (2021),
[arXiv:2109.10995 [nucl-th]].

\bibitem{CobosMartinez:2021bia}
J.~Cobos Martinez, K.~Tsushima, G.~Krein and A.~W.~Thomas,
PoS \textbf{PANIC2021}, 199 (2022),
[arXiv:2111.13820 [nucl-th]].

\bibitem{PDG2024}
S. Navas et al. (Particle Data Group), Phys. Rev. D \textbf{110}, 030001 (2024).



\end{thebibliography}
\end{document}